\documentclass{aa}  

\usepackage{graphicx}
\usepackage{xcolor}
\usepackage{braket}
\usepackage{subcaption}
\usepackage{ctable}
\usepackage[flushleft]{threeparttable}
\usepackage{comment}
\usepackage{placeins}
\usepackage{txfonts}
\usepackage[colorlinks=true,linkcolor=blue,citecolor=blue]{hyperref}%

\newcommand{\NII}{{[N\,{\sc ii}]\ }}

\newcommand{\NIVl}{{N\,{\sc iv}]\,$\lambda$}}
\newcommand{\NIVll}{{N\,{\sc iv}]\,$\lambda\lambda$}}
\newcommand{\NVl}{{N\,{\sc v}\,$\lambda$}}

\newcommand{\NIIll}{{[N\,{\sc ii}]\,$\lambda\lambda$}}
\newcommand{\NeIII}{{[Ne\,{\sc iii}\,]}}

\newcommand{\NeIIIl}{{[Ne\,{\sc iii}]\,$\lambda$}}
\newcommand{\NeIV}{{[Ne\,{\sc iv}]\,}}

\newcommand{\NeIVl}{{[Ne\,{\sc iv}]\,$\lambda$}}
\newcommand{\NeIVll}{{[Ne\,{\sc iv}]\,$\lambda\lambda$}}
\newcommand{\NeV}{{[Ne\,{\sc v}]\,}}

\newcommand{\NeVl}{{[Ne\,{\sc v}]\,$\lambda$}}
\newcommand{\SII}{{[S\,{\sc ii}]\,}}

\newcommand{\SIIll}{{[S\,{\sc ii}]\,$\lambda\lambda$}}

\newcommand{\OIII}{{[O\,{\sc iii}]\,}}

\newcommand{\OIIIl}{{[O\,{\sc iii}]\,$\lambda$}}
\newcommand{\OIIIll}{{[O\,{\sc iii}]\,$\lambda\lambda$}}

\newcommand{\OIIll}{{[O\,{\sc ii}]\,$\lambda\lambda$}}
\newcommand{\OI}{{[O\,{\sc i}]\,}}

\newcommand{\OIl}{{[O\,{\sc i}]\,$\lambda$}}

\newcommand{\CIII}{{C\,{\sc iii}]\,}}
\newcommand{\CIIIs}{{C\,{\sc iii}]\,}}
\newcommand{\CIIIll}{\CIIIs$\lambda\lambda$}
\newcommand{\CIV}{{C\,{\sc iv}\,}}

\newcommand{\CIVll}{{C\,{\sc iv}\,$\lambda\lambda$}\xspace}

\newcommand{\HeII}{{He\,{\sc ii}\,}}
\newcommand{\HeIIl}{{He\,{\sc ii}\,$\lambda$}}

\newcommand{\Ha}{H$\alpha$\,}

\newcommand{\Hb}{H$\beta$\,}

\newcommand{\Hg}{H$\gamma$\,}
\newcommand{\Lya}{Ly$\alpha$\,}

\usepackage{xspace}
\let\oldAA\AA
\renewcommand{\AA}{\text{\oldAA}\xspace}

\begin{document}

   \title{Narrow-line AGN  selection in CEERS: Spectroscopic selection, physical properties, and X-ray and radio analysis}
   \titlerunning{Narrow-line AGN selection in CEERS}

   \author{Giovanni Mazzolari
          \inst{1,2,3}\fnmsep
          \thanks{giovanni.mazzolari@inaf.it}
          \and
          Jan Scholtz\inst{3,4}
          \and 
          Roberto Maiolino\inst{3,4,5}
          \and 
          Roberto Gilli\inst{2}
          \and
          Alberto Traina\inst{2}
          \and
          Ivan E. López \inst{2}
          \and
          Hannah Übler \inst{3,4}
          \and
          Bartolomeo Trefoloni \inst{6,7}
          \and
          Francesco D'Eugenio \inst{3,4}
            \and
          Xihan Ji \inst{3,4}
          \and
          Marco Mignoli \inst{2}
          \and
          Fabio Vito \inst{2}
          \and
          Cristian Vignali \inst{1}
          \and
          Marcella Brusa \inst{1}
          }

\authorrunning{G. Mazzolari et al.}
   \institute{  Dipartimento di Fisica e Astronomia, Università di Bologna, Via Gobetti 93/2, I-40129 Bologna, Italy
\and INAF – Osservatorio di Astrofisica e Scienza dello Spazio di Bologna, Via Gobetti 93/3, I-40129 Bologna, Italy
\and Kavli Institute for Cosmology, University of Cambridge, Madingley Road, Cambridge, CB3 0HA, UK
\and Cavendish Laboratory, University of Cambridge, 19 JJ Thomson Avenue, Cambridge, CB3 0HE, UK
\and Department of Physics and Astronomy, University College London, Gower Street, London WC1E 6BT, UK 
\and Dipartimento di Fisica e Astronomia, Università di Firenze, Via G. Sansone 1, 50019 Sesto Fiorentino, Firenze, Italy
\and INAF – Osservatorio Astrofisico di Arcetri, Largo Enrico Fermi 5, 50125 Firenze, Italy
            }
   \date{}

  \abstract
   {The transformative era opened by the \textit{James Webb Space Telescope} (JWST) on the high-$z$ Universe allows us to investigate the early stages of supermassive black hole (SMBH) evolution, with the first results showing a greater than expected number of active galactic nuclei (AGNs) at very early times. In this work, we spectroscopically select narrow-line AGNs (NLAGNs) among the $\sim 300$ publicly available medium-resolution spectra of the Cosmic Evolution Early Release Science Survey (CEERS). Using both traditional and newly identified emission line NLAGN diagnostics diagrams, we identified 52 NLAGNs at $2\lesssim z \lesssim 9$, on which we performed a detailed multiwavelength analysis. We also identified four new $z\lesssim 2$ broad-line AGNs (BLAGNs), in addition to the eight previously reported $z>4.5$ BLAGNs. 
   We found that the traditional BPT diagnostic diagrams are not suited to identifying high-$z$ AGNs, while most of the high-$z$ NLAGN were selected using the recently proposed AGN diagnostic diagrams based on the \OIIIl4363 auroral line or high-ionization emission lines.
   We compared the emission line velocity dispersion and the obscuration levels of the sample of NLAGNs with those of the parent sample without finding significant differences between the two distributions, suggesting a population of AGNs heavily buried and not significantly impacting the host galaxies' physical properties, as was further confirmed by spectral energy distribution fitting.
   The bolometric luminosities of the high-$z$ NLAGNs selected in this work are $\sim 1.5$dex below the ones sampled by surveys before JWST, potentially explaining the weak impact of these AGNs. 
   Finally, we investigated the X-ray properties of the selected NLAGNs and of the sample of high-$z$ BLAGNs. We found that all but four NLAGNs are undetected in the deep X-ray image of the field, as well as all the high-$z$ BLAGNs. We did not obtain a detection even by stacking the undetected sources, resulting in an X-ray weakness of $\sim 1-2$ dex from what was expected based on their bolometric luminosities.
   To discriminate between a heavily obscured AGN scenario or an intrinsic X-ray weakness of these sources, we performed a radio (1.4GHz) stacking analysis, which did not reveal any detection and left open the questions about the origin of the X-ray weakness.}


   \maketitle
%

\section{Introduction}\label{sec:intro}
Thanks to the successful launch of the \textit{James Webb Space Telescope} \citep[JWST;][]{Gardner23, Rigby23}, we are now able to investigate with a high resolution and an unprecedented sensitivity both the photometric and spectroscopic properties of galaxies up to $z\sim14$ \citep{Carniani24,CurtisLake22, Robertson22}. Within this context, recent studies, exploiting both spectroscopic and imaging data from JWST, have revealed a large population of active galactic nuclei (AGNs) at high redshift \citep{Kocevski23,ubler23,ubler23b,Matthee23,Maiolino23a,Maiolino23c,Greene23,Bogdan23,Goulding23,Kokorev23,Furtak23,Juodzbalis24,Scholtz23b,Chisholm24}, providing a unique opportunity to study the properties of supermassive black hole (SMBH) and AGN-galaxy coevolution since very early times.

Different works, selecting AGNs at high-$z$ using JWST NIR and MIR photometry, have shown that the AGN population at $z>3$ is larger than previously expected and that it is probably dominated by obscured or even heavily obscured sources, as is predicted by some coevolutional models \cite{Hopkins08,hickox18}. \cite{Yang23}, taking advantage of the JWST-MIRI photometry of the Cosmic Evolution Early Release Science Survey \citep[CEERS;][]{Finkelstein22}, investigated the AGN population using spectral energy distribution (SED) modeling and found a black-hole accretion rate density (BHARD) at $z>3$ $\sim 0.5$ dex higher than what was expected from previous X-ray AGN studies \citep{vito16,vito18}. Also, \cite{Lyu23}, performing a similar analysis using the JWST/MIRI data of the SMILES survey, selected a remarkable fraction of AGNs among the MIRI-detected sources ($\sim 7\%$) and found a statistically significant increase in the obscured AGN fraction with redshift, as had previously been found in other works \citep{Signorini23, gilli22, buchner15, aird15}.

The existence of a larger-than-expected AGN population at $z>4$ has also been shown by spectroscopic studies. For example, \citet{Maiolino23c} and \citet{Harikane23}, by selecting broad-line AGNs (BLAGNs, or type I) among JWST/NIRSpec spectra, found a significant AGN excess at $z>4$ with respect to the AGN luminosity functions derived using X-ray data \citep[][]{giallongo19}.  

JWST data not only allow for the identification of a higher fraction of AGNs at high-$z$ but also show some new and unexpected features in this early population of SMBHs. Studies have revealed that early SMBHs tend to be overmassive relative to the host galaxy stellar mass, when compared with the local AGN distribution \citep{Maiolino23c, Bogdan23,Furtak23, Juodzbalis24}, suggesting that the early stages of the coevolution between the SMBH and the host galaxy can be dominated by SMBH growth. This could further suggest that the preferred driving channel for the SMBH formation is the so-called ``direct collapse black hole'' scenario, together with (or alternatively) episodes of super-Eddington accretion \citep{Scholtz_2023_GN-z11}.

Another remarkable feature of high-$z$ newly discovered AGNs is that they seem to be X-ray-weak compared to the low-redshift AGN population. X-rays have traditionally been used to select AGNs because at typical AGN X-ray luminosities ($\log L_{X}\geq 42$), the level of contamination by stellar processes is low, allowing for a generally pure AGN selection. However, even if X-ray photons are able to penetrate through dense environments, they are almost completely absorbed at Compton-thick (CTK) hydrogen column densities ($\log (N_{H}/\rm cm^{-2})\geq 24$), making the X-ray identification of heavily obscured AGNs challenging \citep{vignali15,lanzuisi18,Goulding23,Maiolino24_X}.

Specifically, most of the newly selected AGNs, including BLAGNs, lack any X-ray emission \citep{Maiolino24_X,Yue24}, even if they are located in fields covered by some of the deepest extragalactic X-ray observations ever performed. Given the very few exceptions \citep{Goulding23,Kovacs24,Maiolino24_X} of X-ray AGN detections in the early Universe, it is possible that the X-ray weakness could be due to intrinsic properties of high-$z$ AGNs, as has also been suggested for some low-$z$ AGNs \citep{simmonds18,Zhang23}. In this view, AGNs can be characterized by a different accretion-disk and/or X-ray corona structure that can determine, for example, a larger ratio of the optical to X-ray emission ($\alpha_{OX}$) due to a much lower efficiency of the corona in producing X-ray photons, or simply a lack of corona \citep{Proga05}.
Recently, \cite{Maiolino24_X}, analyzing the lack of X-ray emission in a large sample of high-$z$ sources unambiguously identified as AGNs, suggested that their X-ray weakness could also be ascribed to the presence, in the inner region of the AGNs, of a spherical distribution of clouds with CTK column densities and very low dust content, such as the broad-line region (BLR) clouds. In this hypothesis, high-$z$ BLAGNs do not need to be characterized by any kind of intrinsic X-ray weakness, as their lack of X-ray emission would simply be caused by the inner CTK BLR gas.

JWST spectroscopy can give us information on the nature of the observed sources, allowing us to identify and investigate not only the population of BLAGNs, which is probably only the tip of the iceberg of the entire AGN census, but also the vast and hidden population of narrow-line AGNs (NLAGNs, also called type II).
An additional significant difference between high-$z$ and low-$z$ AGNs lies in the remarkably different physical environments constituted by their host galaxies. Indeed, early galaxies are systematically more metal-poor and characterized by younger stellar populations \citep{Curti23} that increase their ionization parameters in the ISM \citep[$\rm \log U$;][]{Cameron23,Curti23}. These two effects have a significant impact on the selection of NLAGNs, traditionally based on optical emission-line diagnostic diagrams, because the lower metallicities and the larger $\log U$ make star-forming galaxies (SFGs) and AGNs move toward each other and eventually overlap on the traditional diagnostic diagrams \citep{ubler23,Kocevski23, Maiolino23c,Scholtz23b,Mazzolari24b}.

However, the identification of the elusive population of NLAGNs is crucial in order to have a complete census of the AGN population, especially at high-$z$, and to build up statistical AGN samples that can solve the problems in AGN identification and properties at high-$z$. So far, a systematic search for NLAGNs at $z>3$ has only been attempted by a few works. In \cite{Scholtz23b}, the authors selected 42 NLAGNs among $\sim 200$ medium-resolution (MR) spectra of the JADES survey \citep{Eisenstein23,bunker23b}. In particular, they used JADES spectra to select NLAGN through a series of rest-frame optical, rest-frame UV diagnostic diagrams, and by exploiting the detection of high-ionization emission lines, which usually require an AGN as a photoionizing source.
The selected sample allowed the authors to investigate the population of AGNs up to $z\sim 9$, down to bolometric luminosities of $\log L_{bol}\sim 42$ and host galaxy stellar masses of $\rm 10^{7}M_{\odot}$, vastly expanding the region of the parameter space populated by these objects at early times.

In this work, we perform a similar analysis, investigating the NLAGN population hidden among the sample of MR spectra collected in the CEERS survey. The paper is organized as follows. In Sec.~\ref{sec:methods}, we present the data and the sample of spectra analyzed throughout this work and the techniques involved in the subsequent analysis: emission line fitting, spectral stacking, SED-fitting, and X-ray and radio stacking.
In Sec.~\ref{sec:results}, we describe the different emission line diagnostic diagrams involved in the NLAGN selection and the results of the selection. Then, in Sec.~\ref{sec:discussion}, we discuss the physical properties of the selected NLAGNs: the AGN prevalence (Sec.~\ref{sec:agn_prev}), the distribution in velocity dispersion and obscuration (Sec.~\ref{sec:vel}, Sec.~\ref{sec:av}), the bolometric luminosities (Sec.~\ref{sec:lbol}), the host galaxies properties (Sec.~\ref{sec:host_prop}), and finally, in Sec.~\ref{sec:xweak} and Sec.~\ref{sec:rad_res}, the X-ray and radio analysis, respectively. We assume a flat $\Lambda$CDM Universe with $H_{0}=70 ~\rm{km ~s^{-1} ~Mpc^{-1}}$, $\Omega_{m}=0.3$, and $\Omega_{\Lambda}=0.7$.

\section{Data and methods}\label{sec:methods}

\subsection{CEERS observational data}\label{sec:ceers}
We used publicly available MR JWST NIRSpec \citep{Jakobsen22} Micro-Shutter Assembly \citep[MSA;][]{Ferruit23} data from the CEERS program  \citep[Program ID:1345,][]{Finkelstein22}. The CEERS NIRSpec observations consist of six pointings in the `Extended Groth Strip' field \citep[EGS;][]{Rhodes00,Davis07}, all of which utilized the three grating and filter combination of G140M/F100LP, G235M/F170LP, and G395M/F290LP. This provided a spectral resolution of R $\sim 1000$ over the wavelength range of approximately 1–5 $\mu$m \citep{Jakobsen22}. Each grating and filter combination was observed for a total of 3107s for each pointing.\\
In particular, we considered the 313 spectra reduced and published by the CEERS collaboration in their latest data release, DR-0.7\footnote{\url{https://ceers.github.io/dr07.html}}. The data reduction was performed by the CEERS collaboration using the JWST Calibration Pipeline version 1.8.5 \citep{Bushouse22}, using the CRDS context “jwst\_1029.pmap.” The spectra were reduced using standard processing pipeline parameters, with some specific deviation in particular regarding the “jump” parameters and the so-called “snowball” corrections. Nodded background subtraction was employed. The pipeline was instructed that all targets should be treated as point sources, determining each 1D spectrum to be extracted from the 2D spectral data file over a specified range of pixels in the cross-dispersion direction. Extraction apertures on the 2D spectra were determined for each object by interactive visual inspection and were specified explicitly for the pipeline step “extract\_1d.”  Most faint galaxies are compact, and the extraction apertures adopted for nearly all objects have heights ranging from 3 to 6 pixels (scale = 0.10 arcsec/pixel), with a median value of 4 pixels.
The default pipeline path-loss correction was employed. This calibration is based primarily on pre-flight modeling, assuming the targets are point sources and determining these corrections to be incomplete for significantly extended sources, but this is not the case for the large majority of the targets. Flux calibration uses the default reference files for the adopted CRDS context. \\
For this work, we used the spectra with the three single grating spectra combined together. The data from the individual MR gratings were resampled to a common wavelength vector in the overlapping regions, adopting the wavelength grating that in the overlap region has the worst spectral resolution. The flux values in the overlap regions are the average of the individual grating values weighted by the flux errors, excluding pixels affected by masked artefacts.
After a careful visual inspection of the 313 single MR spectra available, we were able to attribute a secure redshift to 217 sources, constituting our parent sample throughout this work.\\
\cite{Harikane23} already investigated a limited sample of these spectra (i.e., only those at $z>3.8$) to look for BLAGN, selecting 10 AGNs at z = $4.015 - 6.936$ (2 of which are marked only as candidates) whose broad component is only seen in the permitted \Ha or \Hb lines and not in the forbidden \OIIIl5007 line. In the rest of the paper we mark the 8 most reliable BLAGNs selected in \cite{Harikane23} as the sample of high-$z$ BLAGNs.

\subsection{Emission line fitting}\label{sec:line_fit}
The NLAGN selection performed in this work is based on emission line diagnostics diagrams, where the AGN selection is provided by demarcation lines taken from the literature or by comparing the distribution of the sources with photoionization models, following the procedure done in \cite{Scholtz23b}\footnote{\url{https://github.com/honzascholtz/Qubespec}}.\\
We used a modified version of the publicly available code \texttt{QubeSpec} \citep{Scholtz23b,DEugenio24} to fit the spectra of the sources. We fit the emission lines considering only small wavelength ranges around the single (or group of) emission lines ($\sim 500\AA$), with each emission line fit using a single Gaussian component. The continuum was fit with a power law model, and in most of the spectra it was not detected. This approach for the fit of the continuum is sufficient for describing a narrow continuum range around an emission line of interest, and we found no instances of a strong continuum that required more sophisticated (e.g., stellar emission) modeling. \texttt{QubeSpec} uses a Bayesian approach, which requires defining prior probability distributions for each model parameter. In our fit, we assume log-uniform priors for the peak of the Gaussian describing each line and also for the continuum normalization. The prior on the lines full-width half maximum (FWHMs) is set to a uniform distribution spanning from the minimum resolution of the NIRSpec/MSA ($\sim 200$km/s) up to a maximum of 700 km/s. For the fit of the high-$z$ BLAGNs identified in \cite{Harikane23} we also include in the fit a broad component in the \Ha (with a FWHM uniform prior between $900-5000$km/s) to carefully disentangle the broad \Ha emission from the narrow one, which is the only component used in the diagnostic diagrams.
The prior on the redshift is a normal distribution centered on the redshift obtained from the visual inspection with a standard deviation of 100 km/s.\\
To estimate the posterior probability distribution of all these parameters, we use a Markov chain Monte Carlo integrator \citep[\texttt{emcee};][]{ForemanMackey13}.\\
To address the NLAGN selection, we performed a fit of the following rest-frame optical and UV emission line blocks:
\begin{itemize}
    \item \Ha + \NIIll6548,83 + \SIIll6716,31
    \item \OIl6300
    \item \Hb + \OIIIll4959,5008 + \HeIIl4686
    \item \Hg + \OIIIl4363
    \item \OIIll3726,29 + \NeIIIl3869
    \item \NeVl3426
    \item \NeIVll2422,24
    \item \CIIIll1907,10
    \item \OIIIll1660,66 + \HeIIl1640 +\CIVll1548,51
    \item \NIVll1483,86
    \item \Lya + \NVl1240
\end{itemize}
For the components of the \OIIIll4959,5008 and \NIIll6548,84 doublets, we fixed the ratios between the two components to be 1:3 given by the relative Einstein coefficients of the two transitions. Given the resolution of the NIRSpec MR spectra,  we fit the \OIIll3726,29, \NeIVll2422,24, \CIVll1548,51, \OIIIll1660,66, \CIIIll1907,10, and \NIVll1483,86 doublets as single Gaussians, centered at the mean wavelength between the two components. The signal-to-noise ratio (S/N) of each line was evaluated considering the same equations reported in \cite{mignoli19} and \cite{Lenz92}:
\begin{equation}\label{eq:SN1}
    \rm (S/N) = C\times \biggl(\frac{FWHM}{\Delta\lambda}\biggr)^{1/2} (S/N)_0,
\end{equation} 
where $\Delta \lambda$ is the sampling interval of the spectrum, $(S/N)_0$ is the peak S/N (i.e., $(S/N)_0=f_0/\sigma_0$, where $f_0, \sigma_0$ are the flux density and the noise level at the central pixel), and C is a proportionality constant whose values are reported in Table 1 of \cite{Lenz92}. We also checked the S/N using the following equation:
\begin{equation}
    \rm S/N=\frac{\sum_i f_i}{(\sum_i \sigma_i^2)^{1/2}},
\end{equation} 
where $f_i$ and $\sigma_i$ are the flux densities and associated errors for flux bins falling inside the Gaussian profile of the fit, and we found results consistent with those derived from Eq.~\ref{eq:SN1}. 
We required at least S/N$>5$ for the detection. Then, we visually inspected all the detected lines to exclude any possible false detection. In case of undetection, we derived the upper limit to the flux of each line as 3$\sigma$ errors (with $\sigma$ the error returned by \texttt{QubeSpec}).\\
While fitting the \Ha line, we found 4 sources at $z\leq2$ for which the residuals from the narrow-line-only fit were too large, and a broad component was required. These sources were not selected as in \cite{Harikane23} because in that work the authors investigated only sources at $z>3.8$. We identify these sources as new BLAGNs at low-$z$; their spectra are presented in Appendix~\ref{sec:app1}. 
In fitting all the BLAGNs, we did not find any significant broad component either in the \Hb or in the \OIIIl5007 lines.

\begin{table}[h!]
 \centering
   \caption{Definitions of line ratios adopted throughout the paper.}
 \begin{tabular}{ll}
  \hline
  Diagnostics & Line Ratio\\
  \hline
  R3 & $\log_{10}($[\ion{O}{III}]$\lambda$5007 / H$\beta)$\\
  N2 & $\log_{10}($[\ion{N}{II}]$\lambda$6583 / H$\alpha)$\\
  S2 & $\log_{10}($[\ion{S}{II}]$\lambda\lambda$6716,31 / H$\alpha)$\\
  O1 & $\log_{10}($[\ion{O}{I}]$\lambda\lambda$6300 / H$\alpha)$\\
  He2 & $\log_{10}($\ion{He}{II}$\lambda$4686 / H$\beta)$\\
  O3Hg & $\log_{10}($[\ion{O}{III}]$\lambda$4363 / H$\gamma)$ \\
  O32 & $\log_{10}($[\ion{O}{III}]$\lambda$5007 / [\ion{O}{II}]$\lambda\lambda$3726,29)\\
  Ne3O2 & $\log_{10}($\NeIIIl3869 / \OIIll3726,29)\\
  C43 &  $\log_{10}($\ion{C}{IV}$\lambda\lambda$1549,51/\ion{C}{III]}$\lambda\lambda$1907,09)\\
  C3He2 & $\log_{10}($\CIII$\lambda\lambda$1907,09 / \ion{He}{II}$\lambda$1640)\\
  Ne43 &  $\log_{10}($\NeIVl$\lambda$2422,24/\NeIIIl3869)\\
  N5C3  &  $\log_{10}($\NVl$\lambda$1239,42/\ion{C}{III]}$\lambda\lambda$1907,09)\\
  N4C3  &  $\log_{10}($\NIVl$\lambda$1483,86/\ion{C}{III]}$\lambda\lambda$1907,09)\\
 \hline
 \end{tabular}
  \label{tab:line_ratios}
\end{table}

\subsection{Spectral stacking} \label{sec:spec_stack}
\begin{figure*}
	\includegraphics[width=2\columnwidth]{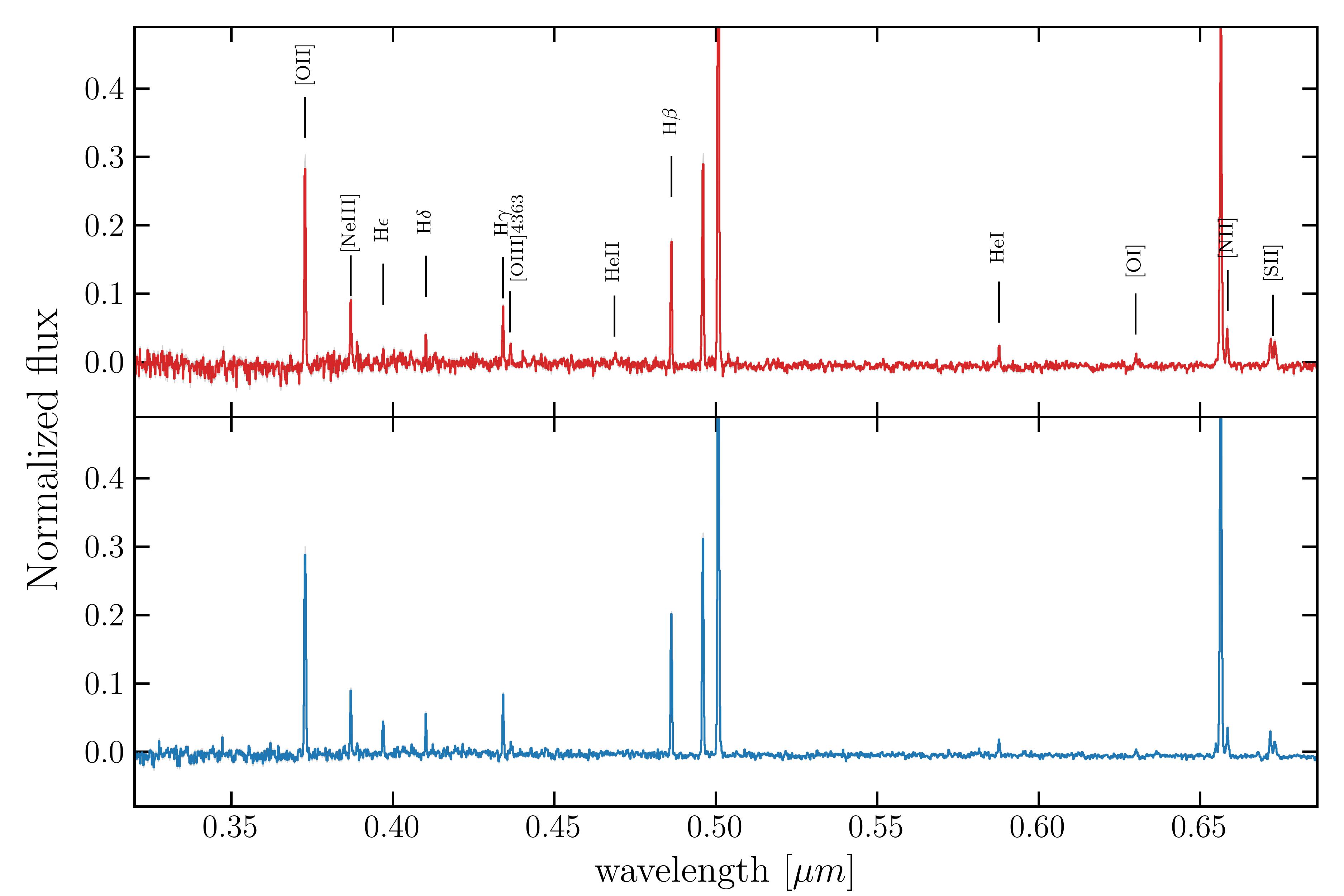}
    \caption{Stacked spectra of the final sample of NLAGNs (red) and non-AGNs (blue) derived in the manner described in Sec.~\ref{sec:spec_stack}. We also marked the positions of some relevant lines.}
    \label{fig:stack_spec}
\end{figure*}

Once the final sample of 52 NLAGNs has been selected, to get the average emission line properties of the SFGs and NLAGN samples, we stacked the MR spectra for each of the emission line complexes investigated in this work. 
The spectral stacking was performed by normalizing all the NLAGN and SFG spectra by the peak flux of the \OIIIl5007 line in order not to be biased toward the brightest objects. The \OIIIl5007 line is the most common and well-detected line among the CEERS MR spectra, but is not available in 8 AGNs and in 14 SFG spectra because for these sources the line falls into a detector gap. Therefore, we excluded them from the stacking. We shifted all the spectra to the rest frame, and then we resampled each of the spectra to a uniform and common wavelength grid, given that the three components of each spectrum (corresponding to the three different gratings) have different resolutions and wavelength bins.
The new and uniform wavelength grid was defined, for each line complex, by the wavelength bin allowing for the best spectral resolution among the different resolutions of the spectra involved in the stacking of the line(s). We checked that the results do not significantly change, taking the wavelength bin corresponding to the worst spectral resolution among the spectra. Before stacking the spectra, we also fit and subtracted the continuum from the rest-frame rebinned spectra. \\
We finally opted for spectral stacking involving a bootstrap procedure, given the number of sources involved. The stacked spectrum and uncertainties of each spectral bin were derived taking the median and standard deviation of 250 bootstrap realizations of the input spectra.
The resulting spectra from the stacking of all NLAGNs and SFGs are presented in Fig.~\ref{fig:stack_spec}.

\subsection{SED fitting}\label{sec:sedfit}
To investigate the physical properties of the host galaxies of the selected NLAGNs and to compare them with the properties of the parent sample, we performed a SED fitting analysis using \texttt{CIGALE} \citep{boquien19}. Starting from the 217 spectra of the MR sample with a secure redshift, we first crossmatch the target coordinates with the HST+JWST publicly available photometric catalog of the CEERS field reported in the DJA archive\footnote{\url{https://s3.amazonaws.com/grizli-v2/JwstMosaics/v7/index.html}}. Considering a 0.5" matching radius we found 117 counterparts. This catalog includes seven HST bands (F435W, F606W, F814W, F105W, F125W, F140W, F160W) and ten JWST/NIRCam and MIRI bands (F115W, F150W, 182M, F200W, 210M, F277W, F356W, F410M, F444W, F770W). Since not the whole CEERS MR sample is covered by JWST imaging, we also crossmatched the remaining sources with the 3D-HST multiwavelength catalog \citep{Momcheva16,Skelton14}, covering the EGS field and using the same crossmatching radius to avoid false matches. We found 93 additional counterparts, bringing the total number of sources with an associated optical/NIR photometry to 210. For the remaining seven spectra, we did not find a counterpart closer than 0.5", but their counterparts are returned when we considered a larger crossmatching radius ($0.5"<r<1.2"$). For sources detected in the JWST+HST catalog, we did not consider the additional photometry of the 3D-HST catalog since most of these sources are undetected in photometric bands bluer than the HST bands (being at high-$z$), while the photometry at longer wavelengths than HST bands is already covered by JWST (the deepest in the near and mid-infrared). \\
The SED fitting was performed considering two different parameter grids, reported in Appendix~\ref{sec:app2}, one for sources at $z<3$ and the other for sources at $z>3$. 
We used delayed star formation history (SFH) models for both redshift groups because they are able to reproduce both early-type and late-type galaxies, with an additional term that allows for a recent (and constant) variation in the star formation rate (SFR) that can be in the burst or in the quench phase. We adopted stellar templates from \cite{bruzual03}, and a Chabrier initial mass function \citep{chabrier03}. We also include the nebular emission module that is extremely important to account for the contribution of emission lines in the broad-band photometry \citep{Schaerer2012,Salvato2019}, in particular for the low-mass and high-$z$ regimes, where emission lines can account for a consistent fraction of the wideband photometric flux. This module is computed by \texttt{CIGALE} based on a grid of user-provided parameters and in a self-consistent manner using a grid of \texttt{Cloudy} photoionization calculations \cite{Ferland13}. For the attenuation of the stellar continuum emission, we considered the \texttt{dustatt\_modified\_CF00} module \citep{Charlot00}, which allows different attenuations for the young and old stellar populations. We also include dust emission in the IR following the empirical templates of \cite{Dale14}. The main differences in the $z>3$ grid with respect to the low-$z$ one consist in the larger parameter space explored for the possible final starburst state of high-$z$ galaxies and in the lower metallicities and higher ionization parameters allowed in the stellar and nebular modules, according with recent JWST results \citep{Endsley23, Tang23}. Indeed, SFGs at high-$z$ are observed to be more frequent in a bursty SF regime \citep{Faisst20, Dressler23, Looser23} and also, the gas metallicities are generally 0.5-1 dex lower than in the local Universe \citep{Curti23b,Nakajima23}.

For sources classified as AGNs based on our NLAGN selection, as well as the BLAGNs selected in \cite{Harikane23}, we included AGN modules in the SED fitting to account for the AGN emission. We employed the \texttt{skirtor2016} module introduced in \cite{yang20}, which is widely used in the community and has demonstrated reliability in studying various aspects of AGN \citep[e.g.,][]{Mountrichas22, LopezIE23, Yang2023}. The SED produced by the AGN combines emissions from the accretion disk, torus, and polar dust. The accretion disk, responsible for the UV-optical emission at the central region, is parameterized according to \cite{Schartmann05}, with a typical delta value of -0.36. Photons from the accretion disk can be obscured and scattered by dust in the vicinity, within the torus and/or in the polar direction. Specifically, for the torus, the \texttt{skirtor2016} module employs a clumpy two-phase model \citep{Stalevski16}, based on the 3D radiative-transfer code SKIRT \citep{SKIRTBaes2011ApJS..196...22B, SKIRTCamps2015A&C.....9...20C}. As our photometry does not cover the restframe mid-IR, we fixed the optical depth at 9.7 $\mu$m at $\tau$ = 3. For the polar dust component, we used the Small Magellanic Cloud (SMC) extinction curve, recommended due to its preference in AGN observations \citep[e.g.,][]{2012MNRAS.427.3103B}. The extinction amplitude, parameterized as $E(B-V)$, is a user-defined free parameter for which we chose a typical value of $E(B-V) = 0.04$. Emission from the polar dust maintains energy conservation, assuming isotropic emission and a “gray body” model \citep{2012MNRAS.425.3094C}. Finally, we allowed inclination values from 50 to 80 degrees and varied the AGN fraction (defined as the ratio between the AGN luminosity and the host galaxy luminosity between 0.15 and 2 microns) from 0.1 (weak AGN contamination) to 0.7 (dominant AGN emission).

\subsection{X-ray and radio stacking}\label{sec:xray_rad_stack}
To investigate the multiwavelength properties of the CEERS NLAGNs selected using the diagnostic diagrams, we looked at their counterparts in the X-rays and Radio images of the EGS field.\\
The \textit{Chandra} AEGIS-XD field is the third deepest X-ray field ever performed, with an exposure that reaches $\sim$800 ks. The X-ray observations and the related X-ray catalog are described in \cite{nandra15}. The X-ray catalog contains 937 X-ray sources, and 553 of these sources (those with enough photon counts) also have an X-ray spectral analysis performed by \cite{buchner15}. By crossmatching the CEERS MR catalog with the X-ray catalog, we found seven matches, involving sources at $0.5<z<3$. The X-ray spectral analysis classified two of these sources as galaxies (CEERS-3060, CEERS-3051), the other 5 (CEERS-2919, CEERS-2808, CEERS-2904, CEERS-2989, CEERS-2900) as AGN. In particular, CEERS-2904, CEERS-2989, and CEERS-2919 are part of the new low-redshift BLAGN sample presented in Appendix~\ref{sec:app1}.\\
Four out of the five X-ray AGNs were selected as NLAGNs by our spectroscopic selection, as we are going to present in Sec.~\ref{sec:results}. However, none of the other 48 NLAGNs selected show any indication of X-ray emission. Therefore, we decided to perform an X-ray stacking analysis of the NLAGNs not X-ray detected using the publicly available code CSTACK v4.5 \cite{miyaji08}. For each target, CSTACK provides the net (background-subtracted) count rate in the soft (0.5–2 keV) and hard (2–8 keV) bands using all the observations from the AEGIS-XD survey and the associated exposure maps. Since the survey is a mosaic, multiple observations at different off-axis angles can cover the same position. Due to the variation in the \textit{Chandra} PSF with the off-axis angle, for each observation of an object, CSTACK defines a circular source extraction region with the size determined by the 90\% encircled counts fraction radius ($r_{90}$) (with a minimum of 1"), to optimize the S/N of the stacked signals. We set the background region for each source to be a $30\times30$ arcsec$^2$ area centered on the object, excluding the region around the object used to compute the net counts. Moreover, the code excludes the photon counts of the background circular regions around all the detected X-ray sources, with radii that depend on the net counts of the X-ray source.  
We derived the count rates in the soft band (0.5-2keV, SB), the band for which the detector is most sensitive and can, therefore, provide tighter constraints on the X-ray analysis. Then we converted the count rates into fluxes assuming an X-ray power spectrum with an intrinsic spectral index $\Gamma=1.9$ and using the \textit{Chandra} tool PIMMS\footnote{\url{https://cxc.harvard.edu/toolkit/pimms.jsp}}.\\

The field targeted by the CEERS survey is also covered by 1.4GHz observations, the AEGIS20 described in \cite{Ivison07}. This radio survey was performed in 2006 with the Very Large Array (VLA) over an area of $\sim0.75$deg$^2$ and reaching, in the central region, a rms of $\sim 10\mu$Jy. The resulting radio catalog \citep{Ivison07} contains 1123 individual radio sources, of which only two are part of the CEERS MR sample, CEERS-2900 and CEERS-3129, the first (that is also X-ray detected) showing the typical morphology of a radio-loud AGN. Both these sources were also identified as AGNs by our selection, but none of the other NLAGNs have any radio counterparts. Therefore, we performed a radio stacking analysis to investigate the radio statistical properties of the NLAGN population below the survey detection threshold.  We applied a pixel-by-pixel median stacking, taking $30\times30$ arcsec$^2$ cutouts around the sources we wanted to stack. This method was proved \citep{white07} to be more robust (compared to the mean stacking, for example) toward systematic effects (i.e., confusion) and also less affected by possible high-flux density contaminants. Since the CEERS survey covers only a very small region of the radio image on which the rms is almost uniform, we did not weigh each radio cutout for the corresponding noise value in the stacking procedure. We also estimated the final rms of the stacking images using a median absolute deviation procedure, following \cite{Keller24}.

\begin{figure*}
	\includegraphics[width=1\columnwidth]{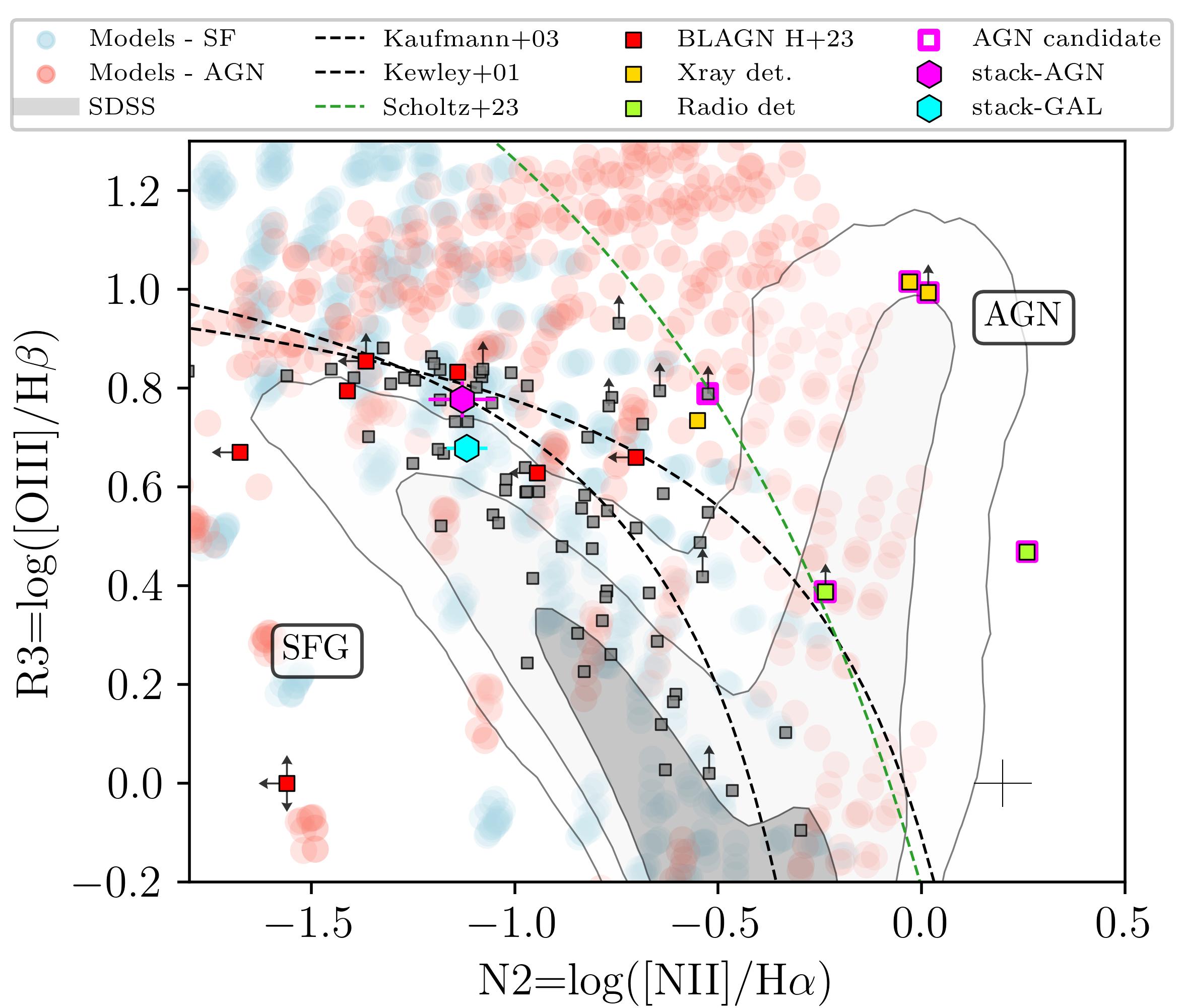}
	\includegraphics[width=1\columnwidth]{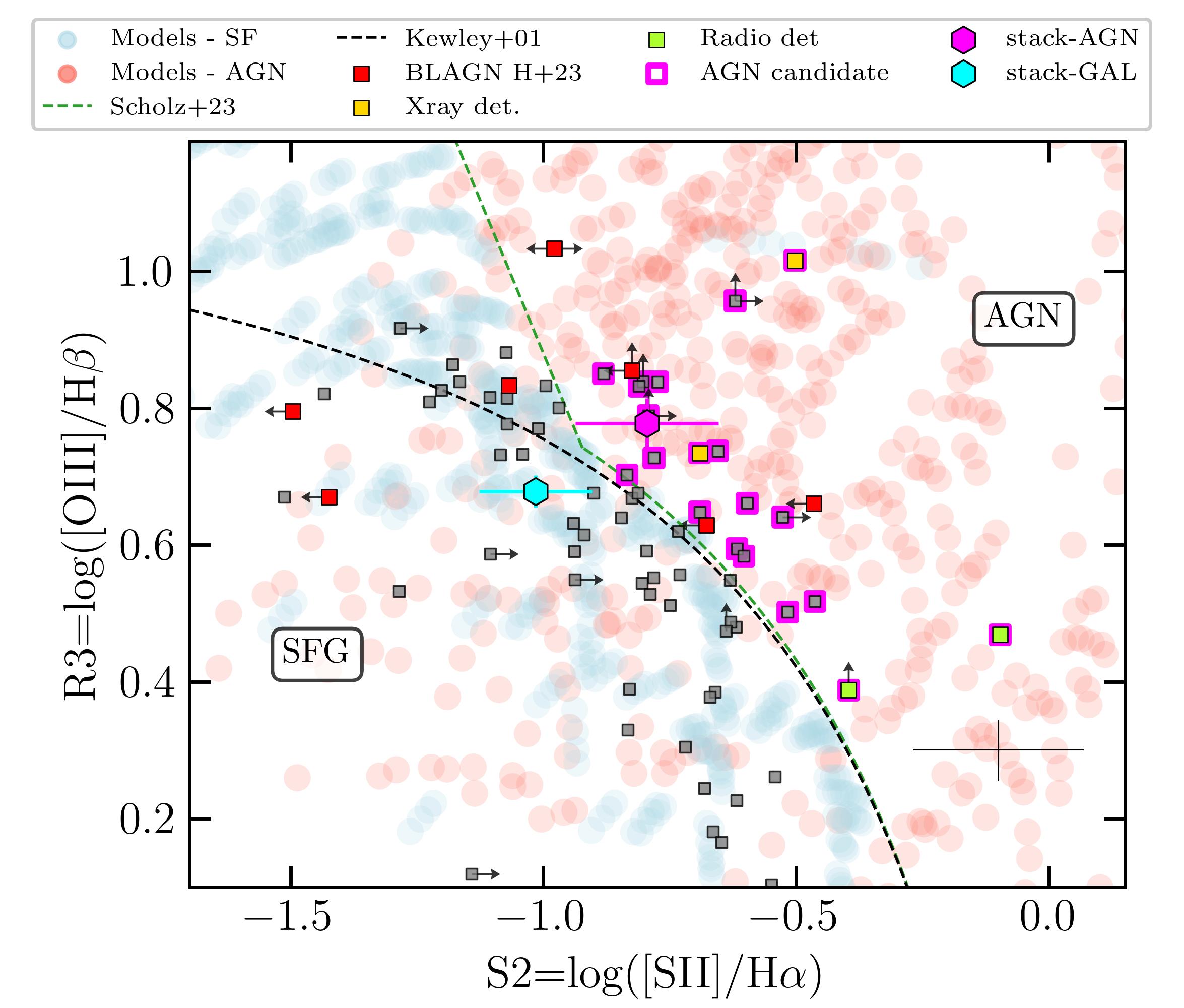}\\
 \centering
 \includegraphics[width=1\columnwidth]{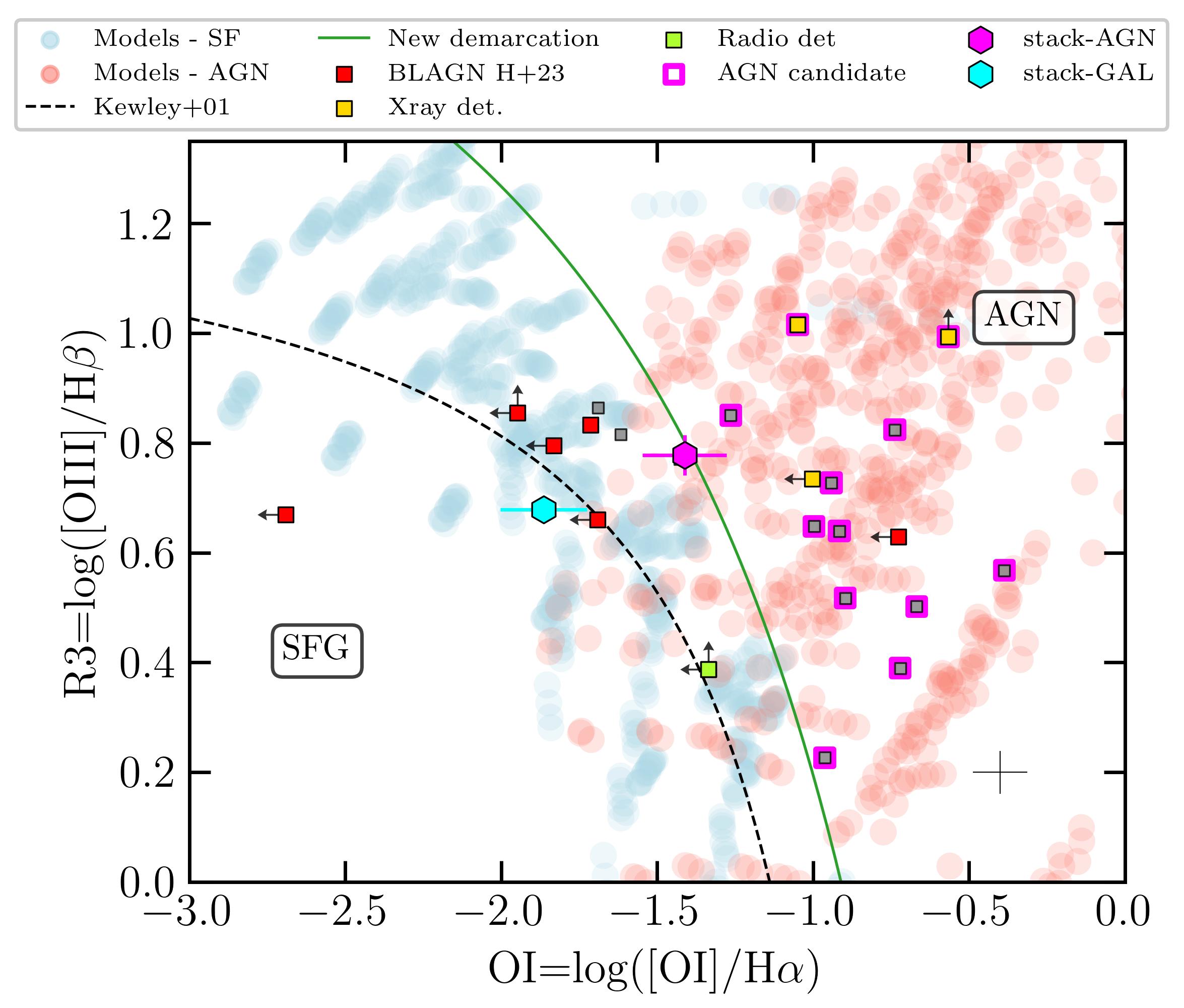}
\caption{\textit{Left:} R3N2 diagnostic diagram (also called BPT). The gray points represent the parent sample of galaxies analyzed in this work. Gold, green and red colors are used, respectively, for X-ray detected sources, radio-detected sources, and the high-$z$ BLAGNs reported in \citep{Harikane23}. The NLAGNs selected using this diagram are marked with magenta edges. In magenta and cyan, we show the line ratios derived from the stacked spectra of all the selected NLAGN (52 sources) and non-AGN sources. The dashed lines refer to three different demarcation lines, as labeled, the one in green is the more conservative demarcation line derived in \cite{Scholtz23b}. The shaded blue and red area represents the regions covered by the SFG and AGN photoionization models computed in \cite{Gutkin16} and \cite{Feltre16}. The gray contours mark the distribution of SDSS sources \citep[taken from SDSS DR7][]{Abazajian09}. In the lower-right corner are reported the median errors of the sample.
    \textit{Right:} R3S2 diagnostic diagram (also called VO87) with the demarcation line originally presented in \cite{Kewley01} (in black) and the new (more conservative) demarcation derived in \cite{Scholtz23b}.
    \textit{Bottom:} R3O1 diagnostic diagram with the demarcation line presented in \cite{Kewley01} (dashed black line) and the new one derived in this work (solid green line).
    }
    \label{fig:BPT}
\end{figure*}

\begin{figure}
	\includegraphics[width=1\columnwidth]{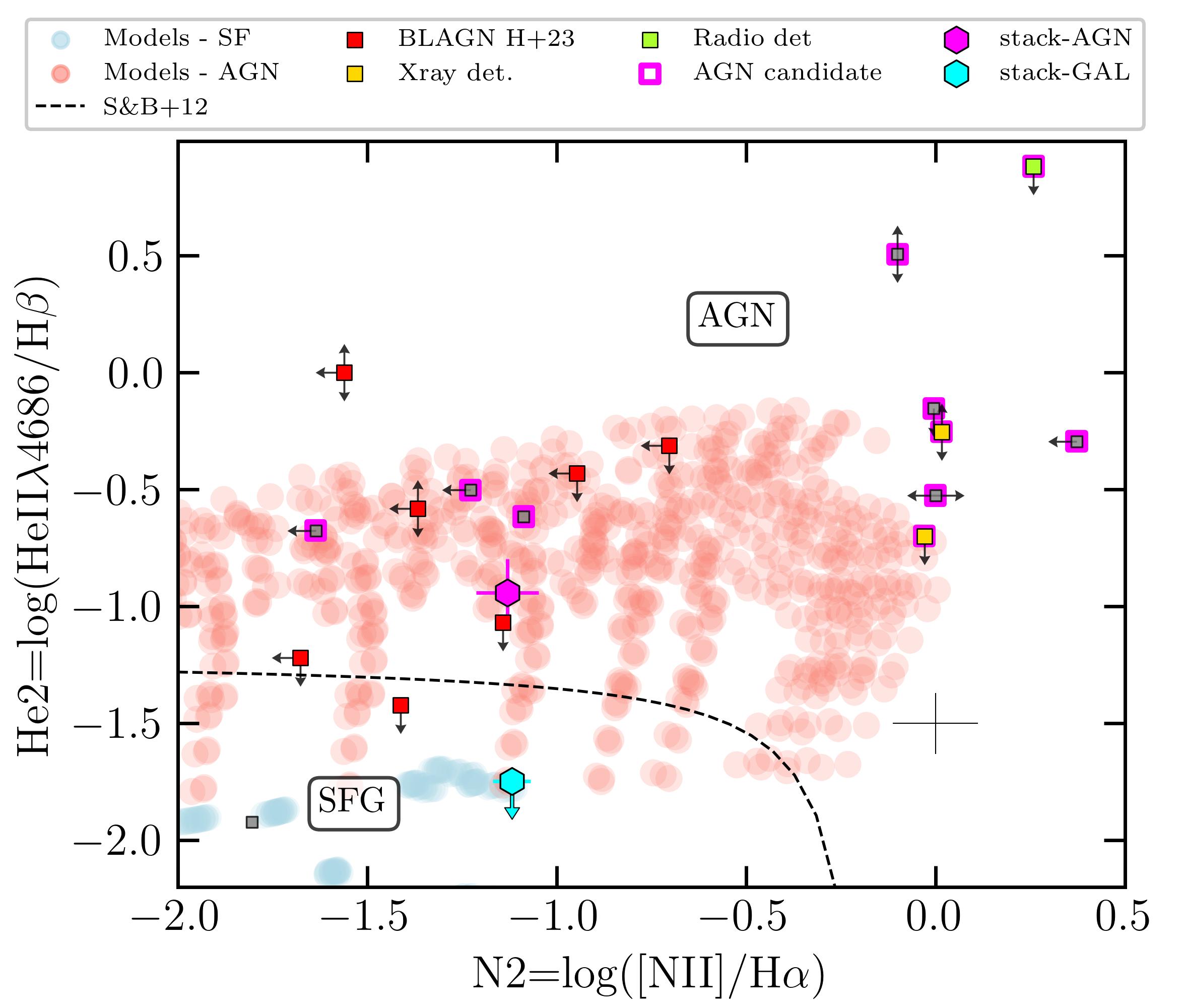}
    \caption{He2N2 diagnostic diagram. The colors follow the same scheme as in Fig.~\ref{fig:BPT}}
    \label{fig:He2_N2}
\end{figure}

\begin{figure*}
	\includegraphics[width=1\columnwidth]{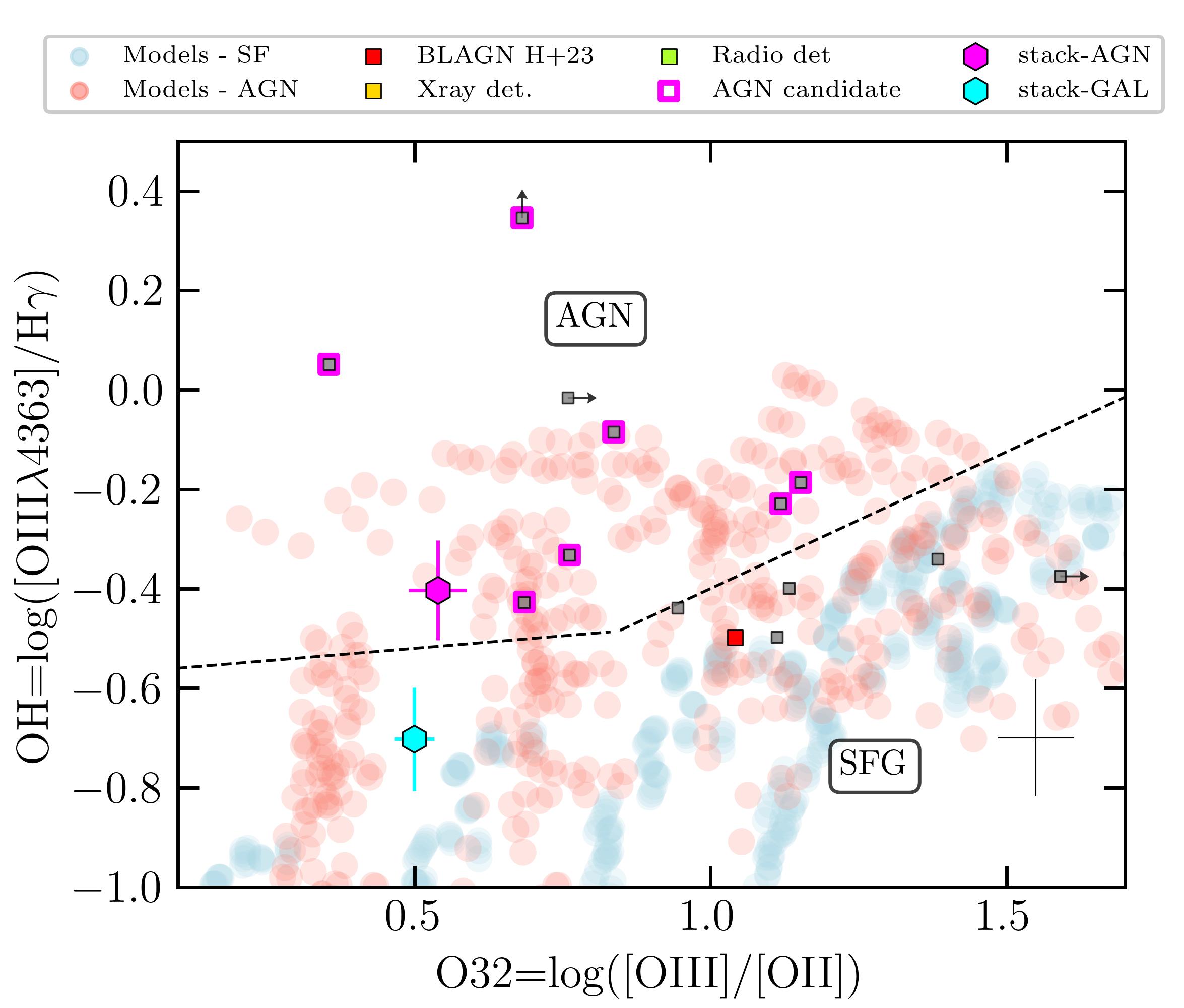}
        \includegraphics[width=1\columnwidth]{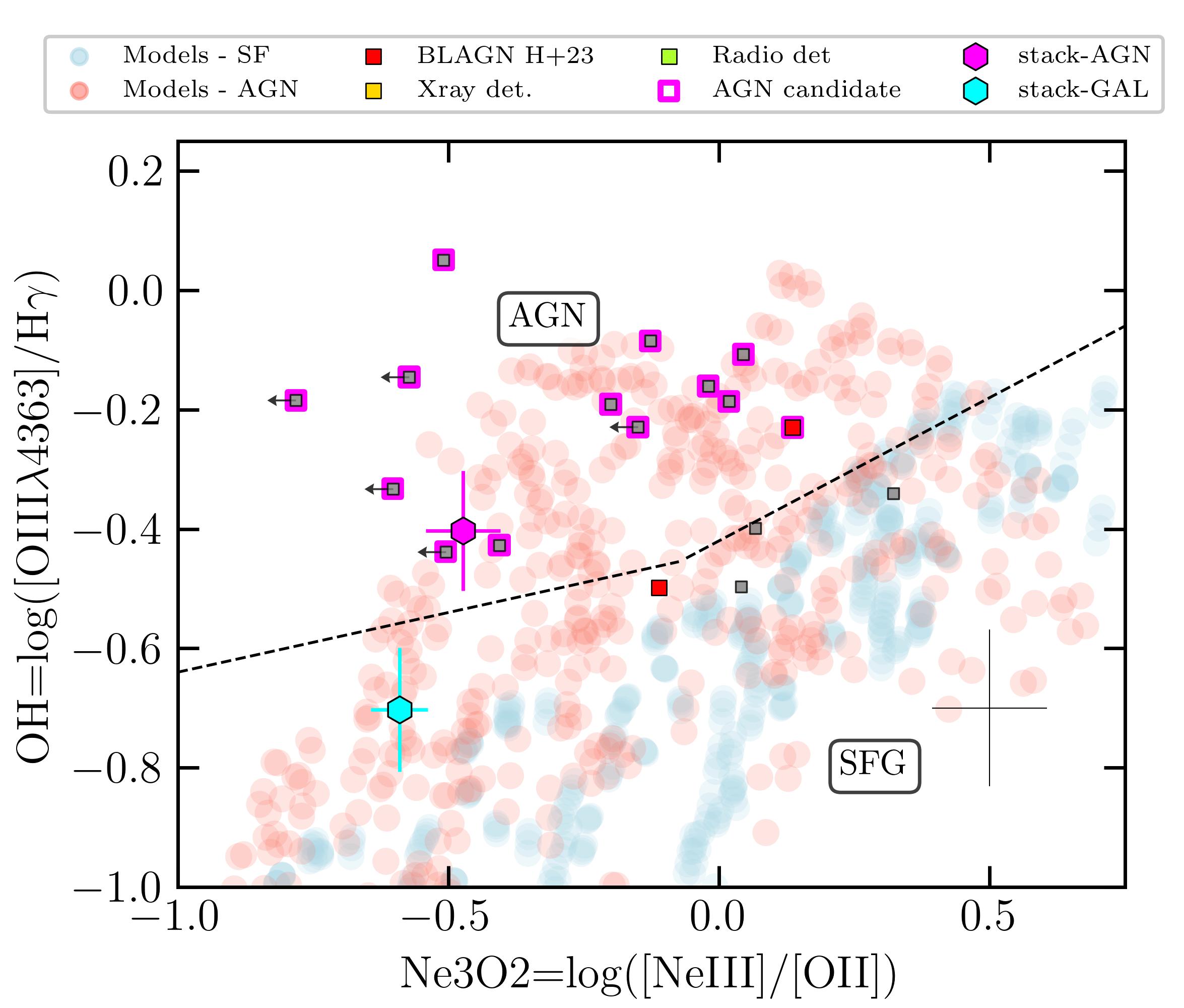}
    \caption{M1 and M2 diagnostic diagrams, firstly presented and discussed in \cite{Mazzolari24b}. The colors follow the same scheme as in Fig.~\ref{fig:BPT}.}
    \label{fig:auroral}
\end{figure*}

\section{Results}\label{sec:results}
Here, we present the result of the NLAGN selection performed on the sample of the 217 MR CEERS spectra with a secure redshift identification. The NLAGN selection was performed using emission line diagnostic diagrams.\\
In all the diagrams, we plot both the observed data points and also the line ratios derived from the photoionization models described in \cite{Feltre16} and \cite{Gutkin16}. These models were computed using the \texttt{Cloudy} code \citep{Ferland13} for SFG and AGN NL regions and for various gas metallicities, ionization parameters, dust content, and ISM densities, and considering a wide range of the parameters space \citep[for the full grid of values covered by the parameter space we refer to Table 1 in][]{Feltre16}. To allow for a sufficiently comprehensive coverage of the SFG models but to avoid physical conditions rarely (or never) found in the general populations of SFG in the local or high-$z$ Universe, we applied to these models the same cut in the parameter space as presented and discussed in \cite{Scholtz23b}. In particular, we selected all models with metallicities between 0.001 and 0.02 (corresponding to 0.06-1.3 solar) and a dust-to-metal mass ratio of 0.3, which is intermediate between the range observed in the most metal-poor absorbers \citep{Konstantopoulou23} and the Milky-Way value of 0.45. We consider all models with carbon to oxygen abundance ratio in the range 0.38-1.00 solar to describe a variety of different or less common star formation grids.
Through the work, photoionization models for AGN and SFG will be shown in red and in blue, respectively.
We also mark with distinctive colors the seven sources that are X-ray detected (gold), the two radio-detected sources (green), and the eight sources that were selected as BLAGNs at $z>4.5$ by \citet[][]{Harikane23} (red). In each diagnostic diagram, the sources with magenta edges are those selected as NLAGN in that specific diagnostic.\\
It is worth noting that the line fluxes involved in the diagnostic diagrams described in the next Sections are not corrected for the effect of dust. This is because we are always considering ratios between lines close in wavelength to each other, hence subjected to almost the same reddening, making the ratio insensible to the presence of dust. The few exceptions are discussed and justified. The acronyms used for the line ratios are reported in Tab.~\ref{tab:line_ratios}.

\subsection{Optical diagnostics}\label{sec:opt_diagn}
In Fig.~\ref{fig:BPT}, we show the traditional BPT diagram \citep{Baldwin81}, the R3S2 (also called VO87), and the R3O1 \citep{Veilleux87} diagnostics diagrams based on the R3-N2, R3-S2, and R3-O1 line ratios, respectively.
The traditional demarcation lines between SFG and AGN provided by \cite{Kauffmann03} and \cite{Kewley01} already proved to be way less effective in selecting high-$z$ AGNs compared to the local Universe \citep{Scholtz23b,ubler23,Maiolino23c}. Indeed, the lower metallicities of high-$z$ sources make \NII (but also \SII) emission lines fainter, shifting the objects (AGN included) toward the left part of the BPT diagram. At the same time, the higher ionization parameter of the high-$z$ SFG \citep{Cameron23,Sanders23, topping24}, due to, on average, the presence of younger stars and lower metallicities compared to the local Universe, moves the SFG toward higher R3 ratios, generating a large overlap between AGN and SFGs. Therefore, in the R3N2 and R3S2 diagnostics, we used the {\it conservative} demarcation lines provided in \cite{Scholtz23b}, derived considering the distribution of the photoionization models of \cite{Gutkin16,Feltre16} and \cite{Nakajima22}.\\
Since the R3O1 diagnostic diagram was not included in the analysis of \cite{Scholtz23b}, we followed the same fitting method to the photoionization models to find more conservative demarcation lines with respect to those presented in \cite{Kewley01}. We derived the following dividing line between the AGN and SFG population:
\begin{equation}
    \text{R3}=2.5 + \frac{2.65}{\text{O1}-0.15}.
\end{equation}
As it is possible to see from the three plots, none of the BLAGNs at high-$z$ are in the AGN region of the three diagnostics, based on the new demarcation lines, and only one would be selected considering the traditional demarcation lines in the R3S2 and R3O1 diagrams (is the same source, CEERS-1665).\\
The five AGNs selected with the R3N2 diagram are all at $z<2.9$. Three of them are X-ray-detected AGNs (one is also radio-detected, i.e., CEERS-2900), and one is another radio-detected source (CEERS-3129). It is worth noting that even the X-ray source at the highest redshift (i.e., CEERS-2808 at $z=3.38$), does not fall in the pure-AGN region of this diagnostic diagram. \\
The R3S2 diagnostic diagram appears effective in selecting AGN also at a higher redshift compared to the BPT. The new demarcation line provided in \cite{Scholtz23b} partially overlaps with the traditional one proposed in \cite{Kewley01}. In this case, we select 21 AGNs, six of which at $z>3$, one at $z=5.27$. None of the BLAGNs is selected as AGN based on this diagnostic, while three X-ray sources and the two radio sources are confirmed as AGN also in this diagnostic.\\
Also the R3O1 diagnostic diagram is more effective in identifying NLAGN compared to the R3N2. In this diagram, we select 12 AGNs, 7 at $z<3$ (including 2 X-ray sources), and 5 at $z>3$, one of these at $z\sim 6$. We further discuss the effectiveness of this diagnostic and of the R3S2 in Sec.~\ref{sec:sel_meth}.\\
In Fig.~\ref{fig:He2_N2}, we show the diagnostic based on the He2 versus N2 lines ratio. In this case, the demarcation line is the original one provided by \cite{Shirazi12}, which still holds even at high-$z$, as also reported in \cite{Dors23} and \cite{Tozzi23}. Overall, we detect \HeIIl4686 in seven sources: all except one of these are classified as AGN. None of the CEERS BLAGNs show detection of the \HeIIl4686 line. In this diagnostic, there are also five sources that are selected as AGN only based on the N2 ratio.\\
The last diagnostics exploiting rest-frame optical lines are the two diagnostics based on the O3Hg ratio and proposed in \cite{Mazzolari24b}. Hereafter, we refer with M1 to the O3Hg versus O32 diagnostic diagram and with M2 to the O3Hg versus Ne3O2 diagram. These diagnostics proved to be effective in selecting those AGN characterized by high O3Hg ratios at a given ratio of O32 or Ne3O2, and were able to select NLAGN also at $z>6$ \citep{Mazzolari24b}.
The \OIIIl4363 line is sensitive to the electron temperature of the ISM. The effectiveness of these diagnostics is related to the fact that the average energy of AGN’s ionizing photons is higher than that of young stars in SFGs, because the AGN ionizing source (the accretion disk) produces a harder SED with respect to stars. Therefore, AGN can more efficiently heat the gas, boosting the \OIIIl4363 line. Using the two diagnostic diagrams reported in Fig.~\ref{fig:auroral}, we were able to select 15 distinct NLAGN, up to $z\sim6$. In the case of the left panel of Fig.~\ref{fig:auroral}, the O32 line ratio can suffer from the effect of dust reddening due to the wavelength distance of the two lines involved. However, the effect of dust attenuation on this diagnostic moves sources toward the right, without contaminating the AGN-only region with SFGs, allowing us to select only pure AGN.

We note that in the diagnostics presented in Fig.~\ref{fig:BPT}, \ref{fig:He2_N2}, \ref{fig:auroral} there are some sources lying close enough to the demarcation lines that their line ratios' error can potentially place them outside from the AGN-only region of the diagrams. We considered all the selected NLAGN whose 1$\sigma$ uncertainties are compatible with SF-driven photoionization, and we double-checked if they were safely identified as NLAGN in other diagnostics (i.e., without 1$\sigma$ errors crossing the demarcation line). We identified four NLAGNs with errors compatible with SFG ionization -- CEERS-2668, CEERS-3535, CEERS-1836, and CEERS-3585 -- and we conservatively decided to mark them only as candidates NLAGNs. CEERS-2900 should also be included in this sample, but given its detection in both the X-ray and radio image of the field it can be safely considered a NLAGN.

\subsection{UV diagnostic}\label{sec:uv_diagn}
\begin{figure}
	\includegraphics[width=1\columnwidth]{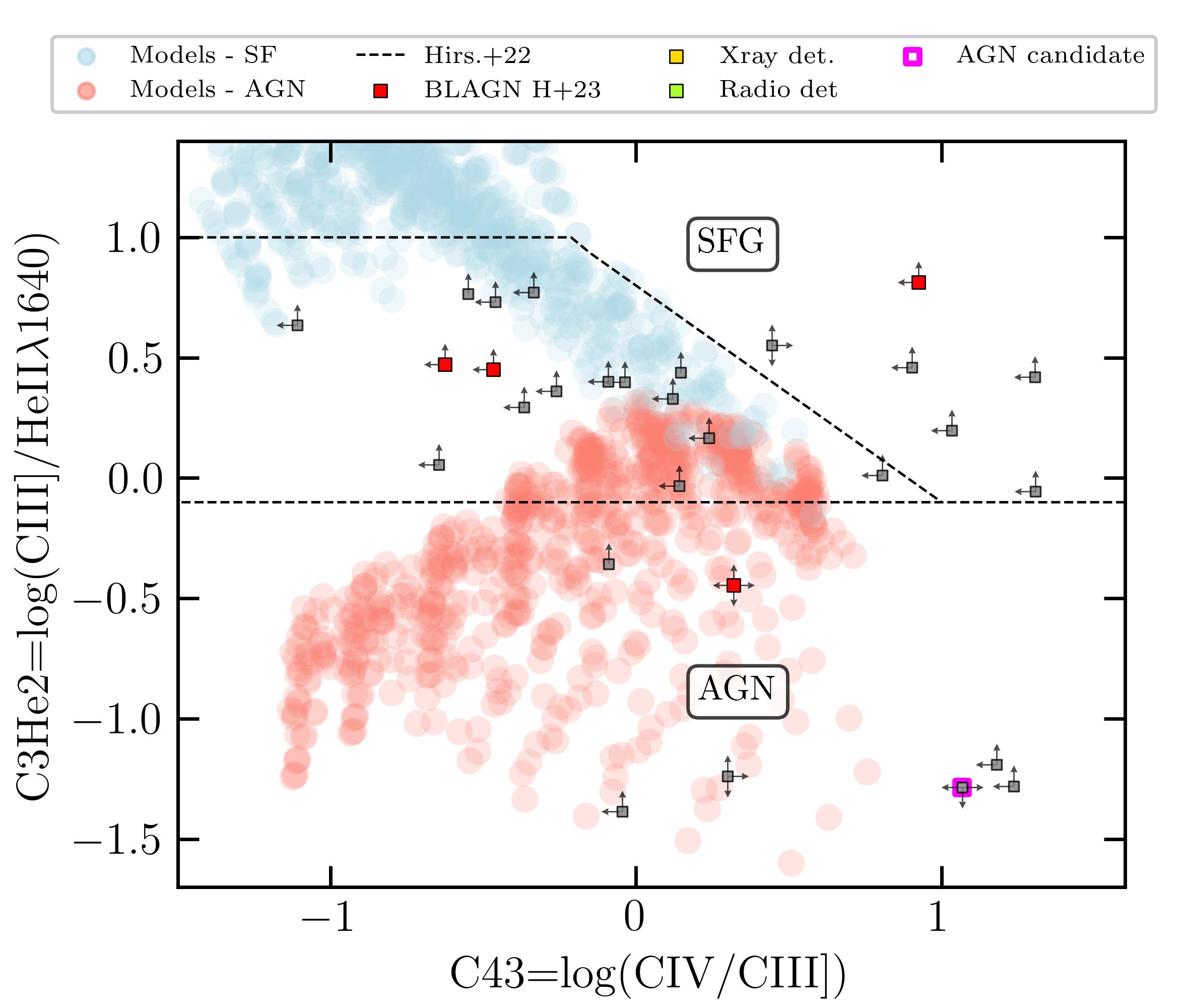}
    \caption{C3He2 versus C43 diagnostic diagrams, selecting only one NLAGN due to the difficulties in detecting rest-UV lines in high-$z$ galaxies with the short exposure times of the CEERS survey. The stacking line ratios and the average errors are not reported because of the poor statistics. The colors follow the same scheme as in Fig.~\ref{fig:BPT}. The demarcation lines are those defined in \cite{Hirschmann22}.}
    \label{fig:UV_diag}
\end{figure}

\begin{figure}
	\includegraphics[width=1\columnwidth]{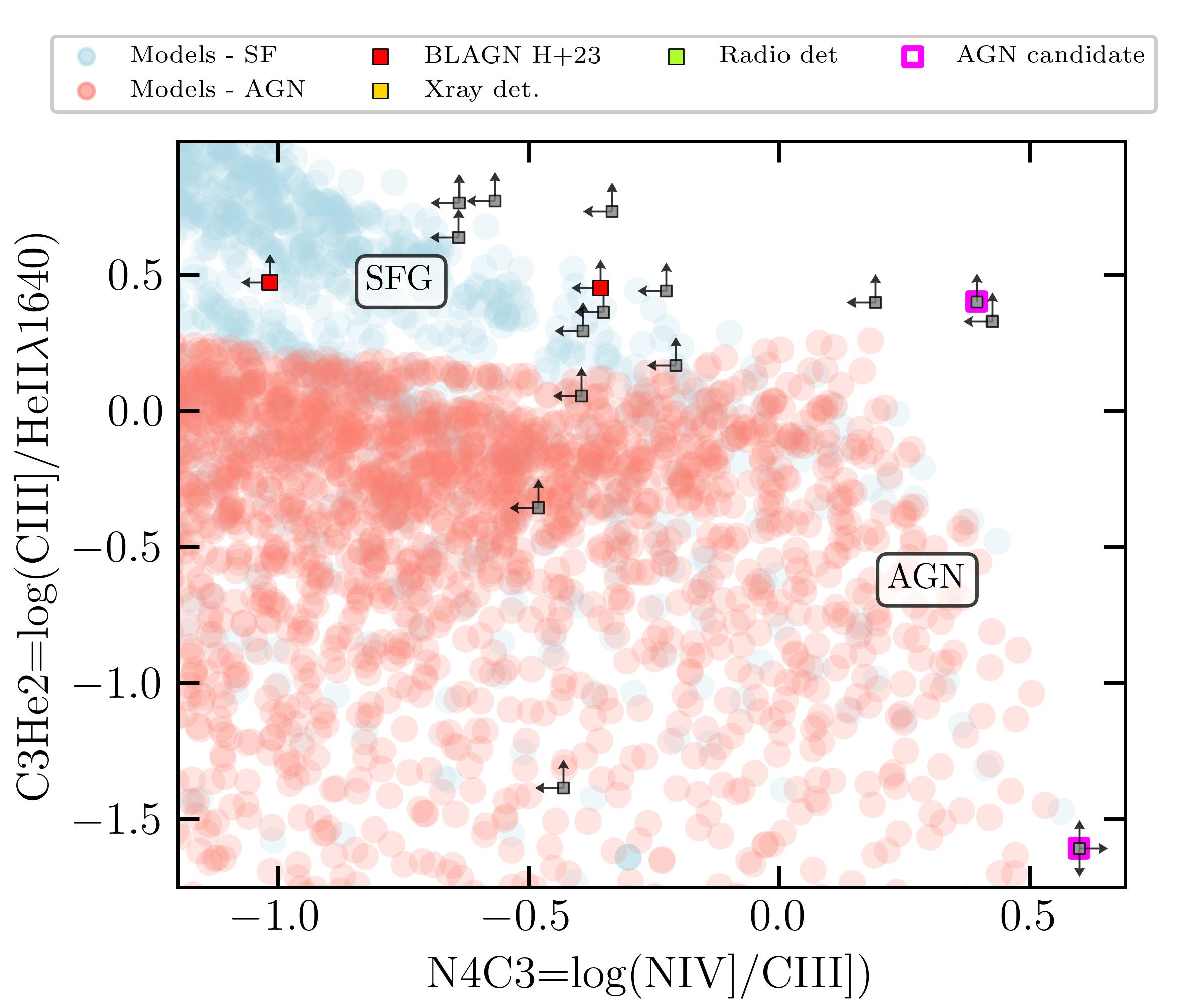}\\
        \includegraphics[width=1\columnwidth]{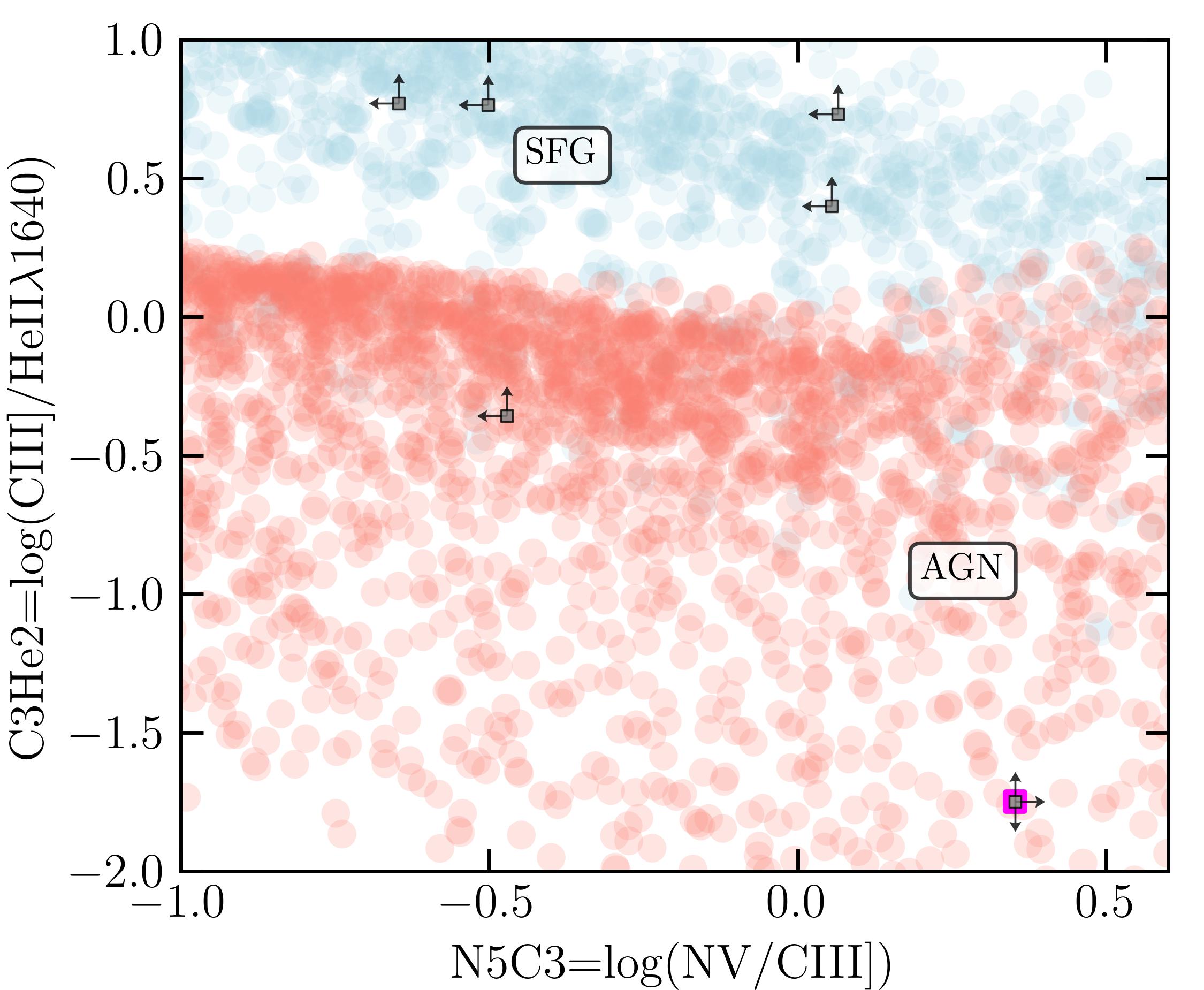}\\
         \includegraphics[width=1\columnwidth]{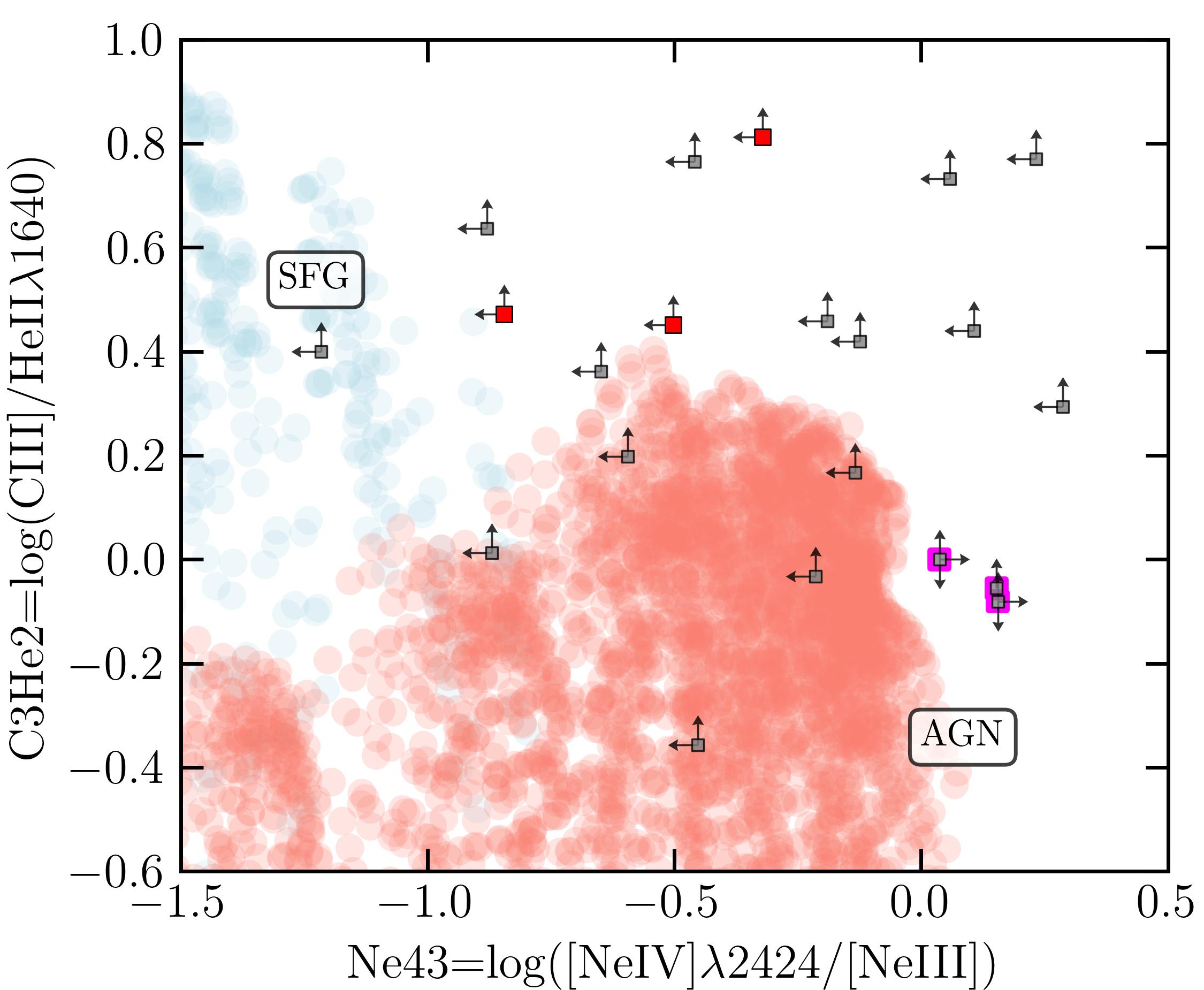}
    \caption{From top to the bottom: C3He2 versus N4C3, C3He2 versus N5C3, C3He2 versus Ne4C3 diagnostic diagrams. The stacking line rations and the average errors are not reported because of the poor statistics. The colors follow the same scheme as in Fig.~\ref{fig:BPT} }
     \label{fig:high-ion}
\end{figure}

In this section, we focus on the diagnostic diagrams involving rest-frame UV lines \citep{Mingozzi24,Mascia24,Scholtz23b,Feltre16}. 
In particular, we considered the UV diagnostic diagram C3He2-C43 and the diagnostic diagrams involving the so-called high-ionization emission lines, \NIVl1486, \NVl1243, and \NeIVl2425.\\
The C3He2-C43 diagnostic diagram reported in Fig.~\ref{fig:UV_diag} allows us to select one single NLAGN (CEERS-613), whose selection is based on a clear detection of the \HeIIl1640 line, which places the source well into the AGN-dominated region according to the demarcation lines presented in \cite{Hirschmann22}. The detection of the UV lines involved in this diagram is challenging with only $\sim 50$min of JWST exposure (the average on-source exposure time of the CEERS survey). Also in \cite{Scholtz23b} the number of detections of these lines is low, even if the exposure time per target of the JADES survey was $\sim 8$ times longer than the CEERS observations.\\
The high-ionization emission lines are characterized by high-photoionization energies that are more likely to be produced in AGNs, due to their harder ionizing radiation, and therefore can be used as indicators of the presence of an AGN, even if hidden. In the diagnostics diagrams involving these lines, we base the identification of NLAGN on the position of the sources compared to the distribution of AGN and SFG photoionization models of \cite{Gutkin16} and \cite{Feltre16}. In the top panel of Fig.~\ref{fig:high-ion} we show the diagnostic diagram involving the \NIVl1486 emission line, which has an ionization potential of 47eV. This diagnostic allows us to select two NLAGN (CEERS-496, CEERS-1019), based on their extreme position with respect to the distribution of both SFG and AGN photoionization models. However, given the unclear detection of the \HeIIl1640 line in both these sources, and given the uncertainties in the extension of the SFG region in the diagnostic, we decided to mark these NLAGNs only as candidates. In particular, the source in the upper part of the diagram is CEERS-1019, whose AGN nature has already been discussed in multiple works in the literature, some supporting the presence of an AGN based on a low-significance broad \Hb detection \citep{Mascia24,Larson23} and some discarding it \citep{Tang23} and attributing its spectral properties to exotic SF. In the middle panel of the same figure is represented the diagnostic diagram involving the \NVl1240 emission line (ionization potential of 77eV). In this case, we could select only one source with a \NVl1240 detection, CEERS-23, which we again marked as a candidate due to the undetection of the \HeIIl1640 line. In the last panel of Fig.~\ref{fig:high-ion}, we show the \NeIV2424 emission line diagnostic, whose ionization potential is 63eV. We detect the \NeIV2424 line in three different sources (CEERS-1029, CEERS-13061, CEERS-4210), and all of them occupy a region of the diagram covered only by the AGN models, even considering the upper limits in \HeIIl1640. In this diagnostic diagram (where the \NeIV and \NeIII lines are $\sim 1000 \AA$ apart one from the other) the dust correction would move the points toward the right, to even more AGN-extreme values.\\
Furthermore, we detect the \NeV3426 emission line (97eV of ionization potential) in one source (CEERS-8299, at $z=2.15$) that we marked as AGN given the extremely high-ionization energy required to produce this line. Even if at high-$z$ it has been proposed that the \NeV3426 could also be associated with SF processes \citep{Cleri23}, it was always related to AGN activity at $z<3$ \cite{Mignoli13,Barchiesi24}. Overall, we selected 52 NLAGNs among the initial 217 sources.

\subsection{Stack of the AGN and non-AGN samples} \label{sec:res_stack}
In each of the diagnostic diagrams reported in Sec.~\ref{sec:results}, we also report the line ratios of the stacked spectra of all the selected NLAGN and of the non-AGN selected sources, derived as presented in Sec.~\ref{sec:spec_stack} and shown in Fig.~\ref{fig:stack_spec}. The two stacked spectra were then fit to measure emission line fluxes using \texttt{Qubspec}, following the same procedure as for all the other single spectra and described in Sec.~\ref{sec:line_fit}. \\
As it is possible to see from the R3N2 diagram, the position of the stacked NLAGN and the non-AGN sources are very close together, with AGN having a slightly larger R3 line ratio. This clearly shows how the BPT diagram and its traditional demarcation lines are no longer useful in effectively separating the AGN population from SFG, probably even for relatively low redshifts ($z>2$). We further discuss this point in Sec.~\ref{sec:sel_meth}. On the contrary, in the R3S2 diagnostic diagram, the line ratios of the stacked NLAGN spectrum clearly lie in the AGN region of the diagnostic, while the non-AGN one is in the SFG-dominated region. The R3S2 line ratios seem to be still informative for the AGN selection also at (relatively) high-$z$, as we discuss in Sec.~\ref{sec:sel_meth}. The same applies to the R3O1 diagnostic diagram. As for the He2N2 diagram, we note that while the NLAGN stacked spectrum shows a clear detection of the \HeII4686 line, which places the final NLAGN stack clearly in the AGN region of the diagnostic, for the non-AGN sample the \HeIIl4686 line is not detected and the line ratios fall in the SFG region, as expected. This is also a strong point in favor of the goodness of our NLAGN selection since the \HeIIl4686 line was detected with high-enough S/N only in 5/52 of the NLAGN, 4 of them at $z<3$. The fact that this line is clearly detected in the NLAGN spectral stack and that the stack line ratios fall in the AGN region of the diagnostic means that signatures of AGN emission are widely present among our NLAGN sample.\\
In both the diagnostic diagrams involving the \OIIIl4363 auroral line, NLAGN are characterized by a larger ($\sim 0.3$ dex) \OIIIl4363/\Hg line ratio compared to the non-AGN, while the two have a similar O3O2 or Ne3O2 line ratios, meaning that the NLAGN selection is mainly driven by a stronger \OIIIl4363 auroral line in AGN, as discussed in \cite{Mazzolari24b}.\\
We did not put the results of the stack in the rest-UV diagnostic diagrams. This is because the \HeIIl1640 and \CIV lines are available only for 33/155 non-AGN and 7/52 NLAGN, and both these lines are not significantly detected in none of the two stacked spectra. This is expected given the faintness of these lines, the possible stronger effect of obscuration at rest-UV wavelengths, and the observing time of the CEERS survey. The same goes for the high-ionization emission line diagnostics: none of the high-ionization emission lines involved in the diagnostics in Fig.~\ref{fig:high-ion} is clearly detected in our stack.

\section{Discussion}\label{sec:discussion}
The IDs and the main physical properties (that will be presented and discussed in the next Sections) of all the 52 NLAGN selected in Sec.~\ref{sec:results} are reported in Table~\ref{table:Sample}.


\subsection{Comparison of AGN selection methods}\label{sec:sel_meth}
\begin{figure}
	\includegraphics[width=1\columnwidth]{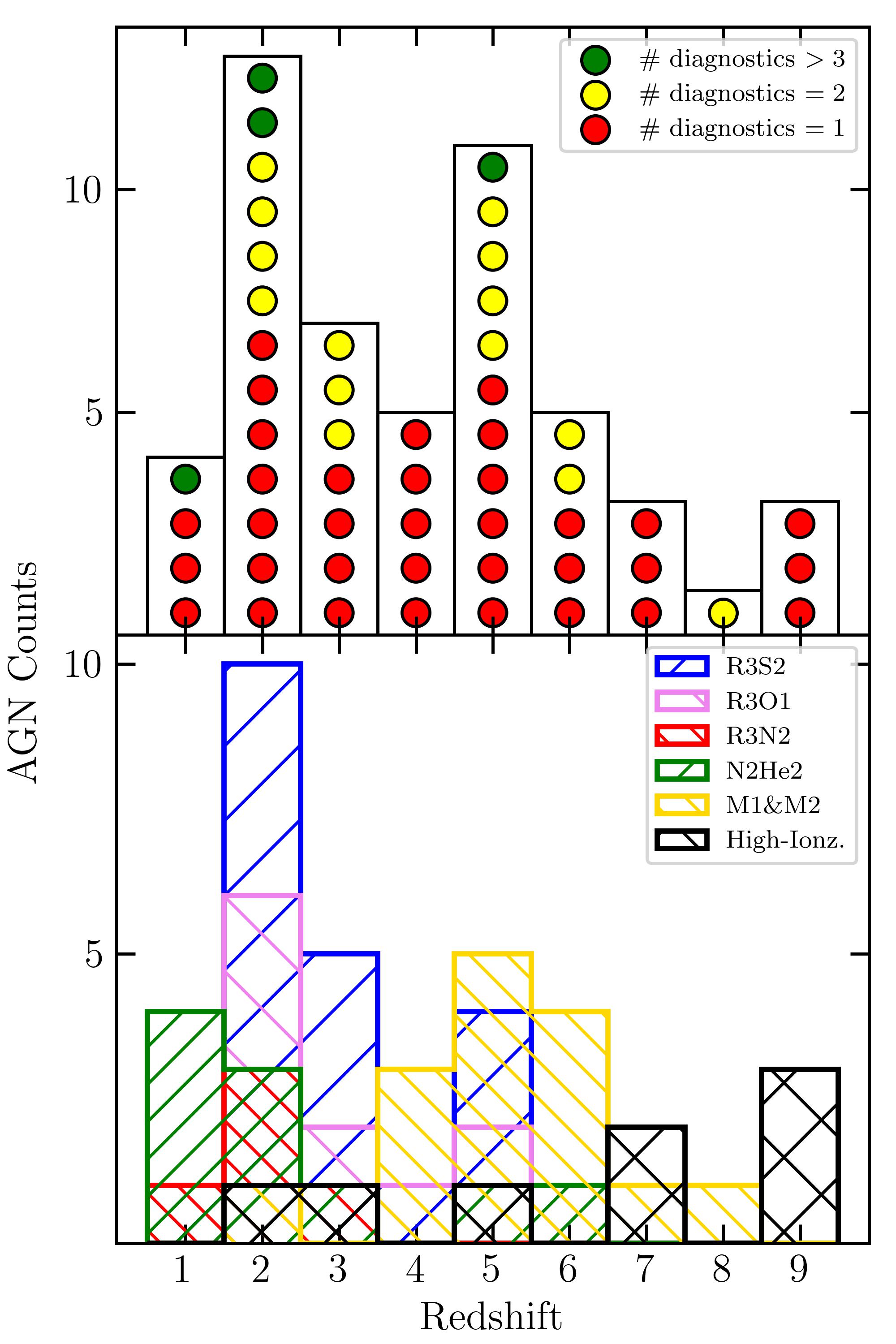}\\
    \caption{Summarizing plots of the NLAGN selection. \textit{Upper:} NLAGN counts in the different redshift bins, color-coded by the number of diagnostics in which each NLAGN has been selected. \textit{Lower:} Distribution of the NLAGN according to the different diagnostic diagrams presented in Sec.~\ref{sec:results}, as labeled.}
    \label{fig:selection}
\end{figure}

In this section, we discuss the effectiveness of the different diagnostic diagrams reported in Sec.~\ref{sec:results}. For the discussion below, the term effectiveness refers to the ability of a diagnostic diagram to efficiently select (i.e., isolate in the diagram) NLAGN among all the considered sources.\\
As presented in the top panel of Fig.~\ref{fig:selection}, most of the NLAGN are selected based on only one single diagnostic diagram (35/52, the 65\%), 13 are selected in two different AGN diagnostics, while 4 are selected based on more than two diagnostics diagrams.  The approach chosen for the NLAGN selection in this work clearly favors the high-completeness instead of the high-purity criterium, selecting as NLAGN all the sources showing the dominance of the AGN photoionization in at least one diagnostic diagram. However, the fact that most of the NLAGN are selected based on one single NLAGN diagnostic is not surprising. The NLAGN diagnostic diagrams used in our work involve emission lines that cover a wide wavelength range (from rest-UV to rest-optical), over which the spectral features of sources at $1\lesssim z\lesssim 9$ can vary a lot depending on many parameters (metallicities, stellar population in the host galaxy, radio or X-ray emission, etc.). Given this, different NLAGN diagnostic diagrams are intrinsically sensitive to different AGN properties and, in turn, can become ineffective (and therefore less useful in separating AGNs from SFGs) under different conditions. Additionally, given the wide redshift range of the sources analyzed in this work, not all of them have in their observed spectra all the lines used for the diagnostic reported in Sec.\ref{sec:opt_diagn},\ref{sec:uv_diagn}, mostly because lines might be redshift out from the observed wavelength range or fall into a spectral gap.\\
Looking at the lower panel of Fig.~\ref{fig:selection}, we see the distribution of the sources selected by the different diagnostic diagrams with respect to their redshift. The diagnostic diagram that selected the larger number of the AGN is the R3S2, in particular at $z<3$, but with a non-negligible number of NLAGN also selected at $z\sim5$. The effectiveness of the R3S2 (but also of the R3O1) diagnostic diagram in the low-metallicity (more frequent at high-$z$) regime was already pointed out in some recent works investigating the effectiveness of traditional NLAGN diagnostic diagrams to select AGN among low-metallicity dwarf galaxies \citep{Polimera22}. In particular, they show that the S2 and the O1 line ratios are less metallicity-sensitive and more successful in identifying AGN in these dwarf environments (more similar to those at high-$z$) than the R3N2 diagnostic. The lower effectiveness of the R3N2 at low metallicities (hence at high redshift) can be a consequence of the nitrogen production channel \citep{Henry00} that can determine nitrogen abundance to scale about quadratically with metallicity (at $Z>0.1~Z_\odot$), hence the [NII]/H$\alpha$ ratio drops strongly. On the contrary, Sulfur and oxygen aboundances, being directly produced by massive stars through $\alpha$-processes, can be less dependent on metallicity \citep{Dopita13}. We also notice that \NII and \SII get redshifted out from the reddest grating of the spectra at $z\sim 6.7$, while the \OI gets redshifted out at $z\sim 7$.\\
We further note that \OI emission can sometimes be excited also by shocks, producing high \OI/\Ha ratios not necessarily driven by a dominant AGN ionization. However, shocks in SFG are generally sub-dominant, as evidenced by the distribution of SDSS galaxies in the BPT and by works selecting this kind of population among SDSS SFG \citep[e.g.][found that shock-dominated SFG are $<1\%$ of the SDSS DR7 SFG]{Alatalo16}. For this reason, it would be unlikely that such a large fraction of SFG at higher redshift would exhibit shock-driven emission.
Shocks can also be an important source of ionizing power in mergers \citep{Medling15} and, of course, in AGN \citep{Best00}. While, after a visual inspection, we did not find strong evidence of ongoing mergers in the majority of the sources selected as NLAGN by the R3O1 diagnostic, we note that 8 out of the 12 NLAGN selected using the R3O1 line ratios are also selected as NLAGN in other diagnostic diagrams, supporting the scenario in which for most of the sources the \OI line should be driven by AGN photoionization. \\
In general, at $z<3$ the most effective AGN diagnostic diagrams are the R3S2, N2He2, and R3N2, while the large majority of the AGN at $z>3$ are selected based on the auroral line diagnostics (both M1 and M2 ), or based on the detection of high-ionization emission lines. This supports the fact that traditional AGN diagnostic diagrams (in particular the R3N2) become less effective at high-$z$, mainly due to metallicity-related effects, while diagnostic based on high-ionization emission lines or on the \OIIIl4363 line proved to be effective in selecting AGN also in the early Universe, as already pointed out by \cite{Maiolino23c}, \cite{ubler23} and \cite{Scholtz23b}.\\
The spectra of the NLAGN selected at $z>6$ (together with the main spectral features that lead to the NLAGN classification) are presented in Appendix~\ref{app:high-z_spec}.

\subsection{Selection of X-ray and radio sources in the diagnostics}
In this section, we want to summarize the results related to the X-rays and radio sources involved in our analysis. Among the seven X-ray sources with a spectrum in the CEERS MR sample, two sources (CEERS-3050 and CEERS-3061) are at $z<0.5$, and their rest-frame optical and UV lines are not available in JWST spectra. These sources were classified as SFG by the X-ray spectral analysis performed in \cite{buchner15}. Among the other five X-ray sources, four are selected as NLAGN in at least one of the diagnostics. In particular, the one at highest redshift (CEERS-2808, $z=3.384$) is selected as NLAGN only in the R3S2 diagnostic diagram and not in the R3N2 \cite[considering the more conservative demarcation line reported in][]{Scholtz23b}. The only three X-ray sources selected as NLAGN in the BPT diagram are CEERS-2919, CEERS-2904, and CEERS-2900, all at $z \lesssim 2$, the first two also showing a broad \Ha emission line (see Appendix \ref{sec:app1}). The only X-ray source not selected in any of the diagnostics is CEERS-2989, at $z=1.433$, whose spectrum shows only \Ha and \NII detections and we could not define the R3 line ratio to place it in the above-mentioned diagnostics. It is worth noting that the lack of \OIIIl5007 and \Hb detections in this source is probably due to high obscuration levels. Considering the analysis in Sec.~\ref{sec:av}, we found a lower limit to the ISM obscuration corresponding to $A_V\sim 6$ mag. The large AGN obscuration of this source is further supported by the CTK obscuration level derived by the X-ray analysis performed in \cite{buchner15}. However, this source shows a faint broad \Ha component, as reported in Fig.~\ref{fig:app_blagn}.\\
The two radio-detected sources are CEERS-2900 (that is also X-ray detected) and CEERS-3129, both selected in the R3N2 diagram as well as in the R3S2 one. They also both show strong \NII emission, in particular CEERS-3129, which could be indicative of a shock \citep{Allen08,Nesvadba17}. Furthermore, CEERS-3129 shows a significant broad \Ha component, but no indication of X-ray emission, as we further explore in Sec.~\ref{sec:xweak}.  

\subsection{Comparison with previous CEERS AGN selections}
\cite{Calabro23} attempted to select AGN among the CEERS sources with a MR spectrum using near-infrared emission line diagnostics. To do so, they restricted the redshift range of the sources to $1<z<3$, therefore considering only 65 sources. Using the R3N2 and R3S2 diagnostic diagrams, the authors selected 8 NLAGN that were classified as NLAGNs also based on at least one NIR diagnostic: CEERS-2919, CEERS-3129, CEERS-2904, CEERS-2754, CEERS-5106, CEERS-12286, CEERS-16406, and CEERS-17496. The first three were marked as BLAGN, as we also found in Appendix~\ref{fig:app_blagn}. Five of the sources selected as NLAGN by \cite{Calabro23} are also selected in our work, while the other three sources were not selected because of the more conservative demarcation lines used in the R3S2 and R3N2 diagnostics or because the \SII line was not considered detected with enough significance by our fit. The authors also selected five NLAGNs using near-infrared diagnostic diagrams that were classified as SFG based on the optical diagnostic diagrams: CEERS-2900, CEERS-8515, CEERS-8588, CEERS-8710, and CEERS-9413. Among these, the only one that is classified as NLAGN also in our selection is CEERS-2900; the other three were selected as SFG also by our optical diagnostics (even if CEERS-8588 shows a tentative \NeV line emission, but we conservatively decided to mark it as a non-detection).\\
In \cite{Davis23} the authors investigate the presence of extreme emission-line galaxies (EELG) at $4<z<9$ using JWST NIRCam photometry in the CEERS program. They used a method to photometrically identify EELGs with \Hb + \OIII or \Ha emission of observed-frame equivalent width $>5000\AA$. Among the photometrically selected EELGs there are 39 sources with a NIRSpec PRISM or MR spectrum, 25 with a match in our MR parent sample. Among these, 6 are selected as NLAGNs in our work, in other words $\sim 25\%$, supporting the non-negligible AGN contamination in the photometric selection of this kind of source, as has already been shown by other works \citep{amorin15}.

\subsection{AGN prevalence}\label{sec:agn_prev}
\begin{figure}
	\includegraphics[width=1\columnwidth]{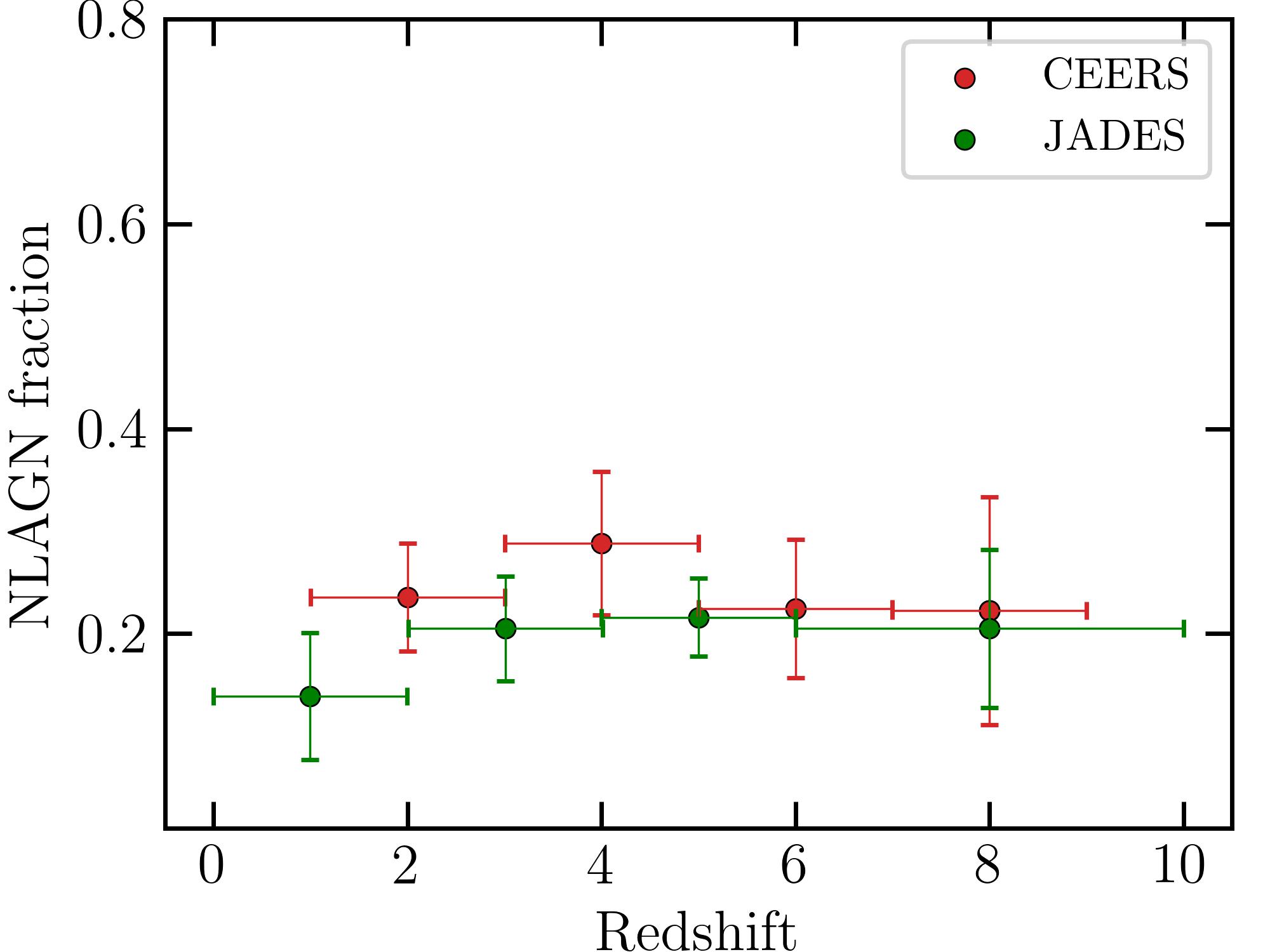}
    \caption{Fraction of the spectroscopically selected NLAGN with respect to the parent sample in this work (CEERS, in red) and in the JADES survey (green) \citep[program ID 1210\&3215,][]{Scholtz23b} in different redshift bins. The errors account for the statistical uncertainties and are computed considering a Poissonian noise.  }
    \label{fig:AGN_prev}
\end{figure}

In Fig.~\ref{fig:AGN_prev}, we show the fraction of the selected NLAGN among the CEERS MR sample. In particular, we consider as parent sample all the galaxies with a MR spectrum and with a secure redshift (217 sources). Then we divided the distribution into four redshift bins of $\Delta z=2$ between $z=0$ and $z=9$, and for each bin, we computed the fraction of AGN compared to the parent sample. The number of sources in the different redshift bins is 122, 85, 59, and 49, going from the lower to the higher redshift bin. We derived an almost constant fraction of AGN among the CEERS MR sources $\sim20\%$. This fraction does not necessarily represent the intrinsic AGN fraction at these redshifts, given the fact that the original selection function of the CEERS spectroscopic survey (necessary to correctly account for completeness corrections) is extremely hard to derive. However, this result agrees well with what was found for the NLAGN selection in the JADES survey by \cite{Scholtz23b}. Given the different sensitivities reached by the CEERS and JADES surveys, one would probably expect a lower fraction of NLAGN selected in the CEERS sample since, in most cases, the AGN selection is based on the detection of faint emission lines. At the same time, as we are going to show in Sec.~\ref{sec:lbol}, the median bolometric luminosity of the NLAGN in the CEERS sample is $\sim 1$ dex higher compared to those selected in JADES, and so the fact that the targets are brighter partially compensate for the shorter exposure times.\\

\subsection{Velocity dispersion}\label{sec:vel}
\begin{figure}
	\includegraphics[width=1\columnwidth]{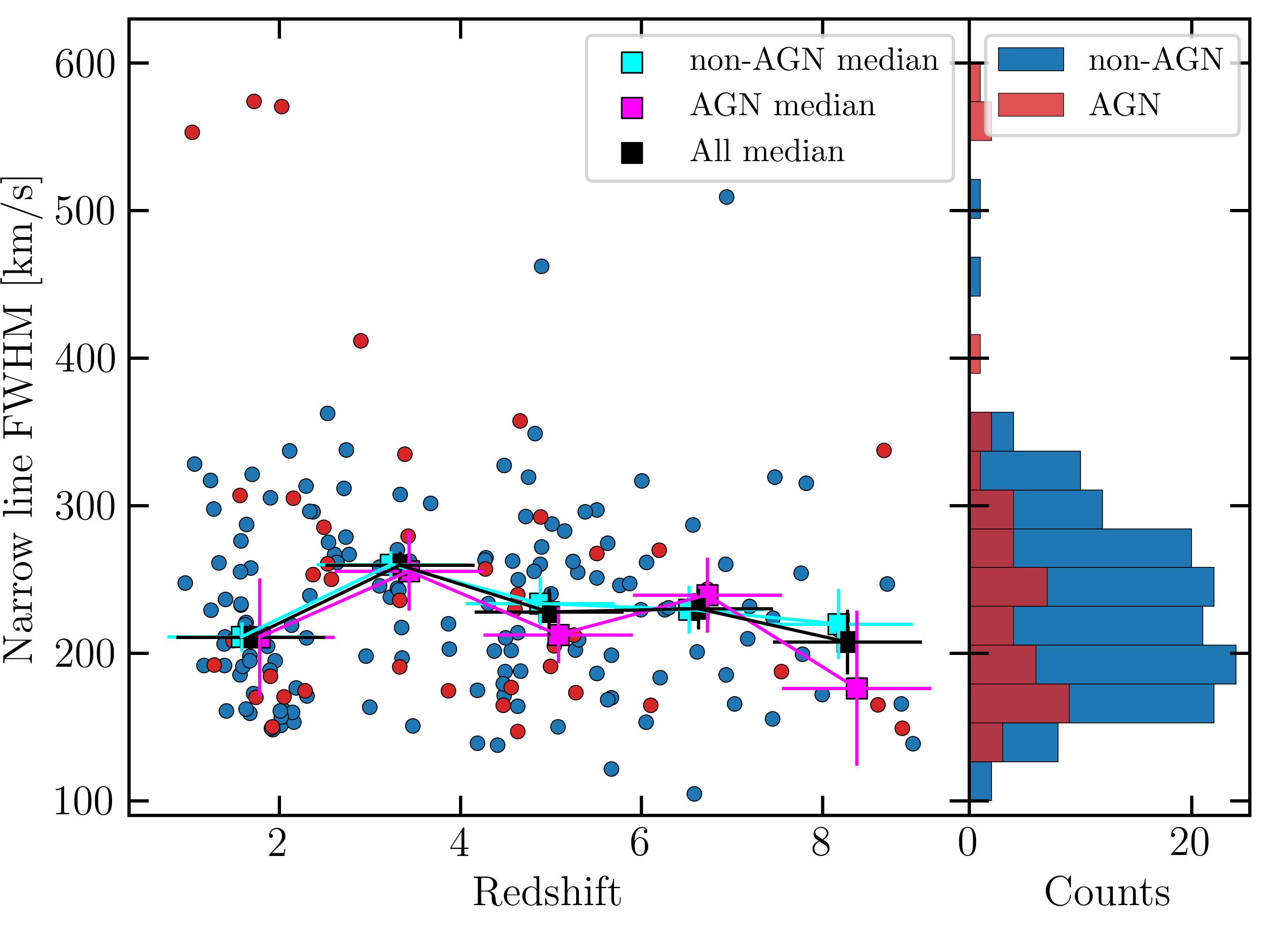}\\
    \caption{ Redshift distribution of the intrinsic \OIIIl5007 line FWHM, resulting from the line-fitting procedure and subtracting the instrumental FWHM. On the left, we also report the median values of the FWHM of NLAGN (magenta), non-AGN (cyan) and of the global population (black) in five different redshift bins. Errors are derived using a bootstrap procedure. There is no significant trend of the FWHM of the sources with redshift. On the right, we report the histogram of the NL-FWHM distribution for AGN (red) and non-AGN (blue).}
    \label{fig:vel}
\end{figure}

In Fig.~\ref{fig:vel}, we show the distribution of the intrinsic narrow line FWHM of the \OIIIl5007 emission line compared to the redshift of the sources. In particular, we considered only sources with the \OIIIl5007 line detected at S/N$>5$. To estimate the intrinsic FWHM, the observed FWHM retrieved by the fit was deconvolved by subtracting in quadrature the instrumental resolution of the grating at the observed wavelength of the \OIIIl5007 line. The instrumental velocity resolution was derived from the [point\_source\_lsf\_f290lp\_g395m\_QD1\_i185\_j85.csv] line spread function file, which was calculated from the instrument model \citep{Ferruit22}, assuming a point-source geometry and a target located in the first MSA quadrant, at the center of shutter (i,j)=(185,85); this procedure is described in \citet{deGraaff24}. The median value of the instrumental resolution is $\sim 160$km/s.\\
The lower and upper boundaries of the observed FWHM distribution are limited by the prior on the width of the line given to \texttt{Qubespec}; that is, $\rm 200km/s<FWHM<700 km/s$. None of the sources, after the subtraction of the instrumental resolution, shows a negative FWHM, but there is a non-negligible number of sources with an FWHM$<200$km/s, meaning that in these cases we are really detecting emission lines at the limits of JWST resolution. Indeed, the cut in S/N and a careful visual inspection confirm beyond any doubt that these lines are real, and, actually, such narrow FWHMs are expected given the trend of decreasing host galaxy stellar mass with redshift and given the fact that with JWST high-resolution spectroscopy were recovered FWHM up to $\sim 100$km/s \citep{Maiolino23c}. On the other hand, there are six sources with an intrinsic FWHM$>400$ km/s. Four out of these six were classified as AGN using the diagnostic diagrams discussed above; in particular, three of these NLAGN are also low-redshift BLAGN reported in Fig.~\ref{fig:app_blagn}. Among the analyzed spectra, we did not find any significant residual from the fit of the narrow \OIIIl5007 line that could be indicative of the presence of an outflow component.\\
In Fig.~\ref{fig:vel}, we also report the median values of the FWHM of the NLAGN population and of the non-AGN in five equally spaced redshift bins.
The distribution of AGN and non-AGN, considering the errors, do not differ significantly in any of the redshift bins. We also perform the Kolmogorov-Smirnov (KS) test on the global marginal distributions in FWHM of the two populations (histogram on the right panel), finding a \textit{p-value}$=0.62$, which does not point toward different parent samples of the two distributions. This indicates that the NLAGN we spectroscopically selected among JWST spectra did not significantly impact their host-galaxy ISM, contrary to what was found in other studies on NLAGN samples, where, however, the NLAGN were selected based on other AGN activity tracers. For example, in the X-ray selected AGN sample of the KASHz survey, investigating sources at Cosmic Noon, \citep{Harrison16} found that $\sim50\%$ of the targets have ionized gas velocities indicative of gas dominated by outflows and/or highly turbulent material, with \OIIIl5007 FWHM$\geq 600$km/s. We further discuss the AGN impact on the sources studied in this work in the next sections.

\subsection{Obscuration}\label{sec:av}
\begin{figure}
	\includegraphics[width=1\columnwidth]{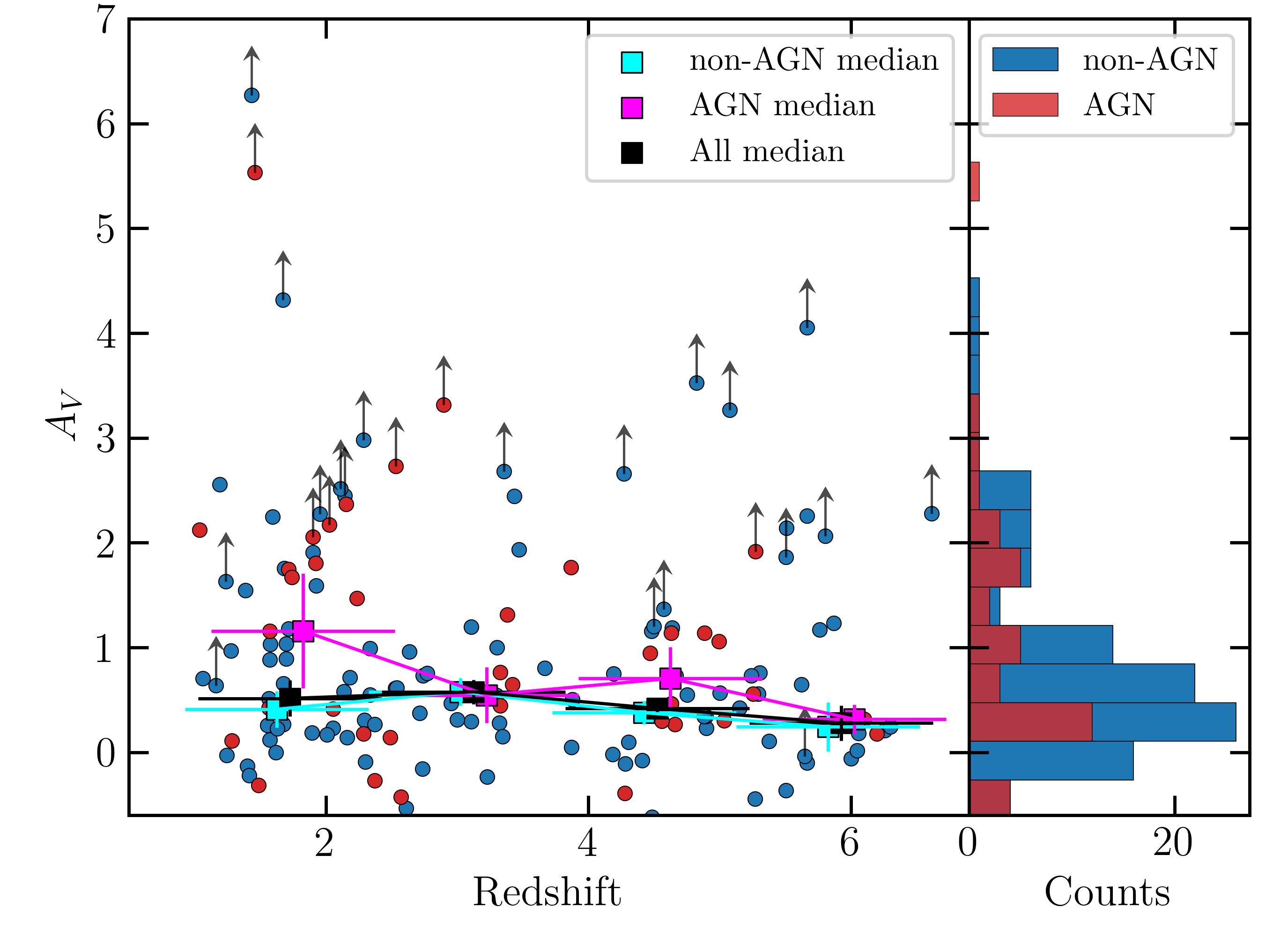}\\
    \caption{Redshift distribution of the values of $A_V$ inferred from the Balmer decrement in sources with detected \Ha and \Hb lines. For sources with upper limits on \Hb we derived lower limits on $A_V$. On the left, we also report the median values of $A_V$ for NLAGN (magenta), non-AGN (cyan) and for the global population (black) in four different redshift bins. Errors are derived using the bootstrap procedure. On the right, we report the histogram of the $A_V$ distribution for AGN (red) and non-AGN (blue).}
        \label{fig:Av}
\end{figure}

In Fig.~\ref{fig:Av}, we show the distribution of the $A_V$ values obtained from the Balmer line decrement. In particular, to derive the $A_V$ we considered the SMC attenuation law \citep{Gordon03}\footnote{We used for the attenuation curve a fit to the average SMC-bar $A_\lambda/A_V$ data points reported in Table 4 of \cite{Gordon03}. We found that the $A_\lambda/A_V$ analytic expression reported in their Eq. 4-5 does not correctly fit their observed data points at $\lambda>3030\AA$ and has $A(\lambda=5500\AA)/A_V\neq1$.} ($R_V=2.74$) for sources at $z>3$ and the \cite{Calzetti00} attenuation law for sources at $z<3$ (with $R_V=3.1$). While the \cite{Calzetti00} attenuation law has been extensively used in the low-$z$ Universe to account for the effects of dust, the choice of the SMC attenuation law is more appropriate for the high-$z$ Universe \citep{Reddy15,Shapley23}, where galaxies are smaller and more compact than in the local Universe \citep{Ono23}. In particular, we select only those sources that have both \Ha and \Hb detected. We assumed CASE B recombination; that is, an intrinsic \Ha/\Hb ratio of 2.86\footnote{$T_e=10^{4}$, $n_e=100\rm cm^{-3}$ \citep{Osterbrock06}}. We also considered sources with upper limits in \Hb\ to derive lower limits in $A_V$.\\
We note that a small number of galaxies in Fig.~\ref{fig:Av} scatter to either surprisingly high values of $A_V$ or else negative values, meaning an \Ha/\Hb line ratio lower than the dust-free minimum value of 2.86. Sources with a dust-free \Ha/\Hb line ratio lower than CASE-B recombination have also been reported recently in the literature \citep{Scarlata24,Yanagisawa24,McClymont24}, but in our case these sources all have \Ha/\Hb line ratios compatible at 1-2$\sigma$ with 2.86. However, there is also the possibility that a small number of these outliers is due to small systematics in the NIRSpec grating-to-grating flux calibration; that is, when \Hb and \Ha are measured in different gratings. A similar result was indeed also observed in \cite{Shapley23} on the same sample analyzed here. Even if present, these calibration issues do not affect our NLAGN selection, because the line ratios of the diagnostics presented above involve lines close to each other, and so generally in the same grating.\\
In Fig.~\ref{fig:Av}, we also plot the median value of the obscuration considering four different redshift bins and taking separately the two populations of NLAGN and sources not identified as AGN. We did not consider for the median values the upper limits in $A_V$.
We did not find a significant evolution of $A_V$ with redshift. We also note that the AGN population is characterized, in the lower redshift bin ($1<z<2.5$), by a $\sim 0.7$mag higher obscuration. At $z<2.5$ the average NLAGN obscuration is 1.15 mag compared to the 0.5 mag of non-AGN. On the contrary, when the whole NLAGN and non-AGN populations are taken into account, the average value of the obscuration across all redshifts is very similar, around $0.3-0.5$ mag.

\subsection{Bolometric luminosities}\label{sec:lbol}
\begin{figure}
	\includegraphics[width=1\columnwidth]{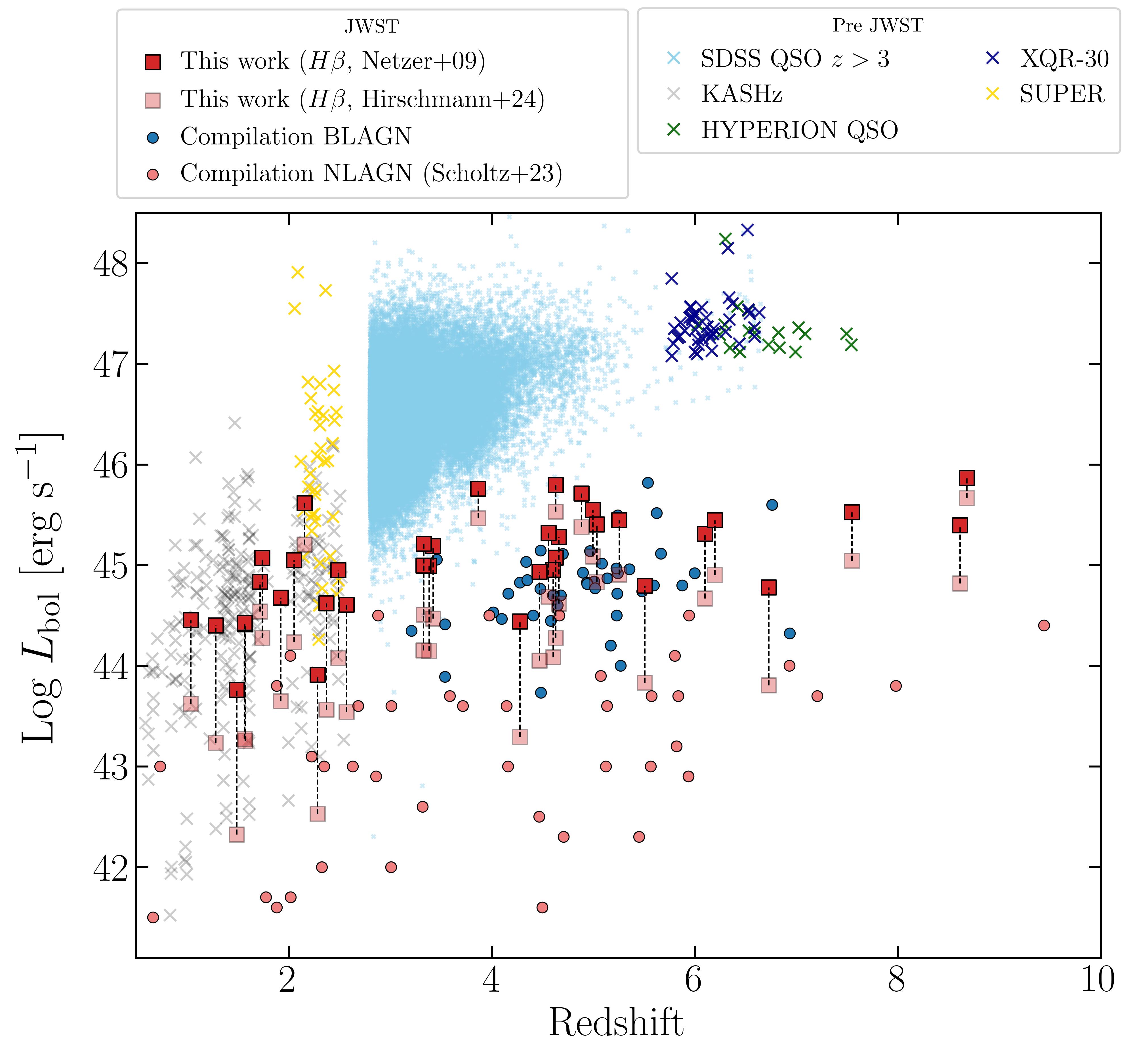}
    \caption{Bolometric luminosities versus redshift of the sample of NLAGN selected in this work (red squares) compared to the bolometric luminosities of other AGN samples selected using JWST spectroscopic observations or pre-JWST, as labeled. The darker red squares represent the baseline bolometric luminosities, derived using \cite{Netzer09} calibration, while the fainter red squares show the values of the bolometric luminosities obtained using the same calibration adopted in \cite{Scholtz23b}. The compilation of JWST selected BLAGN includes the sources taken from \cite{Maiolino23c,Harikane23,Matthee23,ubler23,Kocevski23}. SDSS BLAGN at $z>3$ (light blue crosses) are taken from \cite{Wu22cat}. AGN from the KASHz and SUPER surveys at Cosmic Noon (gray and red gold crosses) are taken from  \cite{Harrison16} and \cite{Kakkad20}, respectively. QSOs samples at Epoch of Reionization are taken from \cite{Zappacosta23} (green crosses, HYPERION sample), and \cite{Mazzucchelli23} (dark-blue crosses, XQR-30).} 
        \label{fig:Lbol}
\end{figure}

In this section, we investigate the bolometric luminosities of the NLAGN selected in this work.
The way to compute the AGN bolometric luminosity is not unique, and it can be done starting from the X-ray luminosity, the luminosity of the BL emission or the UV continuum emission determined by the accretion disk. However, in our case, the majority of the sources are not X-ray detected, they do not have a BL emission, and their continuum is generally undetected. Therefore to estimate the bolometric luminosities, we rely on the dust-corrected narrow line fluxes of the \Hb line, using the calibrations reported in \cite{Netzer09}. These fiducial $\rm L_{bol}$ for the sample of NLAGN (with available \Hb line) are reported with red squares in Fig.~\ref{fig:Lbol}. In the same figure, we also report, with a fainter marker, the value of $\rm L_{bol}$ estimated from the same line fluxes but with the calibration used in \cite{Scholtz23b} and taken from Hirschmann et al. (in prep). In this second case, the bolometric luminosity depends quadratically on the luminosity of the \Hb line, while the \cite{Netzer09} calibration is linear, determining a difference in the estimated $\rm L_{bol}$ that can go up to $\sim 1$ dex (on average the bolometric luminosities computed using Hirschmann et al. calibrations are 0.8 dex lower). To verify the reliability of our fiducial values of $\rm L_{bol}$ we also considered the bolometric luminosity calibrations derived in \cite{Lamastra09} where the NLAGN bolometric luminosities are derived from the dust corrected \OIIIl5007 line. The $\rm L_{bol}$ computed using the calibration in \cite{Netzer09} and those computed using the calibrations in \cite{Lamastra09} agree well, with a median discrepancy of only $\sim 0.25$ dex. \\
However, It is worth noting that all these calibrations assume that the lines used to derive $\rm L_{bol}$ are dominated by the AGN emission. This assumption is not necessarily true for the whole sample of NLAGN, and therefore, our $\rm L_{bol}$ should be generally taken as upper limits. For BLAGN at $z\leq 2$ presented in Fig.~\ref{fig:app_blagn}, we computed the bolometric luminosities considering the same relation used in \cite{Harikane23} to determine the bolometric luminosities of the BLAGN at $z>4.5$ of this sample, i.e the bolometric luminosity calibration derived in \cite{Greene05} from the (dust corrected) broad \Ha emission.\\
In Fig.~\ref{fig:Lbol}, we compare the distribution of $\rm L_{bol}$ of our sample with the AGN bolometric luminosities of other samples at comparable redshifts and taken from the literature. In particular, we report AGN bolometric luminosities derived both from JWST spectroscopic studies and from pre-JWST surveys. Given the bolometric luminosities of $z>3$ AGN detected before the advent of JWST, it is clear that we are now able to sample a completely new regime in the luminosity-redshift space. Comparing the bolometric luminosities of our sample with those of the NLAGN selected sample from \cite{Scholtz23b}, we note that our targets are on average $\sim 1.5$ dex more luminous considering our fiducial calibration, but still more luminous even considering the luminosities derived with the calibration of Hirshman et al. (in prep). On the opposite, the AGN luminosities of our sample are comparable, in the same redshift range, with the distribution of $\rm L_{bol}$ derived from multiple samples of BLAGN detected with JWST.

\subsection{Host galaxies' properties}\label{sec:host_prop}
\begin{figure}
	\includegraphics[width=1\columnwidth]{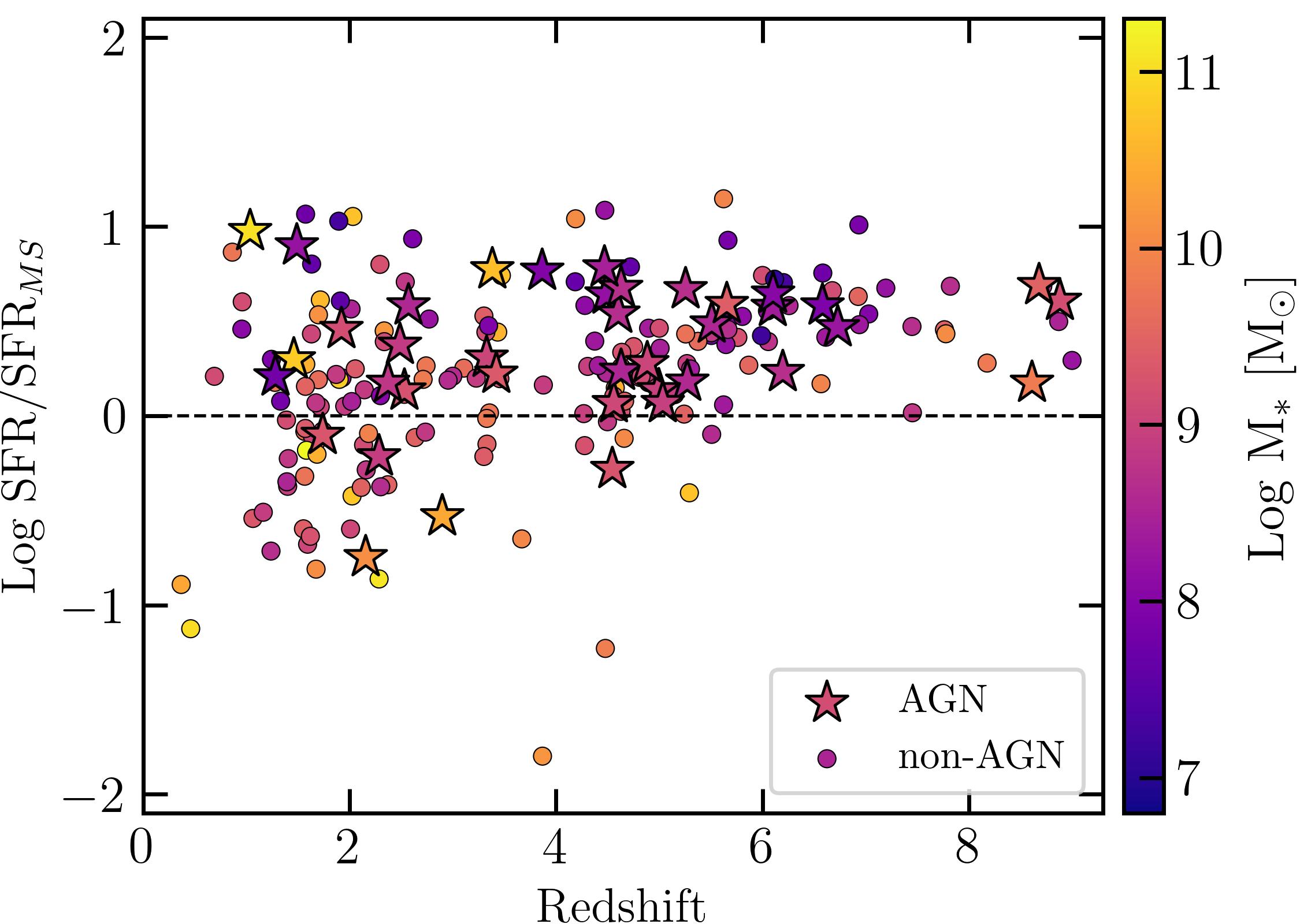}
    \caption{Redshift distribution of the ratio between the SFR derived from the SED-fitting and the SFR computed from the MS relation derived by \cite{Popesso23} at the redshift and $M_*$ of each source. We plot AGN and non-AGN sources with stars and circles, respectively. The sources are color-coded based on the stellar mass, as derived from the SED-fitting.}
    \label{fig:SED}
\end{figure}

In Fig.~\ref{fig:SED}, we show the results from the SED-fitting performed on the whole sample of 217 sources that is described in Sec.~\ref{sec:sedfit}. In particular, we compare the trend with redshift of the ratio between the SFR derived from the \texttt{CIGALE} SED fitting and the one obtained from the main sequence (MS) reported in \cite{Popesso23} at the stellar mass  ($M_{*}$, obtained from the fit) and redshift of each source. We note that sources at $z<4$ are generally distributed almost symmetrically with respect to the MS, with the median value of $\log \rm SFR/SFR_{MS}\sim0.1$. On the contrary, at $z>4$, sources appear to be systematically above the corresponding MS, with a median value of $\log \rm SFR/SFR_{MS}\sim0.4$, in other words with SFRs 2-3 times larger than the respective MS SFR. Sources with larger SFRs are also the sources with the lower $M_{*}$, some of them reaching $\log M_{*}\lesssim 8$. This means that the high-$z$ galaxies analyzed in this sample seem to have, on average, a higher MS normalization, being less massive and more star-forming than expected from the MS. Different works, taking advantage of JWST spectroscopy, have already shown that high-$z$ galaxies are frequently in a state of intense or bursty SF \citep{Dressler23,Looser23,Endsley23}, probably driven by a higher SF efficiency related to the lower metallicities of high-$z$ galaxies. However, this apparent above-MS behavior at high-z could also be due to a selection effect. Indeed, given that we are investigating a spectroscopic sample, we are probably more biased toward high SFR, in particular at high-z. \\
For sources not selected as AGN, we also compared the SFR inferred from \texttt{CIGALE} to the values obtained from the \Ha line \citep[using the measured dust-corrected \Ha luminosity and the relations reported in][]{Shapley23}. On average, we find a good agreement, with $\rm \log SFR_{H\alpha } =0.7 \log SFR_{SED} -0.1$ with 0.25dex of scatter.\\
The AGN distribution in Fig.~\ref{fig:SED} is in agreement with the distribution of the other sources not selected as AGN: considering the median distribution of the ratios of the two SFRs, even in different redshift bin, AGN and SFG are indistinguishable, as already found by other works \citep{Ramasawmy19}. The AGN quenching effect on the host galaxy SF is still a largely debated topic \citep{Bugiani24,Man18,Beckmann17, Scholtz18}, and in this case, we do not note a negative impact of the AGN feedback on SF in the host galaxy. 
This finding is also in line with recent studies based on various cosmological simulations, according to which star formation quenching is not primarily driven by the instantaneous AGN activity but the integrated black hole accretion \citep[as traced by black hole mass ][]{Piotrowska22,Bluck23, Scholtz24}. The AGN contribution to the final SED is measured in \texttt{CIGALE} by the parameter $f_{AGN}$, which corresponds to the fraction of the AGN luminosity with respect to the galaxy one in the rest-frame 0.1-2 $\mu$m. As expected, given the NLAGN nature of our sources, for most of the NLAGN, the fit returned a $f_{AGN}<0.5$ according to the scenario, already discussed in Sec.~\ref{sec:vel} and Sec.~\ref{sec:av}, that the AGN activity of the selected NLAGN is significantly buried by the host galaxy, in particular in the rest-frame optical part of the SED traced by the available photometry. This scenario, and the absence of a significant impact of the AGN activity over the host galaxy properties, is even more supported by the fact that we did not find relevant signatures of outflows in our spectroscopic analysis, in particular in the high-$z$ galaxies.\\
From the NLAGN SED-fitting, we also derived the AGN bolometric luminosities (sum of the disk emission plus the one reprocessed by the tours). However, due to the limited photometric range (the reddest filter is F770W for sources with JWST photometry and IRAC4 for sources with 3D-HST photometry), in most of the cases, we were not able to explore the rest-frame MIR NLAGN emission and the AGN bolometric luminosity is usually largely unconstrained. By selecting only those sources with a reliable bolometric luminosity from the SED-fitting, and comparing these values with the $\rm L_{bol}$ estimated in Sec.~\ref{sec:lbol}, we found that the median distributions agree well: the median value of the bolometric luminosity from the \Hb line is $\sim$0.23dex larger than the one derived from the SED-fitting, but the scatter goes up to $\sim1$ dex.


\subsection{X-ray weakness}\label{sec:xweak}

\begin{figure}
	\includegraphics[width=1\columnwidth]      {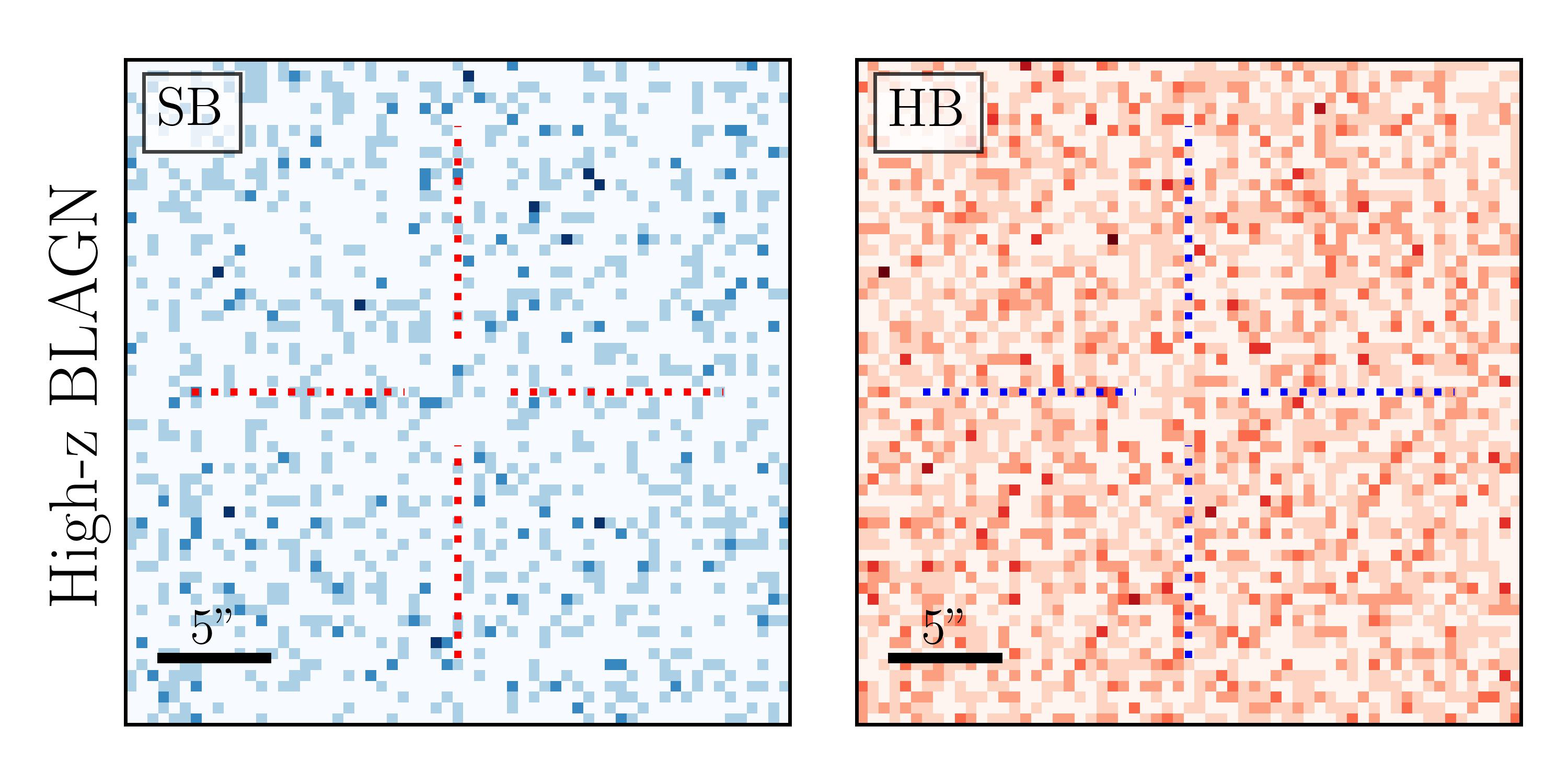}\\
        \includegraphics[width=1\columnwidth]{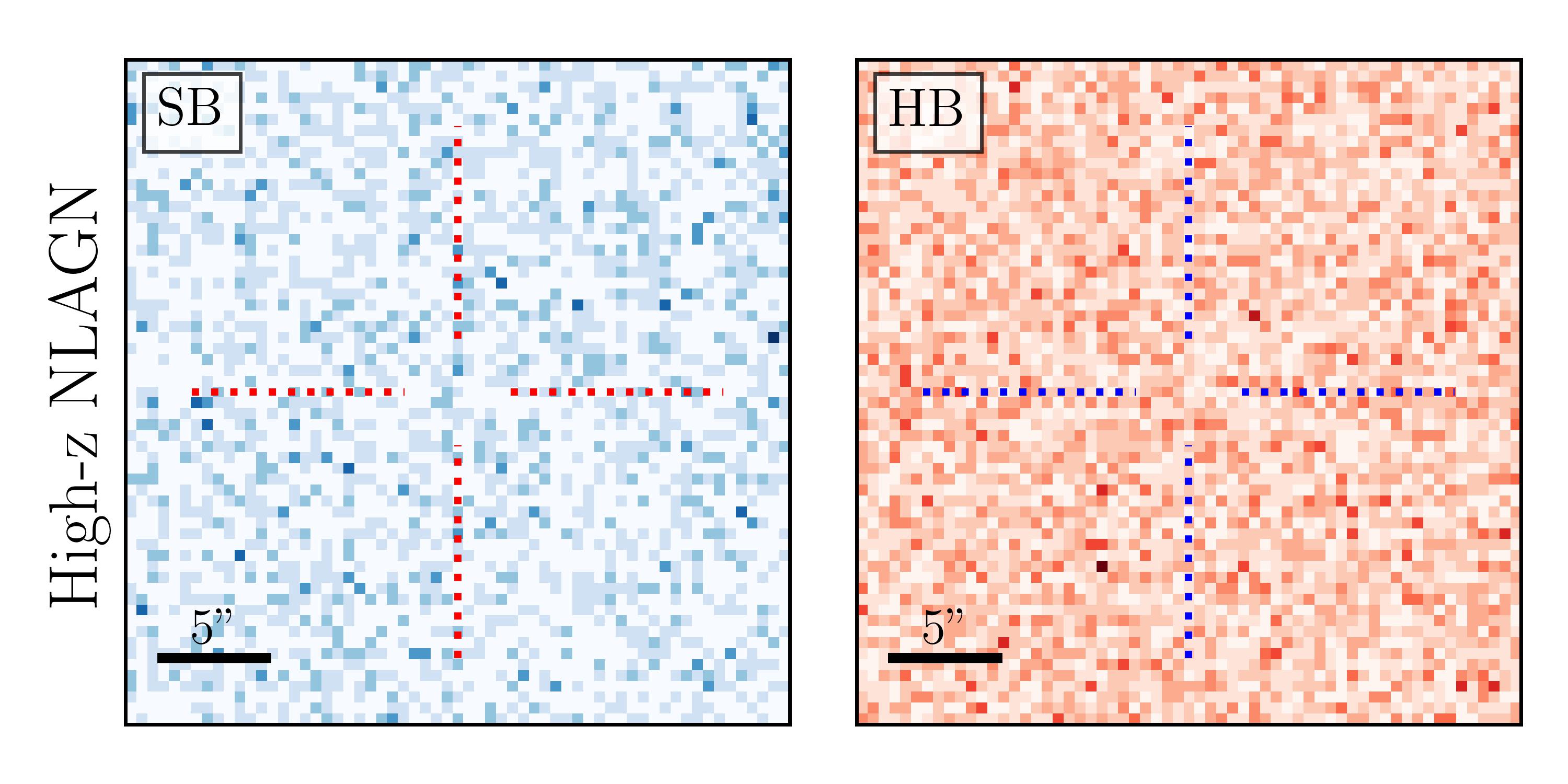}\\
        \includegraphics[width=1\columnwidth]{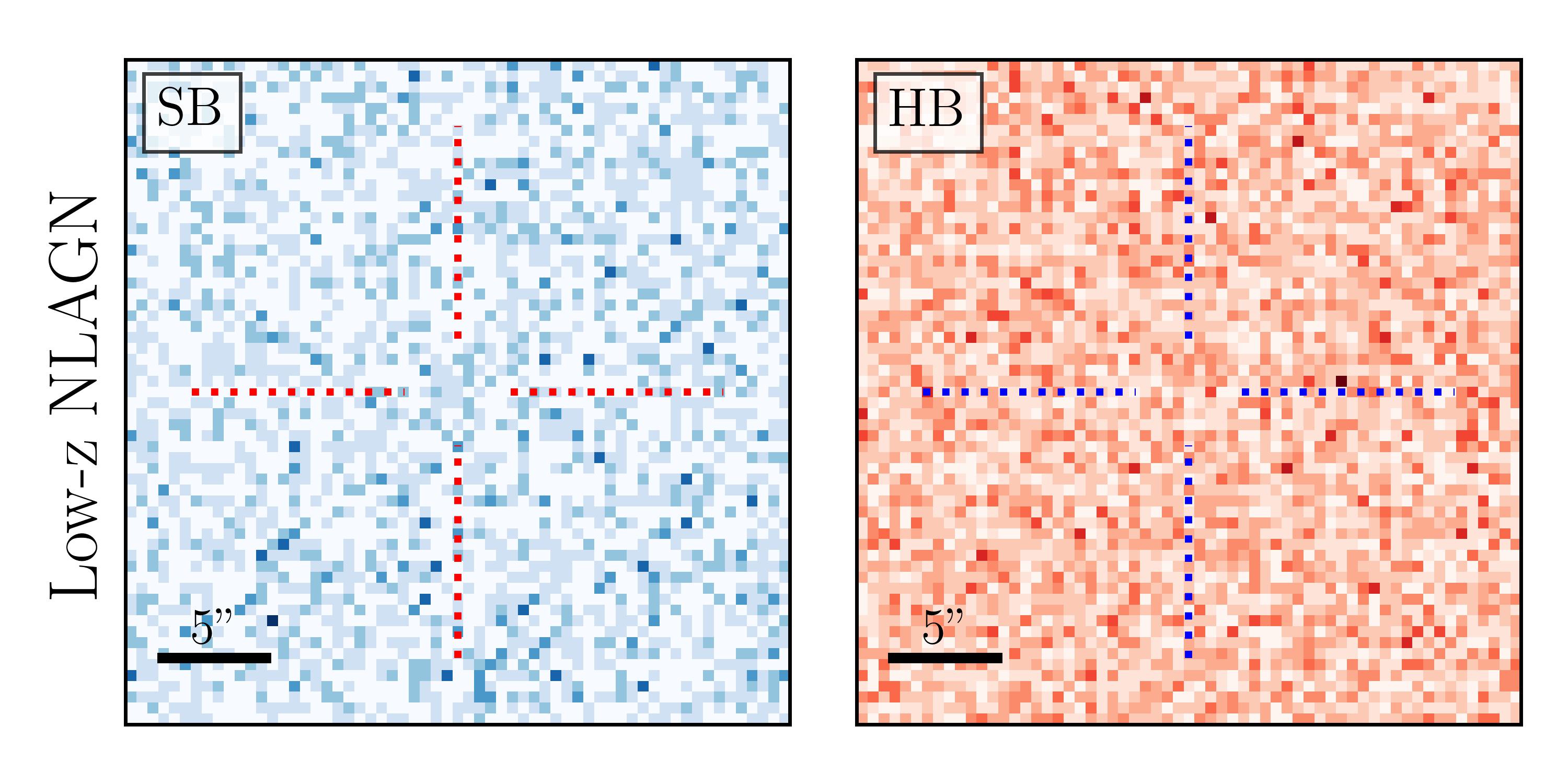}\\
    \caption{30"$\times$30" cutouts of the X-ray stacking image in the observed SB (0.5-2keV) and HB (2-8keV) of the three samples taken into account. From top to bottom: BLAGNs, NLAGNs at $z>4.5$, and NLAGNs at $z<4.5$. None of the samples show a detection in either band.}
    \label{fig:AGN_Xstack}
\end{figure}

\begin{figure*}
        \includegraphics[width=2\columnwidth]{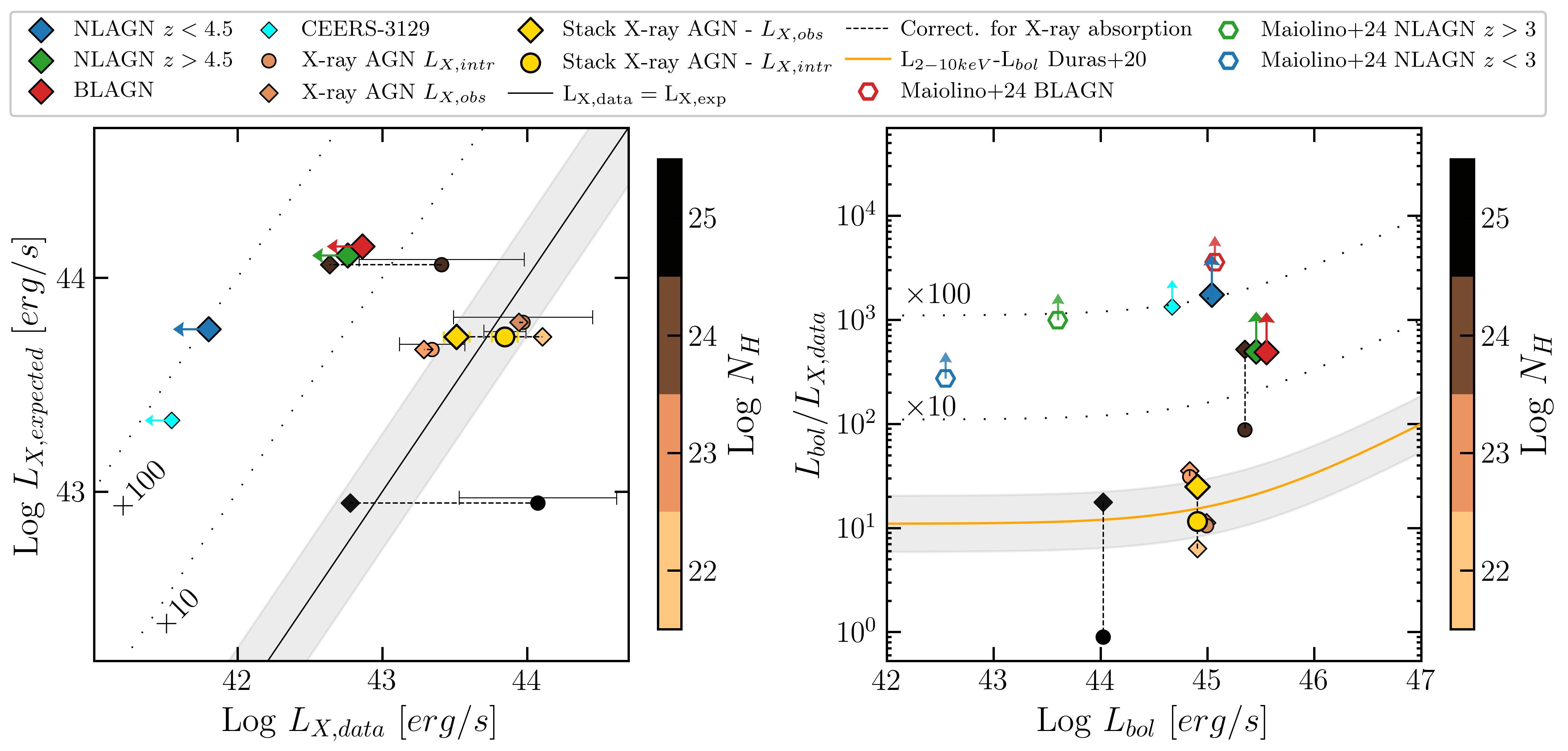}
    \caption{Results coming from the X-ray stacking, showing a significant X-ray weakness of all three samples of NLAGNs at $z<4.5$, NLAGNs at $z>4.5$, and high-$z$ BLAGNs. \textit{Left:} Expected X-ray luminosity versus observed X-ray luminosity (in the 2-10keV band) of the three AGN samples we used for the stacking, as labeled. We also plot the sources in the CEERS MR sample with an X-ray detection color-coded by their obscuration level, as derived from the X-ray spectral analysis \citep{buchner15}. Diamonds refer to observed X-ray luminosities, while circles refer to the intrinsic X-ray luminosities derived from the X-ray analysis. Errors bars mark the error on the intrinsic X-ray luminosities of the sources reported in \cite{buchner15}. The result from the X-ray stacking of these X-ray detected sources is reported as a gold diamond, while the gold circle considers the median of the intrinsic X-ray luminosities of the X-ray sources. The dashed lines correspond to the shift from the observed to intrinsic X-ray luminosities. In cyan, we report the $z=1.037$ BLAGN CEERS-3129, presented in Appendix~\ref{sec:app1}. \textit{Right:} Ratio between the mean bolometric luminosity of the three AGN samples (derived using the values computed in Sec.~\ref{sec:lbol}) and the observed X-ray luminosity upper limits versus the mean bolometric luminosities. The orange line shows the reference $L_{2-10keV}-L_{bol}$ relation derived by \cite{duras20} and used to derive the expected X-ray luminosity of the sources. The empty hexagons represent the lower limits derived from the X-ray stack of the JWST-selected BLAGNs (red) and NLAGNs (green for $z>3$ and blue for $z<3$) on the CDFS and CDFN \citep{Maiolino24_X}.}
    \label{fig:Xstack_lum}
\end{figure*}

As is described in Sec.~\ref{sec:xray_rad_stack}, we performed a detailed analysis of the X-ray and radio counterparts of the selected NLAGN. Four out of five of the X-ray sources classified as AGNs by the X-ray analysis in \cite{buchner15} are also classified as AGNs by our NLAGN diagnostics. All the other 48 NLAGN do not show any X-ray counterpart. Therefore, we decided to perform an X-ray stacking analysis on our sample, using CSTACK as is described in Sec~\ref{sec:xray_rad_stack}. In particular, we decided to divide the sample of NLAGN in two redshift bins, equally populated, using $z=4.5$ as the dividing threshold. We also include in our analysis the BLAGN selected at $z>4.5$ in CEERS by \cite{Harikane23}, since none of them show any indication of X-ray emission and to perform a more complete analysis on the AGN sample among the CEERS MR spectra. We did not stack the sample of low-$z$ BLAGNs presented in Appendix~\ref{sec:app1} because 3 out of 4 of these sources are X-ray detected (except CEERS-3129, which will be discussed later). The median redshifts of the three AGN samples are 2.37 for the low-redshift NLAGNs, 5.50 for the high-$z$ NLAGNs, and 5.43 for the high-$z$ BLAGN sample. In Fig.~\ref{fig:AGN_Xstack}, we show the stacking maps in the SB and HB for the two NLAGN redshift bins and for the BLAGNs, respectively. Given the median redshift of the samples, and considering the SB as the reference observing band since it provides the deepest constraint, we are investigating, on average, the following rest-frame X-ray energy bands: $\sim$1.7-6.5 keV for the low-$z$ NLAGN sample and $\sim3.5-13$keV for the other two samples. The stacking analysis did not reveal any X-ray detection in any of the sub-samples. This is quite surprising given the rest-frame energy range tested and the depth of the X-ray observations. To check whether the non-detections in the stacking of the NLAGN were due to possible galaxy contamination in the NLAGN sample, we also performed the same stacking but considering only NLAGN selected in more than two diagnostics diagrams, but again, the X-ray analysis did not reveal any detection. Therefore, we wanted to test if such X-ray non-detections would be compatible or not with the expected X-ray luminosities of these objects.\\
We used the $\rm L_{bol}-L_{2-10keV}$ luminosity relations derived by \cite{duras20} to compute the expected X-ray luminosities given the baseline dust-corrected bolometric luminosities computed in Sec.~\ref{sec:lbol} for our sample of NLAGN. For the BLAGN, instead, we used the bolometric luminosities computed in \cite{Harikane23} using the \Ha broad emission. The comparison between the expected X-ray luminosities and the upper limits derived from the stacking is shown in Fig.~\ref{fig:Xstack_lum}. For all the AGNs, the upper limits of their observed X-ray luminosities are largely below the expected X-ray luminosities, posing severe constraints on their X-ray production and emission. The X-ray deficit is $\sim 1.3$ dex for BLAGNs and for the sample of NLAGNs at $z>4.5$, and is $\sim 2$ dex for the NLAGNs at $z<4.5$. The X-ray deficit of this last sample is particularly surprising, given the fact that most of these low-$z$ NLAGNs were selected using the traditional BPTs diagnostic diagrams, which are not expected to fail the NLAGN selection at low-$z$. \\
We further check that even taking the bolometric luminosities derived using the calibration of Hirschmann et al. in prep. (shown as faint red squares in Fig.~\ref{fig:Lbol}) the results remain almost unchanged, with the median expected X-ray luminosities of the two NLAGN samples decreasing only of $\sim 0.2$ dex.\\
In Fig.~\ref{fig:Xstack_lum}, we also show the distribution of the X-ray detected sources in our MR spectroscopic sample together with the position of their X-ray stack.
In particular, for each X-ray source, we plot with a diamond the observed X-ray luminosity derived from their X-ray flux assuming $\Gamma=1.9$, namely applying the same procedure as was done for the stack of the other sources. As expected, X-ray sources whose spectral analysis revealed larger column densities are on the left of the 1:1 relation, with a deviation up to $\sim 1.5$ dex for CEER-2989, whose spectral analysis revealed a column density of $\log (N_{H}/ \rm cm^{-2})\sim 25$. This is because the spectral index, $\Gamma=1.9$, does not properly account for the harder spectrum of these heavily obscured sources. For each X-ray source, we also plot with circles the intrinsic, absorption corrected, X-ray luminosity derived in \cite{buchner15}, which generally reduces the deviation from the 1:1 relation, especially if we take into account the large uncertainties on the intrinsic X-ray luminosities of the most obscured X-ray AGN. The same applies to the position of the stack of these X-ray sources: when we considered the observed X-ray luminosity (derived from the X-ray stacking) the point slightly deviates from the median expected X-ray luminosity, but if we correct for the absorption the X-ray luminosities (considering the median intrinsic X-ray luminosity on the x axis), then the golden circle perfectly aligns with the median expected X-ray luminosity of these sources.\\ 
The same conclusions about the different samples can also be drawn by looking at the right panel in Fig.~\ref{fig:Xstack_lum} where we compare the position of the sources with respect to the $\rm L_{bol}-L_{2-10keV}$ relation of \cite{duras20}. In the plot on the right we also show the lower limits derived in \cite{Maiolino24_X} from the X-ray stacking analysis performed on JWST selected BLAGN and NLAGN on the Chandra Deep Field South \citep[CDFS, with 7Ms of \textit{Chandra} observations][]{liu17} and Chandra Deep Field North \citep[CDFN, 2Ms ][]{xue16}. \\
In both figures, we also show in cyan the position of the peculiar source CEERS-3129. This is a BLAGN at $z=1.037$ (whose broad \Ha component is shown in Appendix\ref{sec:app1}), it is not detected in the X-ray image and shows an X-ray deficit of $\sim 2$ dex. This source is also characterized by a consistent dust obscuration: from the Balmer decrement derived in Sec.~\ref{sec:av} we found $A_V=2.12$. 
Since the broad \Hb component is not detected and its upper limit would return an extremely large $A_V$, we assumed for the dust correction of the broad \Ha emission (and therefore for the bolometric luminosity) the value of $A_V$ derived from the NLR. However, this implies that the bolometric luminosity of CEERS-3129 might be underestimated and, consequently, also its expected X-ray luminosity and the amount of its X-ray weakness.
Considering the dust-corrected bolometric luminosity, the lower limit to the X-ray weakness of CEERS-3129 is $\sim$2 dex.\\s
Assuming that these sources are really AGNs, the two possible scenarios that can justify such a lack of X-ray emission are:
\begin{itemize}
    \item Obscuration coming from large column densities, larger than the CTK limit ($\log (N_{H}/ \rm cm^{-2})>24$), and/or
    \item An intrinsic X-ray weakness due to a possible different accretion disk/corona structure. This can determine, for example, a larger ratio between optical and X-ray fluxes due to a lower efficiency of the corona in producing X-ray photons.
\end{itemize}
\cite{Yue24}, analyzing a sample of so-called “little red dots” (LRDs) showing clear AGN signatures (as the detection of a broad \Ha component) and selected from various fields covered by JWST imaging and spectroscopy, found a similar result. The amount of X-ray weakness in the LRD is comparable to what we found in this work for the sample of BLAGN at high-$z$, while they did not investigate the population of NLAGN. In \cite{Yue24}, the authors suggest an intrinsic X-ray weakness as a possible origin of this discrepancy or a wrong AGN classification of these sources; in the latter case, the broad \Ha component might be due to fast galactic outflows (with velocities $\sim 1000$km/s). In our case, however, we did not detect any signature of outflows in any of the forbidden lines of the BLAGN. More evidence and strong arguments against the outflow scenario have been presented by \cite{Juodzbalis24_rosetta}.\\
Also \cite{Maiolino24_X}, considering a large sample of BLAGNs and NLAGNs spectroscopically selected in the JADES survey over the CDFS and CDFN, show that almost all the AGNs discovered by JWST lack any X-ray detection, even in the stack. In this work, the authors investigated different possible origins of this X-ray weakness, focusing in particular on the three possibilities reported above and finding indications in support of one or the other. In particular, they found an enhanced EW(\Ha$_{broad}$) in these sources, which could suggest that the BLR is distributed with a large covering factor around the central source, acting as a dense dust-free absorbing medium that can potentially reach CTK column densities. On the other hand, the authors also find that many of the AGNs newly discovered by JWST have features in common with the population of narrow-line Seyfert 1 (NLS1), which are known to have a steep X-ray spectrum (with $\Gamma>2-3$, and therefore more shifted toward a softer X-ray emission) and/or are characterized by high accretion rates, which are expected to result in a weaker X-ray emission, especially at high redshift, where higher rest frame energies are probed in the \textit{Chandra} bands.

\subsection{Radio stacking to discriminate the nature of X-ray weakness}\label{sec:rad_res}

\begin{figure}
\centering
	\includegraphics[width=0.7\columnwidth]{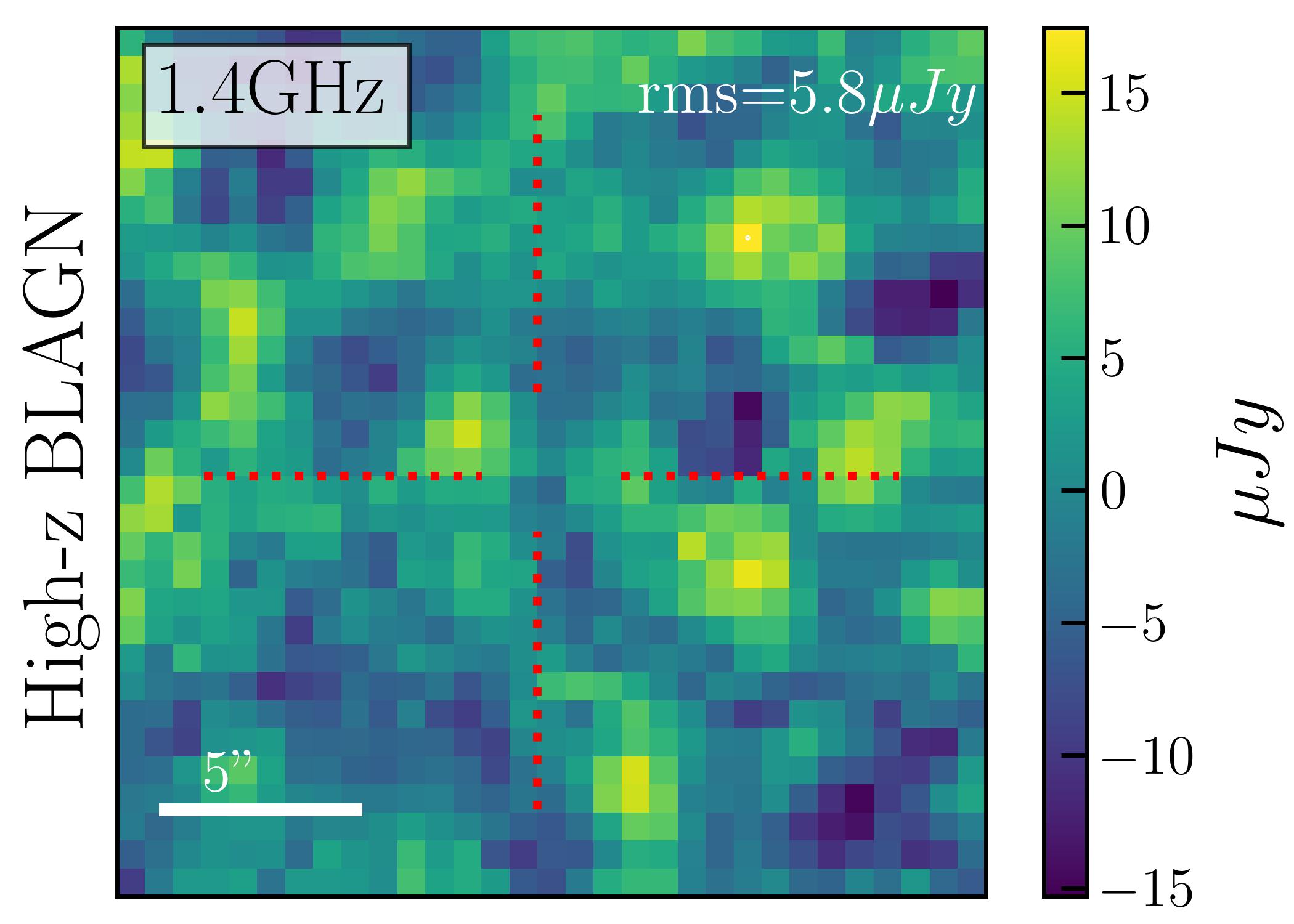}\\
        \includegraphics[width=0.7\columnwidth]{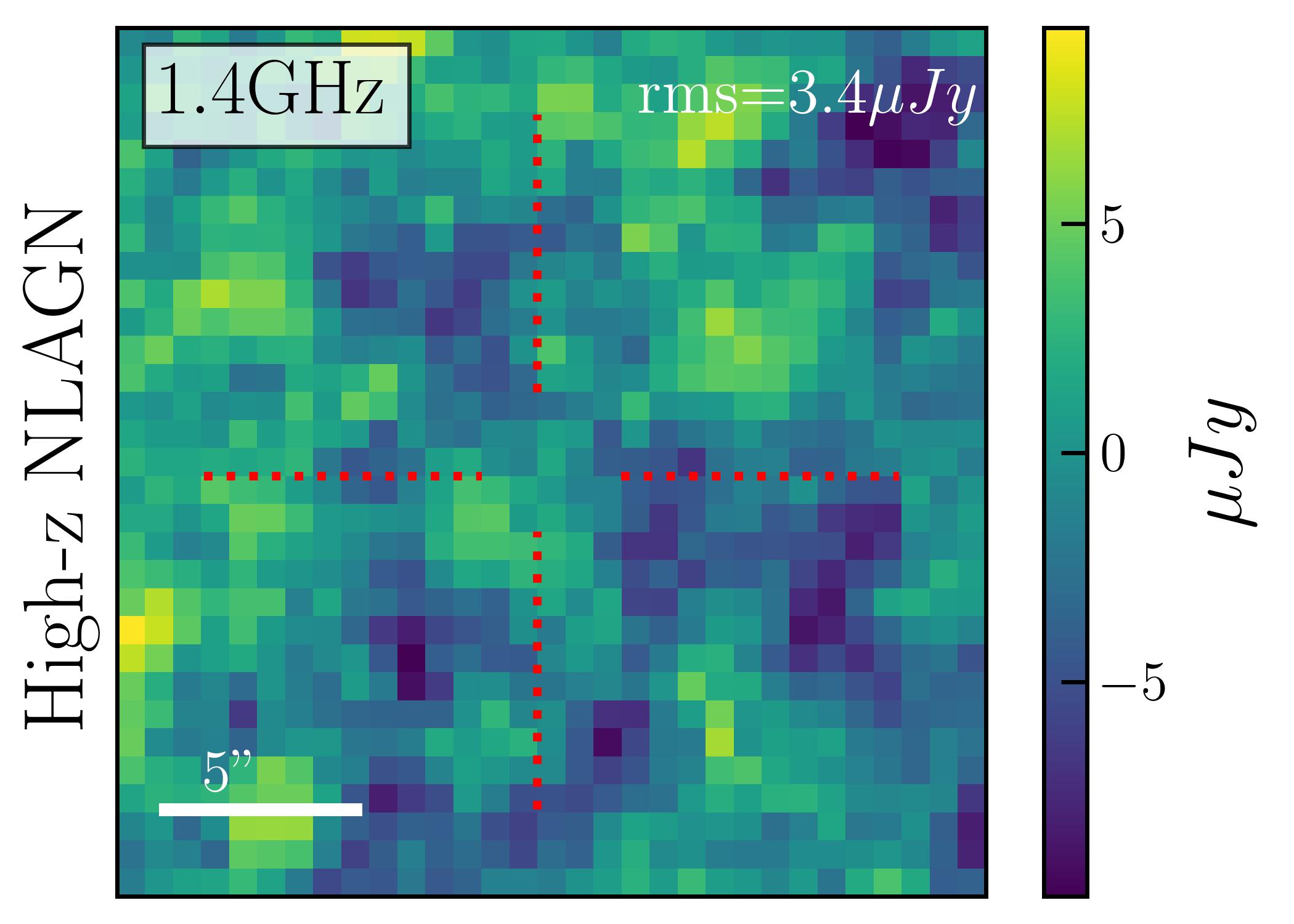}\\
        \includegraphics[width=0.7\columnwidth]{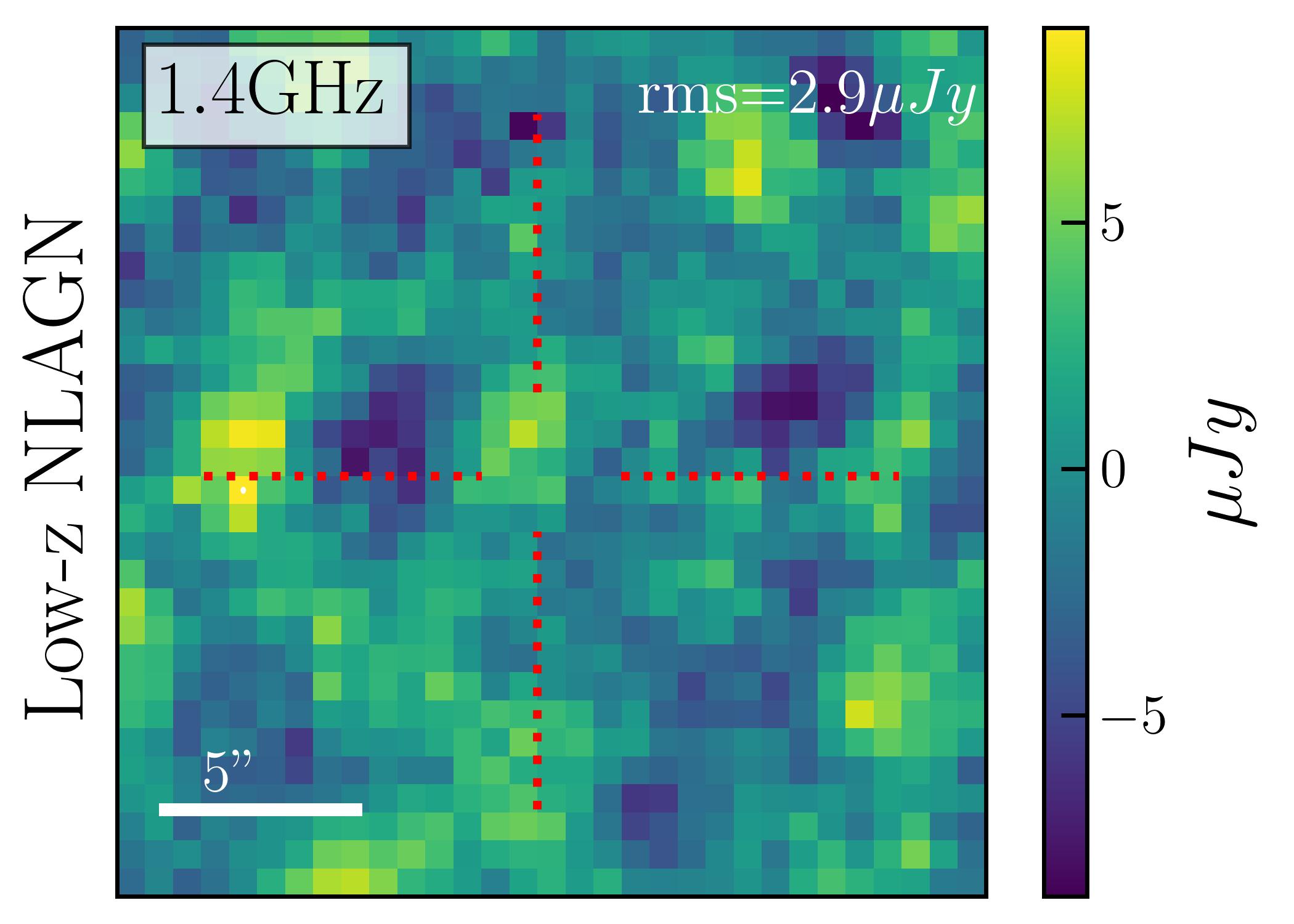}\\
    \caption{Median radio stacking of the three AGN samples taken into account. From top to bottom, we report 25"$\times$25" cutouts of BLAGNs, NLAGNs at $z>4.5$, and NLAGNs at $z<4.5$. None of the samples show radio detection. For each stack, we report the noise (rms) value in the top right corner.}
    \label{fig:AGN_radstack}
\end{figure}

\begin{figure}
	\includegraphics[width=1\columnwidth]{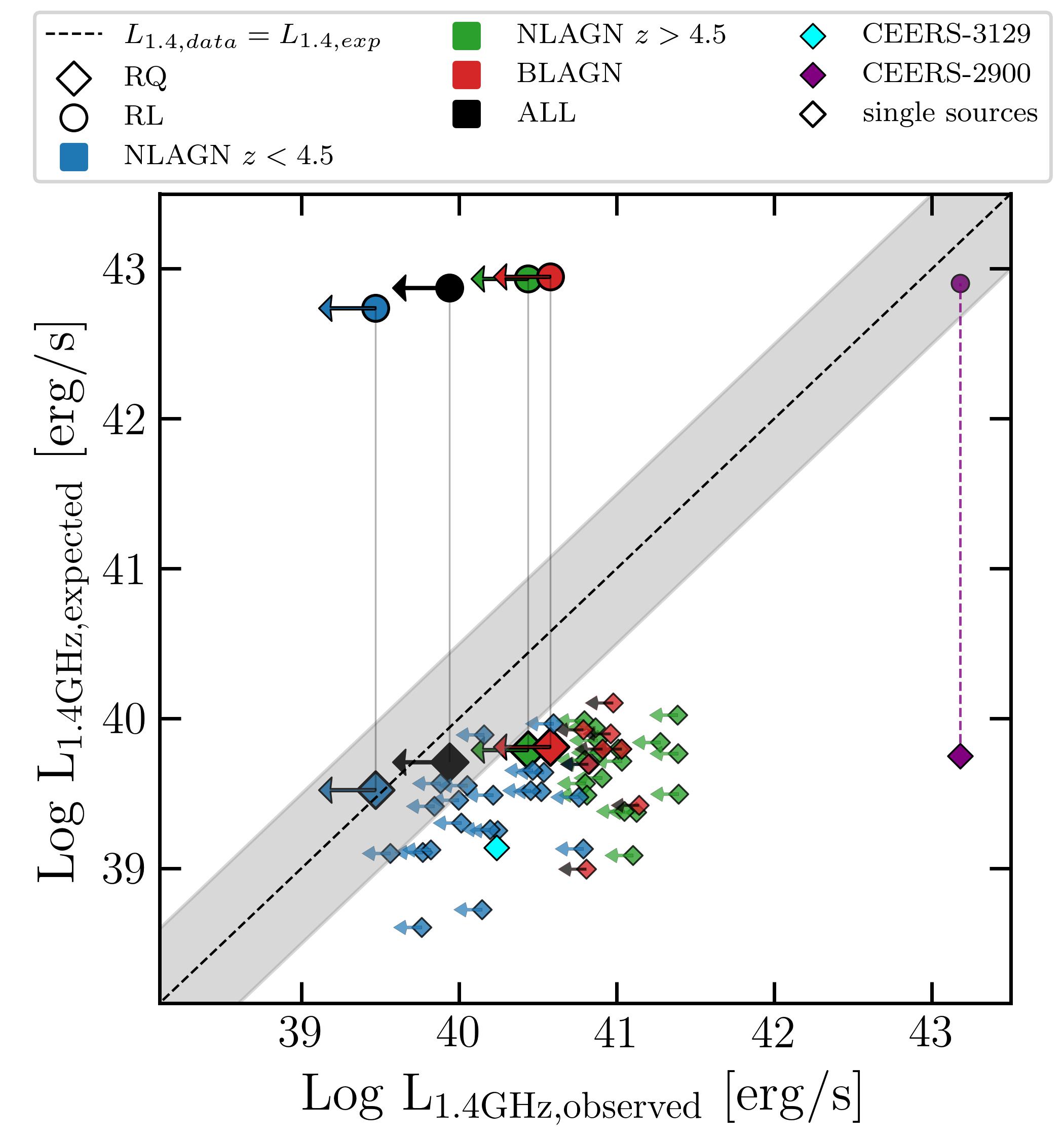}\\
    \caption{Expected 1.4GHz radio luminosity versus observed radio luminosity from the stack of the three AGN samples, as labeled. The diamonds refer to the assumption that AGN are RQ, while circles to the hypothesis that sources are RL. We also show the positions of the single sources according to the color of the respective sample. With a cyan diamond, we report the $z=1.037$ BLAGN CEERS-3129 (presented in Appendix~\ref{sec:app1}). The purple diamond represents the position of CEERS-2900 with the expected $L_{1.4GHz}$ derived using the RQ AGN relation, while the purple circle marks the expected $L_{1.4GHz}$ derived using the RL AGN relation. The shaded gray region represents the dispersion of the $L_{1.4GHz}-L_{2-10keV}$ luminosity relation derived for radio-quiet AGN by \cite{damato22}.}
    \label{fig:radstack_lum}
\end{figure}

To discriminate between the heavily obscured versus intrinsic X-ray weakness scenario and explain the observed X-ray weakness of these AGNs, we took into account the 1.4GHz radio image of the EGS field \citep{Ivison07}. Different works have shown that the physical processes and the magnetic fields at the origin of the hot corona, where the inverse Compton scattering produces the AGN X-ray emission, are also responsible for (at least part of) the synchrotron radiation at the origin of the AGN-related radio emission \citep{Laor08,panessa19}. The connection between these two emissions is also at the basis of the so-called fundamental plane relations \citep{merloni03} or of X-ray - radio luminosity relations reported in different works in the literature \citep{damato22,panessa19,fan16}.
Radio emission is generally largely unaffected by obscuration and can pass freely even from the most obscured environments, contrary to the X-rays that are almost completely absorbed at CTK hydrogen column densities.\\
However, some works show that even the radio emission can be partially or completely absorbed by the BLR clouds if their uniform distribution is also dense enough to determine free-free absorption \citep[][Mazzolari et al. in prep.]{Juodzbalis24_rosetta}. However, it is still not clear whether the X-ray and radio emissions originate and are distributed on the same scale, and, therefore, if the effect of obscuration from BLR clouds can suppress both or not \citep{Paul24}. For example, in \cite{Laor08} the authors found that the smallest possible scale of radio emission ($\nu\sim 1-5\ GHz$) in radio-quiet (RQ) AGN, e.g., coming from a compact radio corona or jet base, should originate from a region $\geq 100$ times the extent of the X-ray emitting core. In this case, the radio emission would probably originate (or be being mainly distributed) outside of the BLR without being absorbed. Therefore, in the following, we assume that there is not a free-free absorption effect on the radio emission of the NLAGN and BLAGN investigated in this work, and we leave a more detailed discussion to Mazzolari et al. in prep.\\
Consequently, if these AGNs are normal AGNs (in terms of bolometric to intrinsic X-ray luminosity), and heavily obscured (but without absorbing also the radio emission), we might expect not to detect their X-ray emission, but we would expect to detect their radio emission at the level returned by the intrinsic X-ray to radio luminosity relations derived in the literature. On the contrary, if the radio image is deep enough to detect the expected AGN radio emission, but none is detected, we would prefer a scenario in which both the radio and X-ray emissions are intrinsically lower with respect to what is observed in normal AGN at lower redshifts.
None of the NLAGNs or BLAGNs are detected in the 1.4GHz image of the field, apart from two sources CEERS-2900 and CEERS-3129. Therefore, we performed a detailed stacking analysis at the position of the sources, as is described in Sec.~\ref{sec:xray_rad_stack}, considering the same three samples as for the X-ray stacking. The radio stacking analysis did not recover any 1.4GHz detection, but allowed us to derive upper limits on the median radio luminosities of the sources in the three samples. We compared these values with the expected AGN radio luminosities. Given the expected (and intrinsic) X-ray luminosities of the sources derived in Sec.~\ref{sec:xweak}, we can estimate the expected 1.4GHz luminosities using the $L_{1.4GHz}-L_{2-10keV}$ luminosity relation derived for radiatively efficient radio-quiet (RQ) AGN by \cite{damato22}. The assumption that the sources are RQ is reasonable given that the fraction of radio-loud (RL) AGN has been found to be $\sim10\%$ of the radio AGN population by different works, and it is not observed to vary with redshift \citep{liu21,williams+15}.\\
It is worth noting that the effectiveness of this test also depends on the fact that the AGN-related radio emission has to be dominant over the radio emission coming from star formation in the host galaxy. To check this, we considered the SFRs derived for these AGNs from the SED fitting in Sec.~\ref{sec:host_prop}, and we converted them into radio luminosities using the relations reported in \cite{novak17,delvecchio21}. For all the three AGN samples the median 1.4GHz luminosities due to star formation are $\sim$1.5 dex lower than the AGN radio luminosity derived from the $L_{1.4GHz}-L_{2-10keV}$ luminosity relation of \cite{damato22}.\\
The comparison between the observed and the expected 1.4GHz luminosities of the three AGN samples is presented in Fig.~\ref{fig:radstack_lum}, together with the upper limits of the single sources in the three samples. Unfortunately, the AEGIS20 radio image is too shallow to give conclusive constraints on the nature of the X-ray weakness of these AGN, being the upper limits of the stack compatible with the expected radio luminosities. We also explore the possibility that the bulk of the sources are instead RL, and, therefore, we computed the expected radio luminosities of the three stacks considering the $L_{1.4GHz}-L_{2-10keV}$ luminosity relation of \cite{fan16}, derived for radiatively inefficient RL AGN. In this case, the observed upper limits on the 1.4GHz luminosities are different orders of magnitudes lower than expected. This excludes that the bulk of our sources is dominated by RL AGN.\\
In Fig.~\ref{fig:radstack_lum}, we also report the position of the two radio-detected sources, CEERS-2900 and CEERS-3129, at $z=1.9$ and $z=1.03$, respectively. The first has a 1.4 GHz radio flux of 58mJy, which corresponds to an observed radio luminosity of $\rm L_{1.4GHz}\sim 10^{43}$erg/s. If we assume that CEERS-2900 is a RQ AGN, the expected radio luminosity given its intrinsic 2-10keV X-ray luminosity reported in \cite{buchner15} (where the source is classified as a CTK AGN) is only $\rm L_{1.4GHz,exp}=6\times 10^{39}$erg/s. This strongly points toward the RL nature of this source. Indeed, considering the RL $\rm L_{1.4GHz}-\log L_{2-10keV}$ luminosity relation reported in \cite{fan16}, the observed and expected radio luminosities correspond almost perfectly.\\
CEERS-3129 has instead a 1.4GHz radio flux of 286 $\mu$Jy that translates into a radio luminosity $\rm L_{1.4GHz}\sim 10^{40}$erg/s. 
The value of the expected $L_{1.4GHz}$, computed starting from the expected $L_{2-10keV}$ luminosity, is almost consistent with the hypothesis of CEERS-3129 being a RQ AGN, since the observed and expected radio luminosities are consistent within 2$\sigma$. This is even truer if we recall that the expected $L_{2-10kev}$, derived from $L_{bol}$ computed using the dust-corrected broad \Ha emission, is probably underestimated because the obscuration of the broad \Ha component is probably larger than the one derived from the narrow-line Balmer decrement (see Sec.~\ref{sec:xweak}).\\
The fact that the expected and observed radio luminosities of CEERS-3129 almost correspond supports the extreme CTK scenario to explain the observed X-ray weakness of this source. However, this scenario is not free from problematics. For example, it requires the CTK envelope around the central SMBH absorbing the X-ray emission to be almost completely dust-free to allow the broad \Ha emission to escape. A natural solution to this problem is that the BLR clouds themselves, with large column densities distributed with approximately a $4\pi$ covering factor, absorb the X-ray photons, as discussed and analyzed in \cite{Maiolino24_X}. 

\section{Conclusions}
In this work, we have analyzed the public \textit{JWST}/NIRSpec MR spectra of the CEERS survey (PID.1340) to identify NLAGNs using multiple emission-line diagnostic diagrams. First, we attributed spectroscopic redshift to 217 out of the 313 targets observed with the \textit{JWST}/NIRSpec MR setup.
We performed spectral line fitting using \texttt{QubeSpec}, and fitting separately different complexes of emission lines, ranging from rest-UV to rest-optical. We considered ten different emission line diagnostic diagrams to perform the NLAGN selection, using both demarcation lines taken from the literature and also exploiting the distribution of SFG and AGN photoionization models taken from \cite{Gutkin16} and \cite{Feltre16}. This process led to the identification of 52 NLAGNs, whose IDs and main physical parameters are reported in Table \ref{table:Sample} to allow the scientific community to perform further studies. The main results are the following:
\begin{itemize}
    \item Traditional AGN diagnostic diagrams, in particular the BPT (R3N2), are less effective in selecting AGNs at $z>3$, because high-$z$ galaxies and AGNs generally overlap in these diagnostics due to the different physical conditions of high-$z$ sources. For the high-$z$ regime, the most effective AGN diagnostic diagrams are the ones based on the \OIIIl4363 line \citep{Mazzolari24b} or on the high-ionization emission lines.
    \item We selected ten NLAGNs at $z>6$ (four of them marked only as candidate NLAGNs), whose spectra are reported in Appendix~\ref{app:high-z_spec}.
    \item The NLAGN selection includes only one out of the eight BLAGNs at $z>4.5$ selected in \citep{Harikane23} -- CEERS-1244 -- four out of five of the X-ray sources with a CEERS MR spectrum and classified as AGNs by the X-ray spectral analysis in \cite{buchner15}, and both the radio detected sources reported in the radio catalog of \cite{Ivison07}.  
    \item NLAGN represent $\sim 20\%$ of the sources analyzed in this work. This fraction is almost constant across different redshifts and agrees well with what was found by \cite{Scholtz23b} for the NLAGN population selected among the spectra of the JADES survey.
    \item We analyzed the distributions of \OIIIl5007 narrow-line FWHMs of the AGN and non-AGN samples without finding significant differences in the distributions of the two populations. This suggests that the NLAGNs identified in this work are not significantly impacting the ISM of their host galaxies. This is probably due to the fact that, thanks to the sensitivity of JWST, we are selecting a fainter NLAGN population that is characterized by having a weaker impact on the host galaxy and lacking powerful outflows (contrary to what has been observed for other NLAGN samples at lower redshifts). Furthermore, in many cases, and in particular for low-mass high-redshift sources, the FWHMs are very close to the resolution limit of JWST.
    \item We also investigated the average obscuration level of the sources by computing the $A_{V}$ values from the Balmer decrements. We did not find a significant evolution with redshift in the value of $A_{V}$. We also note that the AGN population is characterized, in the lower-redshift bin $1<z<2.5$, by a $\sim 0.7$ mag larger obscuration, while in the other redshift bins the median value of obscuration of NLAGNs and non-AGN sources is almost the same. 
    \item We computed the bolometric luminosities of the sources using the dust-corrected \Hb fluxes and the scaling relation reported in \cite{Netzer09}. The range of $L_{bol}$ spanned by our NLAGN sample at $z>3$ is much lower than the AGN bolometric luminosities reported in any pre-JWST samples, but $\sim1$ dex larger than bolometric luminosities of the JWST selected NLAGN reported in \cite{Scholtz23b} selected in the JADES survey (where the exposure times are significantly longer). We tested the robustness of the derived bolometric luminosities by comparing our baseline results with the ones derived using the calibrations reported in \cite{Lamastra09} for the \OIIIl5007 line, finding a good agreement between the two.
    \item We performed a SED-fitting analysis using \texttt{CIGALE} to derive the main physical properties of the host galaxies for the sample of NLAGNs and non-AGNs. We found that, on average, sources at $z>4$ deviate from the main sequence of \cite{Popesso23}, moving toward higher SFR values than what is predicted by the MS. The range of stellar masses extends down to $\sim 10^7 M_{*}$. We did not find any significant difference in the SFR or $M_{*}$ distribution of the NLAGN compared to the ones of the parent sample, suggesting that any potential positive or negative AGN feedback on the host-galaxy SFR is not significantly taking place. This is in agreement with the lack of significant outflow detections in their spectra.
    \item We investigated the X-ray properties of the selected NLAGNs and of the BLAGNs selected in \cite{Harikane23}. We found that none of the BLAGNs is detected in the X-ray image of the field and that only 4 of the 52 NLAGNs are X-ray detected. We performed an X-ray stacking analysis on the AGNs that are not X-ray-detected, finding a significant X-ray deficit between their expected X-ray luminosities (given their bolometric luminosity) and what comes from the X-ray stacking. This observed X-ray weakness is $\sim 1.3$ dex for the high-$z$ BLAGNs and for the sample of NLAGN at $z > 4.5$, and $\sim 2$ dex for the NLAGN at $z < 4.5$. We tried to investigate the origin of the X-ray weakness by studying the radio emission of these sources, which can help to disentangle the heavily obscured AGN scenario from the intrinsic X-ray weak one. Unfortunately, the radio observation covering the region of the CEERS survey is not deep enough to provide a conclusive answer. However, this shows the key role that present and future deep radio continuum observations can have in helping us understand the problem of the X-ray weakness of high-$z$ AGNs.\\
    Interestingly, one of the low-redshift BLAGNs, CEERS-3129, is undetected in the deep X-ray image (showing an X-ray deficit of $\sim2$ dex), but it is detected in the radio image of the field, with a radio luminosity compatible with a RQ AGN nature. The radio detection of CEERS-3129 supports the CTK AGN scenario at the origin of the observed X-ray weakness.
\end{itemize}
JWST is allowing us to reveal the population of heavily obscured AGNs up to very high redshifts ($z>8$) and, at the same time, to investigate the NLAGN population at low redshifts with a spectroscopic sensitivity never explored by pre-JWST surveys. 
This work highlights the necessity of defining new NLAGN diagnostic diagrams and, in general, different and complementary AGN selection techniques suited for the identification of high-redshift AGNs. 
Future works will have to explore in more detail the actual number density and the demography of these objects in the early Universe, carefully accounting for the selection function of each spectroscopic campaign. We also stress that a multiwavelength approach, such as the one outlined in this work, is necessary to explore and potentially provide answers to the peculiar features characterizing high-redshift AGNs. In this sense, the future Square Kilometer Array Observatory (SKAO) and the Advanced X-Ray Imaging Satellite (AXIS) will be crucial to reveal the actual radio and X-ray properties of these objects.

\begin{acknowledgements}
GM acknowledges useful conversations with Andrea Comastri, Roberto Decarli, Sandro Tacchella, William Baker, Callum Witten, Lola Danhaive, Ignas Juodzbalis, Amanda Stoffer, Brian Xing Jiang, and William McClaymont. FDE, JS, RM, and XJ acknowledge support by the Science and Technology Facilities Council (STFC), from the ERC Advanced Grant 695671 ``QUENCH'', and by the
UKRI Frontier Research grant RISEandFALL. RM also acknowledges funding from a research professorship from the Royal Society.
H{\"U} gratefully acknowledges support by the Isaac Newton Trust and by the Kavli Foundation through a Newton-Kavli Junior Fellowship. IEL acknowledges support from the Cassini Fellowship program at INAF-OAS and the European Union’s Horizon 2020 research and innovation program under Marie Sklodowska-Curie grant agreement No. 860744 “Big Data Applications for Black Hole Evolution Studies” (BiD4BESt).
\end{acknowledgements}

%
%

\bibliographystyle{aa}
\bibliography{literature}

@ARTICLE{aird15,
       author = {{Aird}, J. and {Coil}, A.~L. and {Georgakakis}, A. and {Nandra}, K. and
         {Barro}, G. and {P{\'e}rez-Gonz{\'a}lez}, P.~G.},
        title = "{The evolution of the X-ray luminosity functions of unabsorbed and absorbed AGNs out to z̃ 5}",
      journal = {\mnras},
         year = 2015,
        month = aug,
       volume = {451},
       number = {2},
        pages = {1892-1927},
          doi = {10.1093/mnras/stv1062},
archivePrefix = {arXiv},
       eprint = {1503.01120},
 primaryClass = {astro-ph.HE},
       adsurl = {https://ui.adsabs.harvard.edu/abs/2015MNRAS.451.1892A}
}

@ARTICLE{Kokorev23,
       author = {{Kokorev}, Vasily and {Fujimoto}, Seiji and {Labbe}, Ivo and {Greene}, Jenny E. and {Bezanson}, Rachel and {Dayal}, Pratika and {Nelson}, Erica J. and {Atek}, Hakim and {Brammer}, Gabriel and {Caputi}, Karina I. and {Chemerynska}, Iryna and {Cutler}, Sam E. and {Feldmann}, Robert and {Fudamoto}, Yoshinobu and {Furtak}, Lukas J. and {Goulding}, Andy D. and {de Graaff}, Anna and {Leja}, Joel and {Marchesini}, Danilo and {Miller}, Tim B. and {Nanayakkara}, Themiya and {Oesch}, Pascal A. and {Pan}, Richard and {Price}, Sedona H. and {Setton}, David J. and {Smit}, Renske and {Stefanon}, Mauro and {Wang}, Bingjie and {Weaver}, John R. and {Whitaker}, Katherine E. and {Williams}, Christina C. and {Zitrin}, Adi},
        title = "{UNCOVER: A NIRSpec Identification of a Broad-line AGN at z = 8.50}",
      journal = {\apjl},
         year = 2023,
        month = nov,
       volume = {957},
       number = {1},
          eid = {L7},
        pages = {L7},
          doi = {10.3847/2041-8213/ad037a},
archivePrefix = {arXiv},
       eprint = {2308.11610},
 primaryClass = {astro-ph.GA},
       adsurl = {https://ui.adsabs.harvard.edu/abs/2023ApJ...957L...7K}
}

@ARTICLE{panessa19,
       author = {{Panessa}, Francesca and {Baldi}, Ranieri Diego and {Laor}, Ari and {Padovani}, Paolo and {Behar}, Ehud and {McHardy}, Ian},
        title = "{The origin of radio emission from radio-quiet active galactic nuclei}",
      journal = {Nature Astronomy},
         year = 2019,
        month = apr,
       volume = {3},
        pages = {387-396},
          doi = {10.1038/s41550-019-0765-4},
archivePrefix = {arXiv},
       eprint = {1902.05917},
 primaryClass = {astro-ph.GA},
       adsurl = {https://ui.adsabs.harvard.edu/abs/2019NatAs...3..387P}
}

@ARTICLE{boquien19,
       author = {{Boquien}, M. and {Burgarella}, D. and {Roehlly}, Y. and {Buat}, V. and {Ciesla}, L. and {Corre}, D. and {Inoue}, A.~K. and {Salas}, H.},
        title = "{CIGALE: a python Code Investigating GALaxy Emission}",
      journal = {\aap},
         year = 2019,
        month = feb,
       volume = {622},
          eid = {A103},
        pages = {A103},
          doi = {10.1051/0004-6361/201834156},
archivePrefix = {arXiv},
       eprint = {1811.03094},
 primaryClass = {astro-ph.GA},
       adsurl = {https://ui.adsabs.harvard.edu/abs/2019A&A...622A.103B}
}

@ARTICLE{Yue24,
         author = {{Yue}, Minghao and {Eilers}, Anna-Christina and {Ananna}, Tonima Tasnim and {Panagiotou}, Christos and {Kara}, Erin and {Miyaji}, Takamitsu},
        title = "{Stacking X-Ray Observations of ``Little Red Dots'': Implications for Their Active Galactic Nucleus Properties}",
      journal = {\apjl},
         year = 2024,
        month = oct,
       volume = {974},
       number = {2},
          eid = {L26},
        pages = {L26},
          doi = {10.3847/2041-8213/ad7eba},
archivePrefix = {arXiv},
       eprint = {2404.13290},
 primaryClass = {astro-ph.GA},
       adsurl = {https://ui.adsabs.harvard.edu/abs/2024ApJ...974L..26Y}
}

@ARTICLE{Tang23,
       author = {{Tang}, Mengtao and {Stark}, Daniel P. and {Chen}, Zuyi and {Mason}, Charlotte and {Topping}, Michael and {Endsley}, Ryan and {Senchyna}, Peter and {Plat}, Ad{\`e}le and {Lu}, Ting-Yi and {Whitler}, Lily and {Robertson}, Brant and {Charlot}, St{\'e}phane},
        title = "{JWST/NIRSpec spectroscopy of z = 7-9 star-forming galaxies with CEERS: new insight into bright Ly{\ensuremath{\alpha}} emitters in ionized bubbles}",
      journal = {\mnras},
         year = 2023,
        month = dec,
       volume = {526},
       number = {2},
        pages = {1657-1686},
          doi = {10.1093/mnras/stad2763},
archivePrefix = {arXiv},
       eprint = {2301.07072},
 primaryClass = {astro-ph.GA},
       adsurl = {https://ui.adsabs.harvard.edu/abs/2023MNRAS.526.1657T}
}

@ARTICLE{Dors23,
       author = {{Dors}, Oli L. and {Cardaci}, M.~V. and {H{\"a}gele}, G.~F. and {Ilha}, G.~S. and {Oliveira}, C.~B. and {Riffel}, R.~A. and {Riffel}, R. and {Krabbe}, A.~C.},
        title = "{Cosmic metallicity evolution of Active Galactic Nuclei: implications for optical diagnostic diagrams}",
      journal = {\mnras},
         year = 2024,
        month = jan,
       volume = {527},
       number = {3},
        pages = {8193-8212},
          doi = {10.1093/mnras/stad3667},
archivePrefix = {arXiv},
       eprint = {2311.14026},
 primaryClass = {astro-ph.GA},
       adsurl = {https://ui.adsabs.harvard.edu/abs/2024MNRAS.527.8193D}
}

@ARTICLE{Ivison07,
       author = {{Ivison}, R.~J. and {Chapman}, S.~C. and {Faber}, S.~M. and {Smail}, Ian and {Biggs}, A.~D. and {Conselice}, C.~J. and {Wilson}, G. and {Salim}, S. and {Huang}, J. -S. and {Willner}, S.~P.},
        title = "{AEGIS20: A Radio Survey of the Extended Groth Strip}",
      journal = {\apjl},
         year = 2007,
        month = may,
       volume = {660},
       number = {1},
        pages = {L77-L80},
          doi = {10.1086/517917},
archivePrefix = {arXiv},
       eprint = {astro-ph/0607271},
 primaryClass = {astro-ph},
       adsurl = {https://ui.adsabs.harvard.edu/abs/2007ApJ...660L..77I}
}

@ARTICLE{Calabro23,
       author = {{Calabr{\`o}}, Antonello and {Pentericci}, Laura and {Feltre}, Anna and {Arrabal Haro}, Pablo and {Radovich}, Mario and {Seill{\'e}}, Lise-Marie and {Oliva}, Ernesto and {Daddi}, Emanuele and {Amor{\'\i}n}, Ricardo and {Bagley}, Micaela B. and {Bisigello}, Laura and {Buat}, V{\'e}ronique and {Castellano}, Marco and {Cleri}, Nikko J. and {Dickinson}, Mark and {Fern{\'a}ndez}, Vital and {Finkelstein}, Steven L. and {Giavalisco}, Mauro and {Grazian}, Andrea and {Hathi}, Nimish P. and {Hirschmann}, Michaela and {Juneau}, St{\'e}phanie and {Kartaltepe}, Jeyhan S. and {Koekemoer}, Anton M. and {Lucas}, Ray A. and {Papovich}, Casey and {P{\'e}rez-Gonz{\'a}lez}, Pablo G. and {Pirzkal}, Nor and {Santini}, Paola and {Trump}, Jonathan and {de la Vega}, Alexander and {Wilkins}, Stephen M. and {Yung}, L.~Y. Aaron and {Cassata}, Paolo and {Gobat}, Raphael A.~S. and {Mascia}, Sara and {Napolitano}, Lorenzo and {Vulcani}, Benedetta},
        title = "{Near-infrared emission line diagnostics for AGN from the local Universe to z {\ensuremath{\sim}} 3}",
      journal = {\aap},
         year = 2023,
        month = nov,
       volume = {679},
          eid = {A80},
        pages = {A80},
          doi = {10.1051/0004-6361/202347190},
archivePrefix = {arXiv},
       eprint = {2306.08605},
 primaryClass = {astro-ph.GA},
       adsurl = {https://ui.adsabs.harvard.edu/abs/2023A&A...679A..80C}
}

@ARTICLE{Davis23,
       author = {{Davis}, Kelcey and {Trump}, Jonathan R. and {Simons}, Raymond C. and {McGrath}, Elizabeth J. and {Wilkins}, Stephen M. and {Arrabal Haro}, Pablo and {Bagley}, Micaela B. and {Dickinson}, Mark and {Fern{\'a}ndez}, Vital and {Amor{\'\i}n}, Ricardo O. and {Backhaus}, Bren E. and {Cleri}, Nikko J. and {Llerena}, Mario and {Brunker}, Samantha W. and {Barro}, Guillermo and {Bisigello}, Laura and {Brooks}, Madisyn and {Costantin}, Luca and {de La Vega}, Alexander and {Dekel}, Avishai and {Finkelstein}, Steven L. and {Hathi}, Nimish P. and {Hirschmann}, Michaela and {Kartaltepe}, Jeyhan S. and {Koekemoer}, Anton M. and {Lucas}, Ray A. and {Papovich}, Casey and {P{\'e}rez-Gonz{\'a}lez}, Pablo G. and {Pirzkal}, Nor and {Rodighiero}, Giulia and {Rose}, Caitlin and {Yung}, L.~Y. Aaron and {Ceers Collaborators}},
        title = "{A Census from JWST of Extreme Emission-line Galaxies Spanning the Epoch of Reionization in CEERS}",
      journal = {\apj},
         year = 2024,
        month = oct,
       volume = {974},
       number = {1},
          eid = {42},
        pages = {42},
          doi = {10.3847/1538-4357/ad6865},
archivePrefix = {arXiv},
       eprint = {2312.07799},
 primaryClass = {astro-ph.GA},
       adsurl = {https://ui.adsabs.harvard.edu/abs/2024ApJ...974...42D}
}

@ARTICLE{Shapley23,
       author = {{Shapley}, Alice E. and {Sanders}, Ryan L. and {Reddy}, Naveen A. and {Topping}, Michael W. and {Brammer}, Gabriel B.},
        title = "{JWST/NIRSpec Balmer-line Measurements of Star Formation and Dust Attenuation at z   3-6}",
      journal = {\apj},
         year = 2023,
        month = sep,
       volume = {954},
       number = {2},
          eid = {157},
        pages = {157},
          doi = {10.3847/1538-4357/acea5a},
archivePrefix = {arXiv},
       eprint = {2301.03241},
 primaryClass = {astro-ph.GA},
       adsurl = {https://ui.adsabs.harvard.edu/abs/2023ApJ...954..157S}
}

@ARTICLE{Larson23,
       author = {{Larson}, Rebecca L. and {Finkelstein}, Steven L. and {Kocevski}, Dale D. and {Hutchison}, Taylor A. and {Trump}, Jonathan R. and {Arrabal Haro}, Pablo and {Bromm}, Volker and {Cleri}, Nikko J. and {Dickinson}, Mark and {Fujimoto}, Seiji and {Kartaltepe}, Jeyhan S. and {Koekemoer}, Anton M. and {Papovich}, Casey and {Pirzkal}, Nor and {Tacchella}, Sandro and {Zavala}, Jorge A. and {Bagley}, Micaela and {Behroozi}, Peter and {Champagne}, Jaclyn B. and {Cole}, Justin W. and {Jung}, Intae and {Morales}, Alexa M. and {Yang}, Guang and {Zhang}, Haowen and {Zitrin}, Adi and {Amor{\'\i}n}, Ricardo O. and {Burgarella}, Denis and {Casey}, Caitlin M. and {Ch{\'a}vez Ortiz}, {\'O}scar A. and {Cox}, Isabella G. and {Chworowsky}, Katherine and {Fontana}, Adriano and {Gawiser}, Eric and {Grazian}, Andrea and {Grogin}, Norman A. and {Harish}, Santosh and {Hathi}, Nimish P. and {Hirschmann}, Michaela and {Holwerda}, Benne W. and {Juneau}, St{\'e}phanie and {Leung}, Gene C.~K. and {Lucas}, Ray A. and {McGrath}, Elizabeth J. and {P{\'e}rez-Gonz{\'a}lez}, Pablo G. and {Rigby}, Jane R. and {Seill{\'e}}, Lise-Marie and {Simons}, Raymond C. and {de La Vega}, Alexander and {Weiner}, Benjamin J. and {Wilkins}, Stephen M. and {Yung}, L.~Y. Aaron and {Ceers Team}},
        title = "{A CEERS Discovery of an Accreting Supermassive Black Hole 570 Myr after the Big Bang: Identifying a Progenitor of Massive z > 6 Quasars}",
      journal = {\apjl},
         year = 2023,
        month = aug,
       volume = {953},
       number = {2},
          eid = {L29},
        pages = {L29},
          doi = {10.3847/2041-8213/ace619},
archivePrefix = {arXiv},
       eprint = {2303.08918},
 primaryClass = {astro-ph.GA},
       adsurl = {https://ui.adsabs.harvard.edu/abs/2023ApJ...953L..29L}
}

@ARTICLE{Popesso23,
       author = {{Popesso}, P. and {Concas}, A. and {Cresci}, G. and {Belli}, S. and {Rodighiero}, G. and {Inami}, H. and {Dickinson}, M. and {Ilbert}, O. and {Pannella}, M. and {Elbaz}, D.},
        title = "{The main sequence of star-forming galaxies across cosmic times}",
      journal = {\mnras},
         year = 2023,
        month = feb,
       volume = {519},
       number = {1},
        pages = {1526-1544},
          doi = {10.1093/mnras/stac3214},
archivePrefix = {arXiv},
       eprint = {2203.10487},
 primaryClass = {astro-ph.GA},
       adsurl = {https://ui.adsabs.harvard.edu/abs/2023MNRAS.519.1526P}
}

@ARTICLE{Dale14,
       author = {{Dale}, Daniel A. and {Helou}, George and {Magdis}, Georgios E. and {Armus}, Lee and {D{\'\i}az-Santos}, Tanio and {Shi}, Yong},
        title = "{A Two-parameter Model for the Infrared/Submillimeter/Radio Spectral Energy Distributions of Galaxies and Active Galactic Nuclei}",
      journal = {\apj},
         year = 2014,
        month = mar,
       volume = {784},
       number = {1},
          eid = {83},
        pages = {83},
          doi = {10.1088/0004-637X/784/1/83},
archivePrefix = {arXiv},
       eprint = {1402.1495},
 primaryClass = {astro-ph.GA},
       adsurl = {https://ui.adsabs.harvard.edu/abs/2014ApJ...784...83D}
}

@ARTICLE{Lyu23,
       author = {{Lyu}, Jianwei and {Alberts}, Stacey and {Rieke}, George H. and {Shivaei}, Irene and {P{\'e}rez-Gonz{\'a}lez}, Pablo G. and {Sun}, Fengwu and {Hainline}, Kevin N. and {Baum}, Stefi and {Bonaventura}, Nina and {Bunker}, Andrew J. and {Egami}, Eiichi and {Eisenstein}, Daniel J. and {Florian}, Michael and {Ji}, Zhiyuan and {Johnson}, Benjamin D. and {Morrison}, Jane and {Rieke}, Marcia and {Robertson}, Brant and {Rujopakarn}, Wiphu and {Tacchella}, Sandro and {Scholtz}, Jan and {Willmer}, Christopher N.~A.},
        title = "{Active Galactic Nuclei Selection and Demographics: A New Age with JWST/MIRI}",
      journal = {\apj},
         year = 2024,
        month = may,
       volume = {966},
       number = {2},
          eid = {229},
        pages = {229},
          doi = {10.3847/1538-4357/ad3643},
archivePrefix = {arXiv},
       eprint = {2310.12330},
 primaryClass = {astro-ph.GA},
       adsurl = {https://ui.adsabs.harvard.edu/abs/2024ApJ...966..229L}
}

@ARTICLE{DEugenio24,
      author = {{D'Eugenio}, Francesco and {Cameron}, Alex J. and {Scholtz}, Jan and {Carniani}, Stefano and {Willott}, Chris J. and {Curtis-Lake}, Emma and {Bunker}, Andrew J. and {Parlanti}, Eleonora and {Maiolino}, Roberto and {Willmer}, Christopher N.~A. and {Jakobsen}, Peter and {Robertson}, Brant E. and {Johnson}, Benjamin D. and {Tacchella}, Sandro and {Cargile}, Phillip A. and {Rawle}, Tim and {Arribas}, Santiago and {Chevallard}, Jacopo and {Curti}, Mirko and {Egami}, Eiichi and {Eisenstein}, Daniel J. and {Kumari}, Nimisha and {Looser}, Tobias J. and {Rieke}, Marcia J. and {Rodr{\'\i}guez Del Pino}, Bruno and {Saxena}, Aayush and {{\"U}bler}, Hannah and {Venturi}, Giacomo and {Witstok}, Joris and {Baker}, William M. and {Bhatawdekar}, Rachana and {Bonaventura}, Nina and {Boyett}, Kristan and {Charlot}, Stephane and {Danhaive}, A. Lola and {Hainline}, Kevin N. and {Hausen}, Ryan and {Helton}, Jakob M. and {Ji}, Xihan and {Ji}, Zhiyuan and {Jones}, Gareth C. and {Juod{\v{z}}balis}, Ignas and {Maseda}, Michael V. and {P{\'e}rez-Gonz{\'a}lez}, Pablo G. and {Perna}, Michele and {Pusk{\'a}s}, D{\'a}vid and {Shivaei}, Irene and {Silcock}, Maddie S. and {Simmonds}, Charlotte and {Smit}, Renske and {Sun}, Fengwu and {Villanueva}, Natalia C. and {Williams}, Christina C. and {Zhu}, Yongda},
        title = "{JADES Data Release 3: NIRSpec/Microshutter Assembly Spectroscopy for 4000 Galaxies in the GOODS Fields}",
      journal = {\apjs},
         year = 2025,
        month = mar,
       volume = {277},
       number = {1},
          eid = {4},
        pages = {4},
          doi = {10.3847/1538-4365/ada148},
archivePrefix = {arXiv},
       eprint = {2404.06531},
 primaryClass = {astro-ph.GA},
       adsurl = {https://ui.adsabs.harvard.edu/abs/2025ApJS..277....4D}
}

@ARTICLE{Yanagisawa24,
       author = {{Yanagisawa}, Hiroto and {Ouchi}, Masami and {Nakajima}, Kimihiko and {Yajima}, Hidenobu and {Umeda}, Hiroya and {Baba}, Shunsuke and {Nakagawa}, Takao and {Nakane}, Minami and {Matsumoto}, Akinori and {Ono}, Yoshiaki and {Harikane}, Yuichi and {Isobe}, Yuki and {Xu}, Yi and {Zhang}, Yechi},
        title = "{Balmer Decrement Anomalies in Galaxies at z {\ensuremath{\sim}} 6 Found by JWST Observations: Density-bounded Nebulae or Excited H I Clouds?}",
      journal = {\apj},
         year = 2024,
        month = oct,
       volume = {974},
       number = {2},
          eid = {180},
        pages = {180},
          doi = {10.3847/1538-4357/ad7097},
archivePrefix = {arXiv},
       eprint = {2403.20118},
 primaryClass = {astro-ph.GA},
       adsurl = {https://ui.adsabs.harvard.edu/abs/2024ApJ...974..180Y}
}

@ARTICLE{Scholtz24,
       author = {{Scholtz}, Jan and {D'Eugenio}, Francesco and {Maiolino}, Roberto and {P{\'e}rez-Gonz{\'a}lez}, Pablo G. and {Circosta}, Chiara and {Tacchella}, Sandro and {Williams}, Christina C. and {Alberts}, Stacey and {Arribas}, Santiago and {Baker}, William M. and {Bertola}, Elena and {Bunker}, Andrew J. and {Carniani}, Stefano and {Charlot}, Stephane and {Cresci}, Giovanni and {Jones}, Gareth C. and {Kumari}, Nimisha and {Lamperti}, Isabella and {Looser}, Tobias J. and {Rodr{\'\i}guez Del Pino}, Bruno and {Robertson}, Brant and {Parlanti}, Eleonora and {Perna}, Michele and {{\"U}bler}, Hannah and {Venturi}, Giacomo and {Witstok}, Joris},
        title = "{Net-zero gas inflow: deconstructing the gas consumption history of a massive quiescent galaxy with JWST and ALMA}",
      journal = {arXiv e-prints},
         year = 2024,
        month = may,
          eid = {arXiv:2405.19401},
        pages = {arXiv:2405.19401, submitted to Nature},
          doi = {10.48550/arXiv.2405.19401},
archivePrefix = {arXiv},
       eprint = {2405.19401},
 primaryClass = {astro-ph.GA},
       adsurl = {https://ui.adsabs.harvard.edu/abs/2024arXiv240519401S}
}

@ARTICLE{Scholtz18,
       author = {{Scholtz}, J. and {Alexander}, D.~M. and {Harrison}, C.~M. and {Rosario}, D.~J. and {McAlpine}, S. and {Mullaney}, J.~R. and {Stanley}, F. and {Simpson}, J. and {Theuns}, T. and {Bower}, R.~G. and {Hickox}, R.~C. and {Santini}, P. and {Swinbank}, A.~M.},
        title = "{Identifying the subtle signatures of feedback from distant AGN using ALMA observations and the EAGLE hydrodynamical simulations}",
      journal = {\mnras},
         year = 2018,
        month = mar,
       volume = {475},
       number = {1},
        pages = {1288-1305},
          doi = {10.1093/mnras/stx3177},
archivePrefix = {arXiv},
       eprint = {1712.02708},
 primaryClass = {astro-ph.GA},
       adsurl = {https://ui.adsabs.harvard.edu/abs/2018MNRAS.475.1288S}
}

@ARTICLE{Mazzucchelli23,
       author = {{Mazzucchelli}, C. and {Bischetti}, M. and {D'Odorico}, V. and {Feruglio}, C. and {Schindler}, J. -T. and {Onoue}, M. and {Ba{\~n}ados}, E. and {Becker}, G.~D. and {Bian}, F. and {Carniani}, S. and {Decarli}, R. and {Eilers}, A. -C. and {Farina}, E.~P. and {Gallerani}, S. and {Lai}, S. and {Meyer}, R.~A. and {Rojas-Ruiz}, S. and {Satyavolu}, S. and {Venemans}, B.~P. and {Wang}, F. and {Yang}, J. and {Zhu}, Y.},
        title = "{XQR-30: Black hole masses and accretion rates of 42 z {\ensuremath{\gtrsim}} 6 quasars}",
      journal = {\aap},
         year = 2023,
        month = aug,
       volume = {676},
          eid = {A71},
        pages = {A71},
          doi = {10.1051/0004-6361/202346317},
archivePrefix = {arXiv},
       eprint = {2306.16474},
 primaryClass = {astro-ph.GA},
       adsurl = {https://ui.adsabs.harvard.edu/abs/2023A&A...676A..71M}
}

@ARTICLE{Scarlata24,
       author = {{Scarlata}, C. and {Hayes}, M. and {Panagia}, N. and {Mehta}, V. and {Haardt}, F. and {Bagley}, M.},
        title = "{On the universal validity of Case B recombination theory}",
      journal = {arXiv e-prints},
         year = 2024,
        month = apr,
          eid = {arXiv:2404.09015},
        pages = {arXiv:2404.09015, submitted to OJA},
          doi = {10.48550/arXiv.2404.09015},
archivePrefix = {arXiv},
       eprint = {2404.09015},
 primaryClass = {astro-ph.GA},
       adsurl = {https://ui.adsabs.harvard.edu/abs/2024arXiv240409015S}
}

@ARTICLE{Zappacosta23,
       author = {{Zappacosta}, L. and {Piconcelli}, E. and {Fiore}, F. and {Saccheo}, I. and {Valiante}, R. and {Vignali}, C. and {Vito}, F. and {Volonteri}, M. and {Bischetti}, M. and {Comastri}, A. and {Done}, C. and {Elvis}, M. and {Giallongo}, E. and {La Franca}, F. and {Lanzuisi}, G. and {Laurenti}, M. and {Miniutti}, G. and {Bongiorno}, A. and {Brusa}, M. and {Civano}, F. and {Carniani}, S. and {D'Odorico}, V. and {Feruglio}, C. and {Gallerani}, S. and {Gilli}, R. and {Grazian}, A. and {Guainazzi}, M. and {Marinucci}, A. and {Menci}, N. and {Middei}, R. and {Nicastro}, F. and {Puccetti}, S. and {Tombesi}, F. and {Tortosa}, A. and {Testa}, V. and {Vietri}, G. and {Cristiani}, S. and {Haardt}, F. and {Maiolino}, R. and {Schneider}, R. and {Tripodi}, R. and {Vallini}, L. and {Vanzella}, E.},
        title = "{HYPerluminous quasars at the Epoch of ReionizatION (HYPERION): A new regime for the X-ray nuclear properties of the first quasars}",
      journal = {\aap},
         year = 2023,
        month = oct,
       volume = {678},
          eid = {A201},
        pages = {A201},
          doi = {10.1051/0004-6361/202346795},
archivePrefix = {arXiv},
       eprint = {2305.02347},
 primaryClass = {astro-ph.GA},
       adsurl = {https://ui.adsabs.harvard.edu/abs/2023A&A...678A.201Z}
}

@ARTICLE{Kakkad20,
       author = {{Kakkad}, D. and {Mainieri}, V. and {Vietri}, G. and {Carniani}, S. and {Harrison}, C.~M. and {Perna}, M. and {Scholtz}, J. and {Circosta}, C. and {Cresci}, G. and {Husemann}, B. and {Bischetti}, M. and {Feruglio}, C. and {Fiore}, F. and {Marconi}, A. and {Padovani}, P. and {Brusa}, M. and {Cicone}, C. and {Comastri}, A. and {Lanzuisi}, G. and {Mannucci}, F. and {Menci}, N. and {Netzer}, H. and {Piconcelli}, E. and {Puglisi}, A. and {Salvato}, M. and {Schramm}, M. and {Silverman}, J. and {Vignali}, C. and {Zamorani}, G. and {Zappacosta}, L.},
        title = "{SUPER. II. Spatially resolved ionised gas kinematics and scaling relations in z {\ensuremath{\sim}} 2 AGN host galaxies}",
      journal = {\aap},
         year = 2020,
        month = oct,
       volume = {642},
          eid = {A147},
        pages = {A147},
          doi = {10.1051/0004-6361/202038551},
archivePrefix = {arXiv},
       eprint = {2008.01728},
 primaryClass = {astro-ph.GA},
       adsurl = {https://ui.adsabs.harvard.edu/abs/2020A&A...642A.147K}
}

@article{Wu22cat,
  title={A Catalog of Quasar Properties from Sloan Digital Sky Survey Data Release 16},
  author={Wu, Qiaoya and Shen, Yue},
  journal={The Astrophysical Journal Supplement Series},
  volume={263},
  number={2},
  pages={42},
  year={2022},
  publisher={IOP Publishing}
}

@ARTICLE{Bluck23,
       author = {{Bluck}, Asa F.~L. and {Piotrowska}, Joanna M. and {Maiolino}, Roberto},
        title = "{The Fundamental Signature of Star Formation Quenching from AGN Feedback: A Critical Dependence of Quiescence on Supermassive Black Hole Mass, Not Accretion Rate}",
      journal = {\apj},
         year = 2023,
        month = feb,
       volume = {944},
       number = {1},
          eid = {108},
        pages = {108},
          doi = {10.3847/1538-4357/acac7c},
archivePrefix = {arXiv},
       eprint = {2301.03677},
 primaryClass = {astro-ph.GA},
       adsurl = {https://ui.adsabs.harvard.edu/abs/2023ApJ...944..108B}
}

@ARTICLE{Laor08,
       author = {{Laor}, Ari and {Behar}, Ehud},
        title = "{On the origin of radio emission in radio-quiet quasars}",
      journal = {\mnras},
         year = 2008,
        month = oct,
       volume = {390},
       number = {2},
        pages = {847-862},
          doi = {10.1111/j.1365-2966.2008.13806.x},
archivePrefix = {arXiv},
       eprint = {0808.0637},
 primaryClass = {astro-ph},
       adsurl = {https://ui.adsabs.harvard.edu/abs/2008MNRAS.390..847L}
}

@ARTICLE{Allen08,
       author = {{Allen}, Mark G. and {Groves}, Brent A. and {Dopita}, Michael A. and {Sutherland}, Ralph S. and {Kewley}, Lisa J.},
        title = "{The MAPPINGS III Library of Fast Radiative Shock Models}",
      journal = {\apjs},
         year = 2008,
        month = sep,
       volume = {178},
       number = {1},
        pages = {20-55},
          doi = {10.1086/589652},
archivePrefix = {arXiv},
       eprint = {0805.0204},
 primaryClass = {astro-ph},
       adsurl = {https://ui.adsabs.harvard.edu/abs/2008ApJS..178...20A}
}

@ARTICLE{Nesvadba17,
       author = {{Nesvadba}, N.~P.~H. and {Drouart}, G. and {De Breuck}, C. and {Best}, P. and {Seymour}, N. and {Vernet}, J.},
        title = "{Gas kinematics in powerful radio galaxies at z   2: Energy supply from star formation, AGN, and radio jets}",
      journal = {\aap},
         year = 2017,
        month = apr,
       volume = {600},
          eid = {A121},
        pages = {A121},
          doi = {10.1051/0004-6361/201629357},
archivePrefix = {arXiv},
       eprint = {1610.01627},
 primaryClass = {astro-ph.GA},
       adsurl = {https://ui.adsabs.harvard.edu/abs/2017A&A...600A.121N}
}

@ARTICLE{Paul24,
       author = {{Paul}, Jeremiah D. and {Plotkin}, Richard M. and {Brandt}, W.~N. and {Ellis}, Christopher H. and {Gallo}, Elena and {Greene}, Jenny E. and {Ho}, Luis C. and {Kimball}, Amy E. and {Haggard}, Daryl},
        title = "{Radio Scrutiny of the X-Ray-weak Tail of Low-mass Active Galactic Nuclei: A Novel Signature of High-Eddington Accretion?}",
      journal = {\apj},
         year = 2024,
        month = oct,
       volume = {974},
       number = {1},
          eid = {66},
        pages = {66},
          doi = {10.3847/1538-4357/ad67d1},
archivePrefix = {arXiv},
       eprint = {2404.02423},
 primaryClass = {astro-ph.GA},
       adsurl = {https://ui.adsabs.harvard.edu/abs/2024ApJ...974...66P}
}

@ARTICLE{Juodzbalis24_rosetta,
       author = {{Juod{\v{z}}balis}, Ignas and {Ji}, Xihan and {Maiolino}, Roberto and {D'Eugenio}, Francesco and {Scholtz}, Jan and {Risaliti}, Guido and {Fabian}, Andrew C. and {Mazzolari}, Giovanni and {Gilli}, Roberto and {Prandoni}, Isabella and {Arribas}, Santiago and {Bunker}, Andrew J. and {Carniani}, Stefano and {Charlot}, St{\'e}phane and {Curtis-Lake}, Emma and {de Graaff}, Anna and {Hainline}, Kevin and {Parlanti}, Eleonora and {Perna}, Michele and {P{\'e}rez-Gonz{\'a}lez}, Pablo G. and {Robertson}, Brant and {Tacchella}, Sandro and {{\"U}bler}, Hannah and {Williams}, Christina C. and {Willott}, Chris and {Witstok}, Joris},
        title = "{JADES - the Rosetta stone of JWST-discovered AGN: deciphering the intriguing nature of early AGN}",
      journal = {\mnras},
         year = 2024,
        month = nov,
       volume = {535},
       number = {1},
        pages = {853-873},
          doi = {10.1093/mnras/stae2367},
archivePrefix = {arXiv},
       eprint = {2407.08643},
 primaryClass = {astro-ph.GA},
       adsurl = {https://ui.adsabs.harvard.edu/abs/2024MNRAS.535..853J}
}

@ARTICLE{Piotrowska22,
       author = {{Piotrowska}, Joanna M. and {Bluck}, Asa F.~L. and {Maiolino}, Roberto and {Peng}, Yingjie},
        title = "{On the quenching of star formation in observed and simulated central galaxies: evidence for the role of integrated AGN feedback}",
      journal = {\mnras},
         year = 2022,
        month = may,
       volume = {512},
       number = {1},
        pages = {1052-1090},
          doi = {10.1093/mnras/stab3673},
archivePrefix = {arXiv},
       eprint = {2112.07672},
 primaryClass = {astro-ph.GA},
       adsurl = {https://ui.adsabs.harvard.edu/abs/2022MNRAS.512.1052P}
}

@ARTICLE{Mingozzi24,
       author = {{Mingozzi}, Matilde and {James}, Bethan L. and {Berg}, Danielle A. and {Arellano-C{\'o}rdova}, Karla Z. and {Plat}, Adele and {Scarlata}, Claudia and {Aloisi}, Alessandra and {Amor{\'\i}n}, Ricardo O. and {Brinchmann}, Jarle and {Charlot}, St{\'e}phane and {Chisholm}, John and {Feltre}, Anna and {Gazagnes}, Simon and {Hayes}, Matthew and {Heckman}, Timothy and {Hernandez}, Svea and {Kewley}, Lisa J. and {Kumari}, Nimisha and {Leitherer}, Claus and {Martin}, Crystal L. and {Maseda}, Michael and {Nanayakkara}, Themiya and {Ravindranath}, Swara and {Rigby}, Jane R. and {Senchyna}, Peter and {Skillman}, Evan D. and {Sugahara}, Yuma and {Wilkins}, Stephen M. and {Wofford}, Aida and {Xu}, Xinfeng},
        title = "{CLASSY. VIII. Exploring the Source of Ionization with UV Interstellar Medium Diagnostics in Local High-z Analogs}",
      journal = {\apj},
         year = 2024,
        month = feb,
       volume = {962},
       number = {1},
          eid = {95},
        pages = {95},
          doi = {10.3847/1538-4357/ad1033},
archivePrefix = {arXiv},
       eprint = {2306.15062},
 primaryClass = {astro-ph.GA},
       adsurl = {https://ui.adsabs.harvard.edu/abs/2024ApJ...962...95M}
}

@ARTICLE{Mascia24,
       author = {{Mascia}, S. and {Pentericci}, L. and {Calabr{\`o}}, A. and {Santini}, P. and {Napolitano}, L. and {Arrabal Haro}, P. and {Castellano}, M. and {Dickinson}, M. and {Ocvirk}, P. and {Lewis}, J.~S.~W. and {Amor{\'\i}n}, R. and {Bagley}, M. and {Bhatawdekar}, R. and {Cleri}, N.~J. and {Costantin}, L. and {Dekel}, A. and {Finkelstein}, S.~L. and {Fontana}, A. and {Giavalisco}, M. and {Grogin}, N.~A. and {Hathi}, N.~P. and {Hirschmann}, M. and {Holwerda}, B.~W. and {Jung}, I. and {Kartaltepe}, J.~S. and {Koekemoer}, A.~M. and {Lucas}, R.~A. and {Papovich}, C. and {P{\'e}rez-Gonz{\'a}lez}, P.~G. and {Pirzkal}, N. and {Trump}, J.~R. and {Wilkins}, S.~M. and {Yung}, L.~Y.~A.},
        title = "{New insight on the nature of cosmic reionizers from the CEERS survey}",
      journal = {\aap},
         year = 2024,
        month = may,
       volume = {685},
          eid = {A3},
        pages = {A3},
          doi = {10.1051/0004-6361/202347884},
archivePrefix = {arXiv},
       eprint = {2309.02219},
 primaryClass = {astro-ph.GA},
       adsurl = {https://ui.adsabs.harvard.edu/abs/2024A&A...685A...3M}
}

@ARTICLE{Maiolino24_X,
      author = {{Maiolino}, Roberto and {Risaliti}, Guido and {Signorini}, Matilde and {Trefoloni}, Bartolomeo and {Juod{\v{z}}balis}, Ignas and {Scholtz}, Jan and {{\"U}bler}, Hannah and {D'Eugenio}, Francesco and {Carniani}, Stefano and {Fabian}, Andy and {Ji}, Xihan and {Mazzolari}, Giovanni and {Bertola}, Elena and {Brusa}, Marcella and {Bunker}, Andrew J. and {Charlot}, Stephane and {Comastri}, Andrea and {Cresci}, Giovanni and {DeCoursey}, Christa Noel and {Egami}, Eiichi and {Fiore}, Fabrizio and {Gilli}, Roberto and {Perna}, Michele and {Tacchella}, Sandro and {Venturi}, Giacomo},
        title = "{JWST meets Chandra: a large population of Compton thick, feedback-free, and intrinsically X-ray weak AGN, with a sprinkle of SNe}",
      journal = {\mnras},
         year = 2025,
        month = apr,
       volume = {538},
       number = {3},
        pages = {1921-1943},
          doi = {10.1093/mnras/staf359},
archivePrefix = {arXiv},
       eprint = {2405.00504},
 primaryClass = {astro-ph.GA},
       adsurl = {https://ui.adsabs.harvard.edu/abs/2025MNRAS.538.1921M}
}

@ARTICLE{Lamastra09,
       author = {{Lamastra}, A. and {Bianchi}, S. and {Matt}, G. and {Perola}, G.~C. and {Barcons}, X. and {Carrera}, F.~J.},
        title = "{The bolometric luminosity of type 2 AGN from extinction-corrected [OIII]. No evidence of Eddington-limited sources}",
      journal = {\aap},
         year = 2009,
        month = sep,
       volume = {504},
       number = {1},
        pages = {73-79},
          doi = {10.1051/0004-6361/200912023},
archivePrefix = {arXiv},
       eprint = {0905.4439},
 primaryClass = {astro-ph.CO},
       adsurl = {https://ui.adsabs.harvard.edu/abs/2009A&A...504...73L}
}

@ARTICLE{Man18,
       author = {{Man}, Allison and {Belli}, Sirio},
        title = "{Star formation quenching in massive galaxies}",
      journal = {Nature Astronomy},
         year = 2018,
        month = sep,
       volume = {2},
        pages = {695-697},
          doi = {10.1038/s41550-018-0558-1},
archivePrefix = {arXiv},
       eprint = {1809.00722},
 primaryClass = {astro-ph.GA},
       adsurl = {https://ui.adsabs.harvard.edu/abs/2018NatAs...2..695M}
}

@ARTICLE{Barchiesi24,
       author = {{Barchiesi}, L. and {Vignali}, C. and {Pozzi}, F. and {Gilli}, R. and {Mignoli}, M. and {Gruppioni}, C. and {Lapi}, A. and {Marchesi}, S. and {Ricci}, F. and {Urry}, C.~M.},
        title = "{COSMOS2020: Investigating the AGN-obscured accretion phase at z {\ensuremath{\sim}} 1 via [Ne V] selection}",
      journal = {\aap},
         year = 2024,
        month = may,
       volume = {685},
          eid = {A141},
        pages = {A141},
          doi = {10.1051/0004-6361/202245288},
archivePrefix = {arXiv},
       eprint = {2403.03251},
 primaryClass = {astro-ph.GA},
       adsurl = {https://ui.adsabs.harvard.edu/abs/2024A&A...685A.141B}
}

@ARTICLE{Cleri23,
       author = {{Cleri}, Nikko J. and {Olivier}, Grace M. and {Hutchison}, Taylor A. and {Papovich}, Casey and {Trump}, Jonathan R. and {Amor{\'\i}n}, Ricardo O. and {Backhaus}, Bren E. and {Berg}, Danielle A. and {Fern{\'a}ndez}, Vital and {Finkelstein}, Steven L. and {Fujimoto}, Seiji and {Hirschmann}, Michaela and {Kartaltepe}, Jeyhan S. and {Kocevski}, Dale D. and {Simons}, Raymond C. and {Wilkins}, Stephen M. and {Yung}, L.~Y. Aaron},
        title = "{Using [Ne V]/[Ne III] to Understand the Nature of Extreme-ionization Galaxies}",
      journal = {\apj},
         year = 2023,
        month = aug,
       volume = {953},
       number = {1},
          eid = {10},
        pages = {10},
          doi = {10.3847/1538-4357/acde55},
archivePrefix = {arXiv},
       eprint = {2301.07745},
 primaryClass = {astro-ph.GA},
       adsurl = {https://ui.adsabs.harvard.edu/abs/2023ApJ...953...10C}
}

@ARTICLE{Carniani24,
       author = {{Carniani}, Stefano and {Hainline}, Kevin and {D'Eugenio}, Francesco and {Eisenstein}, Daniel J. and {Jakobsen}, Peter and {Witstok}, Joris and {Johnson}, Benjamin D. and {Chevallard}, Jacopo and {Maiolino}, Roberto and {Helton}, Jakob M. and {Willott}, Chris and {Robertson}, Brant and {Alberts}, Stacey and {Arribas}, Santiago and {Baker}, William M. and {Bhatawdekar}, Rachana and {Boyett}, Kristan and {Bunker}, Andrew J. and {Cameron}, Alex J. and {Cargile}, Phillip A. and {Charlot}, St{\'e}phane and {Curti}, Mirko and {Curtis-Lake}, Emma and {Egami}, Eiichi and {Giardino}, Giovanna and {Isaak}, Kate and {Ji}, Zhiyuan and {Jones}, Gareth C. and {Kumari}, Nimisha and {Maseda}, Michael V. and {Parlanti}, Eleonora and {P{\'e}rez-Gonz{\'a}lez}, Pablo G. and {Rawle}, Tim and {Rieke}, George and {Rieke}, Marcia and {Del Pino}, Bruno Rodr{\'\i}guez and {Saxena}, Aayush and {Scholtz}, Jan and {Smit}, Renske and {Sun}, Fengwu and {Tacchella}, Sandro and {{\"U}bler}, Hannah and {Venturi}, Giacomo and {Williams}, Christina C. and {Willmer}, Christopher N.~A.},
        title = "{Spectroscopic confirmation of two luminous galaxies at a redshift of 14}",
      journal = {\nat},
         year = 2024,
        month = sep,
       volume = {633},
       number = {8029},
        pages = {318-322},
          doi = {10.1038/s41586-024-07860-9},
archivePrefix = {arXiv},
       eprint = {2405.18485},
 primaryClass = {astro-ph.GA},
       adsurl = {https://ui.adsabs.harvard.edu/abs/2024Natur.633..318C}
}

@ARTICLE{Ramasawmy19,
       author = {{Ramasawmy}, Joanna and {Stevens}, Jason and {Martin}, Garreth and {Geach}, James E.},
        title = "{A flat trend of star formation rate with X-ray luminosity of galaxies hosting AGN in the SCUBA-2 Cosmology Legacy Survey}",
      journal = {\mnras},
         year = 2019,
        month = jul,
       volume = {486},
       number = {3},
        pages = {4320-4333},
          doi = {10.1093/mnras/stz1093},
archivePrefix = {arXiv},
       eprint = {1904.07880},
 primaryClass = {astro-ph.GA},
       adsurl = {https://ui.adsabs.harvard.edu/abs/2019MNRAS.486.4320R}
}

@ARTICLE{Beckmann17,
       author = {{Beckmann}, R.~S. and {Devriendt}, J. and {Slyz}, A. and {Peirani}, S. and {Richardson}, M.~L.~A. and {Dubois}, Y. and {Pichon}, C. and {Chisari}, N.~E. and {Kaviraj}, S. and {Laigle}, C. and {Volonteri}, M.},
        title = "{Cosmic evolution of stellar quenching by AGN feedback: clues from the Horizon-AGN simulation}",
      journal = {\mnras},
         year = 2017,
        month = nov,
       volume = {472},
       number = {1},
        pages = {949-965},
          doi = {10.1093/mnras/stx1831},
archivePrefix = {arXiv},
       eprint = {1701.07838},
 primaryClass = {astro-ph.GA},
       adsurl = {https://ui.adsabs.harvard.edu/abs/2017MNRAS.472..949B}
}

@ARTICLE{Bugiani24,
       author = {{Bugiani}, Letizia and {Belli}, Sirio and {Park}, Minjung and {Davies}, Rebecca L. and {Mendel}, J. Trevor and {Johnson}, Benjamin D. and {Khoram}, Amir H. and {Benton}, Chlo{\"e} and {Cimatti}, Andrea and {Conroy}, Charlie and {Emami}, Razieh and {Leja}, Joel and {Li}, Yijia and {Maheson}, Gabriel and {Mathews}, Elijah P. and {Naidu}, Rohan P. and {Nelson}, Erica J. and {Tacchella}, Sandro and {Terrazas}, Bryan A. and {Weinberger}, Rainer},
        title = "{Active Galactic Nucleus Feedback in Quiescent Galaxies at Cosmic Noon Traced by Ionized Gas Emission}",
      journal = {\apj},
         year = 2025,
        month = mar,
       volume = {981},
       number = {1},
          eid = {25},
        pages = {25},
          doi = {10.3847/1538-4357/adaeaf},
archivePrefix = {arXiv},
       eprint = {2406.08547},
 primaryClass = {astro-ph.GA},
       adsurl = {https://ui.adsabs.harvard.edu/abs/2025ApJ...981...25B}
}

@ARTICLE{Endsley23,
      author = {{Endsley}, Ryan and {Stark}, Daniel P. and {Whitler}, Lily and {Topping}, Michael W. and {Johnson}, Benjamin D. and {Robertson}, Brant and {Tacchella}, Sandro and {Alberts}, Stacey and {Baker}, William M. and {Bhatawdekar}, Rachana and {Boyett}, Kristan and {Bunker}, Andrew J. and {Cameron}, Alex J. and {Carniani}, Stefano and {Charlot}, Stephane and {Chen}, Zuyi and {Chevallard}, Jacopo and {Curtis-Lake}, Emma and {Danhaive}, A. Lola and {Egami}, Eiichi and {Eisenstein}, Daniel J. and {Hainline}, Kevin and {Helton}, Jakob M. and {Ji}, Zhiyuan and {Looser}, Tobias J. and {Maiolino}, Roberto and {Nelson}, Erica and {Pusk{\'a}s}, D{\'a}vid and {Rieke}, George and {Rieke}, Marcia and {Rix}, Hans-Walter and {Sandles}, Lester and {Saxena}, Aayush and {Simmonds}, Charlotte and {Smit}, Renske and {Sun}, Fengwu and {Williams}, Christina C. and {Willmer}, Christopher N.~A. and {Willott}, Chris and {Witstok}, Joris},
        title = "{The star-forming and ionizing properties of dwarf z 6-9 galaxies in JADES: insights on bursty star formation and ionized bubble growth}",
      journal = {\mnras},
         year = 2024,
        month = sep,
       volume = {533},
       number = {1},
        pages = {1111-1142},
          doi = {10.1093/mnras/stae1857},
archivePrefix = {arXiv},
       eprint = {2306.05295},
 primaryClass = {astro-ph.GA},
       adsurl = {https://ui.adsabs.harvard.edu/abs/2024MNRAS.533.1111E}
}

@ARTICLE{white07,
       author = {{White}, Richard L. and {Helfand}, David J. and {Becker}, Robert H. and {Glikman}, Eilat and {de Vries}, Wim},
        title = "{Signals from the Noise: Image Stacking for Quasars in the FIRST Survey}",
      journal = {\apj},
         year = 2007,
        month = jan,
       volume = {654},
       number = {1},
        pages = {99-114},
          doi = {10.1086/507700},
archivePrefix = {arXiv},
       eprint = {astro-ph/0607335},
 primaryClass = {astro-ph},
       adsurl = {https://ui.adsabs.harvard.edu/abs/2007ApJ...654...99W}
}

@INPROCEEDINGS{miyaji08,
       author = {{Miyaji}, Takamitsu and {Griffiths}, R.~E. and {C-COSMOS Team}},
        title = "{CSTACK: A Web-Based Stacking Analysis Tool for Deep/Wide Chandra Surveys}",
    booktitle = {AAS/High Energy Astrophysics Division \#10},
         year = 2008,
       series = {AAS/High Energy Astrophysics Division},
       volume = {10},
        month = mar,
          eid = {4.01},
        pages = {4.01},
       adsurl = {https://ui.adsabs.harvard.edu/abs/2008HEAD...10.0401M}
}

@ARTICLE{buchner15,
       author = {{Buchner}, Johannes and {Georgakakis}, Antonis and {Nandra}, Kirpal and
         {Brightman}, Murray and {Menzel}, Marie-Luise and {Liu}, Zhu and
         {Hsu}, Li-Ting and {Salvato}, Mara and {Rangel}, Cyprian and
         {Aird}, James and {Merloni}, Andrea and {Ross}, Nicholas},
        title = "{Obscuration-dependent Evolution of Active Galactic Nuclei}",
      journal = {\apj},
         year = 2015,
        month = apr,
       volume = {802},
       number = {2},
          eid = {89},
        pages = {89},
          doi = {10.1088/0004-637X/802/2/89},
archivePrefix = {arXiv},
       eprint = {1501.02805},
 primaryClass = {astro-ph.HE},
       adsurl = {https://ui.adsabs.harvard.edu/abs/2015ApJ...802...89B}
}

@ARTICLE{williams+15,
       author = {{Williams}, W.~L. and {R{\"o}ttgering}, H.~J.~A.},
        title = "{Radio-AGN feedback: when the little ones were monsters}",
      journal = {\mnras},
         year = 2015,
        month = jun,
       volume = {450},
       number = {2},
        pages = {1538-1545},
          doi = {10.1093/mnras/stv692},
archivePrefix = {arXiv},
       eprint = {1503.08927},
 primaryClass = {astro-ph.GA},
       adsurl = {https://ui.adsabs.harvard.edu/abs/2015MNRAS.450.1538W}
}

@ARTICLE{liu21,
       author = {{Liu}, Yuanqi and {Wang}, Ran and {Momjian}, Emmanuel and {Ba{\~n}ados}, Eduardo and {Zeimann}, Greg and {Willott}, Chris J. and {Matsuoka}, Yoshiki and {Omont}, Alain and {Shao}, Yali and {Li}, Qiong and {Li}, Jianan},
        title = "{Constraining the Quasar Radio-loud Fraction at z {\ensuremath{\sim}} 6 with Deep Radio Observations}",
      journal = {\apj},
         year = 2021,
        month = feb,
       volume = {908},
       number = {2},
          eid = {124},
        pages = {124},
          doi = {10.3847/1538-4357/abd3a8},
archivePrefix = {arXiv},
       eprint = {2012.07301},
 primaryClass = {astro-ph.GA},
       adsurl = {https://ui.adsabs.harvard.edu/abs/2021ApJ...908..124L}
}

@ARTICLE{delvecchio21,
       author = {{Delvecchio}, I. and {Daddi}, E. and {Sargent}, M.~T. and {Jarvis}, M.~J. and {Elbaz}, D. and {Jin}, S. and {Liu}, D. and {Whittam}, I.~H. and {Algera}, H. and {Carraro}, R. and {D'Eugenio}, C. and {Delhaize}, J. and {Kalita}, B.~S. and {Leslie}, S. and {Moln{\'a}r}, D. Cs. and {Novak}, M. and {Prandoni}, I. and {Smol{\v{c}}i{\'c}}, V. and {Ao}, Y. and {Aravena}, M. and {Bournaud}, F. and {Collier}, J.~D. and {Randriamampandry}, S.~M. and {Randriamanakoto}, Z. and {Rodighiero}, G. and {Schober}, J. and {White}, S.~V. and {Zamorani}, G.},
        title = "{The infrared-radio correlation of star-forming galaxies is strongly M$_{{\ensuremath{\star}}}$-dependent but nearly redshift-invariant since z {\ensuremath{\sim}} 4}",
      journal = {\aap},
         year = 2021,
        month = mar,
       volume = {647},
          eid = {A123},
        pages = {A123},
          doi = {10.1051/0004-6361/202039647},
archivePrefix = {arXiv},
       eprint = {2010.05510},
 primaryClass = {astro-ph.GA},
       adsurl = {https://ui.adsabs.harvard.edu/abs/2021A&A...647A.123D}
}

@ARTICLE{duras20,
       author = {{Duras}, F. and {Bongiorno}, A. and {Ricci}, F. and {Piconcelli}, E. and
         {Shankar}, F. and {Lusso}, E. and {Bianchi}, S. and {Fiore}, F. and
         {Maiolino}, R. and {Marconi}, A. and {Onori}, F. and {Sani}, E. and
         {Schneider}, R. and {Vignali}, C. and {La Franca}, F.},
        title = "{Universal bolometric corrections for active galactic nuclei over seven luminosity decades}",
      journal = {\aap},
         year = 2020,
        month = apr,
       volume = {636},
          eid = {A73},
        pages = {A73},
          doi = {10.1051/0004-6361/201936817},
archivePrefix = {arXiv},
       eprint = {2001.09984},
 primaryClass = {astro-ph.GA},
       adsurl = {https://ui.adsabs.harvard.edu/abs/2020A&A...636A..73D}
}

@ARTICLE{damato22,
       author = {{D'Amato}, Q. and {Prandoni}, I. and {Gilli}, R. and {Vignali}, C. and {Massardi}, M. and {Liuzzo}, E. and {Jagannathan}, P. and {Brienza}, M. and {Paladino}, R. and {Mignoli}, M. and {Marchesi}, S. and {Peca}, A. and {Chiaberge}, M. and {Mazzolari}, G. and {Norman}, C.},
        title = "{A deep 1.4 GHz survey of the J1030 equatorial field: A new window on radio source populations across cosmic time}",
      journal = {\aap},
         year = 2022,
        month = dec,
       volume = {668},
          eid = {A133},
        pages = {A133},
          doi = {10.1051/0004-6361/202244452},
archivePrefix = {arXiv},
       eprint = {2210.15595},
 primaryClass = {astro-ph.GA},
       adsurl = {https://ui.adsabs.harvard.edu/abs/2022A&A...668A.133D}
}

@ARTICLE{yang23,
       author = {{Yang}, G. and {Caputi}, K.~I. and {Papovich}, C. and {Arrabal Haro}, P. and {Bagley}, M.~B. and {Behroozi}, P. and {Bell}, E.~F. and {Bisigello}, L. and {Buat}, V. and {Burgarella}, D. and {Cheng}, Y. and {Cleri}, N.~J. and {Dav{\'e}}, R. and {Dickinson}, M. and {Elbaz}, D. and {Ferguson}, H.~C. and {Finkelstein}, S.~L. and {Grogin}, N.~A. and {Hathi}, N.~P. and {Hirschmann}, M. and {Holwerda}, B.~W. and {Huertas-Company}, M. and {Hutchison}, T.~A. and {Iani}, E. and {Kartaltepe}, J.~S. and {Kirkpatrick}, A. and {Kocevski}, D.~D. and {Koekemoer}, A.~M. and {Kokorev}, V. and {Larson}, R.~L. and {Lucas}, R.~A. and {P{\'e}rez-Gonz{\'a}lez}, P.~G. and {Rinaldi}, P. and {Shen}, L. and {Trump}, J.~R. and {de la Vega}, A. and {Yung}, L.~Y.~A. and {Zavala}, J.~A.},
        title = "{CEERS Key Paper. VI. JWST/MIRI Uncovers a Large Population of Obscured AGN at High Redshifts}",
      journal = {\apjl},
         year = 2023,
        month = jun,
       volume = {950},
       number = {1},
          eid = {L5},
        pages = {L5},
          doi = {10.3847/2041-8213/acd639},
archivePrefix = {arXiv},
       eprint = {2303.11736},
 primaryClass = {astro-ph.GA},
       adsurl = {https://ui.adsabs.harvard.edu/abs/2023ApJ...950L...5Y}
}

@ARTICLE{gilli22,
       author = {{Gilli}, R. and {Norman}, C. and {Calura}, F. and {Vito}, F. and {Decarli}, R. and {Marchesi}, S. and {Iwasawa}, K. and {Comastri}, A. and {Lanzuisi}, G. and {Pozzi}, F. and {D'Amato}, Q. and {Vignali}, C. and {Brusa}, M. and {Mignoli}, M. and {Cox}, P.},
        title = "{Supermassive black holes at high redshift are expected to be obscured by their massive host galaxies' interstellar medium}",
      journal = {\aap},
         year = 2022,
        month = oct,
       volume = {666},
          eid = {A17},
        pages = {A17},
          doi = {10.1051/0004-6361/202243708},
archivePrefix = {arXiv},
       eprint = {2206.03508},
 primaryClass = {astro-ph.GA},
       adsurl = {https://ui.adsabs.harvard.edu/abs/2022A&A...666A..17G}
}

@ARTICLE{lanzuisi18,
       author = {{Lanzuisi}, G. and {Civano}, F. and {Marchesi}, S. and {Comastri}, A. and
         {Brusa}, M. and {Gilli}, R. and {Vignali}, C. and {Zamorani}, G. and
         {Brightman}, M. and {Griffiths}, R.~E. and {Koekemoer}, A.~M.},
        title = "{The Chandra COSMOS Legacy Survey: Compton thick AGN at high redshift}",
      journal = {\mnras},
         year = 2018,
        month = oct,
       volume = {480},
       number = {2},
        pages = {2578-2592},
          doi = {10.1093/mnras/sty2025},
archivePrefix = {arXiv},
       eprint = {1803.08547},
 primaryClass = {astro-ph.GA},
       adsurl = {https://ui.adsabs.harvard.edu/abs/2018MNRAS.480.2578L}
}

@ARTICLE{hickox18,
       author = {{Hickox}, Ryan C. and {Alexander}, David M.},
        title = "{Obscured Active Galactic Nuclei}",
      journal = {\araa},
         year = 2018,
        month = sep,
       volume = {56},
        pages = {625-671},
          doi = {10.1146/annurev-astro-081817-051803},
archivePrefix = {arXiv},
       eprint = {1806.04680},
 primaryClass = {astro-ph.GA},
       adsurl = {https://ui.adsabs.harvard.edu/abs/2018ARA&A..56..625H}
}

@ARTICLE{Hopkins08,
       author = {{Hopkins}, Philip F. and {Hernquist}, Lars and {Cox}, Thomas J. and {Kere{\v{s}}}, Du{\v{s}}an},
        title = "{A Cosmological Framework for the Co-Evolution of Quasars, Supermassive Black Holes, and Elliptical Galaxies. I. Galaxy Mergers and Quasar Activity}",
      journal = {\apjs},
         year = 2008,
        month = apr,
       volume = {175},
       number = {2},
        pages = {356-389},
          doi = {10.1086/524362},
archivePrefix = {arXiv},
       eprint = {0706.1243},
 primaryClass = {astro-ph},
       adsurl = {https://ui.adsabs.harvard.edu/abs/2008ApJS..175..356H}
}

@ARTICLE{liu17,
       author = {{Liu}, Teng and {Tozzi}, Paolo and {Wang}, Jun-Xian and {Brandt}, William N. and {Vignali}, Cristian and {Xue}, Yongquan and {Schneider}, Donald P. and {Comastri}, Andrea and {Yang}, Guang and {Bauer}, Franz E. and {Paolillo}, Maurizio and {Luo}, Bin and {Gilli}, Roberto and {Wang}, Q. Daniel and {Giavalisco}, Mauro and {Ji}, Zhiyuan and {Alexander}, David M. and {Mainieri}, Vincenzo and {Shemmer}, Ohad and {Koekemoer}, Anton and {Risaliti}, Guido},
        title = "{X-Ray Spectral Analyses of AGNs from the 7Ms Chandra Deep Field-South Survey: The Distribution, Variability, and Evolutions of AGN Obscuration}",
      journal = {\apjs},
         year = 2017,
        month = sep,
       volume = {232},
       number = {1},
          eid = {8},
        pages = {8},
          doi = {10.3847/1538-4365/aa7847},
archivePrefix = {arXiv},
       eprint = {1703.00657},
 primaryClass = {astro-ph.GA},
       adsurl = {https://ui.adsabs.harvard.edu/abs/2017ApJS..232....8L}
}

@ARTICLE{nandra15,
       author = {{Nandra}, K. and {Laird}, E.~S. and {Aird}, J.~A. and {Salvato}, M. and
         {Georgakakis}, A. and {Barro}, G. and {Perez-Gonzalez}, P.~G. and
         {Barmby}, P. and {Chary}, R. -R. and {Coil}, A. and {Cooper}, M.~C. and
         {Davis}, M. and {Dickinson}, M. and {Faber}, S.~M. and {Fazio}, G.~G. and
         {Guhathakurta}, P. and {Gwyn}, S. and {Hsu}, L. -T. and
         {Huang}, J. -S. and {Ivison}, R.~J. and {Koo}, D.~C. and
         {Newman}, J.~A. and {Rangel}, C. and {Yamada}, T. and {Willmer}, C.},
        title = "{AEGIS-X: Deep Chandra Imaging of the Central Groth Strip}",
      journal = {\apjs},
         year = "2015",
        month = "Sep",
       volume = {220},
       number = {1},
          eid = {10},
        pages = {10},
          doi = {10.1088/0067-0049/220/1/10},
archivePrefix = {arXiv},
       eprint = {1503.09078},
 primaryClass = {astro-ph.HE},
       adsurl = {https://ui.adsabs.harvard.edu/abs/2015ApJS..220...10N}
}

@ARTICLE{novak17,
       author = {{Novak}, M. and {Smol{\v{c}}i{\'c}}, V. and {Delhaize}, J. and {Delvecchio}, I. and {Zamorani}, G. and {Baran}, N. and {Bondi}, M. and {Capak}, P. and {Carilli}, C.~L. and {Ciliegi}, P. and {Civano}, F. and {Ilbert}, O. and {Karim}, A. and {Laigle}, C. and {Le F{\`e}vre}, O. and {Marchesi}, S. and {McCracken}, H. and {Miettinen}, O. and {Salvato}, M. and {Sargent}, M. and {Schinnerer}, E. and {Tasca}, L.},
        title = "{The VLA-COSMOS 3 GHz Large Project: Cosmic star formation history since z   5}",
      journal = {\aap},
         year = 2017,
        month = jun,
       volume = {602},
          eid = {A5},
        pages = {A5},
          doi = {10.1051/0004-6361/201629436},
archivePrefix = {arXiv},
       eprint = {1703.09724},
 primaryClass = {astro-ph.GA},
       adsurl = {https://ui.adsabs.harvard.edu/abs/2017A&A...602A...5N}
}

@ARTICLE{merloni03,
       author = {{Merloni}, Andrea and {Heinz}, Sebastian and {di Matteo}, Tiziana},
        title = "{A Fundamental Plane of black hole activity}",
      journal = {\mnras},
         year = 2003,
        month = nov,
       volume = {345},
       number = {4},
        pages = {1057-1076},
          doi = {10.1046/j.1365-2966.2003.07017.x},
archivePrefix = {arXiv},
       eprint = {astro-ph/0305261},
 primaryClass = {astro-ph},
       adsurl = {https://ui.adsabs.harvard.edu/abs/2003MNRAS.345.1057M}
}

@ARTICLE{fan16,
       author = {{Fan}, Xu-Liang and {Bai}, Jin-Ming},
        title = "{The Radio/X-Ray Correlation and Black Hole Fundamental Plane for Young Radio Sources: Implications for X-Ray Origin and Accretion Mode}",
      journal = {\apj},
         year = 2016,
        month = feb,
       volume = {818},
       number = {2},
          eid = {185},
        pages = {185},
          doi = {10.3847/0004-637X/818/2/185},
       adsurl = {https://ui.adsabs.harvard.edu/abs/2016ApJ...818..185F}
}

@ARTICLE{simmonds18,
       author = {{Simmonds}, C. and {Buchner}, J. and {Salvato}, M. and {Hsu}, L. -T. and {Bauer}, F.~E.},
        title = "{XZ: Deriving redshifts from X-ray spectra of obscured AGN}",
      journal = {\aap},
         year = 2018,
        month = oct,
       volume = {618},
          eid = {A66},
        pages = {A66},
          doi = {10.1051/0004-6361/201833412},
archivePrefix = {arXiv},
       eprint = {1807.01782},
 primaryClass = {astro-ph.IM},
       adsurl = {https://ui.adsabs.harvard.edu/abs/2018A&A...618A..66S}
}

@ARTICLE{vignali15,
       author = {{Vignali}, C. and {Iwasawa}, K. and {Comastri}, A. and {Gilli}, R. and {Lanzuisi}, G. and {Ranalli}, P. and {Cappelluti}, N. and {Mainieri}, V. and {Georgantopoulos}, I. and {Carrera}, F.~J. and {Fritz}, J. and {Brusa}, M. and {Brandt}, W.~N. and {Bauer}, F.~E. and {Fiore}, F. and {Tombesi}, F.},
        title = "{The XMM deep survey in the CDF-S. IX. An X-ray outflow in a luminous obscured quasar at z {\ensuremath{\approx}} 1.6}",
      journal = {\aap},
         year = 2015,
        month = nov,
       volume = {583},
          eid = {A141},
        pages = {A141},
          doi = {10.1051/0004-6361/201525852},
archivePrefix = {arXiv},
       eprint = {1509.05413},
 primaryClass = {astro-ph.GA},
       adsurl = {https://ui.adsabs.harvard.edu/abs/2015A&A...583A.141V}
}

@ARTICLE{vito16,
       author = {{Vito}, F. and {Gilli}, R. and {Vignali}, C. and {Brandt}, W.~N. and
         {Comastri}, A. and {Yang}, G. and {Lehmer}, B.~D. and {Luo}, B. and
         {Basu-Zych}, A. and {Bauer}, F.~E. and {Cappelluti}, N. and
         {Koekemoer}, A. and {Mainieri}, V. and {Paolillo}, M. and
         {Ranalli}, P. and {Shemmer}, O. and {Trump}, J. and {Wang}, J.~X. and
         {Xue}, Y.~Q.},
        title = "{The deepest X-ray view of high-redshift galaxies: constraints on low-rate black hole accretion}",
      journal = {\mnras},
         year = "2016",
        month = "Nov",
       volume = {463},
       number = {1},
        pages = {348-374},
          doi = {10.1093/mnras/stw1998},
archivePrefix = {arXiv},
       eprint = {1608.02614},
 primaryClass = {astro-ph.GA},
       adsurl = {https://ui.adsabs.harvard.edu/abs/2016MNRAS.463..348V}
}

@ARTICLE{vito18,
       author = {{Vito}, F. and {Brandt}, W.~N. and {Yang}, G. and {Gilli}, R. and
         {Luo}, B. and {Vignali}, C. and {Xue}, Y.~Q. and {Comastri}, A. and
         {Koekemoer}, A.~M. and {Lehmer}, B.~D. and {Liu}, T. and
         {Paolillo}, M. and {Ranalli}, P. and {Schneider}, D.~P. and
         {Shemmer}, O. and {Volonteri}, M. and {Wang}, J.},
        title = "{High-redshift AGN in the Chandra Deep Fields: the obscured fraction and space density of the sub-L$_{*}$ population}",
      journal = {\mnras},
         year = "2018",
        month = "Jan",
       volume = {473},
       number = {2},
        pages = {2378-2406},
          doi = {10.1093/mnras/stx2486},
archivePrefix = {arXiv},
       eprint = {1709.07892},
 primaryClass = {astro-ph.GA},
       adsurl = {https://ui.adsabs.harvard.edu/abs/2018MNRAS.473.2378V}
}

@ARTICLE{xue16,
       author = {{Xue}, Y.~Q. and {Luo}, B. and {Brandt}, W.~N. and {Alexander}, D.~M. and
         {Bauer}, F.~E. and {Lehmer}, B.~D. and {Yang}, G.},
        title = "{The 2 Ms Chandra Deep Field-North Survey and the 250 ks Extended Chandra Deep Field-South Survey: Improved Point-source Catalogs}",
      journal = {\apjs},
         year = "2016",
        month = "Jun",
       volume = {224},
       number = {2},
          eid = {15},
        pages = {15},
          doi = {10.3847/0067-0049/224/2/15},
archivePrefix = {arXiv},
       eprint = {1602.06299},
 primaryClass = {astro-ph.GA},
       adsurl = {https://ui.adsabs.harvard.edu/abs/2016ApJS..224...15X}
}

@ARTICLE{yang20,
       author = {{Yang}, G. and {Boquien}, M. and {Buat}, V. and {Burgarella}, D. and {Ciesla}, L. and {Duras}, F. and {Stalevski}, M. and {Brandt}, W.~N. and {Papovich}, C.},
        title = "{X-CIGALE: Fitting AGN/galaxy SEDs from X-ray to infrared}",
      journal = {\mnras},
         year = 2020,
        month = jan,
       volume = {491},
       number = {1},
        pages = {740-757},
          doi = {10.1093/mnras/stz3001},
archivePrefix = {arXiv},
       eprint = {2001.08263},
 primaryClass = {astro-ph.GA},
       adsurl = {https://ui.adsabs.harvard.edu/abs/2020MNRAS.491..740Y}
}

@ARTICLE{Liddle07,
       author = {{Liddle}, Andrew R.},
        title = "{Information criteria for astrophysical model selection}",
      journal = {\mnras},
         year = 2007,
        month = may,
       volume = {377},
       number = {1},
        pages = {L74-L78},
          doi = {10.1111/j.1745-3933.2007.00306.x},
archivePrefix = {arXiv},
       eprint = {astro-ph/0701113},
 primaryClass = {astro-ph},
       adsurl = {https://ui.adsabs.harvard.edu/abs/2007MNRAS.377L..74L}
}

@ARTICLE{Goulding23,
       author = {{Goulding}, Andy D. and {Greene}, Jenny E. and {Setton}, David J. and {Labbe}, Ivo and {Bezanson}, Rachel and {Miller}, Tim B. and {Atek}, Hakim and {Bogd{\'a}n}, {\'A}kos and {Brammer}, Gabriel and {Chemerynska}, Iryna and {Cutler}, Sam E. and {Dayal}, Pratika and {Fudamoto}, Yoshinobu and {Fujimoto}, Seiji and {Furtak}, Lukas J. and {Kokorev}, Vasily and {Khullar}, Gourav and {Leja}, Joel and {Marchesini}, Danilo and {Natarajan}, Priyamvada and {Nelson}, Erica and {Oesch}, Pascal A. and {Pan}, Richard and {Papovich}, Casey and {Price}, Sedona H. and {van Dokkum}, Pieter and {Wang}, Bingjie and {Weaver}, John R. and {Whitaker}, Katherine E. and {Zitrin}, Adi},
        title = "{UNCOVER: The Growth of the First Massive Black Holes from JWST/NIRSpec-Spectroscopic Redshift Confirmation of an X-Ray Luminous AGN at z = 10.1}",
      journal = {\apjl},
         year = 2023,
        month = sep,
       volume = {955},
       number = {1},
          eid = {L24},
        pages = {L24},
          doi = {10.3847/2041-8213/acf7c5},
archivePrefix = {arXiv},
       eprint = {2308.02750},
 primaryClass = {astro-ph.GA},
       adsurl = {https://ui.adsabs.harvard.edu/abs/2023ApJ...955L..24G}
}

@ARTICLE{Bogdan23,
       author = {{Bogd{\'a}n}, {\'A}kos and {Goulding}, Andy D. and {Natarajan}, Priyamvada and {Kov{\'a}cs}, Orsolya E. and {Tremblay}, Grant R. and {Chadayammuri}, Urmila and {Volonteri}, Marta and {Kraft}, Ralph P. and {Forman}, William R. and {Jones}, Christine and {Churazov}, Eugene and {Zhuravleva}, Irina},
        title = "{Evidence for heavy-seed origin of early supermassive black holes from a z {\ensuremath{\approx}} 10 X-ray quasar}",
      journal = {Nature Astronomy},
         year = 2023,
        month = nov,
          doi = {10.1038/s41550-023-02111-9},
archivePrefix = {arXiv},
       eprint = {2305.15458},
 primaryClass = {astro-ph.GA},
       adsurl = {https://ui.adsabs.harvard.edu/abs/2023NatAs.tmp..223B}
}

@ARTICLE{Zhang23,
        author = {{Zhang}, Haowen and {Behroozi}, Peter and {Volonteri}, Marta and {Silk}, Joseph and {Fan}, Xiaohui and {Aird}, James and {Yang}, Jinyi and {Wang}, Feige and {Tee}, Wei Leong and {Hopkins}, Philip F.},
        title = "{TRINITY IV: predictions for supermassive black holes at z {\ensuremath{\gtrsim}} 6}",
      journal = {\mnras},
         year = 2024,
        month = jul,
       volume = {531},
       number = {4},
        pages = {4974-4989},
          doi = {10.1093/mnras/stae1447},
archivePrefix = {arXiv},
       eprint = {2309.07210},
 primaryClass = {astro-ph.GA},
       adsurl = {https://ui.adsabs.harvard.edu/abs/2024MNRAS.531.4974Z}
}

@ARTICLE{Ferruit23,
       author = {{Ferruit}, P. and {Jakobsen}, P. and {Giardino}, G. and {Rawle}, T. and {Alves de Oliveira}, C. and {Arribas}, S. and {Beck}, T.~L. and {Birkmann}, S. and {B{\"o}ker}, T. and {Bunker}, A.~J. and {Charlot}, S. and {de Marchi}, G. and {Franx}, M. and {Henry}, A. and {Karakla}, D. and {Kassin}, S.~A. and {Kumari}, N. and {L{\'o}pez-Caniego}, M. and {L{\"u}tzgendorf}, N. and {Maiolino}, R. and {Manjavacas}, E. and {Marston}, A. and {Moseley}, S.~H. and {Muzerolle}, J. and {Pirzkal}, N. and {Rauscher}, B. and {Rix}, H. -W. and {Sabbi}, E. and {Sirianni}, M. and {te Plate}, M. and {Valenti}, J. and {Willott}, C.~J. and {Zeidler}, P.},
        title = "{The Near-Infrared Spectrograph (NIRSpec) on the James Webb Space Telescope. II. Multi-object spectroscopy (MOS)}",
      journal = {\aap},
         year = 2022,
        month = may,
       volume = {661},
          eid = {A81},
        pages = {A81},
          doi = {10.1051/0004-6361/202142673},
archivePrefix = {arXiv},
       eprint = {2202.03306},
 primaryClass = {astro-ph.IM},
       adsurl = {https://ui.adsabs.harvard.edu/abs/2022A&A...661A..81F}
}

@ARTICLE{Scholtz23b,
       author = {{Scholtz}, Jan and {Maiolino}, Roberto and {D'Eugenio}, Francesco and {Curtis-Lake}, Emma and {Carniani}, Stefano and {Charlot}, Stephane and {Curti}, Mirko and {Silcock}, Maddie S. and {Arribas}, Santiago and {Baker}, William and {Bhatawdekar}, Rachana and {Boyett}, Kristan and {Bunker}, Andrew J. and {Cameron}, Alex and {Chevallard}, Jacopo and {Circosta}, Chiara and {Eisenstein}, Daniel J. and {Hainline}, Kevin and {Hausen}, Ryan and {Ji}, Xihan and {Ji}, Zhiyuan and {Johnson}, Benjamin D. and {Kumari}, Nimisha and {Looser}, Tobias J. and {Lyu}, Jianwei and {Maseda}, Michael V. and {Parlanti}, Eleonora and {Perna}, Michele and {Rieke}, Marcia and {Robertson}, Brant and {Rodr{\'\i}guez Del Pino}, Bruno and {Sun}, Fengwu and {Tacchella}, Sandro and {{\"U}bler}, Hannah and {Venturi}, Giacomo and {Williams}, Christina C. and {Willmer}, Christopher N.~A. and {Willott}, Chris and {Witstok}, Joris},
        title = "{JADES: A large population of obscured, narrow line AGN at high redshift}",
      journal = {arXiv e-prints},
         year = 2023,
        month = nov,
          eid = {arXiv:2311.18731},
        pages = {arXiv:2311.18731, submitted to A\&A},
archivePrefix = {arXiv},
       eprint = {2311.18731},
 primaryClass = {astro-ph.GA},
       adsurl = {https://ui.adsabs.harvard.edu/abs/2023arXiv231118731S}
}

@ARTICLE{Eisenstein23,
       author = {{Eisenstein}, Daniel J. and {Willott}, Chris and {Alberts}, Stacey and {Arribas}, Santiago and {Bonaventura}, Nina and {Bunker}, Andrew J. and {Cameron}, Alex J. and {Carniani}, Stefano and {Charlot}, Stephane and {Curtis-Lake}, Emma and {D'Eugenio}, Francesco and {Endsley}, Ryan and {Ferruit}, Pierre and {Giardino}, Giovanna and {Hainline}, Kevin and {Hausen}, Ryan and {Jakobsen}, Peter and {Johnson}, Benjamin D. and {Maiolino}, Roberto and {Rieke}, Marcia and {Rieke}, George and {Rix}, Hans-Walter and {Robertson}, Brant and {Stark}, Daniel P. and {Tacchella}, Sandro and {Williams}, Christina C. and {Willmer}, Christopher N.~A. and {Baker}, William M. and {Baum}, Stefi and {Bhatawdekar}, Rachana and {Boyett}, Kristan and {Chen}, Zuyi and {Chevallard}, Jacopo and {Circosta}, Chiara and {Curti}, Mirko and {Danhaive}, A. Lola and {DeCoursey}, Christa and {de Graaff}, Anna and {Dressler}, Alan and {Egami}, Eiichi and {Helton}, Jakob M. and {Hviding}, Raphael E. and {Ji}, Zhiyuan and {Jones}, Gareth C. and {Kumari}, Nimisha and {L{\"u}tzgendorf}, Nora and {Laseter}, Isaac and {Looser}, Tobias J. and {Lyu}, Jianwei and {Maseda}, Michael V. and {Nelson}, Erica and {Parlanti}, Eleonora and {Perna}, Michele and {Pusk{\'a}s}, D{\'a}vid and {Rawle}, Tim and {Rodr{\'\i}guez Del Pino}, Bruno and {Sandles}, Lester and {Saxena}, Aayush and {Scholtz}, Jan and {Sharpe}, Katherine and {Shivaei}, Irene and {Silcock}, Maddie S. and {Simmonds}, Charlotte and {Skarbinski}, Maya and {Smit}, Renske and {Stone}, Meredith and {Suess}, Katherine A. and {Sun}, Fengwu and {Tang}, Mengtao and {Topping}, Michael W. and {{\"U}bler}, Hannah and {Villanueva}, Natalia C. and {Wallace}, Imaan E.~B. and {Whitler}, Lily and {Witstok}, Joris and {Woodrum}, Charity},
        title = "{Overview of the JWST Advanced Deep Extragalactic Survey (JADES)}",
      journal = {arXiv e-prints},
         year = 2023,
        month = jun,
          eid = {arXiv:2306.02465},
        pages = {arXiv:2306.02465, submitted to ApJ Supplement},
          doi = {10.48550/arXiv.2306.02465},
archivePrefix = {arXiv},
       eprint = {2306.02465},
 primaryClass = {astro-ph.GA},
       adsurl = {https://ui.adsabs.harvard.edu/abs/2023arXiv230602465E}
}

@ARTICLE{Matthee23,
      author = {{Matthee}, Jorryt and {Naidu}, Rohan P. and {Brammer}, Gabriel and {Chisholm}, John and {Eilers}, Anna-Christina and {Goulding}, Andy and {Greene}, Jenny and {Kashino}, Daichi and {Labbe}, Ivo and {Lilly}, Simon J. and {Mackenzie}, Ruari and {Oesch}, Pascal A. and {Weibel}, Andrea and {Wuyts}, Stijn and {Xiao}, Mengyuan and {Bordoloi}, Rongmon and {Bouwens}, Rychard and {van Dokkum}, Pieter and {Illingworth}, Garth and {Kramarenko}, Ivan and {Maseda}, Michael V. and {Mason}, Charlotte and {Meyer}, Romain A. and {Nelson}, Erica J. and {Reddy}, Naveen A. and {Shivaei}, Irene and {Simcoe}, Robert A. and {Yue}, Minghao},
        title = "{Little Red Dots: An Abundant Population of Faint Active Galactic Nuclei at z {\ensuremath{\sim}} 5 Revealed by the EIGER and FRESCO JWST Surveys}",
      journal = {\apj},
         year = 2024,
        month = mar,
       volume = {963},
       number = {2},
          eid = {129},
        pages = {129},
          doi = {10.3847/1538-4357/ad2345},
archivePrefix = {arXiv},
       eprint = {2306.05448},
 primaryClass = {astro-ph.GA},
       adsurl = {https://ui.adsabs.harvard.edu/abs/2024ApJ...963..129M}
}

@ARTICLE{Greene23,
      author = {{Greene}, Jenny E. and {Labbe}, Ivo and {Goulding}, Andy D. and {Furtak}, Lukas J. and {Chemerynska}, Iryna and {Kokorev}, Vasily and {Dayal}, Pratika and {Volonteri}, Marta and {Williams}, Christina C. and {Wang}, Bingjie and {Setton}, David J. and {Burgasser}, Adam J. and {Bezanson}, Rachel and {Atek}, Hakim and {Brammer}, Gabriel and {Cutler}, Sam E. and {Feldmann}, Robert and {Fujimoto}, Seiji and {Glazebrook}, Karl and {de Graaff}, Anna and {Khullar}, Gourav and {Leja}, Joel and {Marchesini}, Danilo and {Maseda}, Michael V. and {Matthee}, Jorryt and {Miller}, Tim B. and {Naidu}, Rohan P. and {Nanayakkara}, Themiya and {Oesch}, Pascal A. and {Pan}, Richard and {Papovich}, Casey and {Price}, Sedona H. and {van Dokkum}, Pieter and {Weaver}, John R. and {Whitaker}, Katherine E. and {Zitrin}, Adi},
        title = "{UNCOVER Spectroscopy Confirms the Surprising Ubiquity of Active Galactic Nuclei in Red Sources at z > 5}",
      journal = {\apj},
         year = 2024,
        month = mar,
       volume = {964},
       number = {1},
          eid = {39},
        pages = {39},
          doi = {10.3847/1538-4357/ad1e5f},
archivePrefix = {arXiv},
       eprint = {2309.05714},
 primaryClass = {astro-ph.GA},
       adsurl = {https://ui.adsabs.harvard.edu/abs/2024ApJ...964...39G}
}

@ARTICLE{Harikane23,
      author = {{Harikane}, Yuichi and {Zhang}, Yechi and {Nakajima}, Kimihiko and {Ouchi}, Masami and {Isobe}, Yuki and {Ono}, Yoshiaki and {Hatano}, Shun and {Xu}, Yi and {Umeda}, Hiroya},
        title = "{A JWST/NIRSpec First Census of Broad-line AGNs at z = 4-7: Detection of 10 Faint AGNs with M $_{BH}$ {}10$^{6}$-{}10$^{8}$ M $_{{\ensuremath{\odot}}}$ and Their Host Galaxy Properties}",
      journal = {\apj},
         year = 2023,
        month = dec,
       volume = {959},
       number = {1},
          eid = {39},
        pages = {39},
          doi = {10.3847/1538-4357/ad029e},
archivePrefix = {arXiv},
       eprint = {2303.11946},
 primaryClass = {astro-ph.GA},
       adsurl = {https://ui.adsabs.harvard.edu/abs/2023ApJ...959...39H}
}

@ARTICLE{ubler23,
       author = {{{\"U}bler}, Hannah and {Maiolino}, Roberto and {Curtis-Lake}, Emma and {P{\'e}rez-Gonz{\'a}lez}, Pablo G. and {Curti}, Mirko and {Perna}, Michele and {Arribas}, Santiago and {Charlot}, St{\'e}phane and {Marshall}, Madeline A. and {D'Eugenio}, Francesco and {Scholtz}, Jan and {Bunker}, Andrew and {Carniani}, Stefano and {Ferruit}, Pierre and {Jakobsen}, Peter and {Rix}, Hans-Walter and {Rodr{\'\i}guez Del Pino}, Bruno and {Willott}, Chris J. and {Boeker}, Torsten and {Cresci}, Giovanni and {Jones}, Gareth C. and {Kumari}, Nimisha and {Rawle}, Tim},
        title = "{GA-NIFS: A massive black hole in a low-metallicity AGN at z {\ensuremath{\sim}} 5.55 revealed by JWST/NIRSpec IFS}",
      journal = {\aap},
         year = 2023,
        month = sep,
       volume = {677},
          eid = {A145},
        pages = {A145},
          doi = {10.1051/0004-6361/202346137},
archivePrefix = {arXiv},
       eprint = {2302.06647},
 primaryClass = {astro-ph.GA},
       adsurl = {https://ui.adsabs.harvard.edu/abs/2023A&A...677A.145U}
}

@ARTICLE{ubler23b,
    author = {{\"U}bler, Hannah and Maiolino, Roberto and Pérez-González, Pablo G and D’Eugenio, Francesco and Perna, Michele and Curti, Mirko and Arribas, Santiago and Bunker, Andrew and Carniani, Stefano and Charlot, Stéphane and Rodríguez Del Pino, Bruno and Baker, William and Böker, Torsten and Cresci, Giovanni and Dunlop, James and Grogin, Norman A and Jones, Gareth C and Kumari, Nimisha and Lamperti, Isabella and Laporte, Nicolas and Marshall, Madeline A and Mazzolari, Giovanni and Parlanti, Eleonora and Rawle, Tim and Scholtz, Jan and Venturi, Giacomo and Witstok, Joris},
    title = "{GA-NIFS: JWST discovers an offset AGN 740 million years after the big bang}",
    journal = {MNRAS},
    volume = {531},
    number = {1},
    pages = {355-365},
    year = {2024},
    month = {05},
    issn = {0035-8711},
    doi = {10.1093/mnras/stae943},
    url = {https://doi.org/10.1093/mnras/stae943},
}

@ARTICLE{Gordon03,
       author = {{Gordon}, Karl D. and {Clayton}, Geoffrey C. and {Misselt}, K.~A. and {Landolt}, Arlo U. and {Wolff}, Michael J.},
        title = "{A Quantitative Comparison of the Small Magellanic Cloud, Large Magellanic Cloud, and Milky Way Ultraviolet to Near-Infrared Extinction Curves}",
      journal = {\apj},
         year = 2003,
        month = sep,
       volume = {594},
       number = {1},
        pages = {279-293},
          doi = {10.1086/376774},
archivePrefix = {arXiv},
       eprint = {astro-ph/0305257},
 primaryClass = {astro-ph},
       adsurl = {https://ui.adsabs.harvard.edu/abs/2003ApJ...594..279G}
}

@ARTICLE{McClymont24,
       author = {{McClymont}, William and {Tacchella}, Sandro and {D'Eugenio}, Francesco and {Witten}, Callum and {Ji}, Xihan and {Smith}, Aaron and {Maiolino}, Roberto and {Scholtz}, Jan and {Simmonds}, Charlotte and {Witstok}, Joris},
        title = "{The density-bounded twilight of starbursts in the early Universe}",
      journal = {arXiv e-prints},
         year = 2024,
        month = may,
          eid = {arXiv:2405.15859},
        pages = {arXiv:2405.15859, submitted to MNRAS},
          doi = {10.48550/arXiv.2405.15859},
archivePrefix = {arXiv},
       eprint = {2405.15859},
 primaryClass = {astro-ph.GA},
       adsurl = {https://ui.adsabs.harvard.edu/abs/2024arXiv240515859M}
}

@ARTICLE{Ferruit22,
       author = {{Ferruit}, P. and {Jakobsen}, P. and {Giardino}, G. and {Rawle}, T. and {Alves de Oliveira}, C. and {Arribas}, S. and {Beck}, T.~L. and {Birkmann}, S. and {B{\"o}ker}, T. and {Bunker}, A.~J. and {Charlot}, S. and {de Marchi}, G. and {Franx}, M. and {Henry}, A. and {Karakla}, D. and {Kassin}, S.~A. and {Kumari}, N. and {L{\'o}pez-Caniego}, M. and {L{\"u}tzgendorf}, N. and {Maiolino}, R. and {Manjavacas}, E. and {Marston}, A. and {Moseley}, S.~H. and {Muzerolle}, J. and {Pirzkal}, N. and {Rauscher}, B. and {Rix}, H. -W. and {Sabbi}, E. and {Sirianni}, M. and {te Plate}, M. and {Valenti}, J. and {Willott}, C.~J. and {Zeidler}, P.},
        title = "{The Near-Infrared Spectrograph (NIRSpec) on the James Webb Space Telescope. II. Multi-object spectroscopy (MOS)}",
      journal = {\aap},
         year = 2022,
        month = may,
       volume = {661},
          eid = {A81},
        pages = {A81},
          doi = {10.1051/0004-6361/202142673},
archivePrefix = {arXiv},
       eprint = {2202.03306},
 primaryClass = {astro-ph.IM},
       adsurl = {https://ui.adsabs.harvard.edu/abs/2022A&A...661A..81F}
}

@ARTICLE{deGraaff24,
       author = {{de Graaff}, Anna and {Rix}, Hans-Walter and {Carniani}, Stefano and {Suess}, Katherine A. and {Charlot}, St{\'e}phane and {Curtis-Lake}, Emma and {Arribas}, Santiago and {Baker}, William M. and {Boyett}, Kristan and {Bunker}, Andrew J. and {Cameron}, Alex J. and {Chevallard}, Jacopo and {Curti}, Mirko and {Eisenstein}, Daniel J. and {Franx}, Marijn and {Hainline}, Kevin and {Hausen}, Ryan and {Ji}, Zhiyuan and {Johnson}, Benjamin D. and {Jones}, Gareth C. and {Maiolino}, Roberto and {Maseda}, Michael V. and {Nelson}, Erica and {Parlanti}, Eleonora and {Rawle}, Tim and {Robertson}, Brant and {Tacchella}, Sandro and {{\"U}bler}, Hannah and {Williams}, Christina C. and {Willmer}, Christopher N.~A. and {Willott}, Chris},
        title = "{Ionised gas kinematics and dynamical masses of z {\ensuremath{\gtrsim}} 6 galaxies from JADES/NIRSpec high-resolution spectroscopy}",
      journal = {\aap},
         year = 2024,
        month = apr,
       volume = {684},
          eid = {A87},
        pages = {A87},
          doi = {10.1051/0004-6361/202347755},
archivePrefix = {arXiv},
       eprint = {2308.09742},
 primaryClass = {astro-ph.GA},
       adsurl = {https://ui.adsabs.harvard.edu/abs/2024A&A...684A..87D}
}

@ARTICLE{Maiolino23a,
       author = {{Maiolino}, Roberto and {Scholtz}, Jan and {Witstok}, Joris and {Carniani}, Stefano and {D'Eugenio}, Francesco and {de Graaff}, Anna and {{\"U}bler}, Hannah and {Tacchella}, Sandro and {Curtis-Lake}, Emma and {Arribas}, Santiago and {Bunker}, Andrew and {Charlot}, St{\'e}phane and {Chevallard}, Jacopo and {Curti}, Mirko and {Looser}, Tobias J. and {Maseda}, Michael V. and {Rawle}, Timothy D. and {Rodr{\'\i}guez del Pino}, Bruno and {Willott}, Chris J. and {Egami}, Eiichi and {Eisenstein}, Daniel J. and {Hainline}, Kevin N. and {Robertson}, Brant and {Williams}, Christina C. and {Willmer}, Christopher N.~A. and {Baker}, William M. and {Boyett}, Kristan and {DeCoursey}, Christa and {Fabian}, Andrew C. and {Helton}, Jakob M. and {Ji}, Zhiyuan and {Jones}, Gareth C. and {Kumari}, Nimisha and {Laporte}, Nicolas and {Nelson}, Erica J. and {Perna}, Michele and {Sandles}, Lester and {Shivaei}, Irene and {Sun}, Fengwu},
        title = "{A small and vigorous black hole in the early Universe}",
      journal = {\nat},
         year = 2024,
        month = mar,
       volume = {627},
       number = {8002},
        pages = {59-63},
          doi = {10.1038/s41586-024-07052-5},
archivePrefix = {arXiv},
       eprint = {2305.12492},
 primaryClass = {astro-ph.GA},
       adsurl = {https://ui.adsabs.harvard.edu/abs/2024Natur.627...59M}
}

@ARTICLE{Maiolino23c,
         author = {{Maiolino}, Roberto and {Scholtz}, Jan and {Curtis-Lake}, Emma and {Carniani}, Stefano and {Baker}, William and {de Graaff}, Anna and {Tacchella}, Sandro and {{\"U}bler}, Hannah and {D'Eugenio}, Francesco and {Witstok}, Joris and {Curti}, Mirko and {Arribas}, Santiago and {Bunker}, Andrew J. and {Charlot}, St{\'e}phane and {Chevallard}, Jacopo and {Eisenstein}, Daniel J. and {Egami}, Eiichi and {Ji}, Zhiyuan and {Jones}, Gareth C. and {Lyu}, Jianwei and {Rawle}, Tim and {Robertson}, Brant and {Rujopakarn}, Wiphu and {Perna}, Michele and {Sun}, Fengwu and {Venturi}, Giacomo and {Williams}, Christina C. and {Willott}, Chris},
        title = "{JADES: The diverse population of infant black holes at 4 < z < 11: Merging, tiny, poor, but mighty}",
      journal = {\aap},
         year = 2024,
        month = nov,
       volume = {691},
          eid = {A145},
        pages = {A145},
          doi = {10.1051/0004-6361/202347640},
archivePrefix = {arXiv},
       eprint = {2308.01230},
 primaryClass = {astro-ph.GA},
       adsurl = {https://ui.adsabs.harvard.edu/abs/2024A&A...691A.145M}
}

@ARTICLE{Chisholm24,
       author = {{Chisholm}, J. and {Berg}, D.~A. and {Endsley}, R. and {Gazagnes}, S. and {Richardson}, C.~T. and {Lambrides}, E. and {Greene}, J. and {Finkelstein}, S. and {Flury}, S. and {Guseva}, N.~G. and {Henry}, A. and {Hutchison}, T.~A. and {Izotov}, Y.~I. and {Marques-Chaves}, R. and {Oesch}, P. and {Papovich}, C. and {Saldana-Lopez}, A. and {Schaerer}, D. and {Stephenson}, M.~G.},
        title = "{[Ne v] emission from a faint epoch of reionization-era galaxy: evidence for a narrow-line intermediate-mass black hole}",
      journal = {\mnras},
         year = 2024,
        month = nov,
       volume = {534},
       number = {3},
        pages = {2633-2652},
          doi = {10.1093/mnras/stae2199},
archivePrefix = {arXiv},
       eprint = {2402.18643},
 primaryClass = {astro-ph.GA},
       adsurl = {https://ui.adsabs.harvard.edu/abs/2024MNRAS.534.2633C}
}

@ARTICLE{Scholtz_2023_GN-z11,
           author = {{Scholtz}, Jan and {Witten}, Callum and {Laporte}, Nicolas and {{\"U}bler}, Hannah and {Perna}, Michele and {Maiolino}, Roberto and {Arribas}, Santiago and {Baker}, William M. and {Bennett}, Jake S. and {D'Eugenio}, Francesco and {Simmonds}, Charlotte and {Tacchella}, Sandro and {Witstok}, Joris and {Bunker}, Andrew J. and {Carniani}, Stefano and {Charlot}, St{\'e}phane and {Cresci}, Giovanni and {Curtis-Lake}, Emma and {Eisenstein}, Daniel J. and {Kumari}, Nimisha and {Robertson}, Brant and {Rodr{\'\i}guez Del Pino}, Bruno and {Smit}, Renske and {Venturi}, Giacomo and {Williams}, Christina C. and {Willmer}, Christopher N.~A.},
        title = "{GN-z11: The environment of an active galactic nucleus at z = 10.603. New insights into the most distant Ly{\ensuremath{\alpha}} detection}",
      journal = {\aap},
         year = 2024,
        month = jul,
       volume = {687},
          eid = {A283},
        pages = {A283},
          doi = {10.1051/0004-6361/202347187},
archivePrefix = {arXiv},
       eprint = {2306.09142},
 primaryClass = {astro-ph.GA},
       adsurl = {https://ui.adsabs.harvard.edu/abs/2024A&A...687A.283S}
}

@ARTICLE{Furtak23,
       author = {{Furtak}, Lukas J. and {Labb{\'e}}, Ivo and {Zitrin}, Adi and {Greene}, Jenny E. and {Dayal}, Pratika and {Chemerynska}, Iryna and {Kokorev}, Vasily and {Miller}, Tim B. and {Goulding}, Andy D. and {de Graaff}, Anna and {Bezanson}, Rachel and {Brammer}, Gabriel B. and {Cutler}, Sam E. and {Leja}, Joel and {Pan}, Richard and {Price}, Sedona H. and {Wang}, Bingjie and {Weaver}, John R. and {Whitaker}, Katherine E. and {Atek}, Hakim and {Bogd{\'a}n}, {\'A}kos and {Charlot}, St{\'e}phane and {Curtis-Lake}, Emma and {van Dokkum}, Pieter and {Endsley}, Ryan and {Feldmann}, Robert and {Fudamoto}, Yoshinobu and {Fujimoto}, Seiji and {Glazebrook}, Karl and {Juneau}, St{\'e}phanie and {Marchesini}, Danilo and {Maseda}, Micheal V. and {Nelson}, Erica and {Oesch}, Pascal A. and {Plat}, Ad{\`e}le and {Setton}, David J. and {Stark}, Daniel P. and {Williams}, Christina C.},
        title = "{A high black-hole-to-host mass ratio in a lensed AGN in the early Universe}",
      journal = {\nat},
         year = 2024,
        month = apr,
       volume = {628},
       number = {8006},
        pages = {57-61},
          doi = {10.1038/s41586-024-07184-8},
archivePrefix = {arXiv},
       eprint = {2308.05735},
 primaryClass = {astro-ph.GA},
       adsurl = {https://ui.adsabs.harvard.edu/abs/2024Natur.628...57F}
}

@ARTICLE{Juodzbalis24,
       author = {{Juod{\v{z}}balis}, Ignas and {Maiolino}, Roberto and {Baker}, William M. and {Tacchella}, Sandro and {Scholtz}, Jan and {D'Eugenio}, Francesco and {Schneider}, Raffaella and {Trinca}, Alessandro and {Valiante}, Rosa and {DeCoursey}, Christa and {Curti}, Mirko and {Carniani}, Stefano and {Chevallard}, Jacopo and {de Graaff}, Anna and {Arribas}, Santiago and {Bennett}, Jake S. and {Bourne}, Martin A. and {Bunker}, Andrew J. and {Charlot}, St{\'e}phane and {Jiang}, Brian and {Koudmani}, Sophie and {Perna}, Michele and {Robertson}, Brant and {Sijacki}, Debora and {{\"U}bler}, Hannah and {Williams}, Christina C. and {Willott}, Chris and {Witstok}, Joris},
        title = "{A dormant, overmassive black hole in the early Universe}",
      journal = {arXiv e-prints},
         year = 2024,
        month = mar,
          eid = {arXiv:2403.03872},
        pages = {arXiv:2403.03872, accepted for publication on Nature},
          doi = {10.48550/arXiv.2403.03872},
archivePrefix = {arXiv},
       eprint = {2403.03872},
 primaryClass = {astro-ph.GA},
       adsurl = {https://ui.adsabs.harvard.edu/abs/2024arXiv240303872J}
}

@ARTICLE{Rigby23,
       author = {{Rigby}, Jane and {Perrin}, Marshall and {McElwain}, Michael and {Kimble}, Randy and {Friedman}, Scott and {Lallo}, Matt and {Doyon}, Ren{\'e} and {Feinberg}, Lee and {Ferruit}, Pierre and {Glasse}, Alistair and {Rieke}, Marcia and {Rieke}, George and {Wright}, Gillian and {Willott}, Chris and {Colon}, Knicole and {Milam}, Stefanie and {Neff}, Susan and {Stark}, Christopher and {Valenti}, Jeff and {Abell}, Jim and {Abney}, Faith and {Abul-Huda}, Yasin and {Acton}, D. Scott and {Adams}, Evan and {Adler}, David and {Aguilar}, Jonathan and {Ahmed}, Nasif and {Albert}, Lo{\"\i}c and {Alberts}, Stacey and {Aldridge}, David and {Allen}, Marsha and {Altenburg}, Martin and {{\'A}lvarez-M{\'a}rquez}, Javier and {Alves de Oliveira}, Catarina and {Andersen}, Greg and {Anderson}, Harry and {Anderson}, Sara and {Argyriou}, Ioannis and {Armstrong}, Amber and {Arribas}, Santiago and {Artigau}, Etienne and {Arvai}, Amanda and {Atkinson}, Charles and {Bacon}, Gregory and {Bair}, Thomas and {Banks}, Kimberly and {Barrientes}, Jaclyn and {Barringer}, Bruce and {Bartosik}, Peter and {Bast}, William and {Baudoz}, Pierre and {Beatty}, Thomas and {Bechtold}, Katie and {Beck}, Tracy and {Bergeron}, Eddie and {Bergkoetter}, Matthew and {Bhatawdekar}, Rachana and {Birkmann}, Stephan and {Blazek}, Ronald and {Blome}, Claire and {Boccaletti}, Anthony and {B{\"o}ker}, Torsten and {Boia}, John and {Bonaventura}, Nina and {Bond}, Nicholas and {Bosley}, Kari and {Boucarut}, Ray and {Bourque}, Matthew and {Bouwman}, Jeroen and {Bower}, Gary and {Bowers}, Charles and {Boyer}, Martha and {Bradley}, Larry and {Brady}, Greg and {Braun}, Hannah and {Breda}, David and {Bresnahan}, Pamela and {Bright}, Stacey and {Britt}, Christopher and {Bromenschenkel}, Asa and {Brooks}, Brian and {Brooks}, Keira and {Brown}, Bob and {Brown}, Matthew and {Brown}, Patricia and {Bunker}, Andy and {Burger}, Matthew and {Bushouse}, Howard and {Cale}, Steven and {Cameron}, Alex and {Cameron}, Peter and {Canipe}, Alicia and {Caplinger}, James and {Caputo}, Francis and {Cara}, Mihai and {Carey}, Larkin and {Carniani}, Stefano and {Carrasquilla}, Maria and {Carruthers}, Margaret and {Case}, Michael and {Catherine}, Riggs and {Chance}, Don and {Chapman}, George and {Charlot}, St{\'e}phane and {Charlow}, Brian and {Chayer}, Pierre and {Chen}, Bin and {Cherinka}, Brian and {Chichester}, Sarah and {Chilton}, Zack and {Chonis}, Taylor and {Clampin}, Mark and {Clark}, Charles and {Clark}, Kerry and {Coe}, Dan and {Coleman}, Benee and {Comber}, Brian and {Comeau}, Tom and {Connolly}, Dennis and {Cooper}, James and {Cooper}, Rachel and {Coppock}, Eric and {Correnti}, Matteo and {Cossou}, Christophe and {Coulais}, Alain and {Coyle}, Laura and {Cracraft}, Misty and {Curti}, Mirko and {Cuturic}, Steven and {Davis}, Katherine and {Davis}, Michael and {Dean}, Bruce and {DeLisa}, Amy and {deMeester}, Wim and {Dencheva}, Nadia and {Dencheva}, Nadezhda and {DePasquale}, Joseph and {Deschenes}, Jeremy and {Hunor Detre}, {\"O}rs and {Diaz}, Rosa and {Dicken}, Dan and {DiFelice}, Audrey and {Dillman}, Matthew and {Dixon}, William and {Doggett}, Jesse and {Donaldson}, Tom and {Douglas}, Rob and {DuPrie}, Kimberly and {Dupuis}, Jean and {Durning}, John and {Easmin}, Nilufar and {Eck}, Weston and {Edeani}, Chinwe and {Egami}, Eiichi and {Ehrenwinkler}, Ralf and {Eisenhamer}, Jonathan and {Eisenhower}, Michael and {Elie}, Michelle and {Elliott}, James and {Elliott}, Kyle and {Ellis}, Tracy and {Engesser}, Michael and {Espinoza}, Nestor and {Etienne}, Odessa and {Etxaluze}, Mireya and {Falini}, Patrick and {Feeney}, Matthew and {Ferry}, Malcolm and {Filippazzo}, Joseph and {Fincham}, Brian and {Fix}, Mees and {Flagey}, Nicolas and {Florian}, Michael and {Flynn}, Jim and {Fontanella}, Erin and {Ford}, Terrance and {Forshay}, Peter and {Fox}, Ori and {Franz}, David and {Fu}, Henry and {Fullerton}, Alexander and {Galkin}, Sergey and {Galyer}, Anthony and {Garc{\'\i}a Mar{\'\i}n}, Macarena and {Gardner}, Jonathan P. and {Gardner}, Lisa and {Garland}, Dennis and {Garrett}, Bruce and {Gasman}, Danny and {Gaspar}, Andras and {Gaudreau}, Daniel and {Gauthier}, Peter and {Geers}, Vincent and {Geithner}, Paul and {Gennaro}, Mario and {Giardino}, Giovanna and {Girard}, Julien and {Giuliano}, Mark and {Glassmire}, Kirk and {Glauser}, Adrian and {Glazer}, Stuart and {Godfrey}, John and {Golimowski}, David and {Gollnitz}, David and {Gong}, Fan and {Gonzaga}, Shireen and {Gordon}, Michael and {Gordon}, Karl and {Goudfrooij}, Paul and {Greene}, Thomas and {Greenhouse}, Matthew and {Grimaldi}, Stefano and {Groebner}, Andrew and {Grundy}, Timothy and {Guillard}, Pierre and {Gutman}, Irvin and {Ha}, Kong Q. and {Haderlein}, Peter and {Hagedorn}, Andria and {Hainline}, Kevin and {Haley}, Craig and {Hami}, Maryam and {Hamilton}, Forrest and {Hammel}, Heidi and {Hansen}, Carl and {Harkins}, Tom and {Harr}, Michael and {Hart}, Jessica and {Hart}, Quyen and {Hartig}, George and {Hashimoto}, Ryan and {Haskins}, Sujee and {Hathaway}, William and {Havey}, Keith and {Hayden}, Brian and {Hecht}, Karen and {Heller-Boyer}, Chris and {Henriques}, Caroline and {Henry}, Alaina and {Hermann}, Karl and {Hernandez}, Scarlin and {Hesman}, Brigette and {Hicks}, Brian and {Hilbert}, Bryan and {Hines}, Dean and {Hoffman}, Melissa and {Holfeltz}, Sherie and {Holler}, Bryan J. and {Hoppa}, Jennifer and {Hott}, Kyle and {Howard}, Joseph M. and {Howard}, Rick and {Hunter}, Alexander and {Hunter}, David and {Hurst}, Brendan and {Husemann}, Bernd and {Hustak}, Leah and {Ilinca Ignat}, Luminita and {Illingworth}, Garth and {Irish}, Sandra and {Jackson}, Wallace and {Jahromi}, Amir and {Jakobsen}, Peter and {James}, LeAndrea and {James}, Bryan and {Januszewski}, William and {Jenkins}, Ann and {Jirdeh}, Hussein and {Johnson}, Phillip and {Johnson}, Timothy and {Jones}, Vicki and {Jones}, Ron and {Jones}, Danny and {Jones}, Olivia and {Jordan}, Ian and {Jordan}, Margaret and {Jurczyk}, Sarah and {Jurling}, Alden and {Kaleida}, Catherine and {Kalmanson}, Phillip and {Kammerer}, Jens and {Kang}, Huijo and {Kao}, Shaw-Hong and {Karakla}, Diane and {Kavanagh}, Patrick and {Kelly}, Doug and {Kendrew}, Sarah and {Kennedy}, Herbert and {Kenny}, Deborah and {Keski-kuha}, Ritva and {Keyes}, Charles and {Kidwell}, Richard and {Kinzel}, Wayne and {Kirk}, Jeff and {Kirkpatrick}, Mark and {Kirshenblat}, Danielle and {Klaassen}, Pamela and {Knapp}, Bryan and {Knight}, J. Scott and {Knollenberg}, Perry and {Koehler}, Robert and {Koekemoer}, Anton and {Kovacs}, Aiden and {Kulp}, Trey and {Kumari}, Nimisha and {Kyprianou}, Mark and {La Massa}, Stephanie and {Labador}, Aurora and {Labiano}, Alvaro and {Lagage}, Pierre-Olivier and {Lajoie}, Charles-Philippe and {Lallo}, Matthew and {Lam}, May and {Lamb}, Tracy and {Lambros}, Scott and {Lampenfield}, Richard and {Langston}, James and {Larson}, Kirsten and {Law}, David and {Lawrence}, Jon and {Lee}, David and {Leisenring}, Jarron and {Lepo}, Kelly and {Leveille}, Michael and {Levenson}, Nancy and {Levine}, Marie and {Levy}, Zena and {Lewis}, Dan and {Lewis}, Hannah and {Libralato}, Mattia and {Lightsey}, Paul and {Link}, Miranda and {Liu}, Lily and {Lo}, Amy and {Lockwood}, Alexandra and {Logue}, Ryan and {Long}, Chris and {Long}, Douglas and {Loomis}, Charles and {Lopez-Caniego}, Marcos and {Lorenzo Alvarez}, Jose and {Love-Pruitt}, Jennifer and {Lucy}, Adrian and {Luetzgendorf}, Nora and {Maghami}, Peiman and {Maiolino}, Roberto and {Major}, Melissa and {Malla}, Sunita and {Malumuth}, Eliot and {Manjavacas}, Elena and {Mannfolk}, Crystal and {Marrione}, Amanda and {Marston}, Anthony and {Martel}, Andr{\'e} and {Maschmann}, Marc and {Masci}, Gregory and {Masciarelli}, Michaela and {Maszkiewicz}, Michael and {Mather}, John and {McKenzie}, Kenny and {McLean}, Brian and {McMaster}, Matthew and {Melbourne}, Katie and {Mel{\'e}ndez}, Marcio and {Menzel}, Michael and {Merz}, Kaiya and {Meyett}, Michele and {Meza}, Luis and {Miskey}, Cherie and {Misselt}, Karl and {Moller}, Christopher and {Morrison}, Jane and {Morse}, Ernie and {Moseley}, Harvey and {Mosier}, Gary and {Mountain}, Matt and {Mueckay}, Julio and {Mueller}, Michael and {Mullally}, Susan and {Murphy}, Jess and {Murray}, Katherine and {Murray}, Claire and {Mustelier}, David and {Muzerolle}, James and {Mycroft}, Matthew and {Myers}, Richard and {Myrick}, Kaila and {Nanavati}, Shashvat and {Nance}, Elizabeth and {Nayak}, Omnarayani and {Naylor}, Bret and {Nelan}, Edmund and {Nickson}, Bryony and {Nielson}, Alethea and {Nieto-Santisteban}, Maria and {Nikolov}, Nikolay and {Noriega-Crespo}, Alberto and {O'Shaughnessy}, Brian and {O'Sullivan}, Brian and {Ochs}, William and {Ogle}, Patrick and {Oleszczuk}, Brenda and {Olmsted}, Joseph and {Osborne}, Shannon and {Ottens}, Richard and {Owens}, Beverly and {Pacifici}, Camilla and {Pagan}, Alyssa and {Page}, James and {Park}, Sang and {Parrish}, Keith and {Patapis}, Polychronis and {Paul}, Lee and {Pauly}, Tyler and {Pavlovsky}, Cheryl and {Pedder}, Andrew and {Peek}, Matthew and {Pena-Guerrero}, Maria and {Penanen}, Konstantin and {Perez}, Yesenia and {Perna}, Michele and {Perriello}, Beth and {Phillips}, Kevin and {Pietraszkiewicz}, Martin and {Pinaud}, Jean-Paul and {Pirzkal}, Norbert and {Pitman}, Joseph and {Piwowar}, Aidan and {Platais}, Vera and {Player}, Danielle and {Plesha}, Rachel and {Pollizi}, Joe and {Polster}, Ethan and {Pontoppidan}, Klaus and {Porterfield}, Blair and {Proffitt}, Charles and {Pueyo}, Laurent and {Pulliam}, Christine and {Quirt}, Brian and {Quispe Neira}, Irma and {Ramos Alarcon}, Rafael and {Ramsay}, Leah and {Rapp}, Greg and {Rapp}, Robert and {Rauscher}, Bernard and {Ravindranath}, Swara and {Rawle}, Timothy and {Regan}, Michael and {Reichard}, Timothy A. and {Reis}, Carl and {Ressler}, Michael E. and {Rest}, Armin and {Reynolds}, Paul and {Rhue}, Timothy and {Richon}, Karen and {Rickman}, Emily and {Ridgaway}, Michael and {Ritchie}, Christine and {Rix}, Hans-Walter and {Robberto}, Massimo and {Robinson}, Gregory and {Robinson}, Michael and {Robinson}, Orion and {Rock}, Frank and {Rodriguez}, David and {Rodriguez Del Pino}, Bruno and {Roellig}, Thomas and {Rohrbach}, Scott and {Roman}, Anthony and {Romelfanger}, Fred and {Rose}, Perry and {Roteliuk}, Anthony and {Roth}, Marc and {Rothwell}, Braden and {Rowlands}, Neil and {Roy}, Arpita and {Royer}, Pierre and {Royle}, Patricia and {Rui}, Chunlei and {Rumler}, Peter and {Runnels}, Joel and {Russ}, Melissa and {Rustamkulov}, Zafar and {Ryden}, Grant and {Ryer}, Holly and {Sabata}, Modhumita and {Sabatke}, Derek and {Sabbi}, Elena and {Samuelson}, Bridget and {Sapp}, Benjamin and {Sappington}, Bradley and {Sargent}, B. and {Sauer}, Arne and {Scheithauer}, Silvia and {Schlawin}, Everett and {Schlitz}, Joseph and {Schmitz}, Tyler and {Schneider}, Analyn and {Schreiber}, J{\"u}rgen and {Schulze}, Vonessa and {Schwab}, Ryan and {Scott}, John and {Sembach}, Kenneth and {Shanahan}, Clare and {Shaughnessy}, Bryan and {Shaw}, Richard and {Shawger}, Nanci and {Shay}, Christopher and {Sheehan}, Evan and {Shen}, Sharon and {Sherman}, Allan and {Shiao}, Bernard and {Shih}, Hsin-Yi and {Shivaei}, Irene and {Sienkiewicz}, Matthew and {Sing}, David and {Sirianni}, Marco and {Sivaramakrishnan}, Anand and {Skipper}, Joy and {Sloan}, G.~C. and {Slocum}, Christine and {Slowinski}, Steven and {Smith}, Erin and {Smith}, Eric and {Smith}, Denise and {Smith}, Corbett and {Snyder}, Gregory and {Soh}, Warren and {Sohn}, Sangmo Tony and {Soto}, Christian and {Spencer}, Richard and {Stallcup}, Scott and {Stansberry}, John and {Starr}, Carl and {Starr}, Elysia and {Stewart}, Alphonso and {Stiavelli}, Massimo and {Straughn}, Amber and {Strickland}, David and {Stys}, Jeff and {Summers}, Francis and {Sun}, Fengwu and {Sunnquist}, Ben and {Swade}, Daryl and {Swam}, Michael and {Swaters}, Robert and {Swoish}, Robby and {Taylor}, Joanna M. and {Taylor}, Rolanda and {Te Plate}, Maurice and {Tea}, Mason and {Teague}, Kelly and {Telfer}, Randal and {Temim}, Tea and {Thatte}, Deepashri and {Thompson}, Christopher and {Thompson}, Linda and {Thomson}, Shaun and {Tikkanen}, Tuomo and {Tippet}, William and {Todd}, Connor and {Toolan}, Sharon and {Tran}, Hien and {Trejo}, Edwin and {Truong}, Justin and {Tsukamoto}, Chris and {Tustain}, Samuel and {Tyra}, Harrison and {Ubeda}, Leonardo and {Underwood}, Kelli and {Uzzo}, Michael and {Van Campen}, Julie and {Vandal}, Thomas and {Vandenbussche}, Bart and {Vila}, Bego{\~n}a and {Volk}, Kevin and {Wahlgren}, Glenn and {Waldman}, Mark and {Walker}, Chanda and {Wander}, Michel and {Warfield}, Christine and {Warner}, Gerald and {Wasiak}, Matthew and {Watkins}, Mitchell and {Weaver}, Andrew and {Weilert}, Mark and {Weiser}, Nick and {Weiss}, Ben and {Weissman}, Sarah and {Welty}, Alan and {West}, Garrett and {Wheate}, Lauren and {Wheatley}, Elizabeth and {Wheeler}, Thomas and {White}, Rick and {Whiteaker}, Kevin and {Whitehouse}, Paul and {Whiteleather}, Jennifer and {Whitman}, William and {Williams}, Christina and {Willmer}, Christopher and {Willoughby}, Scott and {Wilson}, Andrew and {Wirth}, Gregory and {Wislowski}, Emily and {Wolf}, Erin and {Wolfe}, David and {Wolff}, Schuyler and {Workman}, Bill and {Wright}, Ray and {Wu}, Carl and {Wu}, Rai and {Wymer}, Kristen and {Yates}, Kayla and {Yeager}, Christopher and {Yeates}, Jared and {Yerger}, Ethan and {Yoon}, Jinmi and {Young}, Alice and {Yu}, Susan and {Zak}, Dean and {Zeidler}, Peter and {Zhou}, Julia and {Zielinski}, Thomas and {Zincke}, Cristian and {Zonak}, Stephanie},
        title = "{The Science Performance of JWST as Characterized in Commissioning}",
      journal = {\pasp},
         year = 2023,
        month = apr,
       volume = {135},
       number = {1046},
          eid = {048001},
        pages = {048001},
          doi = {10.1088/1538-3873/acb293},
archivePrefix = {arXiv},
       eprint = {2207.05632},
 primaryClass = {astro-ph.IM},
       adsurl = {https://ui.adsabs.harvard.edu/abs/2023PASP..135d8001R}
}

@ARTICLE{Gardner23,
       author = {{Gardner}, Jonathan P. and {Mather}, John C. and {Abbott}, Randy and {Abell}, James S. and {Abernathy}, Mark and {Abney}, Faith E. and {Abraham}, John G. and {Abraham}, Roberto and {Abul-Huda}, Yasin M. and {Acton}, Scott and {Adams}, Cynthia K. and {Adams}, Evan and {Adler}, David S. and {Adriaensen}, Maarten and {Aguilar}, Jonathan Albert and {Ahmed}, Mansoor and {Ahmed}, Nasif S. and {Ahmed}, Tanjira and {Albat}, R{\"u}deger and {Albert}, Lo{\"\i}c and {Alberts}, Stacey and {Aldridge}, David and {Allen}, Mary Marsha and {Allen}, Shaune S. and {Altenburg}, Martin and {Altunc}, Serhat and {Alvarez}, Jose Lorenzo and {{\'A}lvarez-M{\'a}rquez}, Javier and {de Oliveira}, Catarina Alves and {Ambrose}, Leslie L. and {Anandakrishnan}, Satya M. and {Andersen}, Gregory C. and {Anderson}, Harry James and {Anderson}, Jay and {Anderson}, Kristen and {Anderson}, Sara M. and {Aprea}, Julio and {Archer}, Benita J. and {Arenberg}, Jonathan W. and {Argyriou}, Ioannis and {Arribas}, Santiago and {Artigau}, {\'E}tienne and {Arvai}, Amanda Rose and {Atcheson}, Paul and {Atkinson}, Charles B. and {Averbukh}, Jesse and {Aymergen}, Cagatay and {Bacinski}, John J. and {Baggett}, Wayne E. and {Bagnasco}, Giorgio and {Baker}, Lynn L. and {Balzano}, Vicki Ann and {Banks}, Kimberly A. and {Baran}, David A. and {Barker}, Elizabeth A. and {Barrett}, Larry K. and {Barringer}, Bruce O. and {Barto}, Allison and {Bast}, William and {Baudoz}, Pierre and {Baum}, Stefi and {Beatty}, Thomas G. and {Beaulieu}, Mathilde and {Bechtold}, Kathryn and {Beck}, Tracy and {Beddard}, Megan M. and {Beichman}, Charles and {Bellagama}, Larry and {Bely}, Pierre and {Berger}, Timothy W. and {Bergeron}, Louis E. and {Bernier}, Antoine-Darveau and {Bertch}, Maria D. and {Beskow}, Charlotte and {Betz}, Laura E. and {Biagetti}, Carl P. and {Birkmann}, Stephan and {Bjorklund}, Kurt F. and {Blackwood}, James D. and {Blazek}, Ronald Paul and {Blossfeld}, Stephen and {Bluth}, Marcel and {Boccaletti}, Anthony and {Boegner}, Martin E., Jr. and {Bohlin}, Ralph C. and {Boia}, John Joseph and {B{\"o}ker}, Torsten and {Bonaventura}, N. and {Bond}, Nicholas A. and {Bosley}, Kari Ann and {Boucarut}, Rene A. and {Bouchet}, Patrice and {Bouwman}, Jeroen and {Bower}, Gary and {Bowers}, Ariel S. and {Bowers}, Charles W. and {Boyce}, Leslye A. and {Boyer}, Christine T. and {Boyer}, Martha L. and {Boyer}, Michael and {Boyer}, Robert and {Bradley}, Larry D. and {Brady}, Gregory R. and {Brandl}, Bernhard R. and {Brannen}, Judith L. and {Breda}, David and {Bremmer}, Harold G. and {Brennan}, David and {Bresnahan}, Pamela A. and {Bright}, Stacey N. and {Broiles}, Brian J. and {Bromenschenkel}, Asa and {Brooks}, Brian H. and {Brooks}, Keira J. and {Brown}, Bob and {Brown}, Bruce and {Brown}, Thomas M. and {Bruce}, Barry W. and {Bryson}, Jonathan G. and {Bujanda}, Edwin D. and {Bullock}, Blake M. and {Bunker}, A.~J. and {Bureo}, Rafael and {Burt}, Irving J. and {Bush}, James Aaron and {Bushouse}, Howard A. and {Bussman}, Marie C. and {Cabaud}, Olivier and {Cale}, Steven and {Calhoon}, Charles D. and {Calvani}, Humberto and {Canipe}, Alicia M. and {Caputo}, Francis M. and {Cara}, Mihai and {Carey}, Larkin and {Case}, Michael Eli and {Cesari}, Thaddeus and {Cetorelli}, Lee D. and {Chance}, Don R. and {Chandler}, Lynn and {Chaney}, Dave and {Chapman}, George N. and {Charlot}, S. and {Chayer}, Pierre and {Cheezum}, Jeffrey I. and {Chen}, Bin and {Chen}, Christine H. and {Cherinka}, Brian and {Chichester}, Sarah C. and {Chilton}, Zachary S. and {Chittiraibalan}, Dharini and {Clampin}, Mark and {Clark}, Charles R. and {Clark}, Kerry W. and {Clark}, Stephanie M. and {Claybrooks}, Edward E. and {Cleveland}, Keith A. and {Cohen}, Andrew L. and {Cohen}, Lester M. and {Col{\'o}n}, Knicole D. and {Coleman}, Benee L. and {Colina}, Luis and {Comber}, Brian J. and {Comeau}, Thomas M. and {Comer}, Thomas and {Reis}, Alain Conde and {Connolly}, Dennis C. and {Conroy}, Kyle E. and {Contos}, Adam R. and {Contreras}, James and {Cook}, Neil J. and {Cooper}, James L. and {Cooper}, Rachel Aviva and {Correia}, Michael F. and {Correnti}, Matteo and {Cossou}, Christophe and {Costanza}, Brian F. and {Coulais}, Alain and {Cox}, Colin R. and {Coyle}, Ray T. and {Cracraft}, Misty M. and {Crew}, Keith A. and {Curtis}, Gary J. and {Cusveller}, Bianca and {Maciel}, Cleyciane Da Costa and {Dailey}, Christopher T. and {Daugeron}, Fr{\'e}d{\'e}ric and {Davidson}, Greg S. and {Davies}, James E. and {Davis}, Katherine Anne and {Davis}, Michael S. and {Day}, Ratna and {de Chambure}, Daniel and {de Jong}, Pauline and {De Marchi}, Guido and {Dean}, Bruce H. and {Decker}, John E. and {Delisa}, Amy S. and {Dell}, Lawrence C. and {Dellagatta}, Gail and {Dembinska}, Franciszka and {Demosthenes}, Sandor and {Dencheva}, Nadezhda M. and {Deneu}, Philippe and {DePriest}, William W. and {Deschenes}, Jeremy and {Dethienne}, Nathalie and {Detre}, {\"O}rs Hunor and {Diaz}, Rosa Izela and {Dicken}, Daniel and {DiFelice}, Audrey S. and {Dillman}, Matthew and {Disharoon}, Maureen O. and {Dixon}, William V. and {Doggett}, Jesse B. and {Dominguez}, Keisha L. and {Donaldson}, Thomas S. and {Doria-Warner}, Cristina M. and {Santos}, Tony Dos and {Doty}, Heather and {Douglas}, Robert E., Jr. and {Doyon}, Ren{\'e} and {Dressler}, Alan and {Driggers}, Jennifer and {Driggers}, Phillip A. and {Dunn}, Jamie L. and {DuPrie}, Kimberly C. and {Dupuis}, Jean and {Durning}, John and {Dutta}, Sanghamitra B. and {Earl}, Nicholas M. and {Eccleston}, Paul and {Ecobichon}, Pascal and {Egami}, Eiichi and {Ehrenwinkler}, Ralf and {Eisenhamer}, Jonathan D. and {Eisenhower}, Michael and {Eisenstein}, Daniel J. and {El Hamel}, Zaky and {Elie}, Michelle L. and {Elliott}, James and {Elliott}, Kyle Wesley and {Engesser}, Michael and {Espinoza}, N{\'e}stor and {Etienne}, Odessa and {Etxaluze}, Mireya and {Evans}, Leah and {Fabreguettes}, Luce and {Falcolini}, Massimo and {Falini}, Patrick R. and {Fatig}, Curtis and {Feeney}, Matthew and {Feinberg}, Lee D. and {Fels}, Raymond and {Ferdous}, Nazma and {Ferguson}, Henry C. and {Ferrarese}, Laura and {Ferreira}, Marie-H{\'e}l{\'e}ne and {Ferruit}, Pierre and {Ferry}, Malcolm and {Filippazzo}, Joseph Charles and {Firre}, Daniel and {Fix}, Mees and {Flagey}, Nicolas and {Flanagan}, Kathryn A. and {Fleming}, Scott W. and {Florian}, Michael and {Flynn}, James R. and {Foiadelli}, Luca and {Fontaine}, Mark R. and {Fontanella}, Erin Marie and {Forshay}, Peter Randolph and {Fortner}, Elizabeth A. and {Fox}, Ori D. and {Framarini}, Alexandro P. and {Francisco}, John I. and {Franck}, Randy and {Franx}, Marijn and {Franz}, David E. and {Friedman}, Scott D. and {Friend}, Katheryn E. and {Frost}, James R. and {Fu}, Henry and {Fullerton}, Alexander W. and {Gaillard}, Lionel and {Galkin}, Sergey and {Gallagher}, Ben and {Galyer}, Anthony D. and {Garc{\'\i}a Mar{\'\i}n}, Macarena and {Gardner}, Lisa E. and {Garland}, Dennis and {Garrett}, Bruce Albert and {Gasman}, Danny and {G{\'a}sp{\'a}r}, Andr{\'a}s and {Gastaud}, Ren{\'e} and {Gaudreau}, Daniel and {Gauthier}, Peter Timothy and {Geers}, Vincent and {Geithner}, Paul H. and {Gennaro}, Mario and {Gerber}, John and {Gereau}, John C. and {Giampaoli}, Robert and {Giardino}, Giovanna and {Gibbons}, Paul C. and {Gilbert}, Karoline and {Gilman}, Larry and {Girard}, Julien H. and {Giuliano}, Mark E. and {Gkountis}, Konstantinos and {Glasse}, Alistair and {Glassmire}, Kirk Zachary and {Glauser}, Adrian Michael and {Glazer}, Stuart D. and {Goldberg}, Joshua and {Golimowski}, David A. and {Gonzaga}, Shireen P. and {Gordon}, Karl D. and {Gordon}, Shawn J. and {Goudfrooij}, Paul and {Gough}, Michael J. and {Graham}, Adrian J. and {Grau}, Christopher M. and {Green}, Joel David and {Greene}, Gretchen R. and {Greene}, Thomas P. and {Greenfield}, Perry E. and {Greenhouse}, Matthew A. and {Greve}, Thomas R. and {Greville}, Edgar M. and {Grimaldi}, Stefano and {Groe}, Frank E. and {Groebner}, Andrew and {Grumm}, David M. and {Grundy}, Timothy and {G{\"u}del}, Manuel and {Guillard}, Pierre and {Guldalian}, John and {Gunn}, Christopher A. and {Gurule}, Anthony and {Gutman}, Irvin Meyer and {Guy}, Paul D. and {Guyot}, Benjamin and {Hack}, Warren J. and {Haderlein}, Peter and {Hagan}, James B. and {Hagedorn}, Andria and {Hainline}, Kevin and {Haley}, Craig and {Hami}, Maryam and {Hamilton}, Forrest Clifford and {Hammann}, Jeffrey and {Hammel}, Heidi B. and {Hanley}, Christopher J. and {Hansen}, Carl August and {Hardy}, Bruce and {Harnisch}, Bernd and {Harr}, Michael Hunter and {Harris}, Pamela and {Hart}, Jessica Ann and {Hartig}, George F. and {Hasan}, Hashima and {Hashim}, Kathleen Marie and {Hashimoto}, Ryan and {Haskins}, Sujee J. and {Hawkins}, Robert Edward and {Hayden}, Brian and {Hayden}, William L. and {Healy}, Mike and {Hecht}, Karen and {Heeg}, Vince J. and {Hejal}, Reem and {Helm}, Kristopher A. and {Hengemihle}, Nicholas J. and {Henning}, Thomas and {Henry}, Alaina and {Henry}, Ronald L. and {Henshaw}, Katherine and {Hernandez}, Scarlin and {Herrington}, Donald C. and {Heske}, Astrid and {Hesman}, Brigette Emily and {Hickey}, David L. and {Hilbert}, Bryan N. and {Hines}, Dean C. and {Hinz}, Michael R. and {Hirsch}, Michael and {Hitcho}, Robert S. and {Hodapp}, Klaus and {Hodge}, Philip E. and {Hoffman}, Melissa and {Holfeltz}, Sherie T. and {Holler}, Bryan Jason and {Hoppa}, Jennifer Rose and {Horner}, Scott and {Howard}, Joseph M. and {Howard}, Richard J. and {Huber}, Jean M. and {Hunkeler}, Joseph S. and {Hunter}, Alexander and {Hunter}, David Gavin and {Hurd}, Spencer W. and {Hurst}, Brendan J. and {Hutchings}, John B. and {Hylan}, Jason E. and {Ignat}, Luminita Ilinca and {Illingworth}, Garth and {Irish}, Sandra M. and {Isaacs}, John C., III and {Jackson}, Wallace C., Jr. and {Jaffe}, Daniel T. and {Jahic}, Jasmin and {Jahromi}, Amir and {Jakobsen}, Peter and {James}, Bryan and {James}, John C. and {James}, LeAndrea Rae and {Jamieson}, William Brian and {Jandra}, Raymond D. and {Jayawardhana}, Ray and {Jedrzejewski}, Robert and {Jeffers}, Basil S. and {Jensen}, Peter and {Joanne}, Egges and {Johns}, Alan T. and {Johnson}, Carl A. and {Johnson}, Eric L. and {Johnson}, Patricia and {Johnson}, Phillip Stephen and {Johnson}, Thomas K. and {Johnson}, Timothy W. and {Johnstone}, Doug and {Jollet}, Delphine and {Jones}, Danny P. and {Jones}, Gregory S. and {Jones}, Olivia C. and {Jones}, Ronald A. and {Jones}, Vicki and {Jordan}, Ian J. and {Jordan}, Margaret E. and {Jue}, Reginald and {Jurkowski}, Mark H. and {Justis}, Grant and {Justtanont}, Kay and {Kaleida}, Catherine C. and {Kalirai}, Jason S. and {Kalmanson}, Phillip Cabrales and {Kaltenegger}, Lisa and {Kammerer}, Jens and {Kan}, Samuel K. and {Kanarek}, Graham Childs and {Kao}, Shaw-Hong and {Karakla}, Diane M. and {Karl}, Hermann and {Kassin}, Susan A. and {Kauffman}, David D. and {Kavanagh}, Patrick and {Kelley}, Leigh L. and {Kelly}, Douglas M. and {Kendrew}, Sarah and {Kennedy}, Herbert V. and {Kenny}, Deborah A. and {Keski-Kuha}, Ritva A. and {Keyes}, Charles D. and {Khan}, Ali and {Kidwell}, Richard C. and {Kimble}, Randy A. and {King}, James S. and {King}, Richard C. and {Kinzel}, Wayne M. and {Kirk}, Jeffrey R. and {Kirkpatrick}, Marc E. and {Klaassen}, Pamela and {Klingemann}, Lana and {Klintworth}, Paul U. and {Knapp}, Bryan Adam and {Knight}, Scott and {Knollenberg}, Perry J. and {Knutsen}, Daniel Mark and {Koehler}, Robert and {Koekemoer}, Anton M. and {Kofler}, Earl T. and {Kontson}, Vicki L. and {Kovacs}, Aiden Rose and {Kozhurina-Platais}, Vera and {Krause}, Oliver and {Kriss}, Gerard A. and {Krist}, John and {Kristoffersen}, Monica R. and {Krogel}, Claudia and {Krueger}, Anthony P. and {Kulp}, Bernard A. and {Kumari}, Nimisha and {Kwan}, Sandy W. and {Kyprianou}, Mark and {Labador}, Aurora Gadiano and {Labiano}, {\'A}lvaro and {Lafreni{\`e}re}, David and {Lagage}, Pierre-Olivier and {Laidler}, Victoria G. and {Laine}, Benoit and {Laird}, Simon and {Lajoie}, Charles-Philippe and {Lallo}, Matthew D. and {Lam}, May Yen and {LaMassa}, Stephanie Marie and {Lambros}, Scott D. and {Lampenfield}, Richard Joseph and {Lander}, Matthew Ed and {Langston}, James Hutton and {Larson}, Kirsten and {Larson}, Melora and {LaVerghetta}, Robert Joseph and {Law}, David R. and {Lawrence}, Jon F. and {Lee}, David W. and {Lee}, Janice and {Lee}, Yat-Ning Paul and {Leisenring}, Jarron and {Leveille}, Michael Dunlap and {Levenson}, Nancy A. and {Levi}, Joshua S. and {Levine}, Marie B. and {Lewis}, Dan and {Lewis}, Jake and {Lewis}, Nikole and {Libralato}, Mattia and {Lidon}, Norbert and {Liebrecht}, Paula Louisa and {Lightsey}, Paul and {Lilly}, Simon and {Lim}, Frederick C. and {Lim}, Pey Lian and {Ling}, Sai-Kwong and {Link}, Lisa J. and {Link}, Miranda Nicole and {Lipinski}, Jamie L. and {Liu}, XiaoLi and {Lo}, Amy S. and {Lobmeyer}, Lynette and {Logue}, Ryan M. and {Long}, Chris A. and {Long}, Douglas R. and {Long}, Ilana D. and {Long}, Knox S. and {L{\'o}pez-Caniego}, Marcos and {Lotz}, Jennifer M. and {Love-Pruitt}, Jennifer M. and {Lubskiy}, Michael and {Luers}, Edward B. and {Luetgens}, Robert A. and {Luevano}, Annetta J. and {Lui}, Sarah Marie G. Flores and {Lund}, James M., III and {Lundquist}, Ray A. and {Lunine}, Jonathan and {L{\"u}tzgendorf}, Nora and {Lynch}, Richard J. and {MacDonald}, Alex J. and {MacDonald}, Kenneth and {Macias}, Matthew J. and {Macklis}, Keith I. and {Maghami}, Peiman and {Maharaja}, Rishabh Y. and {Maiolino}, Roberto and {Makrygiannis}, Konstantinos G. and {Malla}, Sunita Giri and {Malumuth}, Eliot M. and {Manjavacas}, Elena and {Marini}, Andrea and {Marrione}, Amanda and {Marston}, Anthony and {Martel}, Andr{\'e} R. and {Martin}, Didier and {Martin}, Peter G. and {Martinez}, Kristin L. and {Maschmann}, Marc and {Masci}, Gregory L. and {Masetti}, Margaret E. and {Maszkiewicz}, Michael and {Matthews}, Gary and {Matuskey}, Jacob E. and {McBrayer}, Glen A. and {McCarthy}, Donald W. and {McCaughrean}, Mark J. and {McClare}, Leslie A. and {McClare}, Michael D. and {McCloskey}, John C. and {McClurg}, Taylore D. and {McCoy}, Martin and {McElwain}, Michael W. and {McGregor}, Roy D. and {McGuffey}, Douglas B. and {McKay}, Andrew G. and {McKenzie}, William K. and {McLean}, Brian and {McMaster}, Matthew and {McNeil}, Warren and {De Meester}, Wim and {Mehalick}, Kimberly L. and {Meixner}, Margaret and {Mel{\'e}ndez}, Marcio and {Menzel}, Michael P. and {Menzel}, Michael T. and {Merz}, Matthew and {Mesterharm}, David D. and {Meyer}, Michael R. and {Meyett}, Michele L. and {Meza}, Luis E. and {Midwinter}, Calvin and {Milam}, Stefanie N. and {Miller}, Jay Todd and {Miller}, William C. and {Miskey}, Cherie L. and {Misselt}, Karl and {Mitchell}, Eileen P. and {Mohan}, Martin and {Montoya}, Emily E. and {Moran}, Michael J. and {Morishita}, Takahiro and {Moro-Mart{\'\i}n}, Amaya and {Morrison}, Debra L. and {Morrison}, Jane and {Morse}, Ernie C. and {Moschos}, Michael and {Moseley}, S.~H. and {Mosier}, Gary E. and {Mosner}, Peter and {Mountain}, Matt and {Muckenthaler}, Jason S. and {Mueller}, Donald G. and {Mueller}, Migo and {Muhiem}, Daniella and {M{\"u}hlmann}, Prisca and {Mullally}, Susan Elizabeth and {Mullen}, Stephanie M. and {Munger}, Alan J. and {Murphy}, Jess and {Murray}, Katherine T. and {Muzerolle}, James C. and {Mycroft}, Matthew and {Myers}, Andrew and {Myers}, Carey R. and {Myers}, Fred Richard R. and {Myers}, Richard and {Myrick}, Kaila and {Nagle}, Adrian F., IV and {Nayak}, Omnarayani and {Naylor}, Bret and {Neff}, Susan G. and {Nelan}, Edmund P. and {Nella}, John and {Nguyen}, Duy Tuong and {Nguyen}, Michael N. and {Nickson}, Bryony and {Nidhiry}, John Joseph and {Niedner}, Malcolm B. and {Nieto-Santisteban}, Maria and {Nikolov}, Nikolay K. and {Nishisaka}, Mary Ann and {Noriega-Crespo}, Alberto and {Nota}, Antonella and {O'Mara}, Robyn C. and {Oboryshko}, Michael and {O'Brien}, Marcus B. and {Ochs}, William R. and {Offenberg}, Joel D. and {Ogle}, Patrick Michael and {Ohl}, Raymond G. and {Olmsted}, Joseph Hamden and {Osborne}, Shannon Barbara and {O'Shaughnessy}, Brian Patrick and {{\"O}stlin}, G{\"o}ran and {O'Sullivan}, Brian and {Otor}, O. Justin and {Ottens}, Richard and {Ouellette}, Nathalie N. -Q. and {Outlaw}, Daria J. and {Owens}, Beverly A. and {Pacifici}, Camilla and {Page}, James Christophe and {Paranilam}, James G. and {Park}, Sang and {Parrish}, Keith A. and {Paschal}, Laura and {Patapis}, Polychronis and {Patel}, Jignasha and {Patrick}, Keith and {Pattishall}, Robert A., Jr. and {Paul}, Douglas William and {Paul}, Shirley J. and {Pauly}, Tyler Andrew and {Pavlovsky}, Cheryl M. and {Pe{\~n}a-Guerrero}, Maria and {Pedder}, Andrew H. and {Peek}, Matthew Weldon and {Pelham}, Patricia A. and {Penanen}, Konstantin and {Perriello}, Beth A. and {Perrin}, Marshall D. and {Perrine}, Richard F. and {Perrygo}, Chuck and {Peslier}, Muriel and {Petach}, Michael and {Peterson}, Karla A. and {Pfarr}, Tom and {Pierson}, James M. and {Pietraszkiewicz}, Martin and {Pilchen}, Guy and {Pipher}, Judy L. and {Pirzkal}, Norbert and {Pitman}, Joseph T. and {Player}, Danielle M. and {Plesha}, Rachel and {Plitzke}, Anja and {Pohner}, John A. and {Poletis}, Karyn Konstantin and {Pollizzi}, Joseph A. and {Polster}, Ethan and {Pontius}, James T. and {Pontoppidan}, Klaus and {Porges}, Susana C. and {Potter}, Gregg D. and {Prescott}, Stephen and {Proffitt}, Charles R. and {Pueyo}, Laurent and {Quispe Neira}, Irma Aracely and {Radich}, Armando and {Rager}, Reiko T. and {Rameau}, Julien and {Ramey}, Deborah D. and {Alarcon}, Rafael Ramos and {Rampini}, Riccardo and {Rapp}, Robert and {Rashford}, Robert A. and {Rauscher}, Bernard J. and {Ravindranath}, Swara and {Rawle}, Timothy and {Rawlings}, Tynika N. and {Ray}, Tom and {Regan}, Michael W. and {Rehm}, Brian and {Rehm}, Kenneth D. and {Reid}, Neill and {Reis}, Carl A. and {Renk}, Florian and {Reoch}, Tom B. and {Ressler}, Michael and {Rest}, Armin W. and {Reynolds}, Paul J. and {Richon}, Joel G. and {Richon}, Karen V. and {Ridgaway}, Michael and {Riedel}, Adric Richard and {Rieke}, George H. and {Rieke}, Marcia J. and {Rifelli}, Richard E. and {Rigby}, Jane R. and {Riggs}, Catherine S. and {Ringel}, Nancy J. and {Ritchie}, Christine E. and {Rix}, Hans-Walter and {Robberto}, Massimo and {Robinson}, Gregory L. and {Robinson}, Michael S. and {Robinson}, Orion and {Rock}, Frank W. and {Rodriguez}, David R. and {Rodr{\'\i}guez del Pino}, Bruno and {Roellig}, Thomas and {Rohrbach}, Scott O. and {Roman}, Anthony J. and {Romelfanger}, Frederick J. and {Romo}, Felipe P., Jr. and {Rosales}, Jose J. and {Rose}, Perry and {Roteliuk}, Anthony F. and {Roth}, Marc N. and {Rothwell}, Braden Quinn and {Rouzaud}, Sylvain and {Rowe}, Jason and {Rowlands}, Neil and {Roy}, Arpita and {Royer}, Pierre and {Rui}, Chunlei and {Rumler}, Peter and {Rumpl}, William and {Russ}, Melissa L. and {Ryan}, Michael B. and {Ryan}, Richard M. and {Saad}, Karl and {Sabata}, Modhumita and {Sabatino}, Rick and {Sabbi}, Elena and {Sabelhaus}, Phillip A. and {Sabia}, Stephen and {Sahu}, Kailash C. and {Saif}, Babak N. and {Salvignol}, Jean-Christophe and {Samara-Ratna}, Piyal and {Samuelson}, Bridget S. and {Sanders}, Felicia A. and {Sappington}, Bradley and {Sargent}, B.~A. and {Sauer}, Arne and {Savadkin}, Bruce J. and {Sawicki}, Marcin and {Schappell}, Tina M. and {Scheffer}, Caroline and {Scheithauer}, Silvia and {Scherer}, Ron and {Schiff}, Conrad and {Schlawin}, Everett and {Schmeitzky}, Olivier and {Schmitz}, Tyler S. and {Schmude}, Donald J. and {Schneider}, Analyn and {Schreiber}, J{\"u}rgen and {Schroeven-Deceuninck}, Hilde and {Schultz}, John J. and {Schwab}, Ryan and {Schwartz}, Curtis H. and {Scoccimarro}, Dario and {Scott}, John F. and {Scott}, Michelle B. and {Seaton}, Bonita L. and {Seely}, Bruce S. and {Seery}, Bernard and {Seidleck}, Mark and {Sembach}, Kenneth and {Shanahan}, Clare Elizabeth and {Shaughnessy}, Bryan and {Shaw}, Richard A. and {Shay}, Christopher Michael and {Sheehan}, Even and {Sheth}, Kartik and {Shih}, Hsin-Yi and {Shivaei}, Irene and {Siegel}, Noah and {Sienkiewicz}, Matthew G. and {Simmons}, Debra D. and {Simon}, Bernard P. and {Sirianni}, Marco and {Sivaramakrishnan}, Anand and {Slade}, Jeffrey E. and {Sloan}, G.~C. and {Slocum}, Christine E. and {Slowinski}, Steven E. and {Smith}, Corbett T. and {Smith}, Eric P. and {Smith}, Erin C. and {Smith}, Koby and {Smith}, Robert and {Smith}, Stephanie J. and {Smolik}, John L. and {Soderblom}, David R. and {Sohn}, Sangmo Tony and {Sokol}, Jeff and {Sonneborn}, George and {Sontag}, Christopher D. and {Sooy}, Peter R. and {Soummer}, Remi and {Southwood}, Dana M. and {Spain}, Kay and {Sparmo}, Joseph and {Speer}, David T. and {Spencer}, Richard and {Sprofera}, Joseph D. and {Stallcup}, Scott S. and {Stanley}, Marcia K. and {Stansberry}, John A. and {Stark}, Christopher C. and {Starr}, Carl W. and {Stassi}, Diane Y. and {Steck}, Jane A. and {Steeley}, Christine D. and {Stephens}, Matthew A. and {Stephenson}, Ralph J. and {Stewart}, Alphonso C. and {Stiavelli}, Massimo and {}, Hervey Stockman, Jr. and {Strada}, Paolo and {Straughn}, Amber N. and {Streetman}, Scott and {Strickland}, David Kendal and {Strobele}, Jingping F. and {Stuhlinger}, Martin and {Stys}, Jeffrey Edward and {Such}, Miguel and {Sukhatme}, Kalyani and {Sullivan}, Joseph F. and {Sullivan}, Pamela C. and {Sumner}, Sandra M. and {Sun}, Fengwu and {Sunnquist}, Benjamin Dale and {Swade}, Daryl Allen and {Swam}, Michael S. and {Swenton}, Diane F. and {Swoish}, Robby A. and {Tam Litten}, Oi In and {Tamas}, Laszlo and {Tao}, Andrew and {Taylor}, David K. and {Taylor}, Joanna M. and {Plate}, Maurice te and {Van Tea}, Mason and {Teague}, Kelly K. and {Telfer}, Randal C. and {Temim}, Tea and {Texter}, Scott C. and {Thatte}, Deepashri G. and {Thompson}, Christopher Lee and {Thompson}, Linda M. and {Thomson}, Shaun R. and {Thronson}, Harley and {Tierney}, C.~M. and {Tikkanen}, Tuomo and {Tinnin}, Lee and {Tippet}, William Thomas and {Todd}, Connor William and {Tran}, Hien D. and {Trauger}, John and {Trejo}, Edwin Gregorio and {Vinh Truong}, Justin Hoang and {Tsukamoto}, Christine L. and {Tufail}, Yasir and {Tumlinson}, Jason and {Tustain}, Samuel and {Tyra}, Harrison and {Ubeda}, Leonardo and {Underwood}, Kelli and {Uzzo}, Michael A. and {Vaclavik}, Steven and {Valenduc}, Frida and {Valenti}, Jeff A. and {Van Campen}, Julie and {van de Wetering}, Inge and {Van Der Marel}, Roeland P. and {van Haarlem}, Remy and {Vandenbussche}, Bart and {van Dishoeck}, Ewine F. and {Vanterpool}, Dona D. and {Vernoy}, Michael R. and {Vila Costas}, Maria Bego{\~n}a and {Volk}, Kevin and {Voorzaat}, Piet and {Voyton}, Mark F. and {Vydra}, Ekaterina and {Waddy}, Darryl J. and {Waelkens}, Christoffel and {Wahlgren}, Glenn Michael and {Walker}, Frederick E., Jr. and {Wander}, Michel and {Warfield}, Christine K. and {Warner}, Gerald and {Wasiak}, Francis C. and {Wasiak}, Matthew F. and {Wehner}, James and {Weiler}, Kevin R. and {Weilert}, Mark and {Weiss}, Stanley B. and {Wells}, Martyn and {Welty}, Alan D. and {Wheate}, Lauren and {Wheeler}, Thomas P. and {White}, Christy L. and {Whitehouse}, Paul and {Whiteleather}, Jennifer Margaret and {Whitman}, William Russell and {Williams}, Christina C. and {Willmer}, Christopher N.~A. and {Willott}, Chris J. and {Willoughby}, Scott P. and {Wilson}, Andrew and {Wilson}, Debra and {Wilson}, Donna V. and {Windhorst}, Rogier and {Wislowski}, Emily Christine and {Wolfe}, David J. and {Wolfe}, Michael A. and {Wolff}, Schuyler and {Wondel}, Amancio and {Woo}, Cindy and {Woods}, Robert T. and {Worden}, Elaine and {Workman}, William and {Wright}, Gillian S. and {Wu}, Carl and {Wu}, Chi-Rai and {Wun}, Dakin D. and {Wymer}, Kristen B. and {Yadetie}, Thomas and {Yan}, Isabelle C. and {Yang}, Keith C. and {Yates}, Kayla L. and {Yeager}, Christopher R. and {Yerger}, Ethan John and {Young}, Erick T. and {Young}, Gary and {Yu}, Gene and {Yu}, Susan and {Zak}, Dean S. and {Zeidler}, Peter and {Zepp}, Robert and {Zhou}, Julia and {Zincke}, Christian A. and {Zonak}, Stephanie and {Zondag}, Elisabeth},
        title = "{The James Webb Space Telescope Mission}",
      journal = {\pasp},
         year = 2023,
        month = jun,
       volume = {135},
       number = {1048},
          eid = {068001},
        pages = {068001},
          doi = {10.1088/1538-3873/acd1b5},
archivePrefix = {arXiv},
       eprint = {2304.04869},
 primaryClass = {astro-ph.IM},
       adsurl = {https://ui.adsabs.harvard.edu/abs/2023PASP..135f8001G}
}

@ARTICLE{Tozzi23,
       author = {{Tozzi}, Giulia and {Maiolino}, Roberto and {Cresci}, Giovanni and {Piotrowska}, Joanna M. and {Belfiore}, Francesco and {Curti}, Mirko and {Mannucci}, Filippo and {Marconi}, Alessandro},
        title = "{Unveiling hidden active nuclei in MaNGA star-forming galaxies with He II {\ensuremath{\lambda}}4686 line emission}",
      journal = {\mnras},
         year = 2023,
        month = may,
       volume = {521},
       number = {1},
        pages = {1264-1276},
          doi = {10.1093/mnras/stad506},
archivePrefix = {arXiv},
       eprint = {2302.04282},
 primaryClass = {astro-ph.GA},
       adsurl = {https://ui.adsabs.harvard.edu/abs/2023MNRAS.521.1264T}
}

@ARTICLE{Cameron23,
       author = {{Cameron}, Alex J. and {Saxena}, Aayush and {Bunker}, Andrew J. and {D'Eugenio}, Francesco and {Carniani}, Stefano and {Maiolino}, Roberto and {Curtis-Lake}, Emma and {Ferruit}, Pierre and {Jakobsen}, Peter and {Arribas}, Santiago and {Bonaventura}, Nina and {Charlot}, Stephane and {Chevallard}, Jacopo and {Curti}, Mirko and {Looser}, Tobias J. and {Maseda}, Michael V. and {Rawle}, Tim and {Rodr{\'\i}guez Del Pino}, Bruno and {Smit}, Renske and {{\"U}bler}, Hannah and {Willott}, Chris and {Witstok}, Joris and {Egami}, Eiichi and {Eisenstein}, Daniel J. and {Johnson}, Benjamin D. and {Hainline}, Kevin and {Rieke}, Marcia and {Robertson}, Brant E. and {Stark}, Daniel P. and {Tacchella}, Sandro and {Williams}, Christina C. and {Willmer}, Christopher N.~A. and {Bhatawdekar}, Rachana and {Bowler}, Rebecca and {Boyett}, Kristan and {Circosta}, Chiara and {Helton}, Jakob M. and {Jones}, Gareth C. and {Kumari}, Nimisha and {Ji}, Zhiyuan and {Nelson}, Erica and {Parlanti}, Eleonora and {Sandles}, Lester and {Scholtz}, Jan and {Sun}, Fengwu},
        title = "{JADES: Probing interstellar medium conditions at z {\ensuremath{\sim}} 5.5-9.5 with ultra-deep JWST/NIRSpec spectroscopy}",
      journal = {\aap},
         year = 2023,
        month = sep,
       volume = {677},
          eid = {A115},
        pages = {A115},
          doi = {10.1051/0004-6361/202346107},
archivePrefix = {arXiv},
       eprint = {2302.04298},
 primaryClass = {astro-ph.GA},
       adsurl = {https://ui.adsabs.harvard.edu/abs/2023A&A...677A.115C}
}

@ARTICLE{Veilleux87,
       author = {{Veilleux}, Sylvain and {Osterbrock}, Donald E.},
        title = "{Spectral Classification of Emission-Line Galaxies}",
      journal = {\apjs},
         year = 1987,
        month = feb,
       volume = {63},
        pages = {295},
          doi = {10.1086/191166},
       adsurl = {https://ui.adsabs.harvard.edu/abs/1987ApJS...63..295V}
}

@ARTICLE{Nakajima23,
        author = {{Nakajima}, Kimihiko and {Ouchi}, Masami and {Isobe}, Yuki and {Harikane}, Yuichi and {Zhang}, Yechi and {Ono}, Yoshiaki and {Umeda}, Hiroya and {Oguri}, Masamune},
        title = "{JWST Census for the Mass-Metallicity Star Formation Relations at z = 4-10 with Self-consistent Flux Calibration and Proper Metallicity Calibrators}",
      journal = {\apjs},
         year = 2023,
        month = dec,
       volume = {269},
       number = {2},
          eid = {33},
        pages = {33},
          doi = {10.3847/1538-4365/acd556},
archivePrefix = {arXiv},
       eprint = {2301.12825},
 primaryClass = {astro-ph.GA},
       adsurl = {https://ui.adsabs.harvard.edu/abs/2023ApJS..269...33N}
}

@ARTICLE{Kocevski23,
       author = {{Kocevski}, Dale D. and {Onoue}, Masafusa and {Inayoshi}, Kohei and {Trump}, Jonathan R. and {Arrabal Haro}, Pablo and {Grazian}, Andrea and {Dickinson}, Mark and {Finkelstein}, Steven L. and {Kartaltepe}, Jeyhan S. and {Hirschmann}, Michaela and {Aird}, James and {Holwerda}, Benne W. and {Fujimoto}, Seiji and {Juneau}, St{\'e}phanie and {Amor{\'\i}n}, Ricardo O. and {Backhaus}, Bren E. and {Bagley}, Micaela B. and {Barro}, Guillermo and {Bell}, Eric F. and {Bisigello}, Laura and {Calabr{\`o}}, Antonello and {Cleri}, Nikko J. and {Cooper}, M.~C. and {Ding}, Xuheng and {Grogin}, Norman A. and {Ho}, Luis C. and {Hutchison}, Taylor A. and {Inoue}, Akio K. and {Jiang}, Linhua and {Jones}, Brenda and {Koekemoer}, Anton M. and {Li}, Wenxiu and {Li}, Zhengrong and {McGrath}, Elizabeth J. and {Molina}, Juan and {Papovich}, Casey and {P{\'e}rez-Gonz{\'a}lez}, Pablo G. and {Pirzkal}, Nor and {Wilkins}, Stephen M. and {Yang}, Guang and {Yung}, L.~Y. Aaron},
        title = "{Hidden Little Monsters: Spectroscopic Identification of Low-mass, Broad-line AGNs at z > 5 with CEERS}",
      journal = {\apjl},
         year = 2023,
        month = sep,
       volume = {954},
       number = {1},
          eid = {L4},
        pages = {L4},
          doi = {10.3847/2041-8213/ace5a0},
archivePrefix = {arXiv},
       eprint = {2302.00012},
 primaryClass = {astro-ph.GA},
       adsurl = {https://ui.adsabs.harvard.edu/abs/2023ApJ...954L...4K}
}

@ARTICLE{Sanders23,
         author = {{Sanders}, Ryan L. and {Shapley}, Alice E. and {Topping}, Michael W. and {Reddy}, Naveen A. and {Brammer}, Gabriel B.},
        title = "{Excitation and Ionization Properties of Star-forming Galaxies at z = 2.0-9.3 with JWST/NIRSpec}",
      journal = {\apj},
         year = 2023,
        month = sep,
       volume = {955},
       number = {1},
          eid = {54},
        pages = {54},
          doi = {10.3847/1538-4357/acedad},
archivePrefix = {arXiv},
       eprint = {2301.06696},
 primaryClass = {astro-ph.GA},
       adsurl = {https://ui.adsabs.harvard.edu/abs/2023ApJ...955...54S}
}

@ARTICLE{Curti23b,
       author = {{Curti}, Mirko and {Maiolino}, Roberto and {Curtis-Lake}, Emma and {Chevallard}, Jacopo and {Carniani}, Stefano and {D'Eugenio}, Francesco and {Looser}, Tobias J. and {Scholtz}, Jan and {Charlot}, Stephane and {Cameron}, Alex and {{\"U}bler}, Hannah and {Witstok}, Joris and {Boyett}, Kristian and {Laseter}, Isaac and {Sandles}, Lester and {Arribas}, Santiago and {Bunker}, Andrew and {Giardino}, Giovanna and {Maseda}, Michael V. and {Rawle}, Tim and {Rodr{\'\i}guez Del Pino}, Bruno and {Smit}, Renske and {Willott}, Chris J. and {Eisenstein}, Daniel J. and {Hausen}, Ryan and {Johnson}, Benjamin and {Rieke}, Marcia and {Robertson}, Brant and {Tacchella}, Sandro and {Williams}, Christina C. and {Willmer}, Christopher and {Baker}, William M. and {Bhatawdekar}, Rachana and {Egami}, Eiichi and {Helton}, Jakob M. and {Ji}, Zhiyuan and {Kumari}, Nimisha and {Perna}, Michele and {Shivaei}, Irene and {Sun}, Fengwu},
        title = "{JADES: Insights into the low-mass end of the mass-metallicity-SFR relation at 3 < z < 10 from deep JWST/NIRSpec spectroscopy}",
      journal = {\aap},
         year = 2024,
        month = apr,
       volume = {684},
          eid = {A75},
        pages = {A75},
          doi = {10.1051/0004-6361/202346698},
archivePrefix = {arXiv},
       eprint = {2304.08516},
 primaryClass = {astro-ph.GA},
       adsurl = {https://ui.adsabs.harvard.edu/abs/2024A&A...684A..75C}
}

@ARTICLE{Curti23,
       author = {{Curti}, Mirko and {D'Eugenio}, Francesco and {Carniani}, Stefano and {Maiolino}, Roberto and {Sandles}, Lester and {Witstok}, Joris and {Baker}, William M. and {Bennett}, Jake S. and {Piotrowska}, Joanna M. and {Tacchella}, Sandro and {Charlot}, Stephane and {Nakajima}, Kimihiko and {Maheson}, Gabriel and {Mannucci}, Filippo and {Amiri}, Amirnezam and {Arribas}, Santiago and {Belfiore}, Francesco and {Bonaventura}, Nina R. and {Bunker}, Andrew J. and {Chevallard}, Jacopo and {Cresci}, Giovanni and {Curtis-Lake}, Emma and {Hayden-Pawson}, Connor and {Jones}, Gareth C. and {Kumari}, Nimisha and {Laseter}, Isaac and {Looser}, Tobias J. and {Marconi}, Alessandro and {Maseda}, Michael V. and {Scholtz}, Jan and {Smit}, Renske and {{\"U}bler}, Hannah and {Wallace}, Imaan E.~B.},
        title = "{The chemical enrichment in the early Universe as probed by JWST via direct metallicity measurements at z   8}",
      journal = {\mnras},
         year = 2023,
        month = jan,
       volume = {518},
       number = {1},
        pages = {425-438},
          doi = {10.1093/mnras/stac2737},
archivePrefix = {arXiv},
       eprint = {2207.12375},
 primaryClass = {astro-ph.GA},
       adsurl = {https://ui.adsabs.harvard.edu/abs/2023MNRAS.518..425C}
}

@ARTICLE{CurtisLake22,
       author = {{Curtis-Lake}, Emma and {Carniani}, Stefano and {Cameron}, Alex and {Charlot}, Stephane and {Jakobsen}, Peter and {Maiolino}, Roberto and {Bunker}, Andrew and {Witstok}, Joris and {Smit}, Renske and {Chevallard}, Jacopo and {Willott}, Chris and {Ferruit}, Pierre and {Arribas}, Santiago and {Bonaventura}, Nina and {Curti}, Mirko and {D'Eugenio}, Francesco and {Franx}, Marijn and {Giardino}, Giovanna and {Looser}, Tobias J. and {L{\"u}tzgendorf}, Nora and {Maseda}, Michael V. and {Rawle}, Tim and {Rix}, Hans-Walter and {Rodr{\'\i}guez del Pino}, Bruno and {{\"U}bler}, Hannah and {Sirianni}, Marco and {Dressler}, Alan and {Egami}, Eiichi and {Eisenstein}, Daniel J. and {Endsley}, Ryan and {Hainline}, Kevin and {Hausen}, Ryan and {Johnson}, Benjamin D. and {Rieke}, Marcia and {Robertson}, Brant and {Shivaei}, Irene and {Stark}, Daniel P. and {Tacchella}, Sandro and {Williams}, Christina C. and {Willmer}, Christopher N.~A. and {Bhatawdekar}, Rachana and {Bowler}, Rebecca and {Boyett}, Kristan and {Chen}, Zuyi and {de Graaff}, Anna and {Helton}, Jakob M. and {Hviding}, Raphael E. and {Jones}, Gareth C. and {Kumari}, Nimisha and {Lyu}, Jianwei and {Nelson}, Erica and {Perna}, Michele and {Sandles}, Lester and {Saxena}, Aayush and {Suess}, Katherine A. and {Sun}, Fengwu and {Topping}, Michael W. and {Wallace}, Imaan E.~B. and {Whitler}, Lily},
        title = "{Spectroscopic confirmation of four metal-poor galaxies at z = 10.3-13.2}",
      journal = {Nature Astronomy},
         year = 2023,
        month = may,
       volume = {7},
        pages = {622-632},
          doi = {10.1038/s41550-023-01918-w},
archivePrefix = {arXiv},
       eprint = {2212.04568},
 primaryClass = {astro-ph.GA},
       adsurl = {https://ui.adsabs.harvard.edu/abs/2023NatAs...7..622C}
}

@BOOK{Osterbrock06,
       author = {{Osterbrock}, Donald E. and {Ferland}, Gary J.},
        title = "{Astrophysics of gaseous nebulae and active galactic nuclei}",
         year = 2006,
       adsurl = {https://ui.adsabs.harvard.edu/abs/2006agna.book.....O}
}

\begin{appendix} 
\onecolumn
\section{Broad-line AGN at $z\lesssim2$}\label{sec:app1}
\begin{figure}[h!]
\includegraphics[width=0.49\columnwidth]{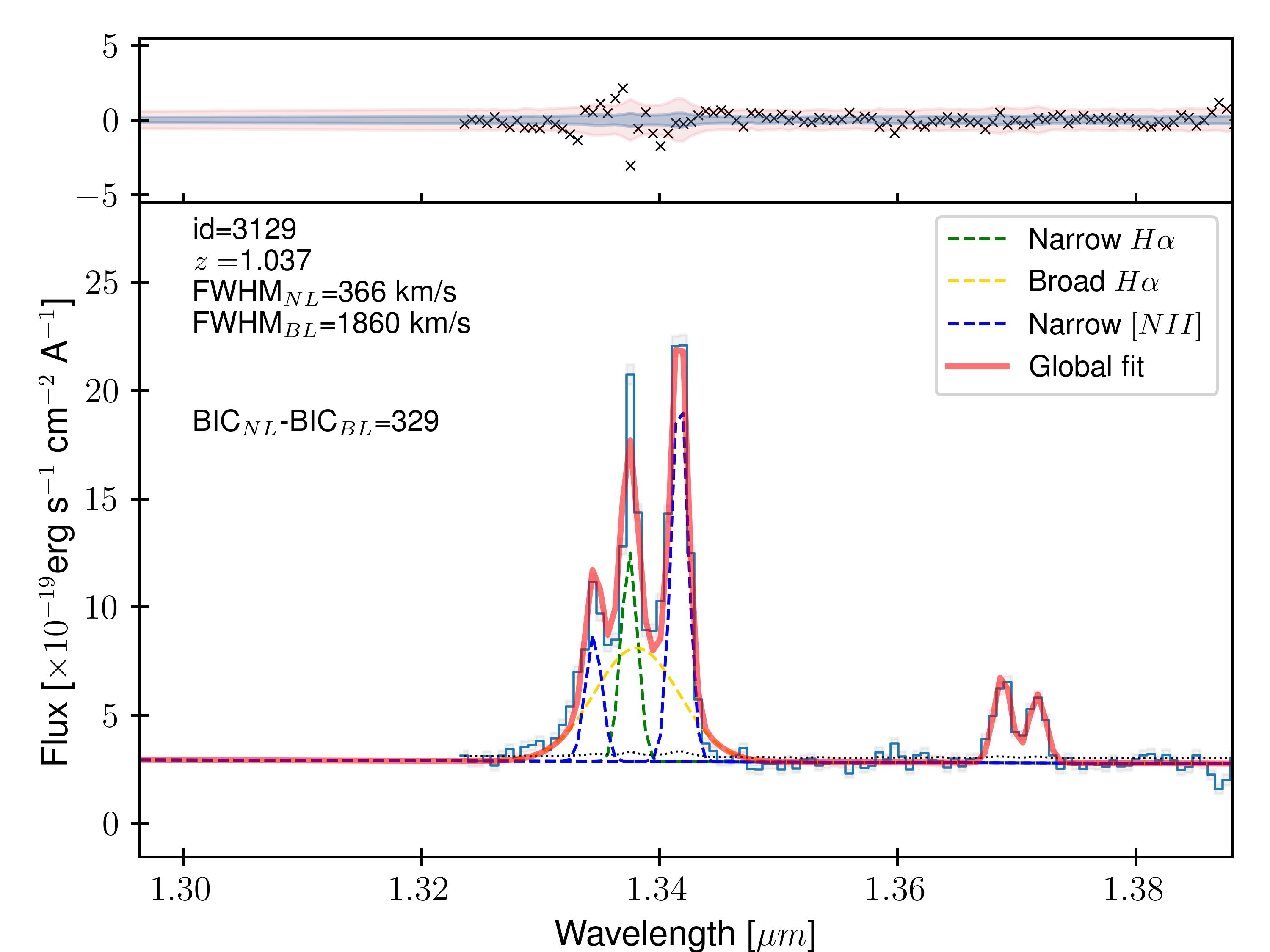}
\includegraphics[width=0.49\columnwidth]{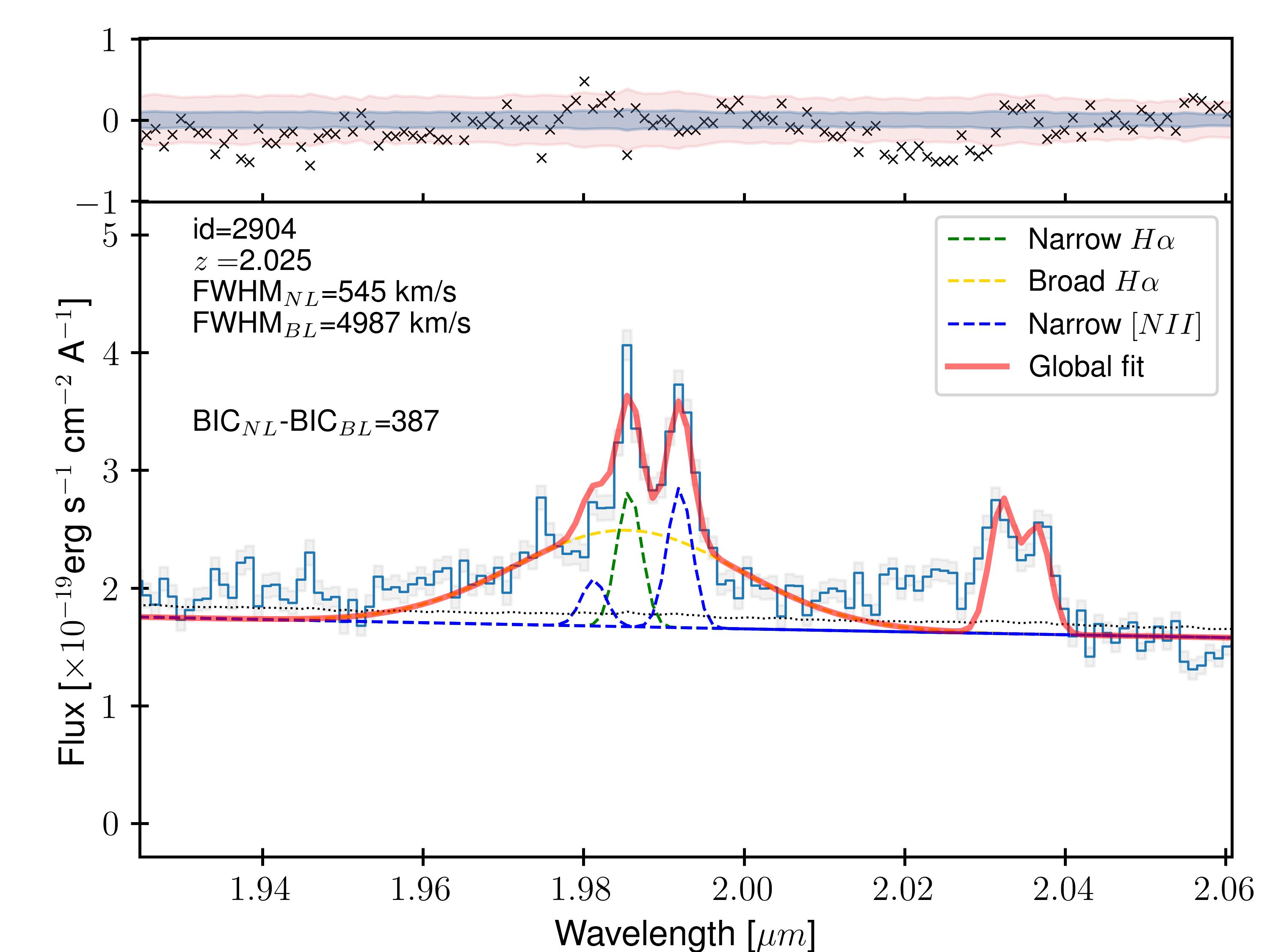}\\
\includegraphics[width=0.49\columnwidth]{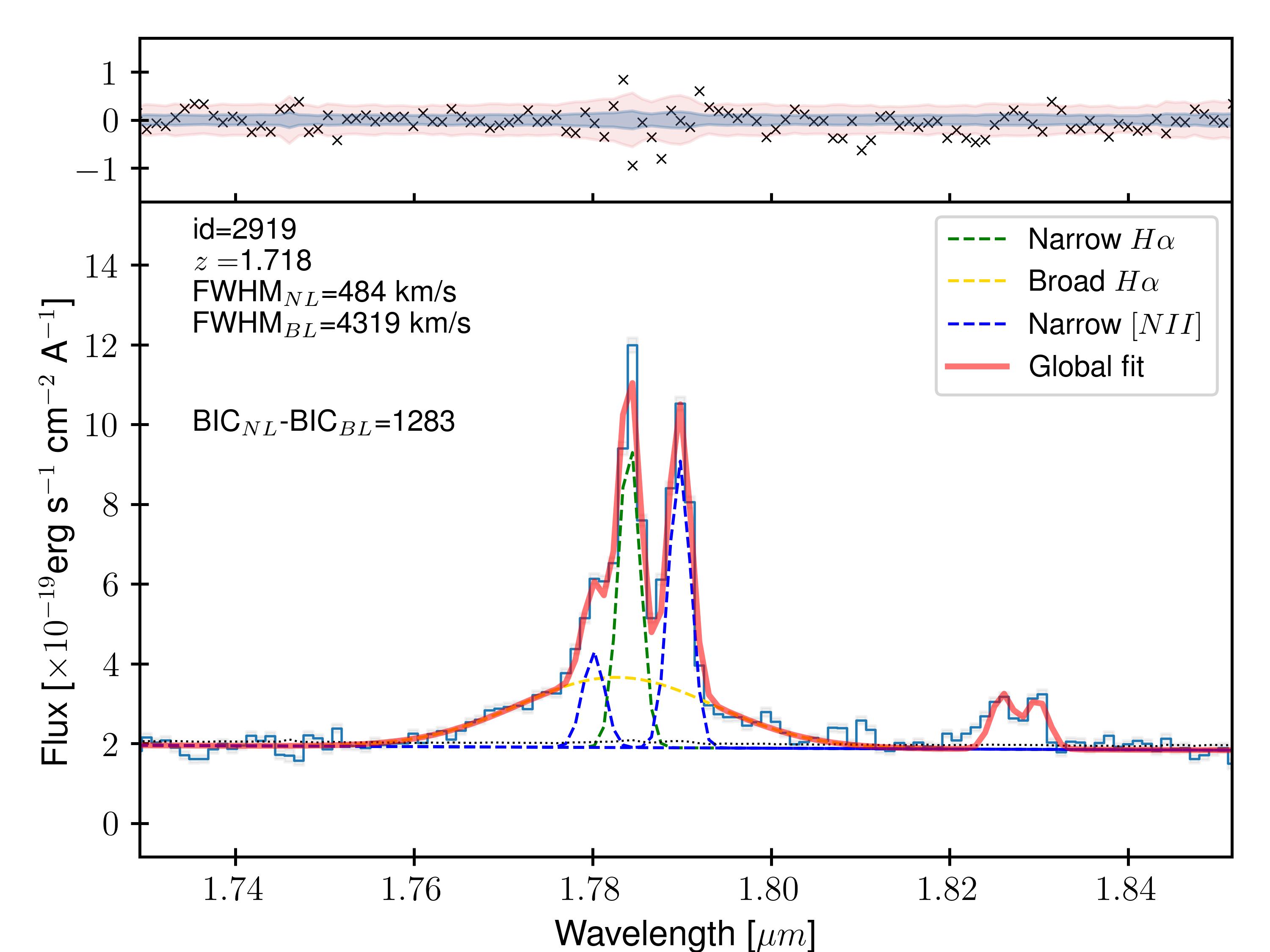}
\includegraphics[width=0.49\columnwidth]{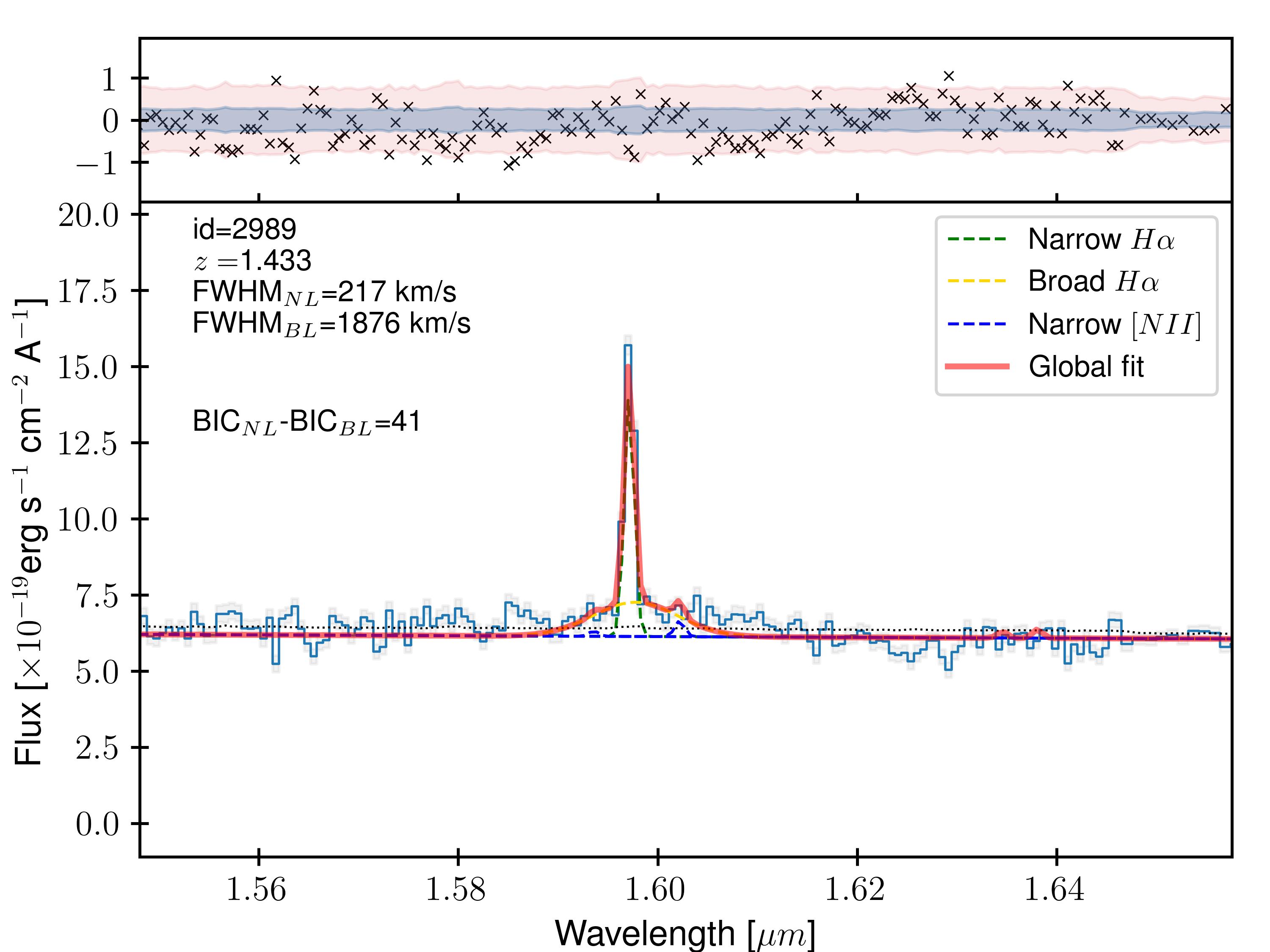}
\caption{Fit of the \Ha, \NII, and \SII complex including both a broad and a narrow components of the \Ha emission line, as is described in Sec.~\ref{sec:line_fit}. The global fit is presented in red, while the narrow and broad \Ha components are in green and yellow, respectively. The blue line represents the fit of the narrow \NII doublet. In the upper panel are shown the residuals of the fit compared with the distribution of the $1\sigma$ (blue) and $3\sigma$ (red) errors on the fluxes. On the upper left part of the plots are reported the id, redshift, FWHMs derived from the fits, and the difference between the BIC value computed from the narrow-only line fit and the one from the fit with the additional broad \Ha component}
    \label{fig:app_blagn}
    
\end{figure}

Here, we present the spectra and line fitting of the four broad-line AGN at $z\leq2$ selected in our work. The four sources are CEERS-2904 (z=2.024), CEERS-2919 (z=1.717), CEERS-3129 (z=1.037), and CEERS-2898 (z=1.433).
Three of these sources are X-ray detected (CEERS-2904, CEERS-2919, CEERS-2989), while one is radio detected (CEERS-3129). These sources were first fit in the range around the \Ha line using a narrow-line only fit (as is described in Sec.~\ref{sec:line_fit}) and then adding a single broad component accounting for the emission of the BLR around the AGN, as presented in Fig.~\ref{fig:app_blagn}. In order to test the significance of the broad \Ha component we used the Bayesian Information Criterion (BIC) parameter \citep{Liddle07}, we found that the BLR model is strongly preferred to the narrow-only fit ($\Delta BIC= BIC_{H\alpha \ NL}- BIC_{H\alpha \ BL}>40$ for all the sources). The $\Delta BIC$ values for the single fit are shown in the right part of each panel in Fig.~\ref{fig:app_blagn}.\\
In fitting the \Hb and \OIIIl5007 lines complex of these sources we did not find any significant broad component in none of the two lines.

\FloatBarrier

\section{\texttt{CIGALE} SED fitting parameters grid}\label{sec:app2}
Here we report the grid of parameters used for the SED-fitting with \texttt{CIGALE} described in Sec.~\ref{sec:sedfit}

\begin{table*}
\centering
\renewcommand{\arraystretch}{1.3}
\caption{\texttt{\texttt{CIGALE}} parameters used for the SED fitting.}

\begin{threeparttable}
\resizebox{\linewidth}{!}{%
\label{tab:CIGALE}
\begin{tabular}{ccc}
\hline
\hline
\multicolumn{1}{l}{}                                                                                                                               & \multicolumn{1}{l}{}                                                                                                                                                                 & \multicolumn{1}{l}{}                                                                                                                                                                                                                                                                                                             \\
\textbf{Parameter}                                                                                                                                 & \textbf{Values}                                                                                                                                                                      & \textbf{Description}                                                                                                                                                                                                                                                                                                             \\
\multicolumn{1}{l}{}                                                                                                                               & \multicolumn{1}{l}{}                                                                                                                                                                 & \multicolumn{1}{l}{}                                                                                                                                                                                                                                                                                                             \\ \hline
\begin{tabular}[c]{@{}c@{}}SFH (\texttt{sfhdelayedbq})\\ $\tau_{\rm main}$\\ \\ age$_{\rm main}$ \\ \\ age$_{\rm bq}$\\ r$_{\rm sfr}$ \end{tabular}& 
\begin{tabular}[c]{@{}c@{}}\\ \\0.2, 0.3, 0.5, 0.7, 1.0, 1.5, 2.0, 3.0, 4.0, 5.0 {[}Gyr{]}\\{\it 0.15, 0.2, 0.3, 0.5, 0.7, 1.0, 1.5} {[}Gyr{]}\\ 0.5, 0.7, 1.0, 1.5, 2.0, 3.0, 4.0, 5.0 {[}Gyr{]}\\ {\it 0.2, 0.3, 0.5, 0.7, 1.0, 1.5} {[}Gyr{]}\\ 0.01, 0.03, 0.05, 0.1 {[}Gyr{]}\\ 0.1, 0.2, 0.5, 1, 2, 5  \\ \textit{0.1, 0.5, 1, 2, 5, 10, 30, 50}\end{tabular} & 
\begin{tabular}[c]{@{}c@{}}\\$e$-folding time of the main stellar population \\ \\ Age of the main stellar population in the galaxy \\ \\ Age of the burst/quench episode\\ Ratio of the SFR after/before age\_bq\end{tabular}\\
\hline
\begin{tabular}[c]{@{}c@{}}Stellar component (\texttt{bc03})\\ IMF\\ Z\\ \\ Separation age\end{tabular}                                                        & \begin{tabular}[c]{@{}c@{}}\\1\\ 0.008, 0.02\\ {\it 0.0004, 0.008, 0.02}\\ 10 {[}Myr{]} \\\end{tabular}                                                                                                                      & \begin{tabular}[c]{@{}c@{}}\\ \\Initial mass function: 1 (Chabrier 2003)\\ Metallicity \\ \\ Age of the separation between \\ the young and the old star populations\end{tabular}                                                                                                      \\ \hline 
\begin{tabular}[c]{@{}c@{}} Nebular emission (\texttt{nebular}) \\ logU\\ \\ zgas\\ \\ ne \\ \end{tabular} & \begin{tabular}[c]{@{}c@{}} \\ \\-3.0, -2.0, -1.0\\\textit{-2.0, -1.0}\\ 0.008, 0.02\\ \textit{0.0004, 0.008, 0.02}\\100\\\textit{100, 1000}\end{tabular} & \begin{tabular}[c]{@{}c@{}}\\ Ionization parameter\\ \\ Gas metallicity\\ \\ Electron density \end{tabular}
\\ \hline
\begin{tabular}[c]{@{}c@{}}Dust attenuation \\ (\texttt{dustatt\_modified\_cf00})\\ A$^{\rm ISM}_{\rm V}$\\ $\mu$\\ slope ISM\\ slope BC\end{tabular} & \begin{tabular}[c]{@{}c@{}}\\ \\0.01, 0.1, 0.3, 0.5, 0.7, 1.0, 1.5, 2.0, 3.0, 4.0\\ 0.44\\ -0.7\\ -1.3\end{tabular}                                                                                        & \begin{tabular}[c]{@{}c@{}}\\ \\V-band attenuation in the interstellar medium\\ Ratio of the BC-to-ISM attenuation\\ Power law slope of the attenuation in the ISM\\ Power law slope of the attenuation in the BC\end{tabular}                                                                                                        \\ \hline
\begin{tabular}[c]{@{}c@{}}Dust emission (\texttt{dale2014})\\ fracAGN\\ $\alpha$\end{tabular}                                       & \begin{tabular}[c]{@{}c@{}}\\0.0\\ 2.0\end{tabular}                                                                                          & \begin{tabular}[c]{@{}c@{}}\\AGN fraction\\ Slope of the dust emission\end{tabular}                                                                                                                                                                                    \\ \hline
\begin{tabular}[c]{@{}c@{}}AGN component (\texttt{skirtor2016})\\ $r_{\rm ratio}$\\ $\tau$\\$i$\\ $f_{\rm AGN}$\\ EBV\end{tabular} & \begin{tabular}[c]{@{}c@{}}\\20.0\\ 3.0\\50, 60, 70, 80 \\ 0.1, 0.2, 0.3, 0.4, 0.5, 0.6, 0.7 \\ 0.04\end{tabular}                                                & \begin{tabular}[c]{@{}c@{}}\\Ratio of the outer to inner radius of the dusty torus\\ Equatorial optical depth at 9.7 $\mu$m\\ Angle between equatorial axis and line of sight\\ AGN fraction at 0.1-1.0 $\mu m$ \\ E(B-V) for the extinction in the polar direction in magnitudes\end{tabular} \\  \hline \hline 
\end{tabular}}

\tablefoot{The first column reports the name of the templates as well as each individual parameter. In the second column, the parameters are listed, and in the third column, the descriptions of the parameters are provided. In cases where the grid is different for galaxies at $z>3$, the values are reported in italics.}
\end{threeparttable}

\end{table*}

\FloatBarrier

\section{Final narrow-line AGN table}\label{sec:app3}
\begin{table}[]
\caption{List of NLAGN selected in this work among CEERS MR spectra.}

    \begin{tabular}{@{}lcccccccc@{}} 
    \hline 
\hline 
 \centering

ID & Redshift & selection & Log L$_{\rm bol}$ & $A_V$& Log M$_{*}$ & SFR   & Notes\\
  &  & method    & ergs s$^{-1}$ & mag & M$_{\odot}$ & M$_{\odot}$ yr$^{-1}$ & &\\
\hline 
CEERS-23* & 8.8802 &         N5   & 44.92 ([OIII]) & x & 9.17 & 36 & \\ 
CEERS-355 & 6.0995 &   R3O1 He2N2        & 45.31 (H$\beta$) & 0.32 & 8.37 & 5 & \\ 
CEERS-428 & 6.1022 &           M2 & x & x & 8.06 & 2 & \\ 
CEERS-496* & 6.5809 &        N4    & 43.93 ([OIII]) & x & 7.93 & 1 & \\ 
CEERS-613* & 6.7288 &     C3He2    & 44.78 (H$\beta$) & x & 8.28 & 3 & \\ 
CEERS-669 & 5.2714 &  R3S2          & 45.22 ([OIII]) & >1.92 & 8.59 & 3 & \\ 
CEERS-689 & 7.5457 &          M1 M2 & 45.53 (H$\beta$) & x & x & x & \\ 
CEERS-1019* & 8.6785 &        N4    & 45.87 (H$\beta$) & x & 9.49 & 88 & \\ 
CEERS-1029 & 8.61 &      Ne4      & 45.4 (H$\beta$) & x & 9.86 & 58 & \\ 
CEERS-1064 & 6.7905 &           M2 & x & x & x & x & \\ 
CEERS-1173 & 4.9955 &  R3S2          & 45.55 (H$\beta$) & 1.06 & 8.97 & 6 & \\ 
CEERS-1212 & 4.2772 &   R3O1         & 44.44 (H$\beta$) & 0 & 9.22 & 5 & \\ 
CEERS-1244 & 4.4778 &           M2 & 45.28 (H$\alpha$) & x & 9.86 & 1 & BLAGN\\ 
CEERS-1400 & 4.4893 &           M2 & x & 0 & 8.19 & 3 & \\ 
CEERS-1477 & 4.6307 &          M1 M2 & 45.8 (H$\beta$) & 1.14 & 8.54 & 3 & \\ 
CEERS-1536 & 5.0338 &          M1 M2 & 45.41 (H$\beta$) & 0.3 & 8.93 & 5 & \\ 
CEERS-1539 & 4.8838 &  R3S2 R3O1         & 45.71 (H$\beta$) & 1.14 & 9.03 & 10 & \\ 
CEERS-1561 & 6.1965 &          M1 M2 & 45.45 (H$\beta$) & 0.18 & 8.68 & 4 & \\ 
CEERS-1605 & 4.6305 &  R3S2        M1 M2 & 45.07 (H$\beta$) & 0.46 & 8.69 & 11 & \\ 
CEERS-1658 & 4.6035 &    He2N2        & 44.95 (H$\beta$) & x & 8.65 & 7 & \\ 
CEERS-1732 & 4.6598 &   R3O1         & 45.28 (H$\beta$) & 0.27 & 10.02 & 33 & \\ 
CEERS-1746 & 4.5602 &          M1 M2 & 45.32 (H$\beta$) & 0.3 & 9.08 & 6 & \\ 
CEERS-1767 & 4.5455 &           M2 & x & x & 9.28 & 4 & \\ 
CEERS-1836* & 4.4694 &  R3S2          & 44.93 (H$\beta$) & 0.95 & 8.44 & 8 & \\ 
CEERS-2168 & 5.6541 &           M2 & x & x & 9.4 & 51 & \\ 
CEERS-2668* & 2.8984 & R3N2 R3S2          & 45.5 ([OIII]) & >3.32 & 10.42 & 19 & \\ 
CEERS-2754 & 2.238 &    He2N2        & x & 1.47 & x & x & \\ 
CEERS-2808 & 3.3834 &  R3S2          & 44.99 (H$\beta$) & 1.31 & 10.68 & 619 & X-ray\\ 
CEERS-2900 & 1.9004 & R3N2 R3S2          & 43.62 ([OIII]) & >2.06 & 11.17 & 0 & X-ray \& Radio \\ 
CEERS-2904 & 2.0249 & R3N2 R3S2 R3O1 He2N2      M1 M2 & 44.9 (H$\alpha$) & >2.17 & 10.78 & 26 & X-ray \& BLAGN\\ 
CEERS-2919 & 1.7177 & R3N2 R3S2 R3O1 He2N2        & 44.84 (H$\alpha$) & 1.75 & 10.62 & 203 &  X-ray \& BLAGN \\ 
CEERS-3129 & 1.0373 & R3N2 R3S2  He2N2        & 44.46 (H$\alpha$) & 2.12 & 11.04 & 309 & Radio \& BLAGN \\ 
CEERS-3187 & 1.4896 &    He2N2        & 43.76 (H$\beta$) & 0 & 8.23 & 3 & \\ 
CEERS-3223 & 2.2862 &  R3S2          & 43.91 (H$\beta$) & 0.18 & 8.9 & 1 & \\ 
CEERS-3535* & 2.4886 &  R3S2          & 44.95 (H$\beta$) & 0.14 & 8.87 & 6 & \\ 
CEERS-3585* & 3.8662 &           M2 & 45.76 (H$\beta$) & 1.77 & 7.97 & 2 & \\ 
CEERS-3838 & 1.9221 &  R3S2 R3O1         & 44.68 (H$\beta$) & 1.8 & 9.16 & 11 & \\ 
CEERS-4210 & 5.2549 &      Ne4      & 45.45 (H$\beta$) & 0.56 & 8.69 & 12 & \\ 
CEERS-4385 & 3.4219 &    He2N2        & 45.19 (H$\beta$) & 0.65 & 9.33 & 15 & \\ 
CEERS-4402 & 3.3293 &   R3O1         & 45.21 (H$\beta$) & 0.44 & 9.82 & 23 & \\ 
CEERS-5117 & 1.5728 &  R3S2 R3O1         & 44.41 (H$\beta$) & 1.16 & 9.26 & 2 & \\ 
CEERS-6689 & 1.4608 &    He2N2        & x & >5.53 & 10.82 & 95 & \\ 
CEERS-8041 & 2.5319 &  R3S2          & 44.96 ([OIII]) & >2.73 & 9.24 & 8 & \\ 
CEERS-8110 & 1.5662 &   R3O1         & 44.43 (H$\beta$) & 0.43 & 9.72 & 5 & \\ 
CEERS-8299 & 2.1567 &       Ne5     & 45.62 (H$\beta$) & 2.37 & 10.1 & 5 & \\ 
CEERS-9156 & 2.0534 &  R3S2          & 45.05 (H$\beta$) & 0.42 & 9.54 & 6 & \\ 
CEERS-9238 & 1.7415 &  R3S2          & 45.07 (H$\beta$) & 1.67 & 9.29 & 3 & \\ 
CEERS-13661 & 3.3279 &  R3S2    Ne4      & 45.0 (H$\beta$) & 0.76 & 9.03 & 9 & \\ 
CEERS-16551 & 2.5714 &  R3S2 R3O1         & 44.61 (H$\beta$) & 0 & 8.57 & 5 & \\ 
CEERS-17496 & 2.3732 &  R3S2 R3O1         & 44.62 (H$\beta$) & 0 & 8.79 & 3 & \\ 
CEERS-31075 & 1.2831 &    He2N2        & 44.4 (H$\beta$) & 0.11 & 7.8 & 0 & \\ 
CEERS-44804 & 5.508 &          M1  & 44.79 (H$\beta$) & x & 8.54 & 5 & \\ 
\hline
\end{tabular} 
    
\tablefoot{Columns: ID, redshift, diagnostic diagram(s) in which the source is selected, AGN bolometric luminosity (and the emission line used to derive it), ISM obscuration $A_V$, stellar mass, SFR, and any other additional notes. For bolometric luminosities derived from the \Hb line we used the \cite{Netzer09} calibration, while for those derived from the \OIIIl5007 we used the \cite{Lamastra09} calibration. Low-z and high-z BLAGN have instead $L_{bol}$ computed using the broad \Ha lines and the calibrations of \cite{Greene05}.  When a 'x' is reported it means that it was not possible to compute that quantity for that source. Asterisks ($^{*}$) mark tentative NLAGN selections.}
\label{table:Sample}

\end{table}

\FloatBarrier

\section{Spectra of the NLAGN at $z>6$}\label{app:high-z_spec}
Here, we report the 1D and 2D spectra of the NLAGN selected at $z>6$.
\begin{figure*}[h!]
    \includegraphics[width=0.49\columnwidth]{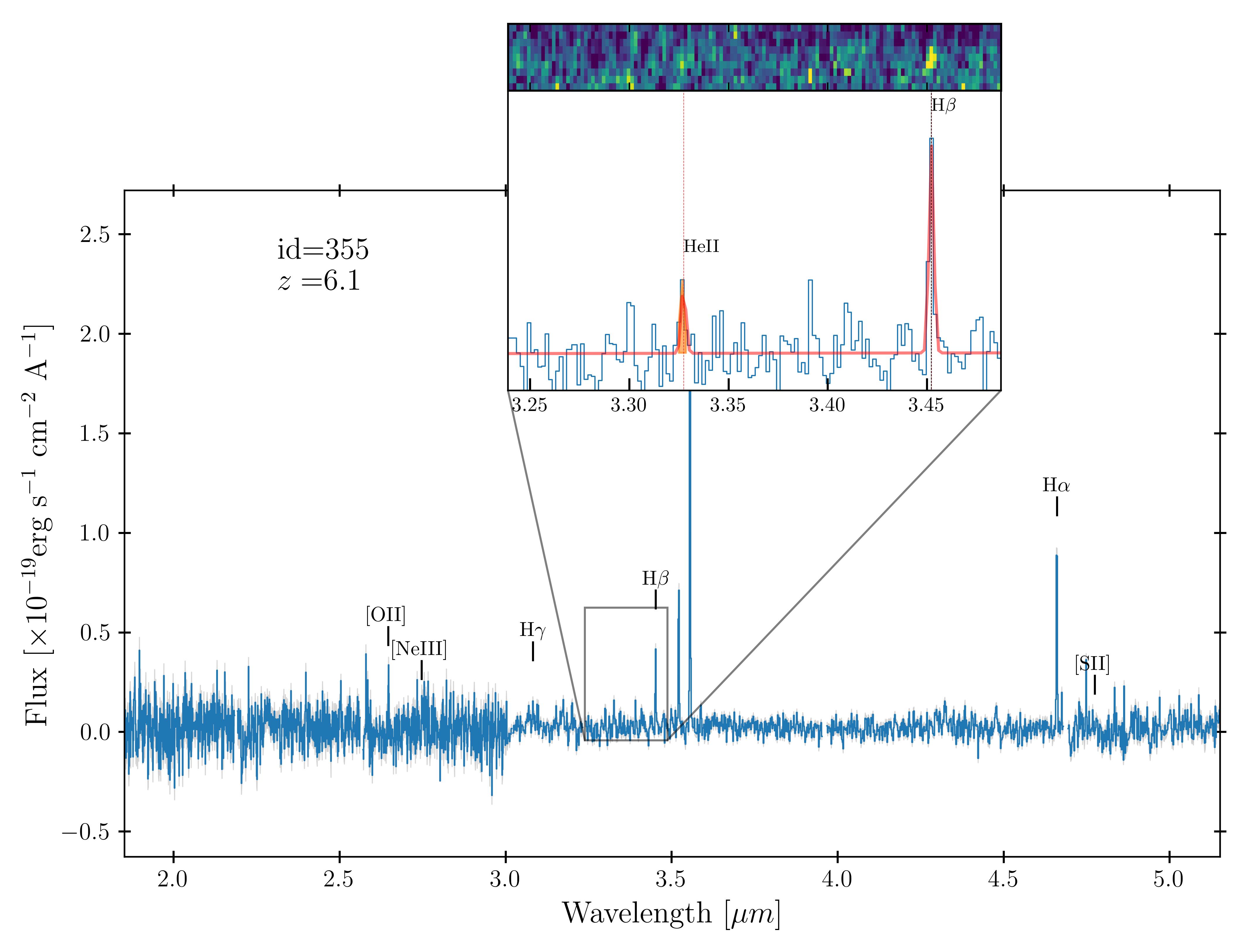}
\includegraphics[width=0.49\columnwidth]{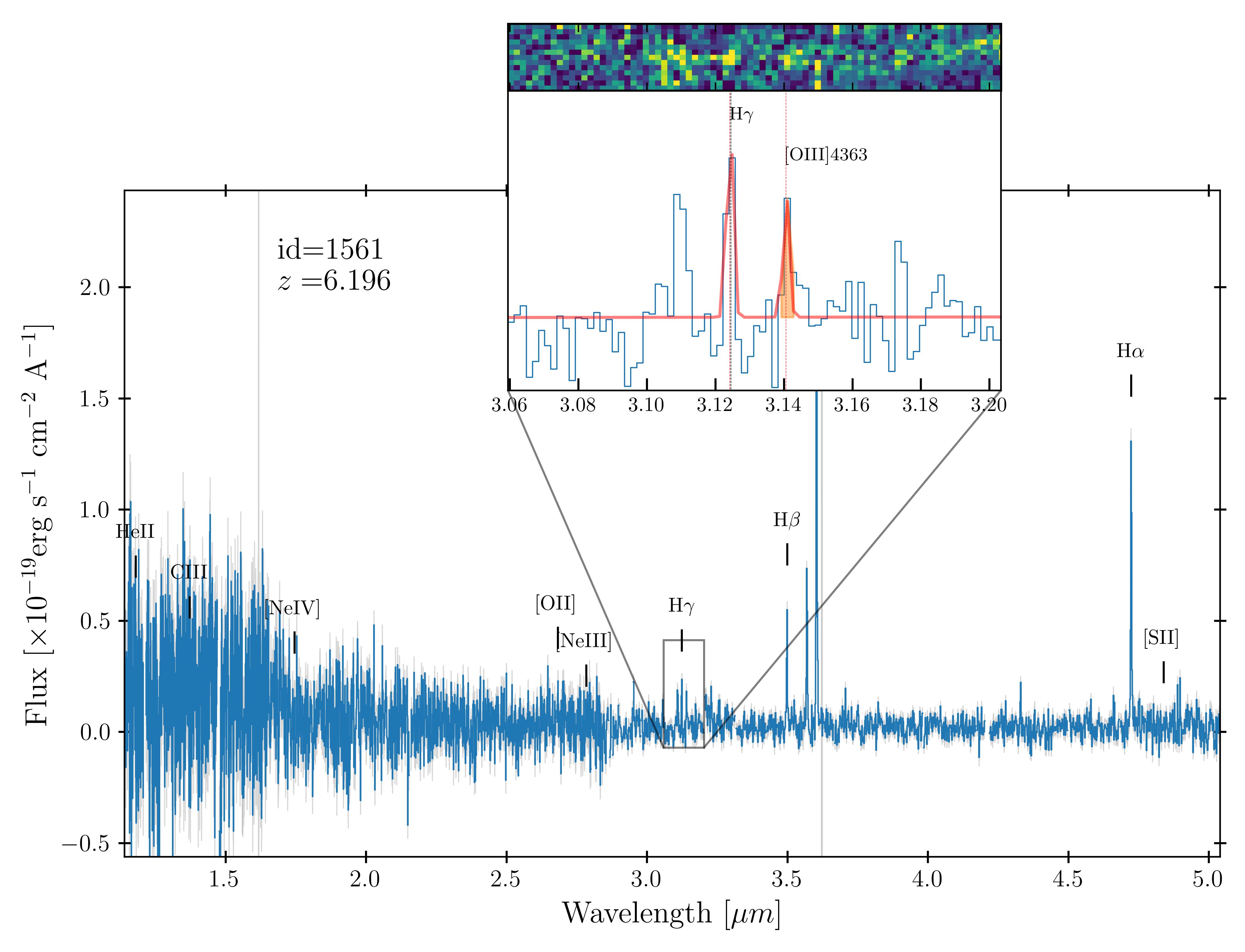}\\
\includegraphics[width=0.49\columnwidth]{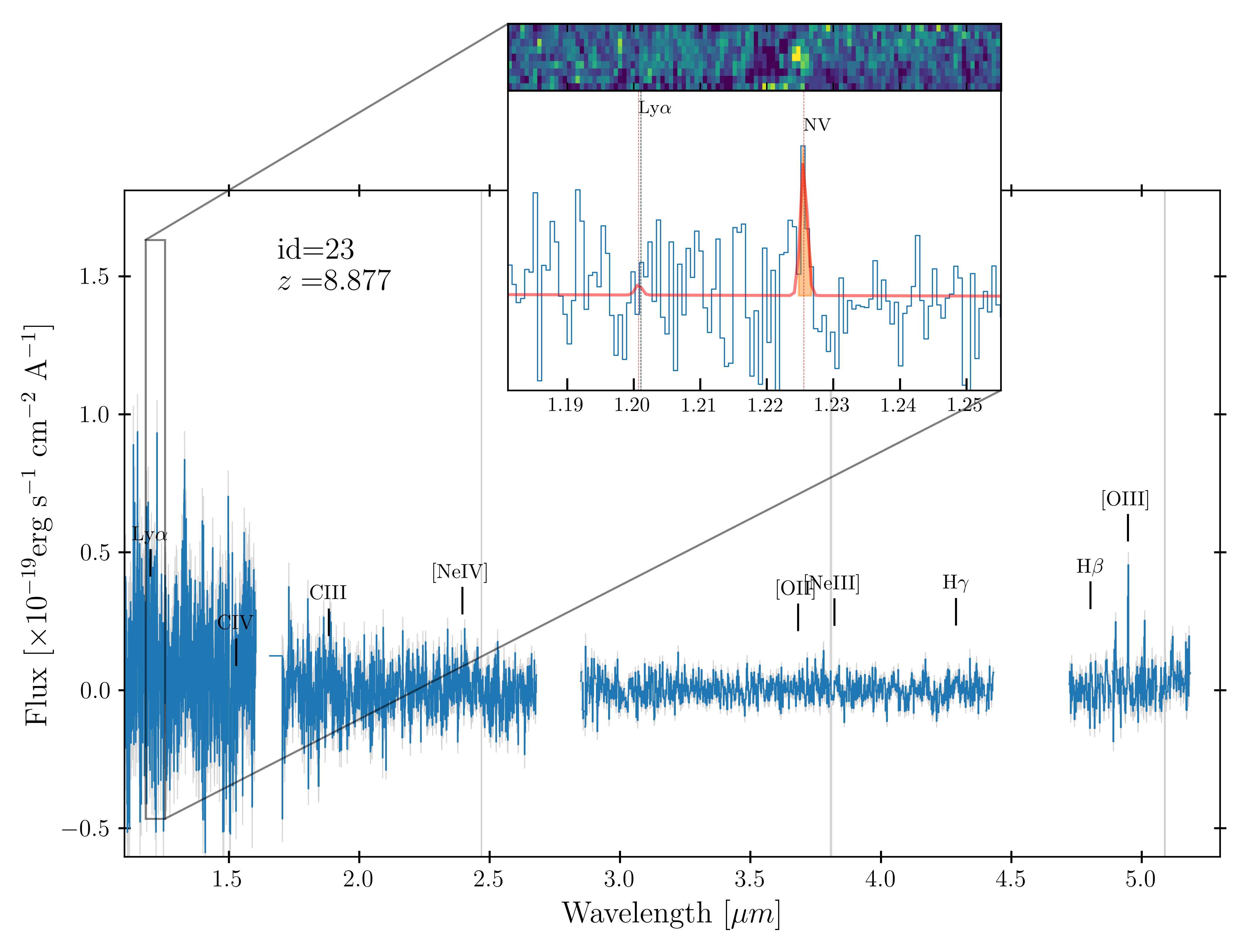}
\includegraphics[width=0.49\columnwidth]{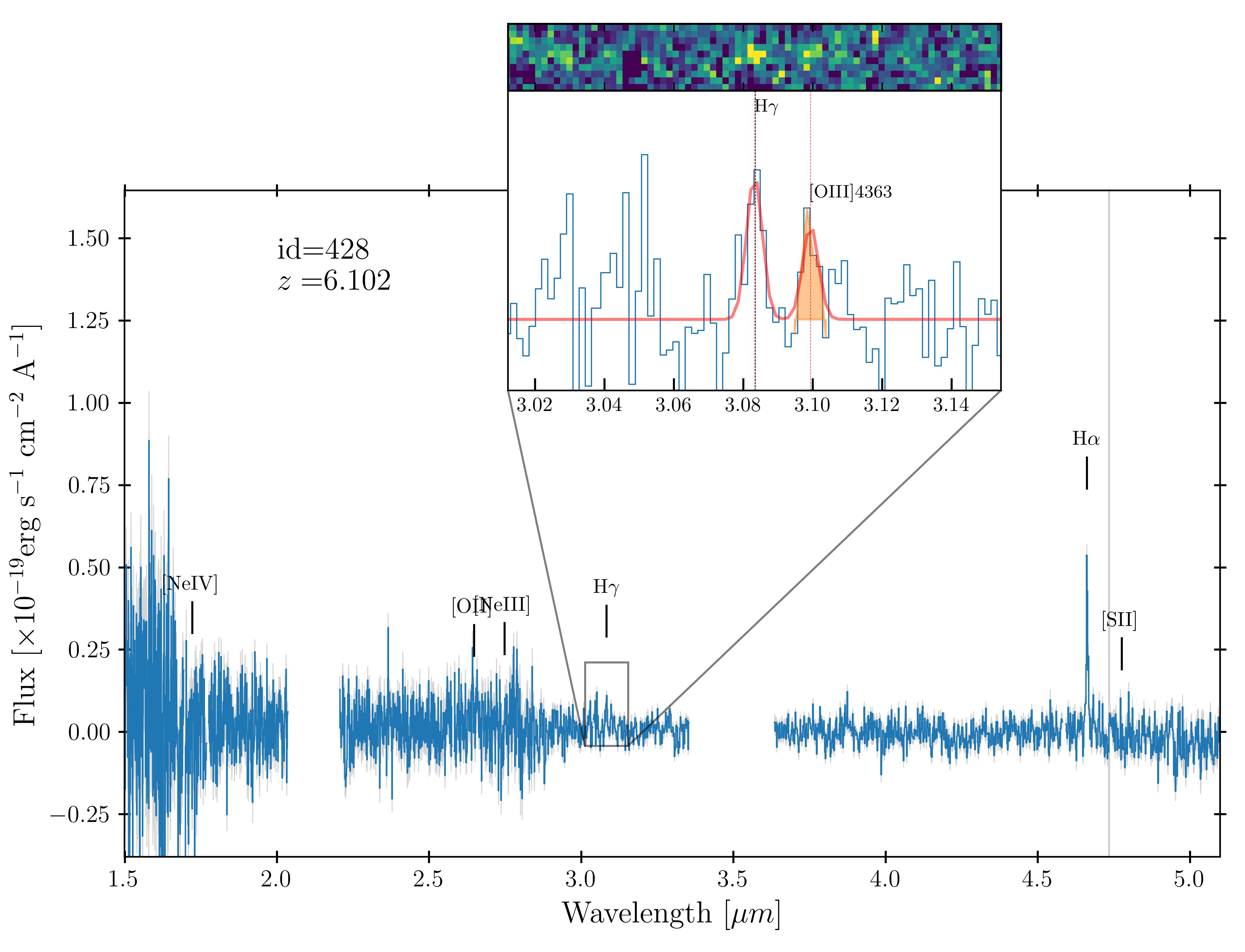}
\includegraphics[width=0.49\columnwidth]{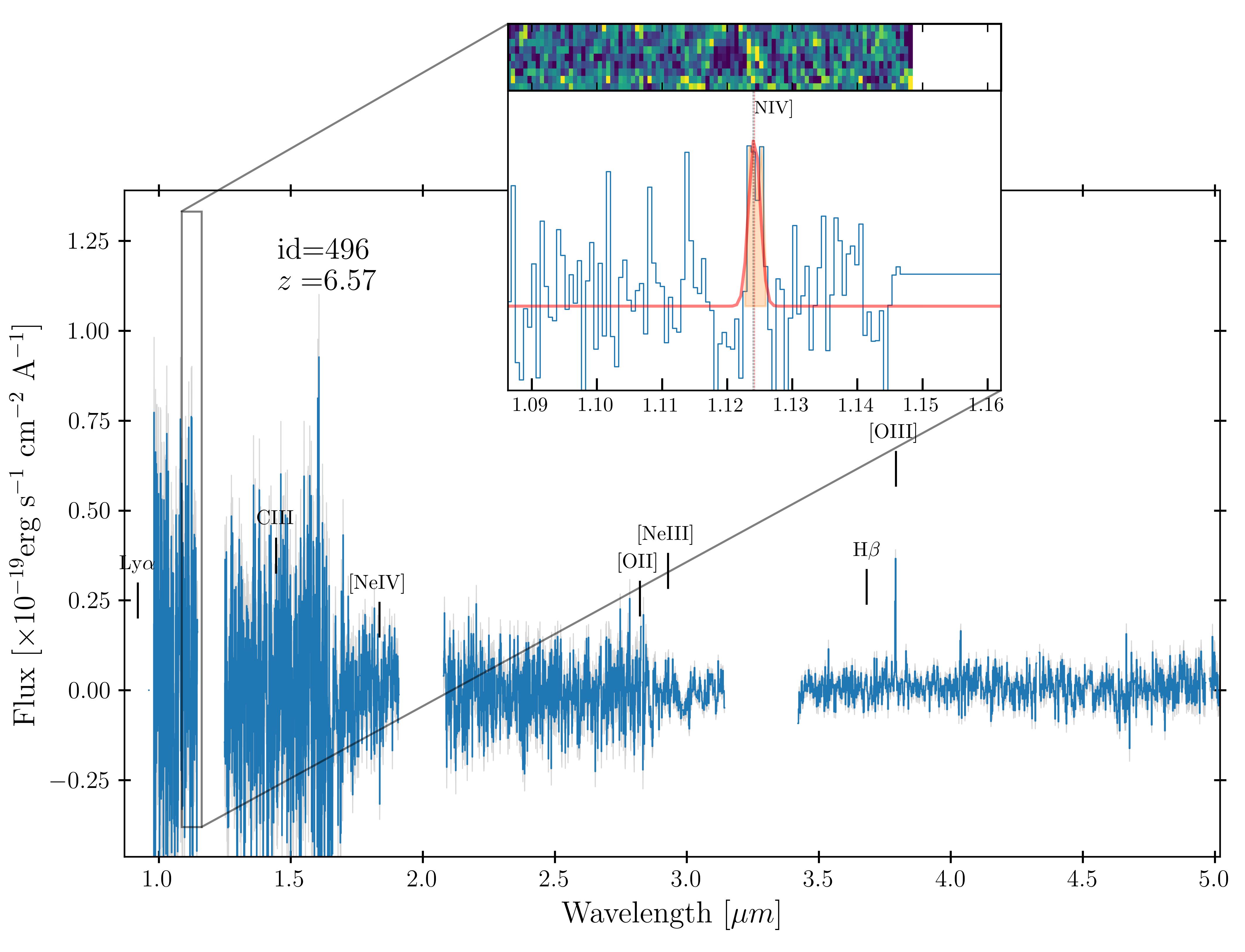}
\includegraphics[width=0.49\columnwidth]{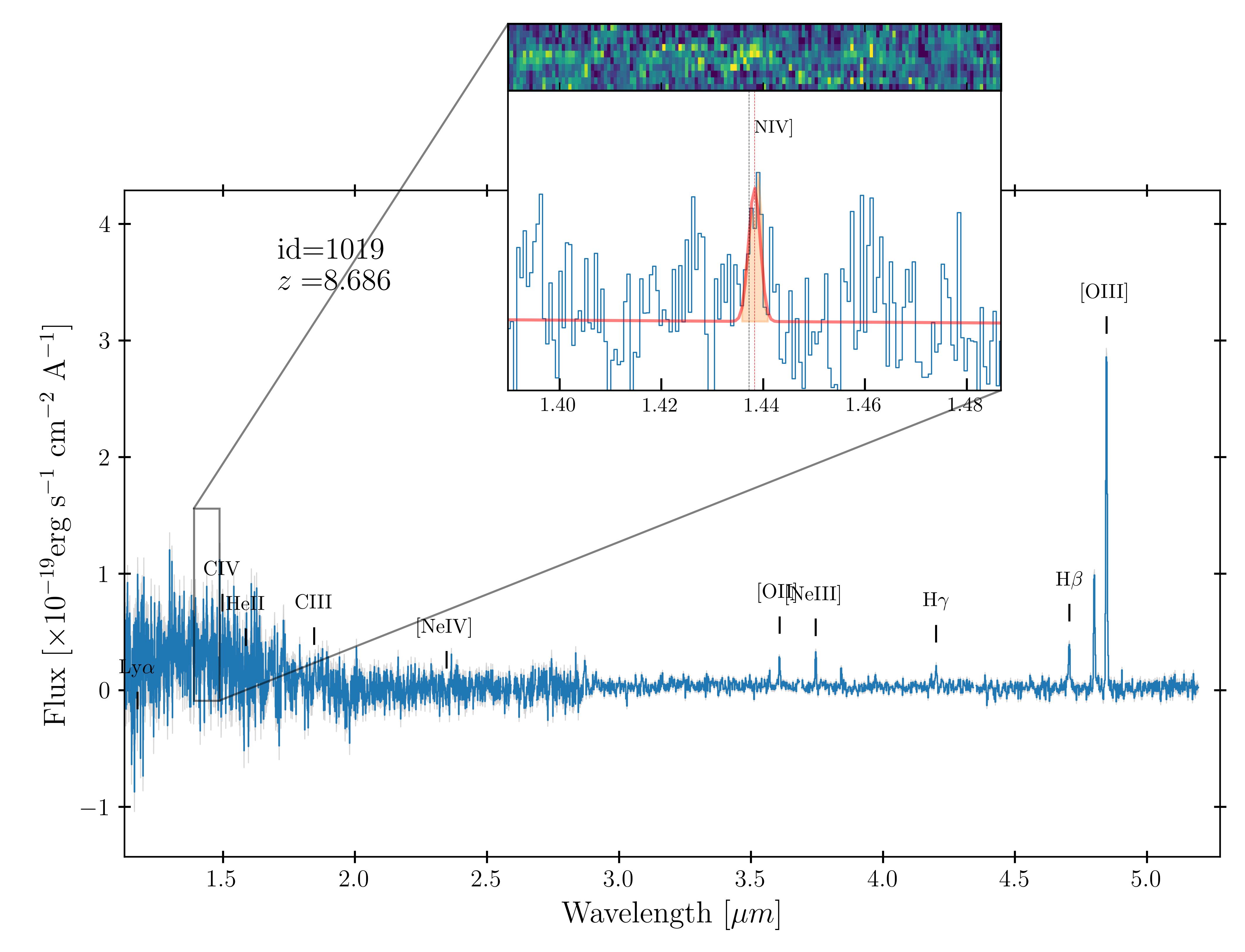}
\caption{Spectra of the sources selected as NLAGN at $z>6$. In each plot, we highlight in the box the main spectral features that led to the identification as a NLAGN (showing both the 1D and 2D spectrum). CEERS-613, CEERS-1019, CEERS-23 and CEERS-496 are marked only as tentative selections. The first because of the dubious \HeIIl1640 line in the 2D spectrum, while the other because of their dubious position in the diagnostic diagrams reported in Sec.~\ref{sec:results}}
\label{fig:high-z_spec}
\end{figure*}

\begin{figure*}
    \includegraphics[width=0.49\columnwidth]{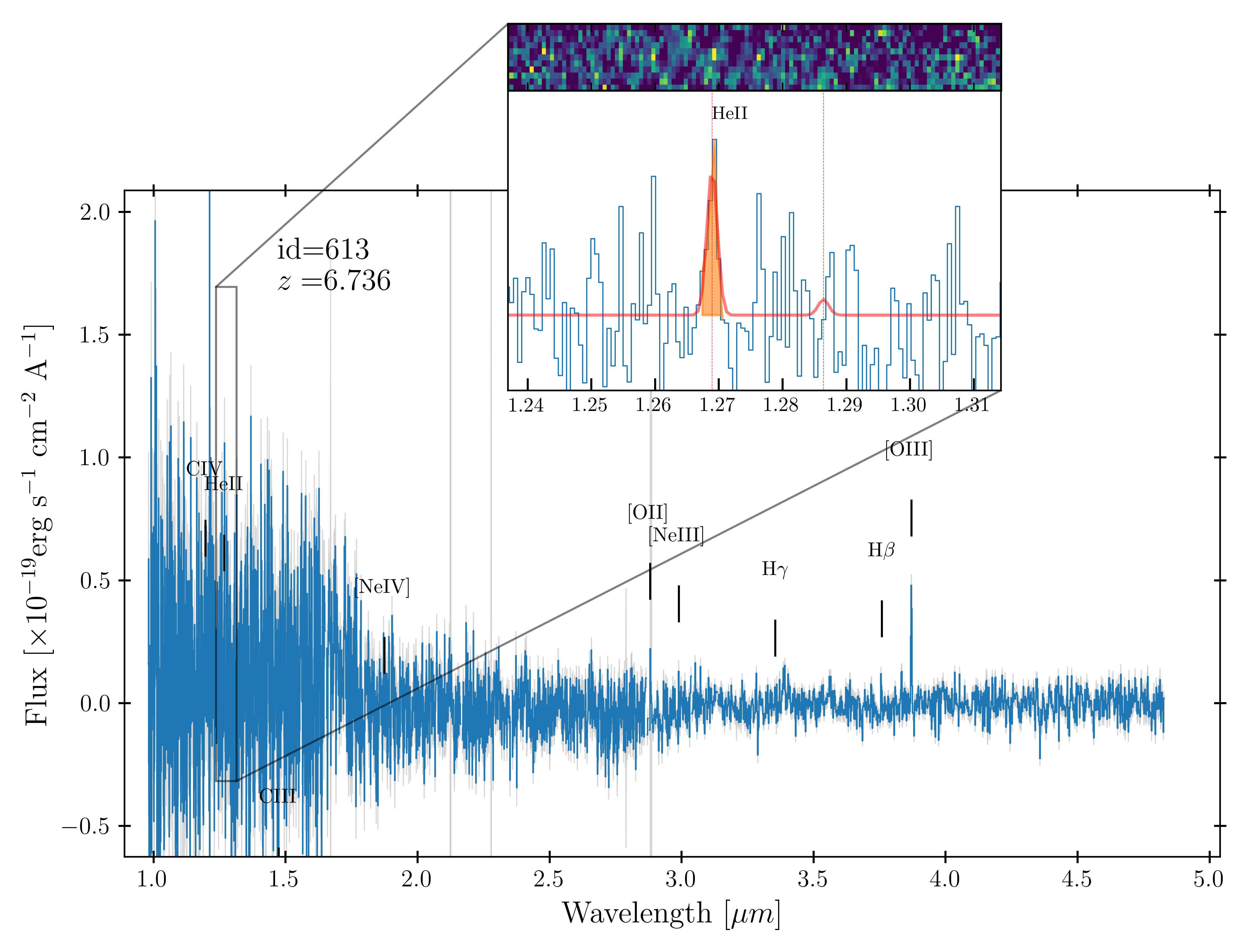}
\includegraphics[width=0.49\columnwidth]{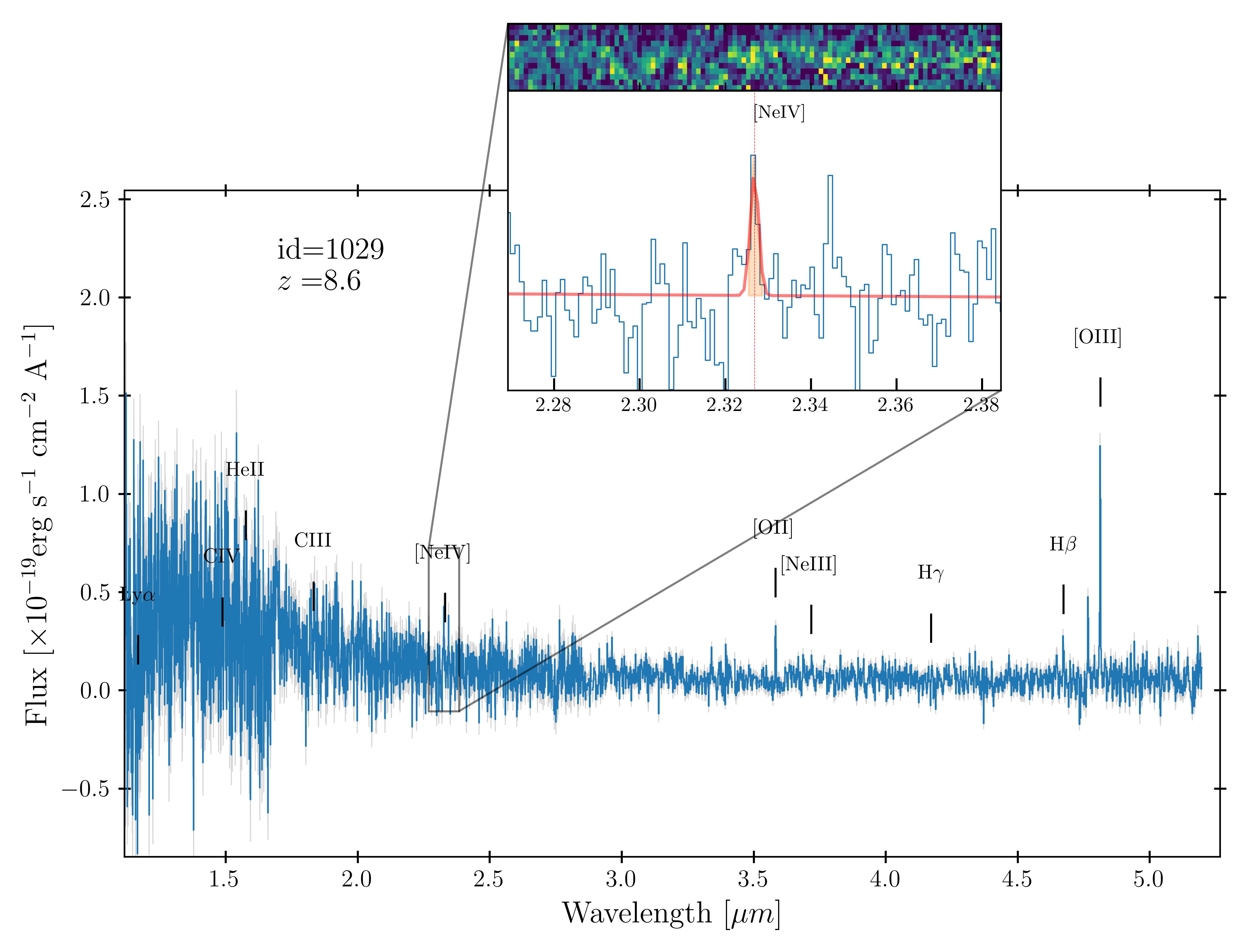}\\
\includegraphics[width=0.49\columnwidth]{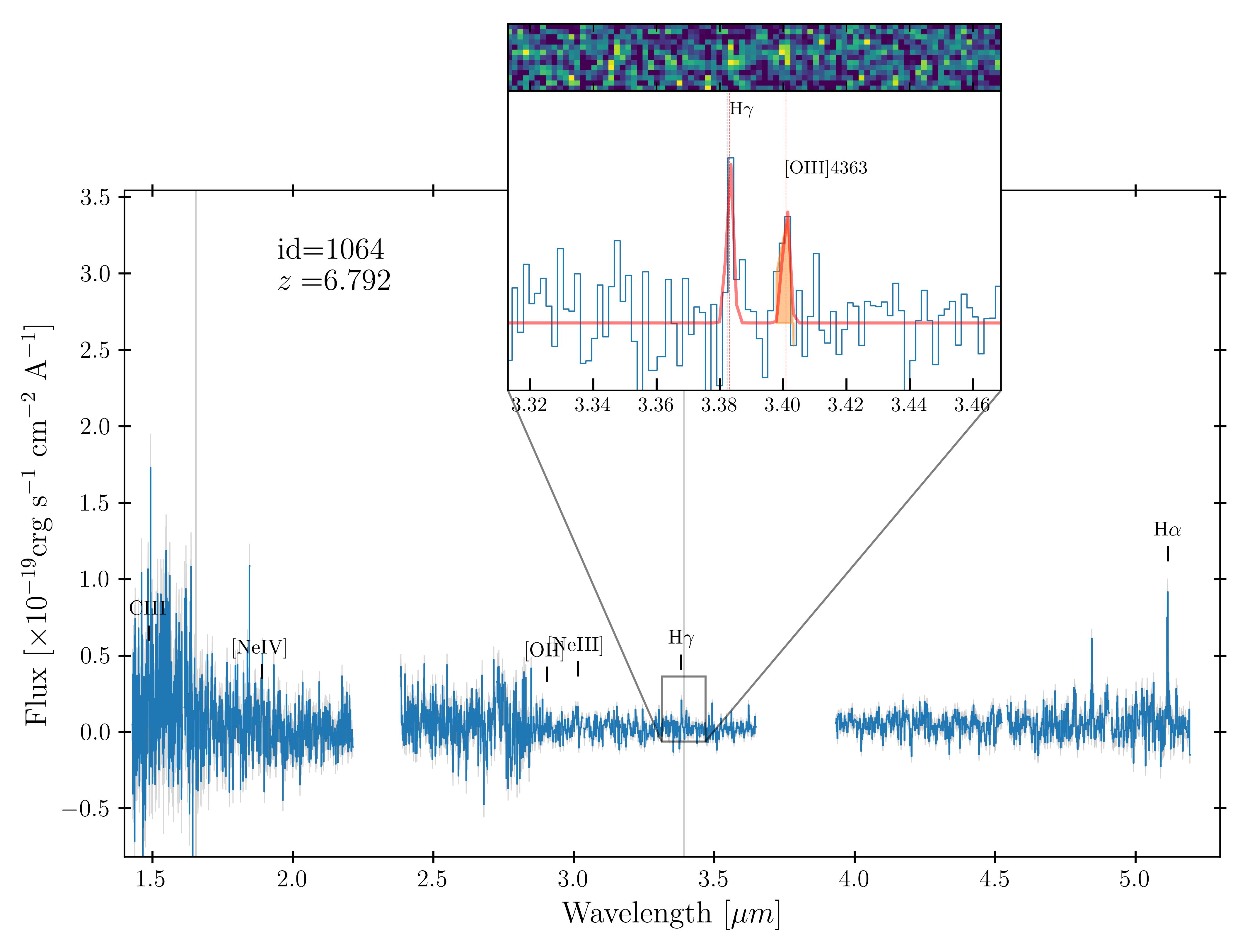}
\includegraphics[width=0.49\columnwidth]{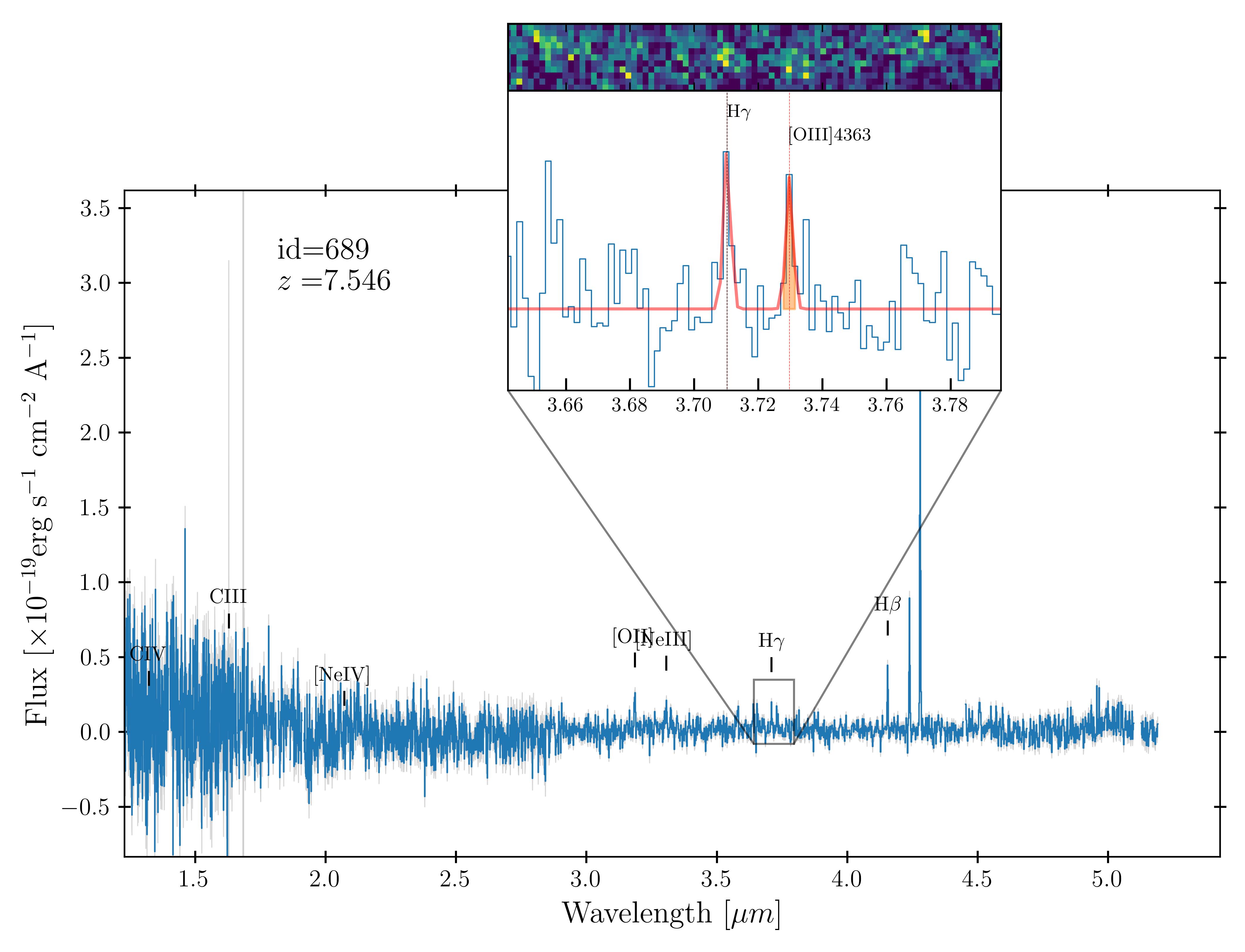}
\caption{Continuation of Figure~\ref{fig:high-z_spec}}
\end{figure*}

\end{appendix}
\end{document}